\documentclass[11pt,a4paper,oneside]{article}
\pdfoutput=1
\usepackage{jheppub}

\usepackage{verbatim}
\usepackage{mathrsfs}
\usepackage{appendix}
\usepackage{caption}
\usepackage{float}
\usepackage{subfig}
\usepackage[vcentermath]{youngtab}
\usepackage{alltt}
\usepackage{multirow}
\usepackage{arydshln}
\usepackage{lscape}
\usepackage{tabu}
\usepackage{booktabs}
\usepackage{amssymb,amsmath,amsfonts,mathrsfs,enumerate,wrapfig,latexsym,wasysym}
\usepackage[table]{xcolor}
\usepackage{placeins}
\usepackage{bbold}
\usepackage{graphicx}
\usepackage{adjustbox}
\usepackage{xcolor}
\usepackage{slashed}    
\usepackage{url}
\usepackage{hyperref}
\hypersetup{
	colorlinks = true,
	linkcolor = blue,
	anchorcolor = blue,
	citecolor = blue,
	filecolor = blue,
	urlcolor = blue,
	bookmarks=true,
	linktocpage=true
}
\usepackage{framed}
\usepackage[numbers,sort&compress]{natbib}
\usepackage{fancyvrb}
\allowdisplaybreaks
\usepackage{multirow}
\usepackage{framed}
\usepackage{makecell}
\usepackage{color}
\usepackage{pifont}
\usepackage{cancel}
\usepackage{import}
\usepackage{subfiles}
\usepackage{rotating}
\usepackage{threeparttable,tablefootnote}\usepackage{mathtools}
\usepackage{pbox}
\usepackage{setspace}
\usepackage{lscape}
\usepackage{color, colortbl}
\usepackage{pdflscape}
\usepackage{fontawesome}

\usepackage{threeparttable}
\usepackage{slashed}
\usepackage{tikz}
\def\checkmark{\tikz\fill[scale=0.4](0,.35) -- (.25,0) -- (1,.7) -- (.25,.15) -- cycle;}
\usepackage[normalem]{ulem}

\definecolor{darkgreen}{cmyk}{1,0,1,0.4}

\definecolor{pink}{cmyk}{0.4,1,0.3,0}

\usepackage{xcolor}
\definecolor{gerua}{rgb}{0.74, 0.2, 0.64}
\definecolor{beguni}{rgb}{0.9, 0.17, 0.31}
\definecolor{darkgreen}{rgb}{0.07, 0.53, 0.03}
\definecolor{skyblue}{rgb}{0.0, 0.53, 0.74}

\makeatletter
\g@addto@macro{\endtabular}{\rowfont{}}% Clear row font
\makeatother
\newcommand{\rowfonttype}{}% Current row font
\newcommand{\rowfont}[1]{% Set current row font
	\gdef\rowfonttype{#1}#1\ignorespaces%
}
\makeatother
\def\tr{\text{tr}}
\usepackage{bbm}

\def\m{\mu}
\def\n{\nu}
\def\s{\sigma}

\def\tr{\text{tr}}

\newcommand{\ope}{Q}
\newcommand{\gev}{\text{ GeV}}
\newcommand{\tev}{\text{ TeV}}

\newcommand{\ifb}{\text{ fb}^{-1}}

\newcommand{\wc}{\mathcal{C}}
\graphicspath{{plots/}}

\definecolor{Gray}{gray}{0.9}

\preprint{CERN-TH-2021-190, IPPP/21/48}
\title{Effective limits on single scalar extensions in the light of recent LHC data}

\author[1,5]{Anisha,}
\author[1,6]{Supratim Das Bakshi,}
\author[2]{Shankha Banerjee,}
\author[3]{Anke Biek\"otter,}
\author[1]{Joydeep Chakrabortty,}
\author[4]{Sunando Kumar Patra,}
\author[3]{Michael Spannowsky} \author{\\}

\affiliation[1]{Department of Physics, Indian Institute of Technology, Kanpur-208016, India.}
\affiliation[2]{CERN, Theoretical Physics Department, CH-1211 Geneva 23, Switzerland.}
\affiliation[3]{Institute for Particle Physics Phenomenology, Department of Physics, Durham University, Durham DH1 3LE, U.K.}
\affiliation[4]{Department of Physics, Bangabasi Evening College, 19 Rajkumar Chakraborty Sarani, Kolkata 700009, West Bengal, India.}
\affiliation[5]{School of Physics \& Astronomy, University of Glasgow, Glasgow G12 8QQ, United Kingdom}
\affiliation[6]{CAFPE and Departamento de F\'isica Te\'orica y del Cosmos,
Universidad de Granada, Campus de Fuentenueva, E--18071 Granada, Spain}

\emailAdd{anisha@glasgow.ac.uk, sdb@ugr.es, shankha.banerjee@cern.ch,
anke.biekoetter@durham.ac.uk, joydeep@iitk.ac.in, sunando.patra@gmail.com, michael.spannowsky@durham.ac.uk}

\abstract{In this paper, we work with 16 different single scalar particle extensions of the Standard Model. We present the sets of dimension-6 effective operators and the associated Wilson coefficients as functions of model parameters after integrating out the heavy scalars up to 1-loop, including the heavy-light mixing, for each such scenario. Using the correspondence between the effective operators and the  observables at electroweak scale, and employing Bayesian statistics, we compute the allowed ranges of new physics parameters that are further translated and depicted in 2-dimensional Wilson coefficient space in the light of the latest CMS and ATLAS data up to $137 \ifb$ and $139\ifb$, respectively. We also adjudge the status of those new physics extensions that offer similar sets of relevant effective operators. In addition, we provide a model-independent fit of $23$ Standard Model effective field theory Wilson coefficients using electroweak precision observables, single and di-Higgs data as well as kinematic distributions of di-boson production.}

\DeclareUnicodeCharacter{2212}{-}
\begin{document}
	\maketitle
	\enlargethispage{10mm}
%%%%%%%%%%%%%%%%%%%%%%%%%%%%%%%%%%%%%%%%%%%%
\section{Introduction}\label{sec:intro}

Despite the successful journey of the Standard Model of particle physics (SM), its inadequacy to explain many experimental observations and also the lack of direct evidence of new physics beyond the SM (BSM) in the high energy experiments compel us to look for indirect signatures of new physics (NP). Low-energy observables, such as the electroweak precision observables (EWPOs), the measurement of SM Higgs production and decay modes as well as di-boson production~\cite{ALEPH:2005ab, Berthier:2016tkq, Khachatryan:2016vau, Aad:2015gba,  Aad:2020plj, Aad:2020xfq, ATLAS-CONF-2021-044, ATLAS:2020bhl, ATLAS:2020syy, ATLAS:2020naq, ATLAS:2021upe, CMS:2020gsy, CMS:2020dvg, CMS:2020xwi, CMS:2020mpn, CMS:2021ixs, CMS:2020dvp, CMS:2021kom, CMS:2021ugl, Aaboud:2019gxl, ATLAS:2020nzk, ATLAS:2019rob, ATLAS:2018dpp, ATLAS:2018rnh, ATLAS:2018uni, CMS:2020tkr,  CMS:2021ssj, CMS:2017hea, Buchmuller:1985jz, Giudice:2007fh, Grzadkowski:2010es, Gupta:2011be,Gupta:2012mi,Banerjee:2012xc, Gupta:2012fy, Banerjee:2013apa, Gupta:2013zza, Elias-Miro:2013eta, Contino:2013kra, Falkowski:2014tna, Englert:2014cva, Gupta:2014rxa, Amar:2014fpa, Buschmann:2014sia, Craig:2014una, Ellis:2014dva, Ellis:2014jta, Banerjee:2015bla, Englert:2015hrx, Ghosh:2015gpa, Degrande:2016dqg, Cohen:2016bsd, Ge:2016zro, Contino:2016jqw, Biekotter:2016ecg, deBlas:2016ojx, Denizli:2017pyu, Barklow:2017suo, Brivio:2017vri, Barklow:2017awn, Khanpour:2017cfq, Englert:2017aqb, Panico:2017frx, Franceschini:2017xkh, Banerjee:2018bio, Grojean:2018dqj,Biekotter:2018rhp, Goncalves:2018ptp,Gomez-Ambrosio:2018pnl, Freitas:2019hbk, Banerjee:2019pks, Banerjee:2019twi, Biekotter:2020flu, Araz:2020zyh, Ellis:2020unq, Banerjee:2020vtm, Almeida:2021asy, Chatterjee:2021nms}, play a crucial role in constraining the SM and leave little room for BSM physics. To understand the present status of NP under the lamp post of experimental data in terms of the non-SM parameters, we need to establish an effective connection between the BSM physics residing at a relatively high scale and observables at the electroweak scale. Effective Field Theories (EFTs) are considered to be elegant tools to extract indirect effects of NP from the experimental data, if any, by bridging the gap between scales.

In a bottom-up approach, the complete and independent set of effective operators for any given mass dimension is computed relying on the symmetry and particle content of the SM -- leading to the most popular notion of the Standard Model Effective Field Theory (SMEFT)~\cite{Grzadkowski:2010es, BUCHMULLER1986621, Brivio_2019}. At dimension six, the commonly used bases of the SMEFT operators include the Warsaw basis~\footnote{The Warsaw basis
accommodates a set of complete and non-redundant effective operators of mass dimension six.}\cite{Grzadkowski:2010es, BUCHMULLER1986621} and the SILH basis~\cite{Giudice:2007fh, Elias-Miro:2013mua}. Generally, the associated Wilson coefficients (WCs) are independent and not pertinent to any specific UV theory. Parametrising low-energy observables in terms of these WCs and analysing them in global fits allows us to constrain potential NP in a (rather) model-independent way. Global analyses in the SMEFT framework have been performed for the EWPO~\cite{Han:2004az}, Higgs~\cite{Corbett:2012ja, Ellis:2014jta, Dumont:2013wma, Corbett:2013pja,Chang:2013cia, Elias-Miro:2013mua, Boos:2013mqa, Ellis:2014dva}, di-boson~\cite{Panico:2017frx, Baglio:2018bkm, Grojean:2018dqj, Baglio:2019uty, Azatov:2019xxn}, the top sector~\cite{Buckley:2015lku, Brivio:2019ius, Bissmann:2019gfc, Hartland:2019bjb,Durieux:2019rbz, vanBeek:2019evb, CMS:2020pnn}, and for combinations of these sectors~\cite{Ellis:2014jta, Contino:2016jqw, Butter:2016cvz, Biekoetter:2018ypq, Ellis:2018gqa, Ellis:2020unq, Anisha:2020ggj, Dawson:2020oco, Ethier:2021bye}. In this work, we consider constraints from EWPO, single and di-Higgs data, as well as distributions from the di-boson production channels.

In a top-down approach, the SMEFT lays the platform to bring different BSM scenarios on the same footing by integrating out heavy non-SM degrees of freedom (DOFs). Each effective operator that emerges in the process is accompanied by a WC which is a function of the model parameters and captures the footprints of NP interactions~\cite{Henning:2014wua, Drozd:2015rsp, Henning:2016lyp, Ellis:2016enq,  Fuentes-Martin:2016uol, Zhang:2016pja, Ellis:2017jns, de_Blas_2018, Bakshi:2018ics, Kramer:2019fwz, DasBakshi:2020ejz, DasBakshi:2020pbf, Haisch_2020, Dawson:2021jcl, Dittmaier:2021fls, Brivio:2021alv, Dedes:2021abc}. It is important to note that most of the phenomenologically interesting BSM scenarios do not induce the complete set of dimension-6 SMEFT operators. In addition, the Wilson coefficients of effective operators computed by integrating out heavy DOFs from a specific NP Lagrangian are functions of model parameters and thus, unlike the model-independent SMEFT case, they are related to each other. As a result of these relations and the typically smaller number of free parameters, the SMEFT parameter space in a top-down analysis is usually much more stringently constrained compared to bottom-up analyses. Top-down SMEFT fits hence play a crucial role in pinning down the nature of BSM physics.

The primary motivation of this work is to estimate the allowed BSM parameter space for different SM extensions and to perform a comparative analysis among those scenarios that are degenerate effective theories upon non-SM DOFs being integrated out. In our analysis, we consider 16 different single scalar extensions of the SM without extending the gauge symmetry. We embed each of the NP Lagrangians in our {\em Mathematica$^\text{\tiny \textregistered}$} program \texttt{CoDEx} \cite{Bakshi:2018ics} to perform the integration up to 1-loop, including the heavy-light mixing, and provide the exhaustive sets of the dimension-6 effective operators in Warsaw basis and their associated WCs as functions of model parameters~\footnote{We provide the sets of effective operators in both Warsaw and SILH bases for all these BSM scenarios in a Mathematica notebook file here \href{https://github.com/effExTeam/Precision-Observables-and-Higgs-Signals-Effective-passageto-select-BSM}{\faGithub}.}. The effective operators generated for these individual models can be verified using the complementary diagrammatic method introduced in ref.~\cite{DasBakshi:2021xbl}. We work within the Bayesian framework to draw the statistical inference and parameter estimation using the {\em Mathematica$^\text{\tiny \textregistered}$} package \texttt{OptEx}~\cite{sunando_patra_2019_3404311}.

In this paper we constrain the SMEFT parameter space in both bottom-up as well as  top-down approaches to provide an up-to-date map of the new physics parameter space of scalar extensions of the SM. 
First, we perform a model-independent SMEFT fit based on the latest CMS and ATLAS data, including many distributions and simplified template cross sections.
Second, we choose 16 different heavy scalars that are being used frequently in literature for different phenomenological reasons, \textit{e.g.},~neutrino mass generation, to explain IceCube, LHCb data, etc.~\cite{Dey:2015eaa, Pilaftsis:1999qt, Arhrib:2011uy, Babu:2009aq, Bambhaniya:2013yca, Buchmuller:1986zs, Arnold:2013cva, Bauer:2015knc, Bandyopadhyay:2016oif, Davidson_2010,Chen:2008hh}. For simplicity, we assume that SM is extended by a single heavy scalar at a time. We compute the effective dimension-6 operators and the associated WCs by performing the integration out of individual heavy fields up to 1-loop, including heavy-light mixing. These matching results incorporating scalar heavy-light mixing are presented in the Warsaw basis. We also provide the results in SILH basis. The obtained results will be useful even when there are multiple heavy scalars. In that case, there will be additional contributions due to mutual interactions among the heavy fields. We further constrain the non-SM parameter space for each individual model using electroweak precision data and Higgs data. We highlight the impact of the individual data sets and their cumulative impact. This helps us to identify which data sets play a crucial role for which model while constraining the non-SM parameter space. We also pin-point the directions in non-SM parameter space that cannot be constrained by our chosen set of data and comment on the observables necessary to constrain these models completely.

\begin{figure}[!tbh]
    \centering
    \includegraphics[height=8cm,width=15cm]{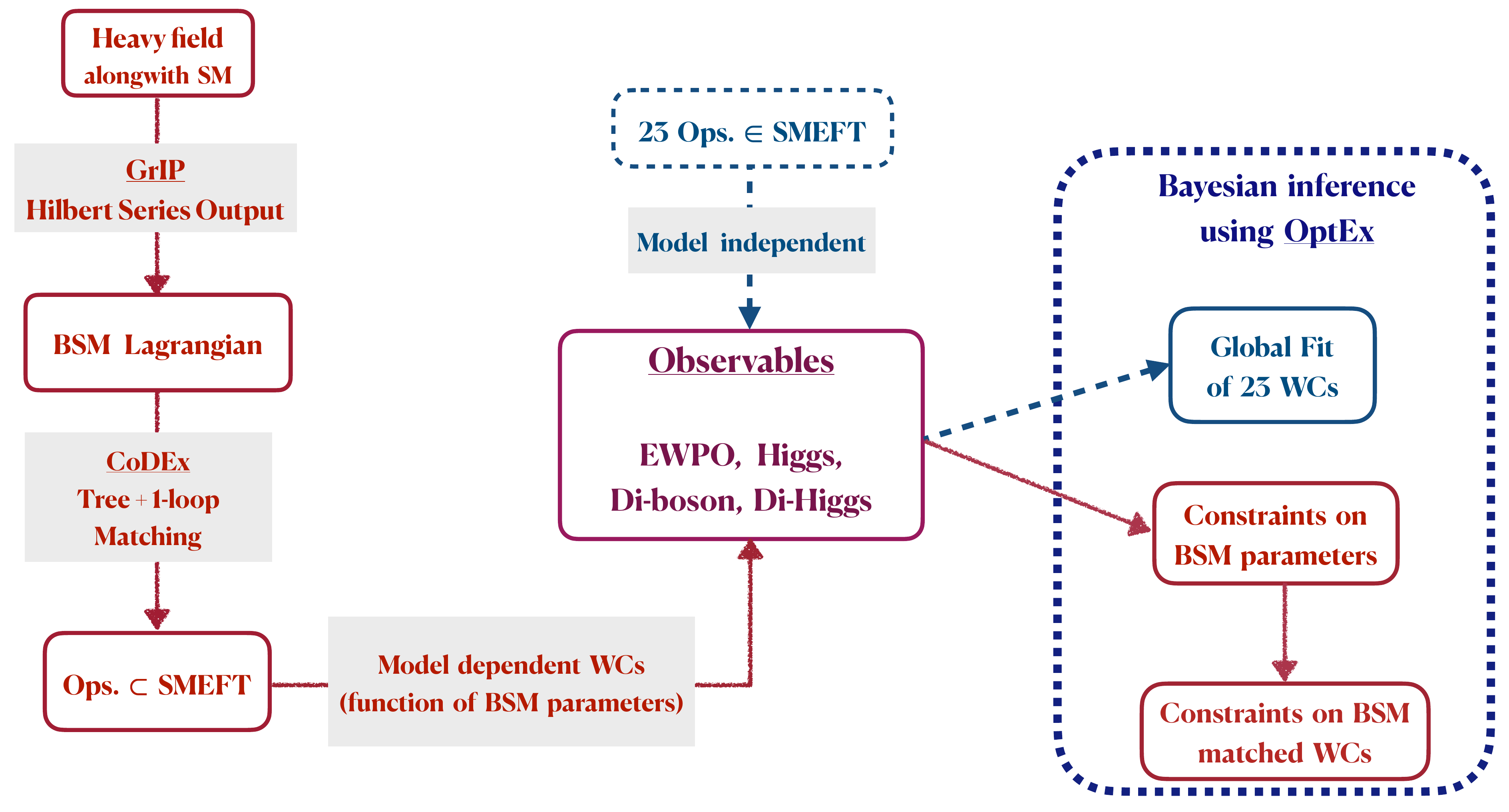}
    \caption{Flowchart depicting the work-flow of this article. Details on the observables are discussed in section~\ref{sec:obs}. The global fit consists of 23 WCs, see fig.~\ref{fig:CompModIndep}. The constraints on the BSM parameters of specific models are discussed in section~\ref{sec:BSMs} and appendix~\ref{sec:remainingmodels}. We use the Hilbert series output from GrIP~\cite{Banerjee:2020bym} to construct the BSM Lagrangian given the heavy field information. The BSM Lagrangians are then implemented in CoDEx~\cite{Bakshi:2018ics} to compute the tree- and the 1-loop-level WCs (including the heavy-light mixed WCs). We perform the statistical analysis based on Bayesian inference using the package OptEx~\cite{sunando_patra_2019_3404311}.}
    \label{fig:flowchart}
\end{figure}

The paper is organised as follows: We provide details on the SMEFT parametrisation of the used observables in section~\ref{sec:param} and describe the corresponding experimental inputs in section~\ref{sec:obs}. In section~\ref{sec:modindep}, we perform a model-independent analysis including 23 WCs contributing to the EWPO, single Higgs, di-Higgs, and di-boson sectors, and present the results for individual one-parameter fits as well as for a global analysis. In section~\ref{sec:BSMs}, we introduce the 16 single scalar extensions of the SM and reduce them into effective theories $\subset$ SMEFT. For each case, we tabulate the sets of emerged effective operators and associated WCs in the Warsaw basis as functions of the model parameters. We then employ similar fitting methodology for our chosen two NP models and display the results in terms of model parameters as well as in terms of the WCs, in section~\ref{sec:BSMs}. We discuss the Renormalisation Group Equations (RGEs) of the SMEFT operators for a specific model and captures its impact on the allowed BSM parameters and WCs in section~\ref{sec:RGE_effects}. We summarise our results and discuss possible future directions in section~\ref{sec:conclusion}. In the appendix~\ref{sec:remainingmodels}, we tabulate the computed effective operators and the WCs for the remaining 14 models adopted in this work. The working principle of this work is depicted in a flowchart, see fig.~\ref{fig:flowchart}.

\section{SMEFT parametrisation}\label{sec:param}

The SMEFT describes new physics with higher-dimensional (in mass with dimensions $\ge$ 5) operators consisting of only the SM fields and respecting the SM symmetries. The effective Lagrangian is given by
\begin{align}\label{eq:bsmlagr}
	\mathcal{L}_{_\text{eff}} = \mathcal{L}_{_\text{SM}}^{d\leq 4} +  \mathcal{L}_{_\text{SM}}^{\text{EFT}} 
	= \mathcal{L}_{_\text{SM}}^{d\leq 4} + \sum\limits_{d=5,...} \sum\limits_{i} \left( 
	\frac{\wc^{(d)}_{i}}{\Lambda^{d-4}}\right) \ope_{i}^{(d)} \, ,
\end{align}
where $\ope_{i}^{(d)}$ are the effective operators of mass dimension~$d$ and $\wc^{(d)}_{i}$  are the accompanying WCs, respectively. The index~$i$ runs over the number of independent effective operators. At dimension six, there are $2499$ independent operators in the most flavour general case. This number reduces to $59$ when assuming minimal flavour violation (MFV)~\cite{DAmbrosio:2002vsn} and real WCs. On the other hand, in a top-down approach, the number of operators is determined by integrating out the heavy BSM fields. In this work, we have confined ourselves to dimension-6 operators in the Warsaw basis~\cite{Grzadkowski:2010es} and parametrise observables up to linear order in the WCs. The renormalisable part of the  SM Lagrangian is given for the sake of completeness as
\begin{align}\label{eq:smlagr}
	\mathcal{L}_{\text{SM}}^{d\leq 4} &= -\frac{1}{4} G^a_{\mu\nu}G^{a,{\mu\nu}}-\frac{1}{4} W^I_{\mu\nu}W^{I,{\mu\nu}}-\frac{1}{4}B_{\mu\nu}B^{\mu\nu}+ |D_{\mu}H|^2 - \mu_{_H}^2 |H|^2 -\frac{1}{2} \lambda_{H}^{\text{SM}} |H|^4 \nonumber\\
	&+\bar{l}_{_L}i \slashed{D} \,l_{_L} + \bar{q}_{_L}i \slashed{D} \,q_{_L} + \bar{e}_{_R} i \slashed{D} \, e_{_R} + \bar{u}_{_R}\, i \slashed{D} \, u_{_R} + \bar{d}_{_R} \, i \slashed{D}\, d_{_R}\nonumber\\
	&-\left\lbrace Y^{\text{SM}}_{e}  H^\dagger \bar{e}_{_R} \, l_{_L}		 \, +  Y^{\text{SM}}_{u} \widetilde{H}^\dagger \bar{u}_{_R}  \, q_{_L} \,  + \, Y^{\text{SM}}_{d} H^\dagger	 \bar{d}_{_R}  \, q_{_L}		 \, + \,\text{h.c.} \right\rbrace,
\end{align}
where $B_{\mu\nu}, W^I_{\mu\nu}, \text{ and } G^a_{\mu\nu}$ are the field strength tensors of the SM gauge groups  $U(1)_{_Y}$, $SU(2)_{_L}$, and $SU(3)_{_C}$, respectively, with $a=1,\dots,8$. The adopted conventions for the quantum numbers of the SM fields are shown in tab.~\ref{tab:fieldconvention}. $D$ denotes the covariant derivative, $Y^{\text{SM}}_{u,d,e}$ are the SM Yukawa couplings, and $\widetilde{H} = i  \,  \sigma^2 \, H^\ast$ is the conjugate Higgs field.

Our SMEFT predictions are computed at linear order in the WCs in the electroweak $\{ \alpha_{EW}, \, G_F, \, m_Z\}$ input scheme with the following input values 
\begin{equation}\label{eq:SMinputval}
    \begin{split}
        &\alpha^{-1}_\text{EW} = 127.95 , \quad G_F = 1.6638 \times 10^{-5} \gev^{-2} , \\
        m_Z = &91.1876 \gev, \quad m_H = 125.09\gev , \quad m_t = 173.2 \gev \, .
    \end{split}
\end{equation}
We generally assume MFV. However, for the dimension-6 operators contributing to the Yukawa couplings, we distinguish between flavours and introduce different WCs for muons $\wc_{\mu H}$ and tau leptons $\wc_{\tau H}$, as well as the charm $\wc_{cH}$ and top quarks $\wc_{tH}$. There is of course no direct measurement of the charm Yukawa coupling apart from the $h\to J/\psi\gamma$ searches~\cite{Perez:2015aoa} and in our analysis $\wc_{cH}$ can be considered as a proxy for a modification of the total Higgs decay width which can be constrained from a global fit of all Higgs signal strengths. It is important to note that with more data we will be able to explore the structure of the SMEFT operators which may not always follow MFV. Refs.~\cite{Falkowski:2020znk, Efrati:2015eaa} study various such cases where the requirement for MFV is relaxed.

In total, we include $23$ WCs in our model-independent analysis. A graphical summary of the WCs and the observables that they contribute to is given in fig.~\ref{fig:WC_coef_summary}. The corresponding operators are listed in tab.~\ref{table:22operators} in the appendix~\ref{app:operators}.

\begin{figure}[thb]
    \centering
    \includegraphics[width=0.6\textwidth]{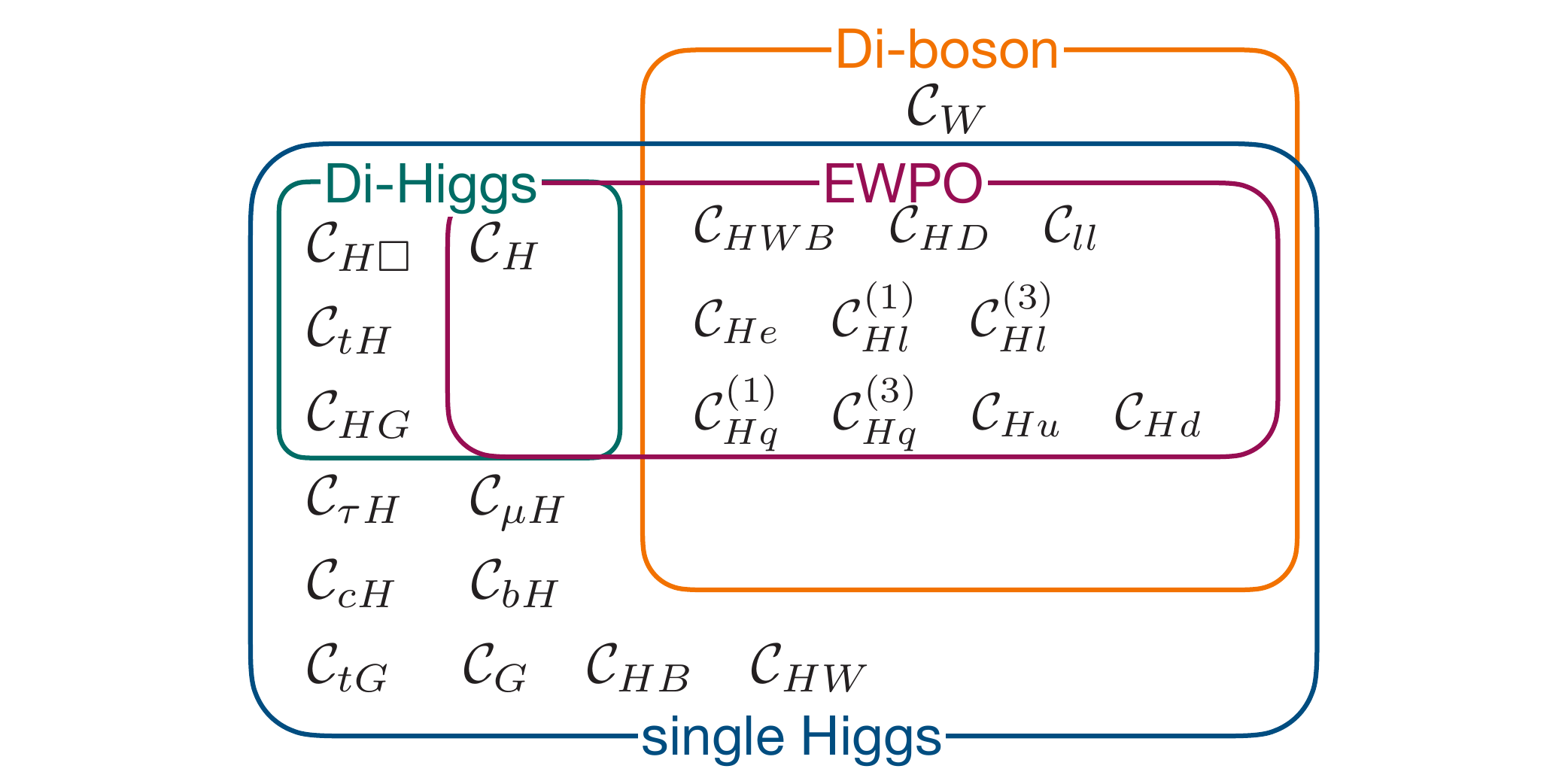}
    \caption{Graphical summary of the considered Wilson coefficients in our model-independent analysis and the observables they contribute to.}
    \label{fig:WC_coef_summary}
\end{figure}

\paragraph{Higgs signal strengths and simplified template cross sections:} For the Higgs sector, most of the theory predictions are generated at leading order (LO) using the \texttt{SMEFTsim} model~\cite{Brivio:2017btx} in \texttt{MadGraph~2.7.3}~\cite{Alwall:2014hca} with the default \texttt{NNPDF23\_NLO} parton distribution functions~\cite{Ball:2013hta}. Where available, SMEFT predictions are taken directly from \texttt{fitmaker}~\cite{Ellis:2020unq}, which uses the same tools and settings as outlined above.

For the gluon fusion STXS categories, we instead use the predictions from the ATLAS combination in ref.~\cite{ATLAS:2020naq} which include matching, merging and parton showering. LO predictions for the operator $\ope_G$, which are not present in the ATLAS reference, are added using the \texttt{SMEFT@NLO} model~\cite{Degrande:2020evl}.  
For all other channels, we cannot use the ATLAS predictions due to different input parameter scheme choices and instead rely on the predictions at parton level generated with~\texttt{MadGraph}. Overall, we describe our Higgs data sets in terms of the following WCs.
\begin{equation}
\begin{split}
    \wc_{H\square}, \quad
    &\wc_{HW}, \quad 
    \wc_{HB}, \quad 
    \wc_{HG}, \quad
    \wc_{tH}, \quad
    \wc_{cH}, \quad
    \wc_{bH}, \quad
    \wc_{\tau H}, \quad
    \wc_{\mu H}, \quad
    \wc_{G}, \quad
    \wc_{tG}, 
    \\
    \wc_{H}, \quad
    &\wc_{HWB}, \quad 
    \wc_{HD}, \quad 
    \wc_{ll}, \quad 
    \wc_{Hl}^{(1)}, \quad 
    \wc_{Hl}^{(3)}, \quad 
    \wc_{He}, \quad 
    \wc_{Hq}^{(1)}, \quad 
    \wc_{Hq}^{(3)}, \quad 
    \wc_{Hd}, \quad 
    \wc_{Hu}.
\end{split}
\end{equation}
Currently, we only include $\wc_{H}$ in the predictions of total Higgs signal strength measurements and not for STXS measurements. The corresponding contributions of $\ope_{H}$ to the predictions of different Higgs production and decay channels are used from refs.~\cite{Degrassi:2016wml, Degrassi:2019yix}. Notice that four-quark operators, which could contribute, for instance, to the $t\bar{t}h$ and $th$ categories, are not included in our study at the moment. The inclusion of four-quark operators in global fits with top data has, however, not lead to a weakening of constraints on operators relevant for the top-Higgs sector such as $\wc_{tH}$, $\wc_{G}$ and $\wc_{HG}$~\cite{Ellis:2020unq, Ethier:2021bye}. We therefore assume that our approach does not overestimate the constraints on these operators.

\paragraph{Di-Higgs production cross section:} The SMEFT predictions for the di-Higgs production cross section are generated using the \texttt{ggHH} model~\cite{Heinrich:2020ckp} within the \texttt{POWHEG-BOX-V2}~\cite{Nason:2004rx, Frixione:2007vw, Alioli:2010xd} framework which implements BSM effects in a non-linear EFT framework including full NLO QCD corrections with massive top quarks. The Warsaw basis WCs $\wc_i$
are related to the five anomalous couplings $c_j$ in the \texttt{ggHH} model via
\begin{equation}
\begin{split}
c_t &= 1 + \frac{v^2}{\Lambda^2} \wc_{H \square} -  \frac{v^2}{\Lambda^2 y_t} \wc_{tH} \, , 
\qquad 
c_{tt} =  \frac{v^2}{\Lambda^2} \left( \frac{1}{2} \wc_{H \square} - \frac{3}{4 y_t} \wc_{tH} \right) \\
c_{hhh} &= 1 + 3  \frac{v^2}{\Lambda^2} \wc_{H \square}-  \frac{v^2}{\lambda \Lambda^2} \wc_H \, ,  
\qquad
c_{ggh} = 2 \, c_{gghh} = (16\pi)^2 \,  \frac{2v^2}{\Lambda^2 g_s^2} \wc_{H G} \, ,
\end{split}
\end{equation}
where we use ref.~\cite{FalkowskiTranlation} to translate from the SILH to the Warsaw basis. The effect of the operator $\ope_{tG}$ on the di-Higgs cross section is not taken into account. The di-Higgs cross section is thus parametrised in terms of the operators 
\begin{equation}
    \wc_H, \quad \wc_{H \square}, \quad \wc_{tH}, \quad \wc_{HG} \, .
\end{equation}

\paragraph{Di-boson distributions:} We include distributions of $WZ$ and $WW$ production. As shown in ref.~\cite{Baglio:2019uty}, the $k$~factors for the SM and the SMEFT interference in di-boson distributions can be very different. The SMEFT predictions for these channels are generated using the \texttt{WZanomal} and \texttt{WWanomal} models~\cite{Baglio:2019uty} within the \texttt{POWHEG-BOX-V2} which includes NLO QCD corrections~\cite{Dixon:1999di, Baur:1994aj, Azatov:2019xxn}. We shower and hadronise our results using \texttt{Pythia 8.2}~\cite{Sjostrand:2014zea}.

For the $WZ$ production with leptonic decays, we include the ATLAS $m_T^{WZ}$ production distribution~\cite{ATLAS:2019bsc}. For the last bin of this distribution, which is an overflow bin, we have cut off the SMEFT prediction at $m_T^{WZ} = 1 \tev$ to ensure the validity of the EFT. We implement the cuts provided in ref.~\cite{ATLAS:2019bsc} in \texttt{Rivet}~\cite{Buckley:2010ar} and validate the analysis by comparing our \texttt{POWHEG} + \texttt{Pythia 8.2} SM predictions to the ones in the experimental reference. The SM predictions agree within $5\,\%$ in each bin.

For $WW$ production in the $e^{\pm}\nu\mu^{\mp}\nu$ final state, we include the leading lepton's $p_T$ distribution, $p_T^{\ell_1}$, from the ATLAS study~\cite{ATLAS:2019rob}. In order to consider the most sensitive bins, we take the predictions of bins 8 to 14. Similar to the $WZ$ analysis, we utilise the cuts from the ATLAS study~\cite{ATLAS:2019rob} in \texttt{Rivet}~\cite{Buckley:2010ar} and validate our \texttt{POWHEG} + \texttt{Pythia 8.2} SM predictions to the ones in the experimental paper. The SM predictions agree within $6\,\%$.

The SMEFT predictions for the ATLAS $\Delta \phi_{jj}$ distribution in the electroweak $Zjj$ production~\cite{ATLAS:2020nzk} are taken directly from \texttt{fitmaker}. This distribution tightly constrains anomalous triple gauge couplings induced by the operator $\ope_W$, which is generated in several of the considered SM extensions. 

For LEP $WW$ data, we use the SM and SMEFT parameterisations as well as the theoretical uncertainties from ref.~\cite{Berthier:2016tkq} for the total and differential angular cross-sections at different energies. 

The di-boson predictions are expressed in terms of the WCs 
\begin{equation}
    \wc_{W}, \quad
    \wc_{HWB}, \quad 
    \wc_{HD}, \quad 
    \wc_{Hl}^{(1)}, \quad 
    \wc_{ll}, \quad 
    \wc_{Hl}^{(3)}, \quad 
    \wc_{He}, \quad
    \wc_{Hq}^{(1)}, \quad 
    \wc_{Hq}^{(3)}, \quad 
    \wc_{Hd}, \quad 
    \wc_{Hu}
    \, .
\end{equation}

\paragraph{Electroweak precision observables:} We calculate the SMEFT parametrisations of the EWPO based on refs.~\cite{Berthier:2015oma, Brivio:2017vri, Dawson:2019clf}, see also appendix~A2 of ref.~\cite{Anisha:2020ggj} for the explicit expressions and tabs.~10.4 and 10.5 of ref.~\cite{10.1093/ptep/ptaa104} for the most accurate SM predictions. The EWPO predictions are calculated in terms of the WCs 
\begin{equation}
    \wc_{HWB}, \quad \wc_{HD}, \quad 
    \wc_{ll}, \quad 
    \wc_{Hl}^{(1)}, \quad 
    \wc_{Hl}^{(3)}, \quad 
    \wc_{He}, \quad
    \wc_{Hq}^{(1)}, \quad 
    \wc_{Hq}^{(3)}, \quad 
    \wc_{Hd}, \quad 
    \wc_{Hu}, \quad
    \wc_{H}
    \, ,
\end{equation}
where the contribution of $\wc_H$ is only included in the SMEFT parameterisation of $m_W$ from ref.~\cite{Degrassi:2021uik}.

\section{Observables}\label{sec:obs}
Our fits include data from EWPO, LEP-$WW$ measurements as well as LHC data for single Higgs, di-Higgs, di-boson, and electroweak $Zjj$ production processes. 

\begin{itemize}
    \item We include a total of $15$ electroweak precision observables. They consist of pseudo-measurements on the $Z$~resonance~\cite{ALEPH:2005ab}, as well as a combination of the $W$~mass measurements at LEP~\cite{ALEPH:2013dgf}, Tevatron~\cite{Group:2012gb} and ATLAS~\cite{ATLAS:2017rzl}, the LEP and Tevatron combination of the decay width of the $W$~boson~\cite{Zyla:2020zbs} and the Tevatron $\sin^2 \theta_\text{eff}$~measurement~\cite{CDF:2018cnj}
    \begin{equation}
\begin{split}
    \Gamma_{Z}, \;
    \sigma^{0}_{\text{had}}, \; 
    R^{0}_{l}, \; 
    A_{l}, \; 
    &A_{l}(\text{SLD}), \; 
    A^{l}_{FB}, \; 
    R^{0}_{c}, \;
    A_{c}, \; 
    A^{c}_{FB}, \;
    R^{0}_{b}, 
    A_{b}, \; 
    A^{0,b}_{FB}, \\
    &m_{W}, \; 
    \Gamma_{W}, \;
    \sin^2\theta_{\text{eff}}^{l}(\text{Tev}) \, .
\end{split}
\end{equation}

\item We include LHC single-Higgs data from ATLAS and CMS. For LHC Run~I, we incorporate the combination of ATLAS and CMS results in ref.~\cite{Khachatryan:2016vau}. For ATLAS Run~II, we add the STXS results from refs.~\cite{ATLAS:2020naq, ATLAS:2021upe} as well as the measurements of the signal strengths in the $h \to Z \gamma$, $h\to \mu \mu$, $h\to \tau \tau$ and $h \to b \bar{b}$ (VBF and $t \bar{t}h$ only) decay channels. For CMS Run~II, we make use of the signal strength measurements~\cite{CMS:2020gsy, CMS:2020dvg, CMS:2020xwi, CMS:2020mpn} as well as the STXS results~\cite{CMS:2021ixs, CMS:2020dvp, CMS:2021kom, CMS:2021ugl}. For the STXS $h \to ZZ \to 4 \ell$ analysis~\cite{CMS:2021ugl} we neglect the \texttt{qqH-3j} category due to its very large uncertainty. For the $h \to WW$ channel signal strength measurement~\cite{CMS:2020dvg}, we only include the \texttt{0-jet} category and assume that the signal contribution comes from gluon fusion only (we discard the $5\%$ contribution from the other production modes). For the CMS $t\bar{t}h$ analysis~\cite{CMS:2020mpn}, we select only the most sensitive three channels for which there is one single dominant decay mode. These are the following three final states: 2 same-sign (SS) leptons with no hadronic $\tau$, $\tau_h$, ensuing from $h \to WW^*$, 2 SS leptons and 1 $\tau_h$ from $h \to \tau\tau$, and 1 lepton and 2 $\tau_h$ also from $h \to \tau \tau$. Once CMS combines the signal regions and provides the signal strength for different decay channels separately, we will be able to incorporate these in our analysis.

\item For di-Higgs production, the total cross section signal-strength measurements in the $4b$, $2b 2\tau$ and $2b 2\gamma$ decay channels are taken into account~\cite{ATLAS:2018dpp, ATLAS:2018rnh, ATLAS:2018uni, CMS:2020tkr, CMS:2021ssj, CMS:2017hea}. These measurements include $36.1 \ifb$ of data for ATLAS and up to $137\ifb$ for CMS. We have translated the upper limits given in the experimental references into signal strengths measurements as listed in tab.~\ref{tab:diHiggs_data} in appendix~\ref{app:diHiggsData}.

\item We include momentum-dependent di-boson distributions as well as the $\Delta \phi_{jj}$~distribution for electroweak $Zjj$ production~\cite{ATLAS:2020nzk}. For $WZ$ production with leptonic decays, we include the ATLAS $m_T^{WZ}$ production distribution with $36.1\ifb$~\cite{ATLAS:2019bsc}. For $WW \to e \mu \nu \nu$ production, we include the $p_T^{\ell_1}$ distribution of the leading lepton in ATLAS~\cite{ATLAS:2019rob}. We only include bins 8-14 of the distributions since we observed some discrepancies between our SM prediction and the ATLAS SM prediction in the low-$p_T^{\ell_1}$ regime.
We include the measurements and correlation matrices as provided on \texttt{Hepdata}~\cite{Maguire:2017ypu}.

For LEP $WW$ data, we consider the cross-section measurements  for the process $e^{+} e^{-} \rightarrow W^{+} W^{+} \rightarrow l \nu l \nu / l \nu q q /qqqq $ at different centre of mass energies and angular distributions from tabs.~12-15 of ref.~\cite{Berthier:2016tkq}  The actual experimental measurements are from refs.~\cite{L3:2004lwm, OPAL:2007ytu, ALEPH:2013dgf, ALEPH:2004dmh}.

\end{itemize}

The full list of observables included in our fits is given in tab.~\ref{tab:obset}. 
To highlight the constraining power of recent analyses and compare with previous work, we split our set of observables into two sets called ``2020 dataset" and ``this analysis" in the following.
The 2020 dataset is used to compare and crosscheck with Ref.~\cite{Ellis:2020unq}. The set called ``this analysis" contains updated versions of some experimental analyses as well as additional data. To avoid overlap of experimental analyses, some data used in the 2020 dataset has been removed in the ``this analysis" set, see tab.~\ref{tab:obset}.

\begin{table}[h]
	\centering
	\renewcommand{\arraystretch}{2.0}
	\caption{\small {Observables included in the fit. The rightmost column specifies which observables were part of the 2020 dataset used for comparison with previous work. The observables in \textcolor{teal}{teal} are \textbf{exclusively} used in the 2020 dataset. They are not part of our full dataset as they overlap with other observables.}}
	\begin{adjustbox}{width=0.9\textwidth}
		\label{tab:obset}
		\begin{tabular}{|c|c|c|c|c|}
			\hline
			\multicolumn{2}{|c|}{Observables} & no. of measurements	 &  References & 2020 \\
			\hline
			\multicolumn{2}{|c|}{\bf{Electroweak Precision Observables (EWPO)}}& \multirow{3}{*}{15} & \multirow{3}{*}{tab.~1 of ref.~\cite{Haller:2018nnx}} &  \\
			\multicolumn{2}{|c|}{ $\Gamma_{Z}$, $\sigma^{0}_{had}$, $R^{0}_{l}$, $A_{l}$, $A_{l}$(SLD), $A^{l}_{FB}$, 
			\text{sin}$^2\theta_{\text{eff}}^{l}$(Tev),} & \multirow{1}{*}{} & \multirow{1}{*}{} & \checkmark \\
			\multicolumn{2}{|c|}{ $R^{0}_{c}$, $A_{c}$, $A^{c}_{FB}$, $R^{0}_{b}$, $A_{b}$, $A^{b}_{FB}$, $m_{W}$, $\Gamma_{W}$} 	& \multirow{1}{*}{} & \multirow{1}{*}{correlations in ref.~\cite{ALEPH:2005ab}} & \checkmark \\
			\hline
			\multicolumn{2}{|c|}{\bf{LEP-2 WW data}} & \multirow{1}{*}{74} & \multirow{1}{*}{tabs.~12-15 of ref.~\cite{Berthier:2016tkq}} & \checkmark \\ 
			\cline{1-2}
			\hline
			\multicolumn{2}{|c|}{\bf{Higgs Data}} & \multirow{1}{*}{} & \multirow{1}{*}{} & \\ 
			\cline{1-2}
			\multirow{3}{*}{7 and 8 TeV } & ATLAS \& CMS combination  & \multirow{1}{*}{20} & \multirow{1}{*}{tab.~8 of ref.~\cite{Khachatryan:2016vau}} & \checkmark\\
			\cline{2-5}
			\multirow{3}{*}{Run-I data }& ATLAS \& CMS combination $\mu( h \to \mu \mu)$ &  \multirow{1}{*}{1} &  \multirow{1}{*}{tab.~13 of ref.~\cite{Khachatryan:2016vau}} & \checkmark \\ \cline{2-5}
			& ATLAS $\mu (h \to Z \gamma)$ & \multirow{1}{*}{1} & \multirow{1}{*}{ fig.~1 of ref.~\cite{Aad:2015gba}} & \checkmark \\
			\hline
			\multirow{4}{*}{13 TeV ATLAS} &  $\mu ( h \to Z \gamma )$ at 139 $\ifb$ & 1 &  \cite{Aad:2020plj} & \checkmark\\
			& $\mu ( h \to \mu \mu)$ at 139 $\ifb$ & 1 & \cite{Aad:2020xfq} & \checkmark\\
			Run-II data  & $\mu(h \to \tau \tau)$ at 139 $\ifb$ & 4 & fig.~14 of ref.~\cite{ATLAS-CONF-2021-044} &\\
		    & $\mu( h \to bb)$ in VBF and ${ttH}$ at 139 $\ifb$ & 1+1 & \cite{ATLAS:2020bhl,ATLAS:2020syy} & \\
		    \cline{2-5} 
		    & \textcolor{teal}{STXS Higgs combination} & \textcolor{teal}{25} & \textcolor{teal}{figs. 20/21 of ref.~\cite{ATLAS:2019nkf}} & \checkmark \\
		    & STXS $h \to \gamma \gamma/ZZ/b \bar{b}$ at $139\ifb$ & 42 & figs.~1 and 2 of ref.~\cite{ATLAS:2020naq} &\\
			& STXS $ h \rightarrow$ $W W$ in ggF, VBF at $139\ifb$ & 11 & figs.~12 and 14 of ref.~\cite{ATLAS:2021upe} &\\
			\hline
			&  \textcolor{teal}{CMS combination at up to $137\ifb$} & \textcolor{teal}{23}  &   \textcolor{teal}{tab.~4 of ref.~\cite{CMS:2020gsy}} & \checkmark\\
			 &  $\mu(h \to b \bar{b})$ in $Vh$ at $35.9/41.5\ifb$ & 2  &   entries from tab.~4 of ref.~\cite{CMS:2020gsy} &\\
			 &  $\mu(h \to W W)$ in ggF at $137\ifb$ & 1  & \cite{CMS:2020dvg} &\\
			 13 TeV CMS &  $\mu (h\to \mu \mu)$ at  $137\ifb$ & 4  & fig.~11 of ref.~\cite{CMS:2020xwi} &\\
			Run-II data & $\mu (h \to \tau \tau/WW)$ in  $t\bar{t}h$ at $137\ifb$ & 3  & fig.~14 of ref.~\cite{CMS:2020mpn} &\\
			\cline{2-5} 
			 & STXS $h\to WW$ at $137\ifb$ in $Vh$  & 4 &  tab.~9 of ref.~\cite{CMS:2021ixs} &
			\\
			& STXS $h \to \tau \tau$ at  $137\ifb$  & 11 &  figs.~11/12 of ref.~\cite{CMS:2020dvp} &
			\\
			& STXS $h \to \gamma \gamma$ at  $137\ifb$ & 27 & tab.~13 and fig.~21 of ref.~\cite{CMS:2021kom} &
			\\
			& STXS $h \to ZZ $ at  $137\ifb$ & 18 & tab.~6 and fig.~15 of ref.~\cite{CMS:2021ugl} &
			\\
			\hline
			\multicolumn{2}{|c|}{{\bf{ATLAS  $WZ$  13 TeV $m_T^{WZ}$}} at $36.1\ifb$}&6 bins & fig.~4(c) of ref.~\cite{Aaboud:2019gxl} & \checkmark \\
			\hline
			\multicolumn{2}{|c|}{{\bf{ ATLAS $Zjj$  13 TeV $\Delta \phi_{jj}$}} at $139\ifb$}&12 bins & fig.~7(d) of ref.~\cite{ATLAS:2020nzk} & \checkmark \\
			\hline
			\multicolumn{2}{|c|}{{\bf{ ATLAS $WW$  13 TeV $p_T^{\ell 1}$}} at $36.1\ifb$}&7 bins & bins 8-14 of fig.~7(a) of ref.~\cite{ATLAS:2019rob} & \checkmark \\
			\hline
			\multicolumn{2}{|c|}{\bf{Di-Higgs signal strengths ATLAS \& CMS 13 TeV data}}&\multirow{2}{*}{6}  &   \multirow{2}{*}{\cite{ATLAS:2018dpp,ATLAS:2018rnh,ATLAS:2018uni,CMS:2020tkr, CMS:2021ssj,CMS:2017hea}} & \multirow{2}{*}{}  \\
			\multicolumn{2}{|c|}{$\mu_{_{HH}}^{b\bar{b}b \bar{b}}$, $\mu_{_{HH}}^{b \bar{b}\tau \bar{\tau}}$, $\mu_{_{HH}}^{b \bar{b} \gamma \gamma}$} & & & \\
			\hline
		\end{tabular}
	\end{adjustbox}
\end{table}

\section{Model independent SMEFT: Bayesian analysis}\label{sec:modindep}

The statistical analysis in this work is performed within a Bayesian framework. The priors, used in this work for free parameters, follow uniform distribution with some definite ranges. For all numerical results, samples from the un-normalised posterior distributions are used, each of which are generated from a Markov Chain Monte Carlo (MCMC) process. The MCMC algorithm followed here is Metropolis-Hastings~\cite{hastings1970} and all runs come from a single long Markov-chain. We make sure that all obtained samples are independent and identically distributed ($iid$) and the chains converge to desired quantiles. This is ensured through diagnostic checks and sequential runs following the prescriptions of Raftery and Lewis~\cite{raftery1992}. Point estimates are almost always quoted in terms of Medians and fixed quantiles around them.
Though not used, we keep track of all corresponding frequentist maximum likelihood estimates (MLEs), which both help us to track the fit-quality as well as a good choice for the start of the Markov-chains.
As for priors, we use uniform distributions with specific (conservative) ranges for the free fit parameters. As we use the SM theoretical predictions for all observables from several sources, there are no nuisance-type parameters (e.g.~SM input parameters) in this analysis with `informative' priors.
Appendix~\ref{subsec:priors} contains detailed information about the priors for the free parameters here, i.e.~the dimension-6 SMEFT WCs. 

\begin{figure}
    \centering
    \subfloat[]{\includegraphics[width=.95\textwidth]{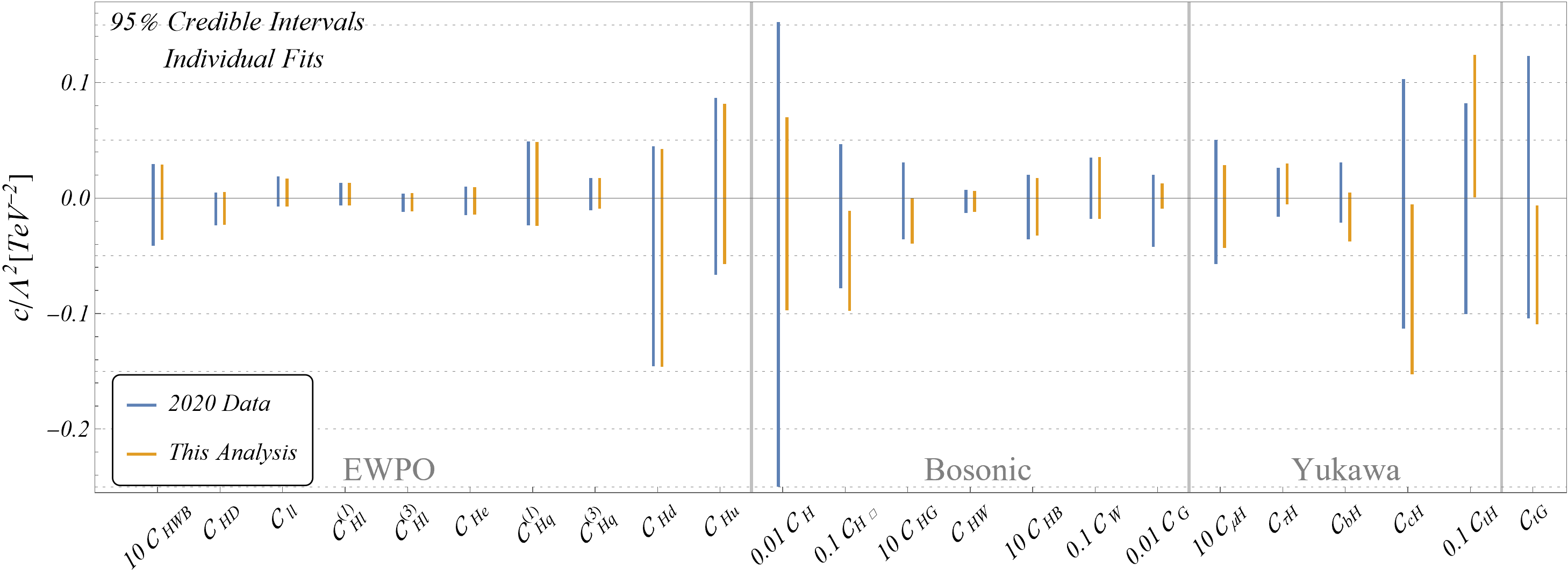}\label{fig:CompModIndepIndiv}}\\
    \subfloat[]{\includegraphics[width=.95\textwidth]{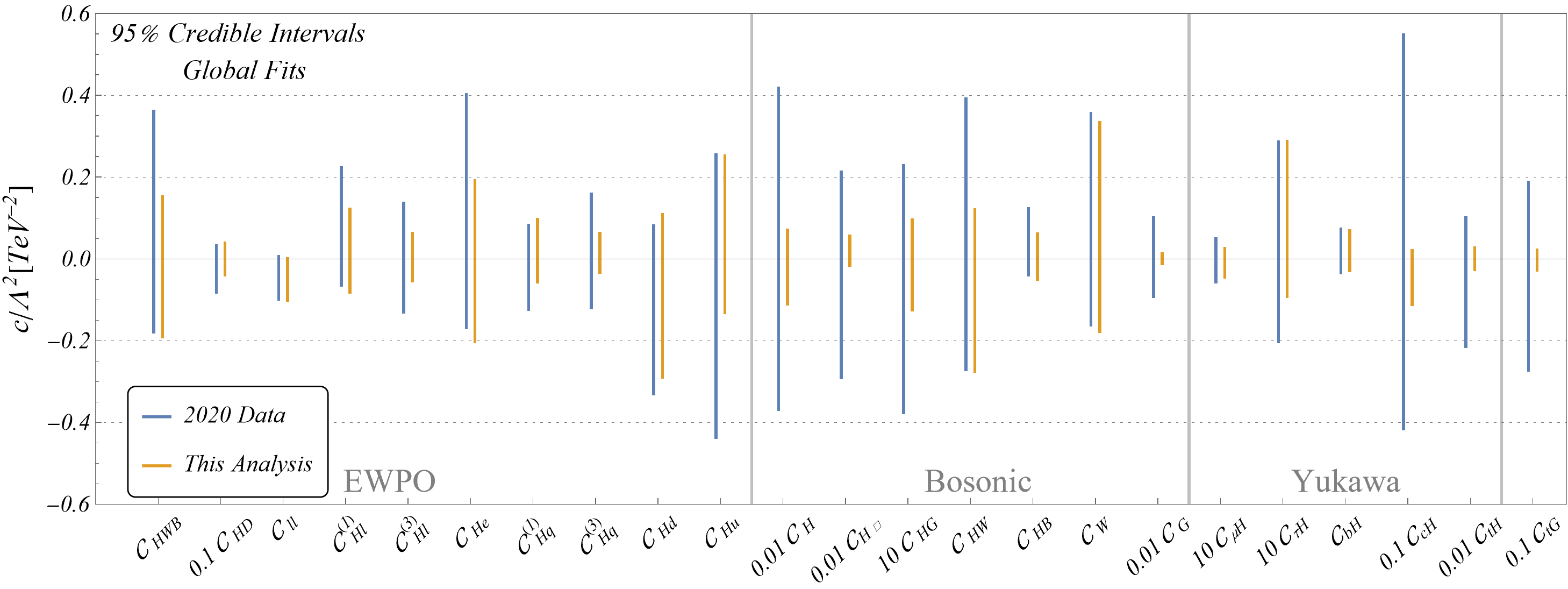}\label{fig:CompModIndepGlob}}
    \caption{\small Individual (top) and global (bottom) 95\% CI limits on the WCs. 
    We compare a fit involving  our full dataset (orange), with a reduced set containing LHC Higgs measurements up to year 2020 (blue), see tab.~\ref{tab:obset} for the dataset definitions. Note that some bounds have been scaled by factors of ten to fit all results on the same $y$-axis.}
    \label{fig:CompModIndep}
\end{figure}

We start the discussion of our model-independent fits by showing the results of one-parameter fits of the WCs in the upper panel of fig.~\ref{fig:CompModIndep}, see also tab.~\ref{tab:fit_results} for numerical results. In the figure, we contrast the results including our full dataset with a reduced set based on the one used in the most recent \texttt{fitmaker} analysis~\cite{Ellis:2020unq}. See tab.~\ref{tab:obset} for the exact dataset definitions. Overall, the WCs constrained through the EWPO typically receive much stronger individual bounds than those constrained through Higgs and di-boson observables alone. The most weakly constrained WC is $\wc_H$ which is most strongly constrained through di-Higgs production in our fit. Comparing the two fits, we find that operators constrained through EWPO have not benefited from the addition of new datasets at the level of one-parameter fits. Several of the limits on the bosonic as well as Yukawa-like operators, on the other hand, have improved significantly. Deviations from the SM for some WCs, for instance for $\wc_{tH}$, are caused by large correlations in the CMS $h \to ZZ \to 4 \ell$ STXS analysis~\cite{CMS:2021ugl},
specifically between the \texttt{ggF 1jet} ($p_T^H \in [0, \, 60]$~GeV), \texttt{ggF 1jet} ($p_T^H \in [60, \, 120]$~GeV), and 
\texttt{qqH 2jet} ($m_{jj} \in [0, \, 350]$~GeV, $Vh$ veto) regions. 
However, in the global analysis with more DOF none of these deviations persist. 
 
In the lower panel of fig.~\ref{fig:CompModIndep}, we display the results of our global fit of $23$ WCs after marginalisation. Numerical results can be found in tab.~\ref{tab:fit_results}. To highlight the restriction power of recent STXS measurements, we again contrast our full data set with LHC data up to 2020. 
We have explicitly checked that the limits from this reduced dataset agree are in good agreement with previous literature~\cite{Ellis:2020unq}. 
The improvements of the limits with the addition of more data is even more visible in the global fit than it was in the one-parameter fits. We find relatively mild improvement for WCs describing modified Higgs Yukawa couplings to leptons or bottom quarks. Many of the other limits, however, have significantly decreased. 

Comparing the limits from one-parameter fits with the global analysis, we find that the limits of six WCs weaken by a factor of ten or more compared to the individual fit limits: $\wc_{HWB}$, $\wc_{HD}$, $\wc_{HW}$, $\wc_{HB}$, $\wc_{He}$, $\wc_{Hl}^{(1)}$. This is the result of the strong correlation of these coefficients as displayed in tab.~\ref{tab:correlations_23wcs}. Many pairs of these WCs have absolute correlation coefficients of $\gtrsim 0.8$ as a result of their joint contribution to the shifts of SM parameters.

\begin{figure}[htb]
    \centering
    \subfloat[]{\includegraphics[width=.95\textwidth]{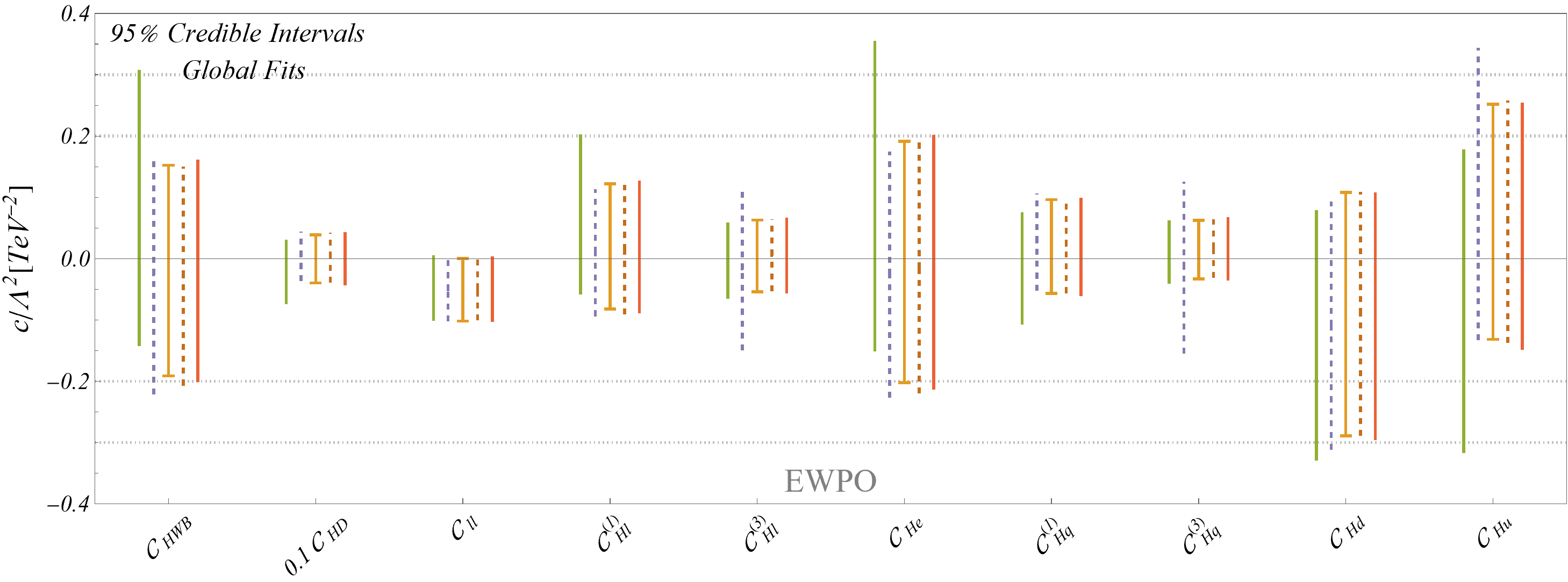}\label{fig:CompModIndepPart1}}\\
    \subfloat[]{\includegraphics[width=.95\textwidth]{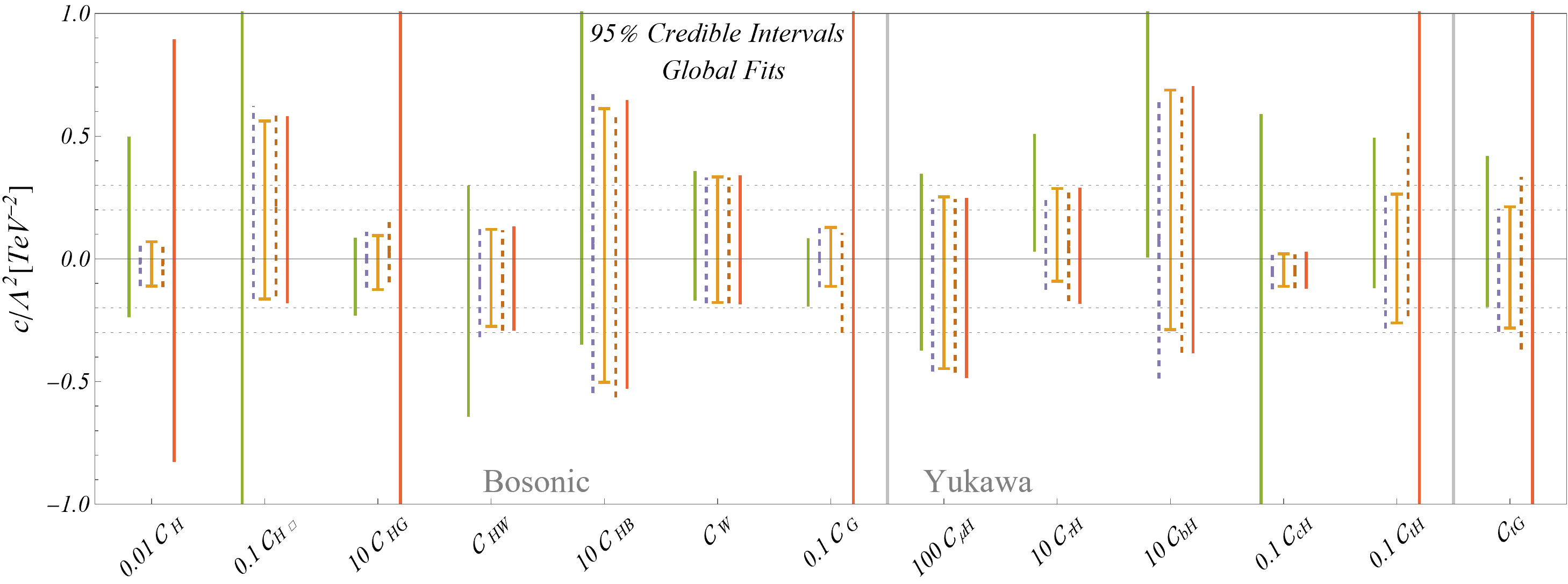}\label{fig:CompModIndepPart2}}\\
    \subfloat[]{\includegraphics[width=.35\textwidth]{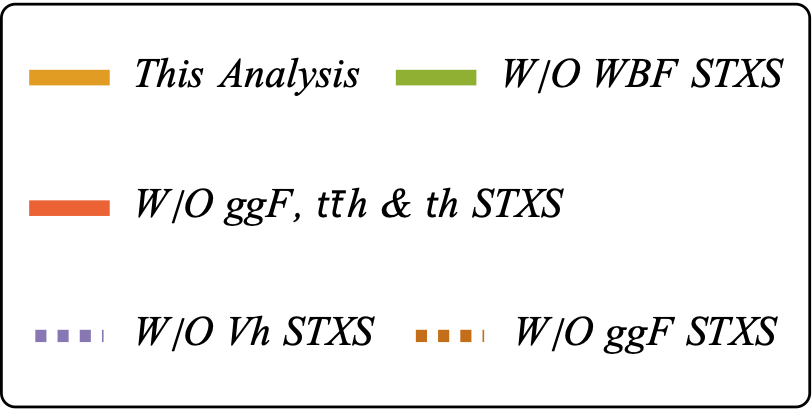}\label{fig:CompModIndepPartLeg}}
    \caption{\small Figs.~\ref{fig:CompModIndepPart1} and \ref{fig:CompModIndepPart2} show the $95\%$ CIs  on 23 WCs in a global fit using different datasets. We constrast the corresponding intervals obtained using the full dataset (orange) with datasets excluding certain sets of STXS measurements. We individually exclude STXS measurements of WBF production (green), $Vh$ measurements (purple), ggF measurements (red dashed), and a combination of ggF with $t\bar{t}h $ and $th$ (red solid).
    Note that some bounds have been scaled by factors of ten to fit all results on the same $y$-axis.}
    \label{fig:CompModIndepPart}
\end{figure}

To gain a better understanding for the relevance of different Higgs production channels for the constraints on different WCs, we display in fig.~\ref{fig:CompModIndepPart} the global fit limits on the WCs when removing certain STXS channels from the analysis. In four additional fits, we have removed the STXS channels for associated production of a Higgs with a vector boson ($Vh$), weak boson fusion (WBF), gluon fusion (ggF), and a combination of gluon fusion as well as top associated production modes (ggF+$t\bar{t}h$+$tH$).
As expected, removing gluon fusion STXS channels from the analysis only weakens the limits on $\wc_{tG}$, $\wc_{tH}$, $\wc_{G}$ and $\wc_{HG}$, highlighting the constraining power of ggF STXS measurements up to high transverse momenta of the Higgs. The highest-momentum STXS regions included in our analysis are $p_T^H \in [200, \, 300, \, 450] \,\text{GeV}$ as well as $p_T^H > 450\,\text{GeV}$~\cite{CMS:2021kom}.
When in addition to removing the gluon fusion STXS channels we also neglect top associated Higgs production, the extreme weakening of the limits on $\wc_{HG}$ and $\wc_{tH}$ leads to looser constraints on $\wc_{H}$. This is because of correlations of $\wc_{H}$ with other WCs ($\wc_{tH}$ and $\wc_{HG}$) which are, however, irrelevant at the current level of constraints from ggF and top associated Higgs production.

STXS measurements of $Vh$ production mainly affect the limits of $\wc_{Hq}^{(3)}$, $\wc_{Hu}$ and $\wc_{Hl}^{(3)}$. While $\wc_{Hq}^{(3)}$ and $\wc_{Hu}$ directly profit from $Vh$ measurements at high $p_T^V$, $\wc_{Hl}^{(3)}$ improves indirectly as it is highly correlated with $\wc_{Hq}^{(3)}$.
WBF STXS measurements influence several bosonic operators such as $\wc_{H\square}$ and $\wc_{HWB}$. Other operators which do not directly contribute to WBF, for instance $\wc_{Hl}^{(1)}$ are affected by WBF STXS measurements through their correlation with other operators. 
$\wc_{cH}$, which only appears in the parametrisation of the Higgs width, is highly affected by WBF STXS data because of its correlation with $\wc_{H \square}$.

\section{BSMs $\xrightarrow {\texttt{CoDEx}}\; \subset$ SMEFT: Effective Operators, WCs and Bayesian Analysis}\label{sec:BSMs}

\begin{table}[htb]
	\caption{\small SM and BSM fields and their spin and gauge quantum numbers. The BSM Lagrangians are constructed using these fields.}
	\label{tab:fieldconvention}
	\scriptsize
	\centering
	\renewcommand{\arraystretch}{1.5}
		\subfloat{\begin{threeparttable}
			\begin{tabular}{|c|c|c|c|c|}
				\hline\hline
				\multirow{2}{*}{SM field} & \multirow{2}{*}{Spin} &\multicolumn{3}{c|}{SM quantum numbers}\\
				\cline{3-5}
				&&$SU(3)_{_{C}}$&$SU(2)_{_{L}}$&$U(1)_{_Y}$\tnote{1}\\
				\hline \hline
				$q_{_L}$ & $\frac{1}{2}$ & 3 & 2 & $\frac{1}{6}$\\
				\hline
				$l_{_L}$ & $\frac{1}{2}$ & 1 & 2 & -$\frac{1}{2}$\\
				\hline
				$u_{_R}$ & $\frac{1}{2}$ & 3 & 1 & $\frac{2}{3}$\\
				\hline
				$d_{_R}$ & $\frac{1}{2}$ & 3 & 1 & -$\frac{1}{3}$\\
				\hline
				$e_{_R}$ & $\frac{1}{2}$ & 1 & 1 & -1\\
				\hline
				$H$ 	 & 0 			 & 1 & 2 & $\frac{1}{2}$\\
				\hline
				$B_{\mu\nu}$ 	 & 1 			 & 1 & 1 & 0\\
				\hline
				$W_{\mu\nu}$ 	 & 1 			 & 1 & 3 & 0\\
				\hline
				$G_{\mu\nu}$ 	 & 1 			 & 8 & 1 & 0\\
				\hline\hline
			\end{tabular}
			\begin{tablenotes}
				\item[1] \footnotesize{Hypercharge convention: $Q_{\text{em}}$ = $T_{3}$ + Y, where $Q_{\text{em}}$, $T_{3}$ and Y are electro-magnetic charge, third component of isospin quantum number and hypercharge respectively.}
			\end{tablenotes}
	\end{threeparttable}}\hspace{.1pt}
	\quad
	\subfloat{\begin{threeparttable}
		\begin{tabular}{|c|c|c|c|c|c|}
			\hline\hline
			\multirow{2}{*}{BSM field} & \multirow{2}{*}{Spin} &\multicolumn{3}{c|}{SM quantum numbers}&\multirow{2}{*}{Mass}\\
			\cline{3-5}
			&&$SU(3)_{_{C}}$&$SU(2)_{_{L}}$&$U(1)_{_Y}$&\\
			\hline \hline
			$\mathcal{S}$ & 0 & 1 & 1 & 0 & $m_{\mathcal{S}}$ \\
			\hline
			$\Delta$ & 0 & 1 & 3 & 0 & $m_{\Delta}$ \\
			\hline \hline
			$\mathcal{S}_1$ & 0 & 1 & 1 & 1 & $m_{\mathcal{S}_1}$ \\
			\hline
			$\mathcal{S}_2$ & 0 & 1 & 1 & 2 & $m_{\mathcal{S}_2}$ \\
			\hline
			$\Delta_1$ & 0 & 1 & 3 & 1 & $m_{\Delta_1}$ \\
			\hline
			$\mathcal{H}_2$ & 0 & 1 & 2 &  $ -\frac{1}{2} $  & $m_{\mathcal{H}_2}$ \\
			\hline
			$\Sigma$ & 0 & 1 & 4 & $\frac{1}{2}$ & $m_{_\Sigma}$ \\
			\hline \hline
			$\varphi_1$ & 0 & 3 & 1 & $ -\frac{1}{3} $ & $m_{\varphi_1}$ \\
			\hline
			$\varphi_2$ & 0 & 3 & 1 & $ -\frac{4}{3} $ & $m_{\varphi_2}$ \\
			\hline  \hline
			$\Theta_1$ & 0 & 3 & 2 & $ \frac{1}{6} $ & $m_{\Theta_1}$ \\
			\hline
			$\Theta_2$ & 0 & 3 & 2 & $ \frac{7}{6} $ & $m_{\Theta_2}$ \\
			\hline
			$\Omega$ & 0 & 3 & 3 & $ -\frac{1}{3} $ & $m_{\Omega}$ \\
			\hline \hline
			$\chi_{_1}$ & 0 & 6 & 3 & $ \frac{1}{3} $ & $m_{\chi_{_1}}$ \\
			\hline
			$\chi_{_2}$ & 0 & 6 & 1 & $ \frac{4}{3} $ & $m_{\chi_{_2}}$ \\
			\hline
			$\chi_{_3}$ & 0 & 6 & 1 & $ -\frac{2}{3} $ & $m_{\chi_{_3}}$ \\
			\hline
			$\chi_{_4}$ & 0 & 6 & 1 & $ \frac{1}{3} $ & $m_{\chi_{_4}}$ \\
			\hline\hline
		\end{tabular}
	\end{threeparttable}}
\end{table}

\begin{figure}[htb]
	\centering
	\subfloat[$\lambda_{\mathcal{H}_{2},1}$ - $\lambda_{\mathcal{H}_{2},2}$]
	{\includegraphics[width=0.325\textwidth, height=6cm]{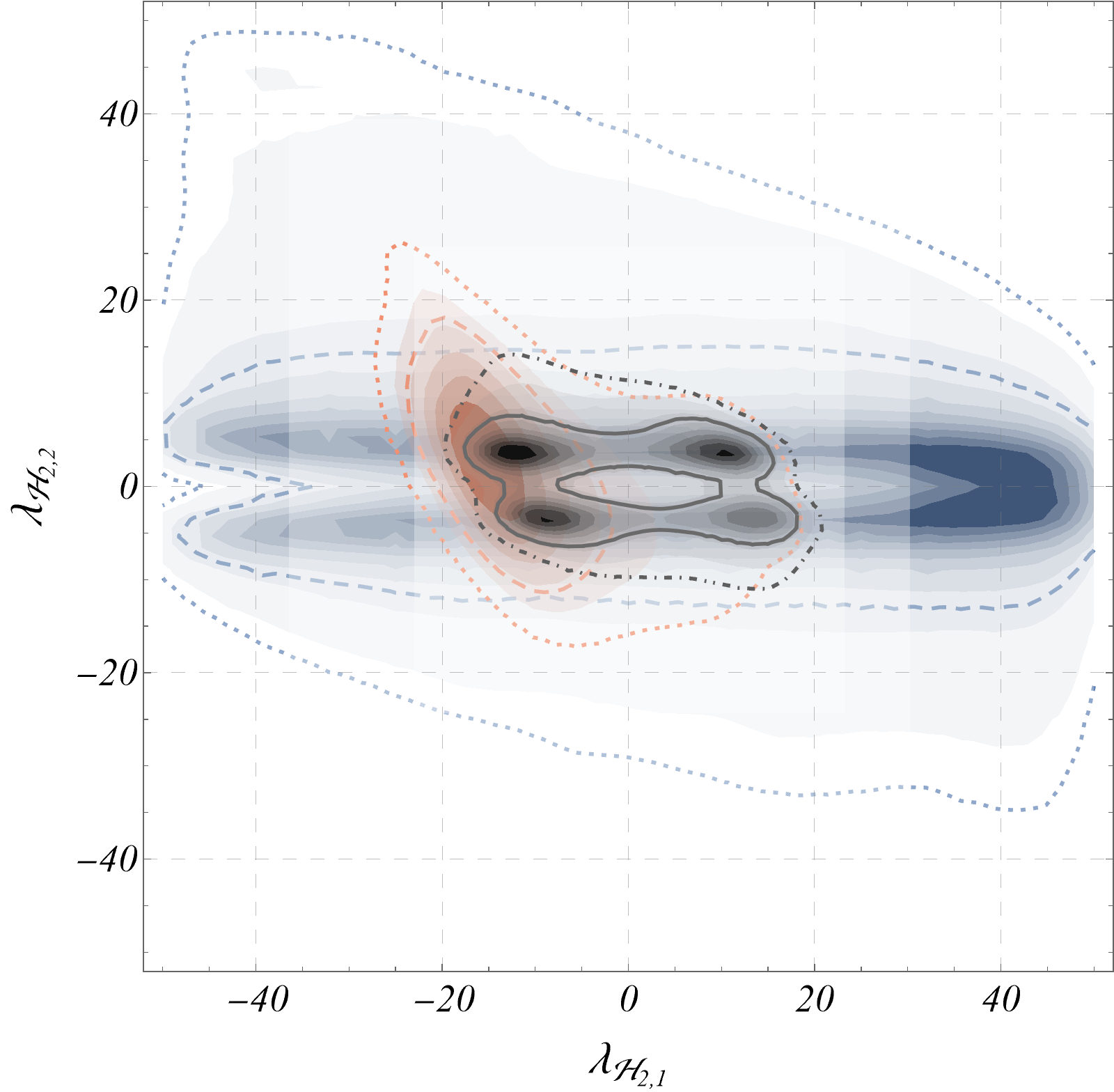}\label{fig:2HDM_param12}}~
	\subfloat[$\lambda_{\mathcal{H}_{2},2}$ - $\lambda_{\mathcal{H}_{2},3}$]
	{\includegraphics[width=0.325\textwidth, height=6cm]{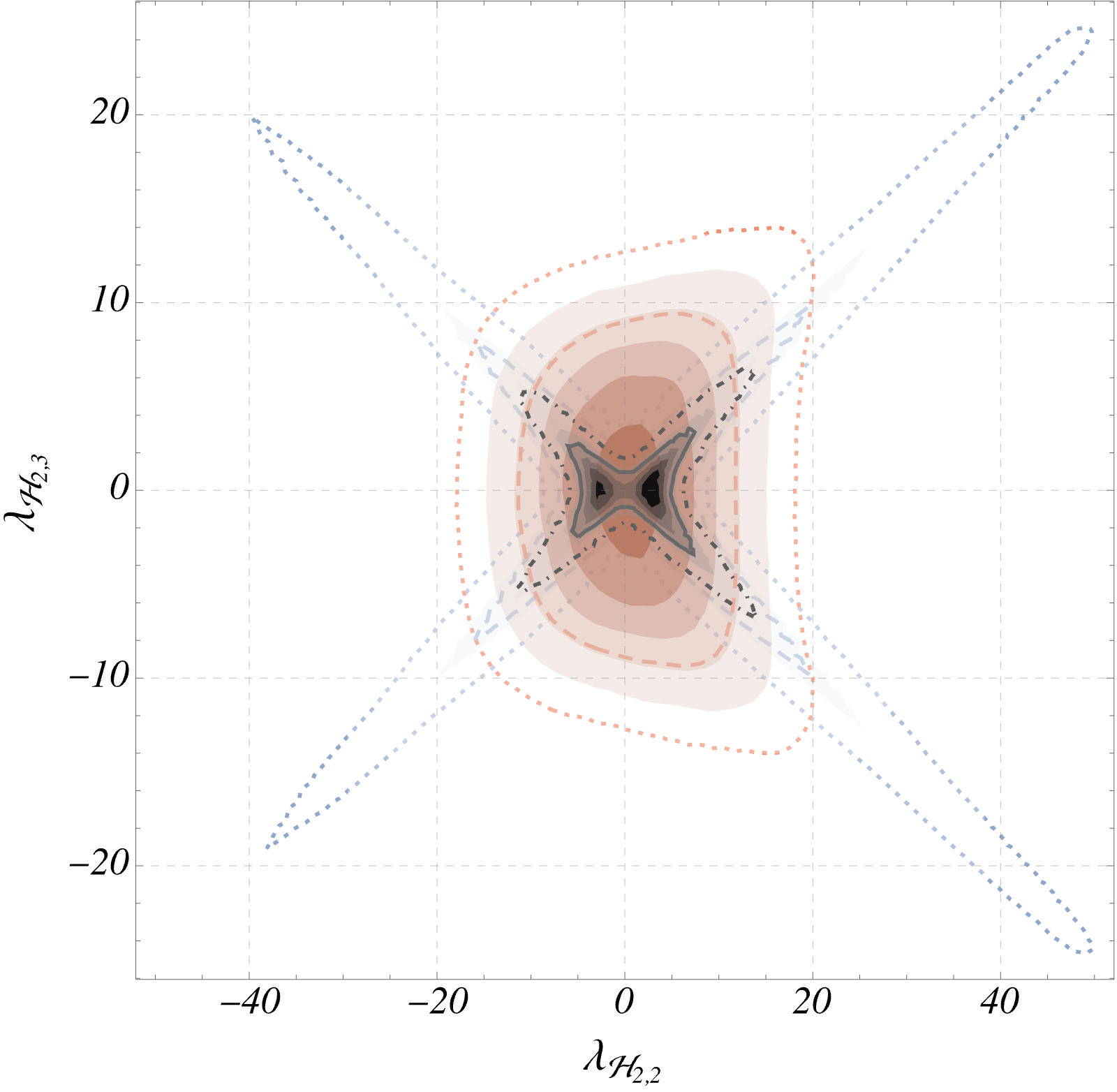}\label{fig:2HDM_param23}}~
	\subfloat[$\lambda_{\mathcal{H}_{2},1}$ - $\lambda_{\mathcal{H}_{2},3}$]
	{\includegraphics[width=0.325\textwidth, height=6cm]{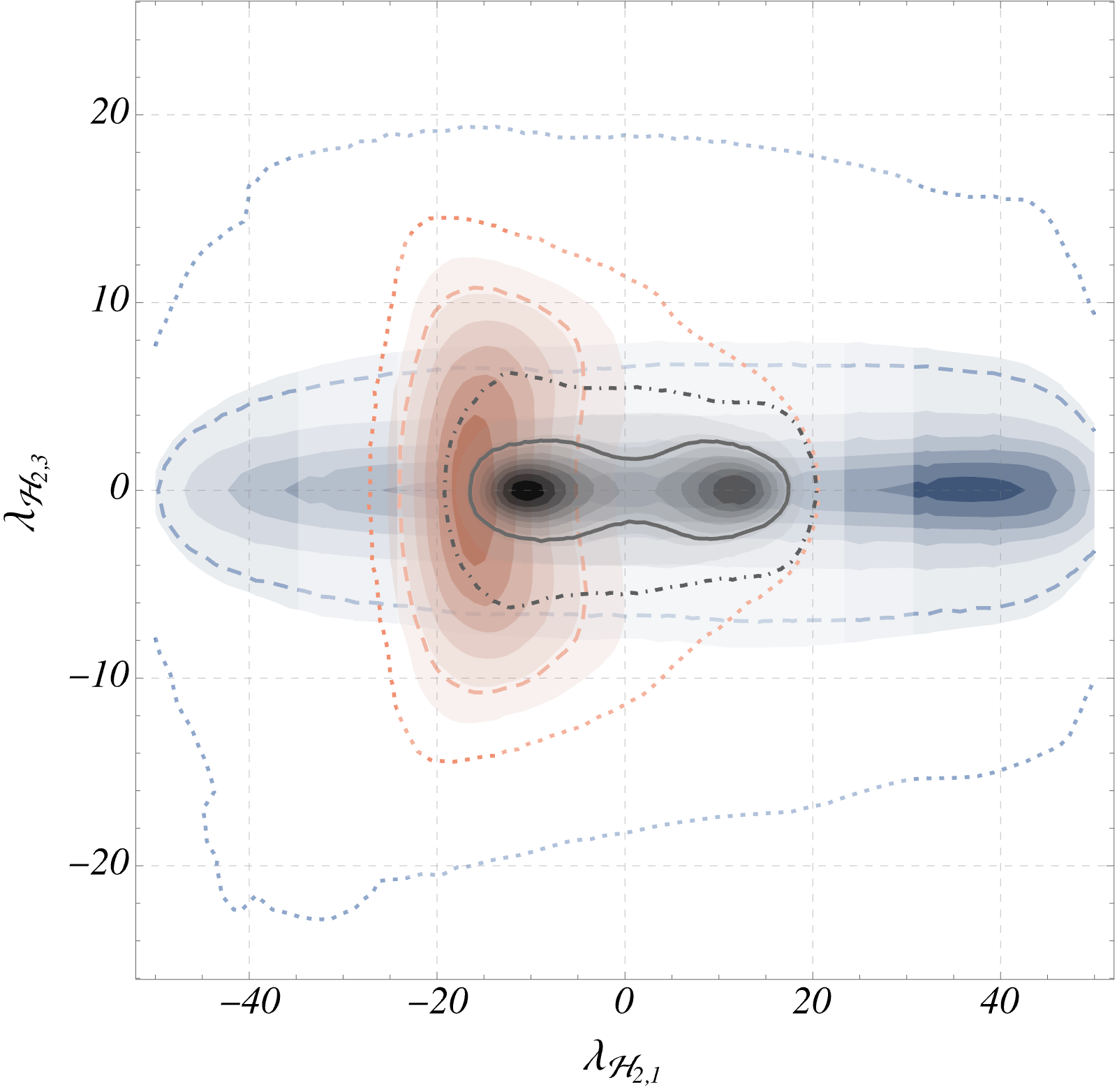}\label{fig:2HDM_param13}}\\
	\subfloat[$\eta_{\Theta_1}^{(1)}$ - $\eta_{\Theta_1}^{(2)}$]
	{\includegraphics[width=0.325\textwidth, height=6cm]{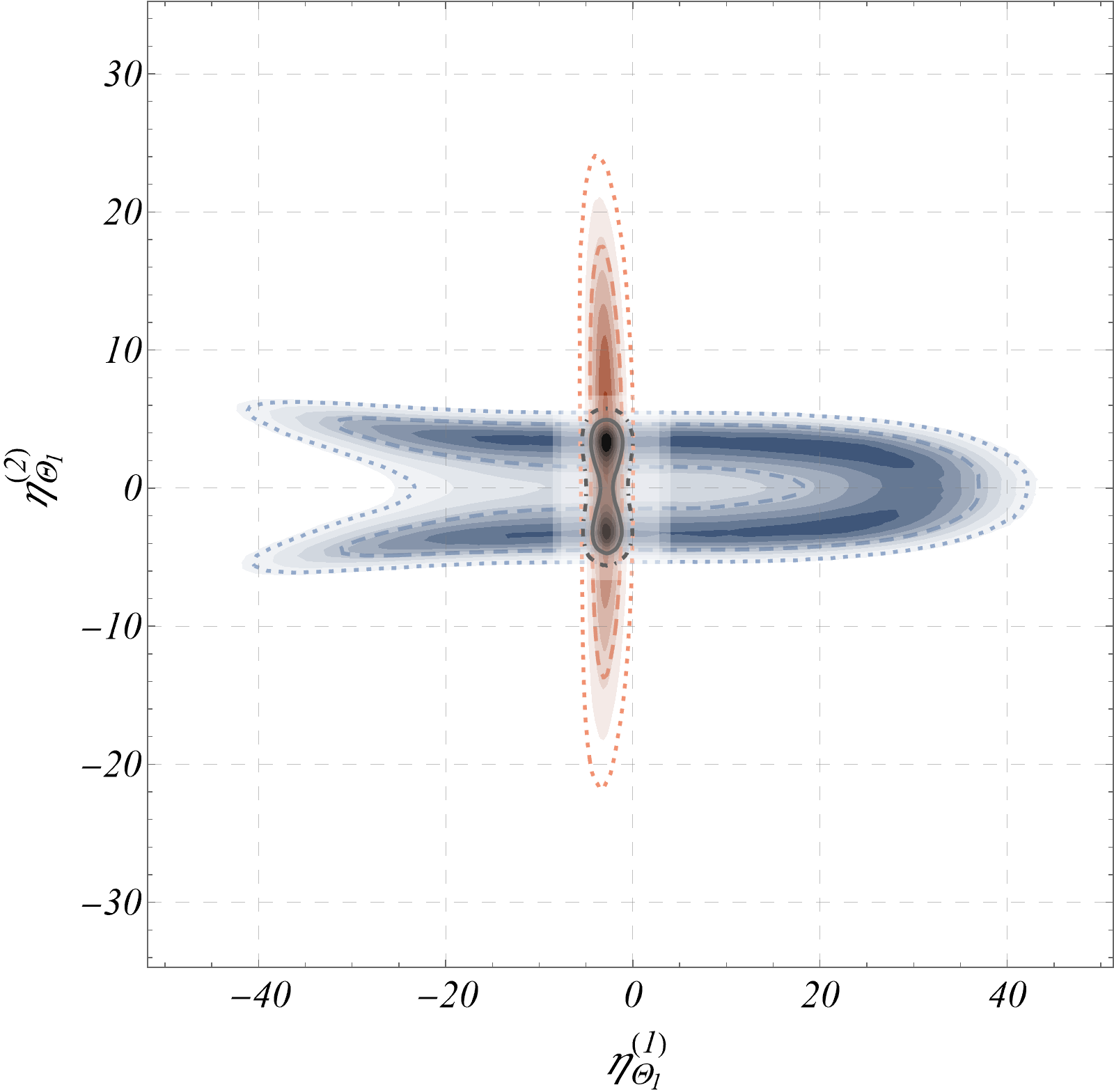}\label{fig:Theta1_param}}~
	\subfloat[Legend]
	{\includegraphics[width=3cm, height=3cm]{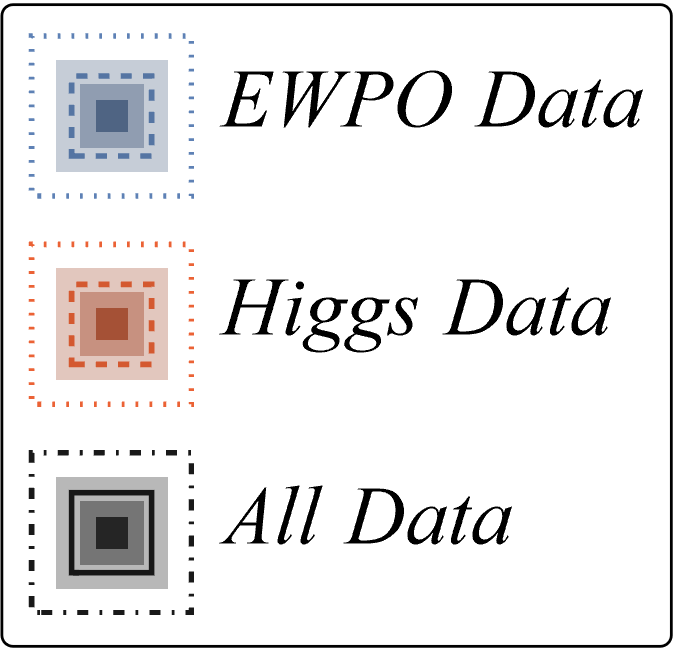}\label{}}
	\caption{\small 
    Two-dimensional marginalised posteriors among the BSM parameters for $\mathcal{H}_{2}$ (top row) and $\Theta_{1}$. The line contours represent the 68\% and 95\% credible intervals (CIs) and the filled contours with changing opacity show the high-probability regions with decreasing probabilities (darker to lighter).  
    We show the results from a fit of ``EWPO" data only (blue), ``Higgs" data only (red) as well as for ``All"  (black). }
	\label{fig:model_param_plots}
\end{figure}

%\newpage
\begin{table*}[!htb]
	\caption{\small Warsaw basis  effective operators and the associated WCs that emerge after integrating-out the heavy field $\mathcal{H}_2 : (1,2,-\frac{1}{2})$.  Operators highlighted in red do not affect our current set of observables and are thus absent from our analysis. Operators highlighted in blue are functions of SM parameters only, while the red coloured ones do not contribute to our observables.}
	\label{tab:H2}
	\centering
	\scriptsize
	\renewcommand{\arraystretch}{1.8}
	\subfloat{
		\begin{tabular}{|*{2}{>{\rowfonttype}c|}}
			\hline \hline
			Dim-6 Ops.&Wilson coefficients\\
			\hline \hline
			$Q_{\text{dH}}$  &  $\frac{\eta _H^2 Y_d^{\text{SM}}}{16 \pi ^2 m_{\mathcal{H}_2}^2}-\frac{3 \eta _H \eta _{\mathcal{H}_2} Y_d^{\text{SM}}}{16 \pi ^2 m_{\mathcal{H}_2}^2}-\frac{\eta _H Y_{\mathcal{H}_2}^{(d)}}{m_{\mathcal{H}_2}^2}$  \\
			&  $-\frac{3 \eta _H \lambda _{\mathcal{H}_2} Y_{\mathcal{H}_2}^{(d)}}{32 \pi ^2 m_{\mathcal{H}_2}^2}+\frac{3 \eta _H \lambda _{\mathcal{H}_2,1} Y_{\mathcal{H}_2}^{(d)}}{16 \pi ^2 m_{\mathcal{H}_2}^2}-\frac{3 \eta _{\mathcal{H}_2} \lambda _{\mathcal{H}_2,1} Y_{\mathcal{H}_2}^{(d)}}{16 \pi ^2 m_{\mathcal{H}_2}^2}$  \\
			&  $\frac{\eta _H \lambda _{\mathcal{H}_2,2} Y_{\mathcal{H}_2}^{(d)}}{4 \pi ^2 m_{\mathcal{H}_2}^2}-\frac{3 \eta _{\mathcal{H}_2} \lambda _{\mathcal{H}_2,2} Y_{\mathcal{H}_2}^{(d)}}{16 \pi ^2 m_{\mathcal{H}_2}^2}+\frac{\lambda _{\mathcal{H}_2,2}^2 Y_d^{\text{SM}}}{192 \pi ^2 m_{\mathcal{H}_2}^2}$  \\
			&  $\frac{5 \eta _H \lambda _{\mathcal{H}_2,3} Y_{\mathcal{H}_2}^{(d)}}{8 \pi ^2 m_{\mathcal{H}_2}^2}+\frac{\lambda _{\mathcal{H}_2,3}^2 Y_d^{\text{SM}}}{48 \pi ^2 m_{\mathcal{H}_2}^2}$  \\
			\hline
			$Q_{\text{eH}}$  &  $\frac{\eta _H^2 Y_e^{\text{SM}}}{16 \pi ^2 m_{\mathcal{H}_2}^2}-\frac{3 \eta _H \eta _{\mathcal{H}_2} Y_e^{\text{SM}}}{16 \pi ^2 m_{\mathcal{H}_2}^2}-\frac{\eta _H Y_{\mathcal{H}_2}^{(e)}}{m_{\mathcal{H}_2}^2}$  \\
			&  $-\frac{3 \eta _H \lambda _{\mathcal{H}_2} Y_{\mathcal{H}_2}^{(e)}}{32 \pi ^2 m_{\mathcal{H}_2}^2}+\frac{3 \eta _H \lambda _{\mathcal{H}_2,1} Y_{\mathcal{H}_2}^{(e)}}{16 \pi ^2 m_{\mathcal{H}_2}^2}-\frac{3 \eta _{\mathcal{H}_2} \lambda _{\mathcal{H}_2,1} Y_{\mathcal{H}_2}^{(e)}}{16 \pi ^2 m_{\mathcal{H}_2}^2}$  \\
			&  $\frac{\eta _H \lambda _{\mathcal{H}_2,2} Y_{\mathcal{H}_2}^{(e)}}{4 \pi ^2 m_{\mathcal{H}_2}^2}-\frac{3 \eta _{\mathcal{H}_2} \lambda _{\mathcal{H}_2,2} Y_{\mathcal{H}_2}^{(e)}}{16 \pi ^2 m_{\mathcal{H}_2}^2}+\frac{\lambda _{\mathcal{H}_2,2}^2 Y_e^{\text{SM}}}{192 \pi ^2 m_{\mathcal{H}_2}^2}$  \\
			&  $\frac{5 \eta _H \lambda _{\mathcal{H}_2,3} Y_{\mathcal{H}_2}^{(e)}}{8 \pi ^2 m_{\mathcal{H}_2}^2}+\frac{\lambda _{\mathcal{H}_2,3}^2 Y_e^{\text{SM}}}{48 \pi ^2 m_{\mathcal{H}_2}^2}$  \\
			\hline
			$Q_{\text{uH}}$  &  $\frac{\eta _H^2 Y_u^{\text{SM}}}{16 \pi ^2 m_{\mathcal{H}_2}^2}+\frac{3 \eta _H \lambda _{\mathcal{H}_2} Y_{\mathcal{H}_2}^{(u)}}{32 \pi ^2 m_{\mathcal{H}_2}^2}+\frac{\eta _H Y_{\mathcal{H}_2}^{(u)}}{m_{\mathcal{H}_2}^2}$  \\
			&  $-\frac{3 \eta _H \eta _{\mathcal{H}_2} Y_u^{\text{SM}}}{16 \pi ^2 m_{\mathcal{H}_2}^2}-\frac{3 \eta _H \lambda _{\mathcal{H}_2,1} Y_{\mathcal{H}_2}^{(u)}}{16 \pi ^2 m_{\mathcal{H}_2}^2}+\frac{3 \eta _{\mathcal{H}_2} \lambda _{\mathcal{H}_2,1} Y_{\mathcal{H}_2}^{(u)}}{16 \pi ^2 m_{\mathcal{H}_2}^2}$  \\
			&  $-\frac{\eta _H \lambda _{\mathcal{H}_2,2} Y_{\mathcal{H}_2}^{(u)}}{4 \pi ^2 m_{\mathcal{H}_2}^2}+\frac{3 \eta _{\mathcal{H}_2} \lambda _{\mathcal{H}_2,2} Y_{\mathcal{H}_2}^{(u)}}{16 \pi ^2 m_{\mathcal{H}_2}^2}+\frac{\lambda _{\mathcal{H}_2,2}^2 Y_u^{\text{SM}}}{192 \pi ^2 m_{\mathcal{H}_2}^2}$  \\
			&  $\frac{\lambda _{\mathcal{H}_2,3}^2 Y_u^{\text{SM}}}{48 \pi ^2 m_{\mathcal{H}_2}^2}-\frac{5 \eta _H \lambda _{\mathcal{H}_2,3} Y_{\mathcal{H}_2}^{(u)}}{8 \pi ^2 m_{\mathcal{H}_2}^2}$  \\
			\hline
			$Q_H$  &  $\frac{3 \eta _H^2 \lambda _{\mathcal{H}_2}}{32 \pi ^2 m_{\mathcal{H}_2}^2}+\frac{17 \eta _H^2 \lambda _H^{\text{SM}}}{16 \pi ^2 m_{\mathcal{H}_2}^2}+\frac{\eta _H^2}{m_{\mathcal{H}_2}^2}$  \\
			&  $-\frac{3 \eta _H^2 \lambda _{\mathcal{H}_2,1}}{4 \pi ^2 m_{\mathcal{H}_2}^2}-\frac{3 \eta _H \eta _{\mathcal{H}_2} \lambda _H^{\text{SM}}}{8 \pi ^2 m_{\mathcal{H}_2}^2}+\frac{3 \eta _H \eta _{\mathcal{H}_2} \lambda _{\mathcal{H}_2,1}}{8 \pi ^2 m_{\mathcal{H}_2}^2}$  \\
			&  $-\frac{13 \eta _H^2 \lambda _{\mathcal{H}_2,2}}{16 \pi ^2 m_{\mathcal{H}_2}^2}+\frac{3 \eta _H \eta _{\mathcal{H}_2} \lambda _{\mathcal{H}_2,2}}{8 \pi ^2 m_{\mathcal{H}_2}^2}-\frac{\lambda _{\mathcal{H}_2,1}^3}{48 \pi ^2 m_{\mathcal{H}_2}^2}$  \\
			&  $\frac{\lambda _H^{\text{SM}} \lambda _{\mathcal{H}_2,2}^2}{96 \pi ^2 m_{\mathcal{H}_2}^2}-\frac{\lambda _{\mathcal{H}_2,1}^2 \lambda _{\mathcal{H}_2,2}}{32 \pi ^2 m_{\mathcal{H}_2}^2}-\frac{\lambda _{\mathcal{H}_2,1} \lambda _{\mathcal{H}_2,2}^2}{32 \pi ^2 m_{\mathcal{H}_2}^2}$  \\
			&  $-\frac{7 \eta _H^2 \lambda _{\mathcal{H}_2,3}}{4 \pi ^2 m_{\mathcal{H}_2}^2}+\frac{\lambda _H^{\text{SM}} \lambda _{\mathcal{H}_2,3}^2}{24 \pi ^2 m_{\mathcal{H}_2}^2}-\frac{\lambda _{\mathcal{H}_2,2}^3}{96 \pi ^2 m_{\mathcal{H}_2}^2}$  \\
			&  $-\frac{\lambda _{\mathcal{H}_2,1} \lambda _{\mathcal{H}_2,3}^2}{8 \pi ^2 m_{\mathcal{H}_2}^2}-\frac{\lambda _{\mathcal{H}_2,2} \lambda _{\mathcal{H}_2,3}^2}{8 \pi ^2 m_{\mathcal{H}_2}^2}$  \\
			\hline
			$Q_{H\square }$  &  $-\frac{g_W^4}{7680 \pi ^2 m_{\mathcal{H}_2}^2}-\frac{3 \eta _H^2}{32 \pi ^2 m_{\mathcal{H}_2}^2}-\frac{\lambda _{\mathcal{H}_2,1}^2}{96 \pi ^2 m_{\mathcal{H}_2}^2}$  \\
			&  $-\frac{\lambda _{\mathcal{H}_2,1} \lambda _{\mathcal{H}_2,2}}{96 \pi ^2 m_{\mathcal{H}_2}^2}+\frac{\lambda _{\mathcal{H}_2,3}^2}{48 \pi ^2 m_{\mathcal{H}_2}^2}$  \\
			\hline
			$Q_{\text{HD}}$  &  $-\frac{g_Y^4}{1920 \pi ^2 m_{\mathcal{H}_2}^2}-\frac{\lambda _{\mathcal{H}_2,2}^2}{96 \pi ^2 m_{\mathcal{H}_2}^2}+\frac{\lambda _{\mathcal{H}_2,3}^2}{24 \pi ^2 m_{\mathcal{H}_2}^2}$  \\
			\hline
			$Q_{\text{HB}}$  &  $\frac{g_Y^2 \lambda _{\mathcal{H}_2,1}}{384 \pi ^2 m_{\mathcal{H}_2}^2}+\frac{g_Y^2 \lambda _{\mathcal{H}_2,2}}{768 \pi ^2 m_{\mathcal{H}_2}^2}$  \\
			\hline
			$Q_{\text{HW}}$  &  $\frac{g_W^2 \lambda _{\mathcal{H}_2,1}}{384 \pi ^2 m_{\mathcal{H}_2}^2}+\frac{g_W^2 \lambda _{\mathcal{H}_2,2}}{768 \pi ^2 m_{\mathcal{H}_2}^2}$  \\
			\hline
			$Q_{\text{HWB}}$  &  $\frac{g_W g_Y \lambda _{\mathcal{H}_2,2}}{384 \pi ^2 m_{\mathcal{H}_2}^2}$  \\
			\hline \rowfont{\color{blue}}
			$Q_{\text{Hl}}{}^{(1)}$  &  $\frac{g_Y^4}{3840 \pi ^2 m_{\mathcal{H}_2}^2}$  \\
			\hline
			$Q_{\text{Hq}}{}^{(1)}$  &  $-\frac{g_Y^4}{11520 \pi ^2 m_{\mathcal{H}_2}^2}$  \\
			\hline \hline
	\end{tabular}}
	\subfloat{
		\begin{tabular}{|*{2}{>{\rowfonttype}c|}}
			\hline \hline
			Dim-6 Ops.&Wilson coefficients\\
			\hline \hline \rowfont{\color{blue}} 
			$Q_{\text{Hd}}$  &  $\frac{g_Y^4}{5760 \pi ^2 m_{\mathcal{H}_2}^2}$  \\
			\hline
			$Q_{\text{He}}$  &  $\frac{g_Y^4}{1920 \pi ^2 m_{\mathcal{H}_2}^2}$  \\
			\hline
			$Q_{\text{Hu}}$  &  $-\frac{g_Y^4}{2880 \pi ^2 m_{\mathcal{H}_2}^2}$  \\
			\hline
			$Q_{\text{Hl}}{}^{(3)}$  &  $-\frac{g_W^4}{1920 \pi ^2 m_{\mathcal{H}_2}^2}$  \\
			\hline
			$Q_{\text{Hq}}{}^{(3)}$  &  $-\frac{g_W^4}{1920 \pi ^2 m_{\mathcal{H}_2}^2}$  \\
			\hline
			$Q_W$  &  $\frac{g_W^3}{5760 \pi ^2 m_{\mathcal{H}_2}^2}$  \\
			\hline
			$Q_{\text{ll}}$  &  $-\frac{g_W^4}{7680 \pi ^2 m_{\mathcal{H}_2}^2}-\frac{g_Y^4}{7680 \pi ^2 m_{\mathcal{H}_2}^2}$  \\
			\hline \rowfont{\color{red}} 
			$Q_{\text{ud}}{}^{(1)}$  &  $\frac{g_Y^4}{4320 \pi ^2 m_{\mathcal{H}_2}^2}$  \\
			\hline
			$Q_{\text{lq}}{}^{(3)}$  &  $-\frac{g_W^4}{3840 \pi ^2 m_{\mathcal{H}_2}^2}$  \\
			\hline
			$Q_{\text{qq}}{}^{(3)}$  &  $-\frac{g_W^4}{7680 \pi ^2 m_{\mathcal{H}_2}^2}$  \\
			\hline
			$Q_{\text{dd}}$  &  $-\frac{g_Y^4}{17280 \pi ^2 m_{\mathcal{H}_2}^2}$  \\
			\hline
			$Q_{\text{ed}}$  &  $-\frac{g_Y^4}{2880 \pi ^2 m_{\mathcal{H}_2}^2}$  \\
			\hline
			$Q_{\text{ee}}$  &  $-\frac{g_Y^4}{1920 \pi ^2 m_{\mathcal{H}_2}^2}$  \\
			\hline
			$Q_{\text{eu}}$  &  $\frac{g_Y^4}{1440 \pi ^2 m_{\mathcal{H}_2}^2}$  \\
			\hline
			$Q_{\text{uu}}$  &  $-\frac{g_Y^4}{4320 \pi ^2 m_{\mathcal{H}_2}^2}$  \\
			\hline
			$Q_{\text{lu}}$  &  $\frac{g_Y^4}{2880 \pi ^2 m_{\mathcal{H}_2}^2}$  \\
			\hline
			$Q_{\text{qe}}$  &  $\frac{g_Y^4}{5760 \pi ^2 m_{\mathcal{H}_2}^2}$  \\
			\hline
			$Q_{\text{ld}}$  &  $-\frac{g_Y^4}{5760 \pi ^2 m_{\mathcal{H}_2}^2}$  \\
			\hline 
			$Q_{\text{qq}}{}^{(1)}$  &  $-\frac{g_Y^4}{69120 \pi ^2 m_{\mathcal{H}_2}^2}$  \\
			\hline
			$Q_{\text{le}}$  &  $-\frac{g_Y^4}{1920 \pi ^2 m_{\mathcal{H}_2}^2}-\frac{3 \lambda _{\mathcal{H}_2} Y_{\mathcal{H}_2}^{(e)}{}^2}{128 \pi ^2 m_{\mathcal{H}_2}^2}-\frac{Y_{\mathcal{H}_2}^{(e)}{}^2}{4 m_{\mathcal{H}_2}^2}$  \\
			\hline
			$Q_{\text{qd}}{}^{(1)}$  &  $\frac{g_Y^4}{17280 \pi ^2 m_{\mathcal{H}_2}^2}-\frac{3 \lambda _{\mathcal{H}_2} Y_{\mathcal{H}_2}^{(d)}{}^2}{128 \pi ^2 m_{\mathcal{H}_2}^2}-\frac{Y_{\mathcal{H}_2}^{(d)}{}^2}{4 m_{\mathcal{H}_2}^2}$  \\
			\hline
			$Q_{\text{qu}}{}^{(1)}$  &  $-\frac{g_Y^4}{8640 \pi ^2 m_{\mathcal{H}_2}^2}-\frac{3 \lambda _{\mathcal{H}_2} Y_{\mathcal{H}_2}^{(u)}{}^2}{128 \pi ^2 m_{\mathcal{H}_2}^2}-\frac{Y_{\mathcal{H}_2}^{(u)}{}^2}{4 m_{\mathcal{H}_2}^2}$  \\
			\hline
			$Q_{\text{quqd}}{}^{(1)}$  &  $-\frac{3 \lambda _{\mathcal{H}_2} Y_{\mathcal{H}_2}^{(d)} Y_{\mathcal{H}_2}^{(u)}}{64 \pi ^2 m_{\mathcal{H}_2}^2}-\frac{Y_{\mathcal{H}_2}^{(d)} Y_{\mathcal{H}_2}^{(u)}}{2 m_{\mathcal{H}_2}^2}$  \\
			\hline
			$Q_{\text{lequ}}{}^{(1)}$  &  $\frac{3 \lambda _{\mathcal{H}_2} Y_{\mathcal{H}_2}^{(e)} Y_{\mathcal{H}_2}^{(u)}}{64 \pi ^2 m_{\mathcal{H}_2}^2}+\frac{Y_{\mathcal{H}_2}^{(e)} Y_{\mathcal{H}_2}^{(u)}}{2 m_{\mathcal{H}_2}^2}$  \\
			\hline
			$Q_{\text{lq}}{}^{(1)}$  &  $\frac{g_Y^4}{11520 \pi ^2 m_{\mathcal{H}_2}^2}$  \\
			\hline
			$Q_{\text{ledq}}$  &  $\frac{3 \lambda _{\mathcal{H}_2} Y_{\mathcal{H}_2}^{(d)} Y_{\mathcal{H}_2}^{(e)}}{64 \pi ^2 m_{\mathcal{H}_2}^2}+\frac{Y_{\mathcal{H}_2}^{(d)} Y_{\mathcal{H}_2}^{(e)}}{2 m_{\mathcal{H}_2}^2}$  \\
			\hline \hline
	\end{tabular}}
\end{table*}

In this work, we consider 16 different BSM scenarios which are extensions of the SM by single scalar representations. This choice is guided by our phenomenological interest and to encapsulate the features of various new physics interactions.~\footnote{The impact of the new physics parameters on renormalisation of the SM parameters are ignored in our analysis.} Colour singlet scalars with higher hypercharges are used to explain the origin of light neutrino masses through higher dimensional operators. The coloured  leptoquark scalars are used to explain icecube data, LHCb observations, etc.~\cite{Dey:2015eaa, Buchmuller:1986zs, Arnold:2013cva, Bauer:2015knc, Bandyopadhyay:2016oif, Davidson_2010, Chen:2008hh}. We chose BSM extensions that encapsulate the features of these heavy scalars which are frequently used in the literature. Even when the SM is extended by multiple heavy scalars, one can immediately get to know which effective operators will emerge after integrating out those heavy fields. Of course there will be additional contributions from the mutual interactions among the heavy fields. We start with the individual complete BSM Lagrangian and integrate out the heavy non-SM scalar multiplets using the automated program Mathematica based package \texttt{CoDEx}~\cite{Bakshi:2018ics}. The BSM scenarios are then expressed in terms of the SM-renormalisable Lagrangian accompanied by the effective operators as given in eq.~\eqref{eq:bsmlagr}.

\begin{figure}[htb]
	\centering
	\subfloat[$\mathcal{C}_{dH}$ - $\mathcal{C}_{eH}$]
	{\includegraphics[width=0.325\textwidth, height=5.8cm]{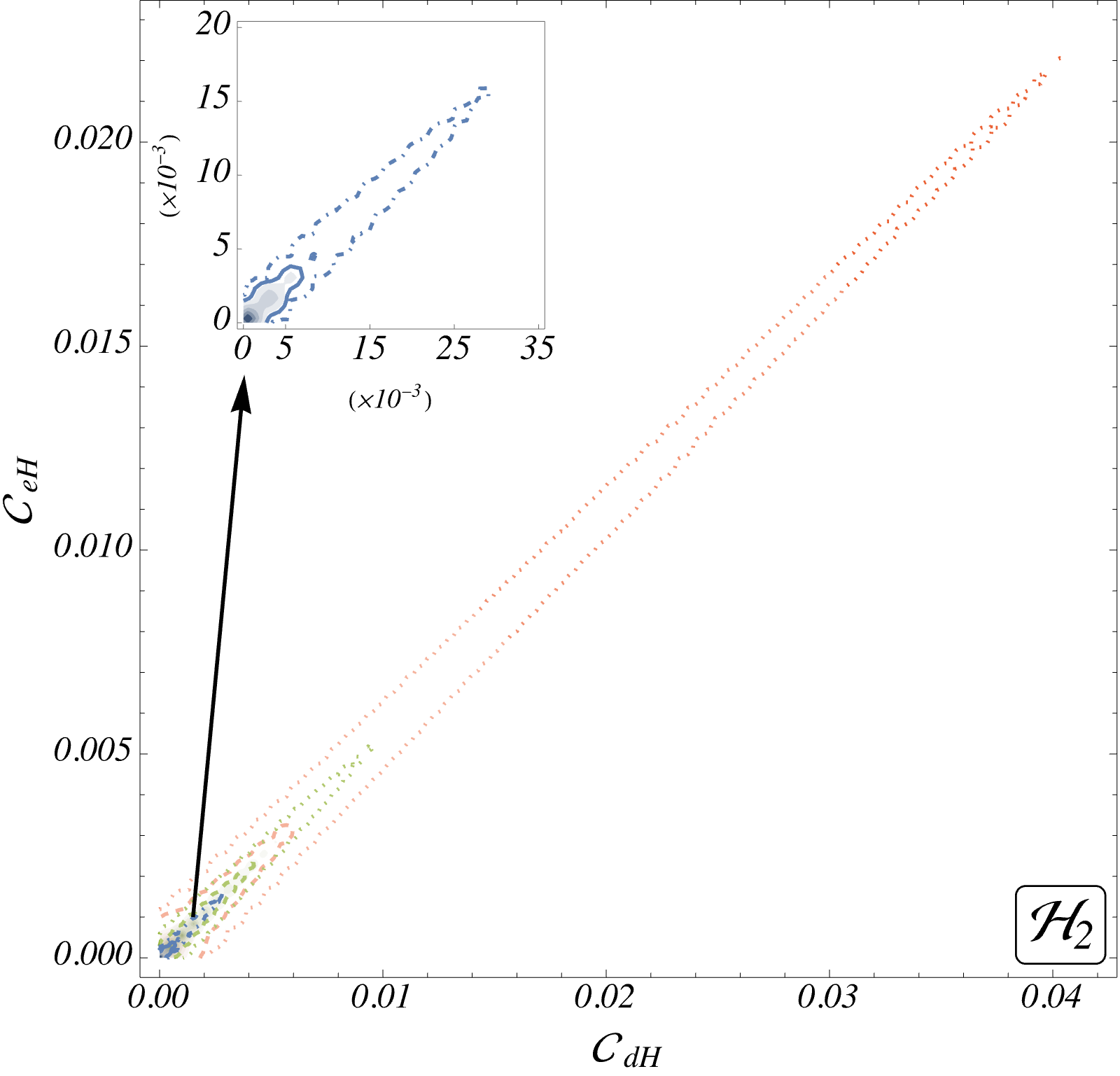}\label{fig:2HDM_CdHvsCeH}}~ 
	\subfloat[$\mathcal{C}_{H}$ - $\mathcal{C}_{HB}$]
	{\includegraphics[width=0.325\textwidth, height=5.8cm]{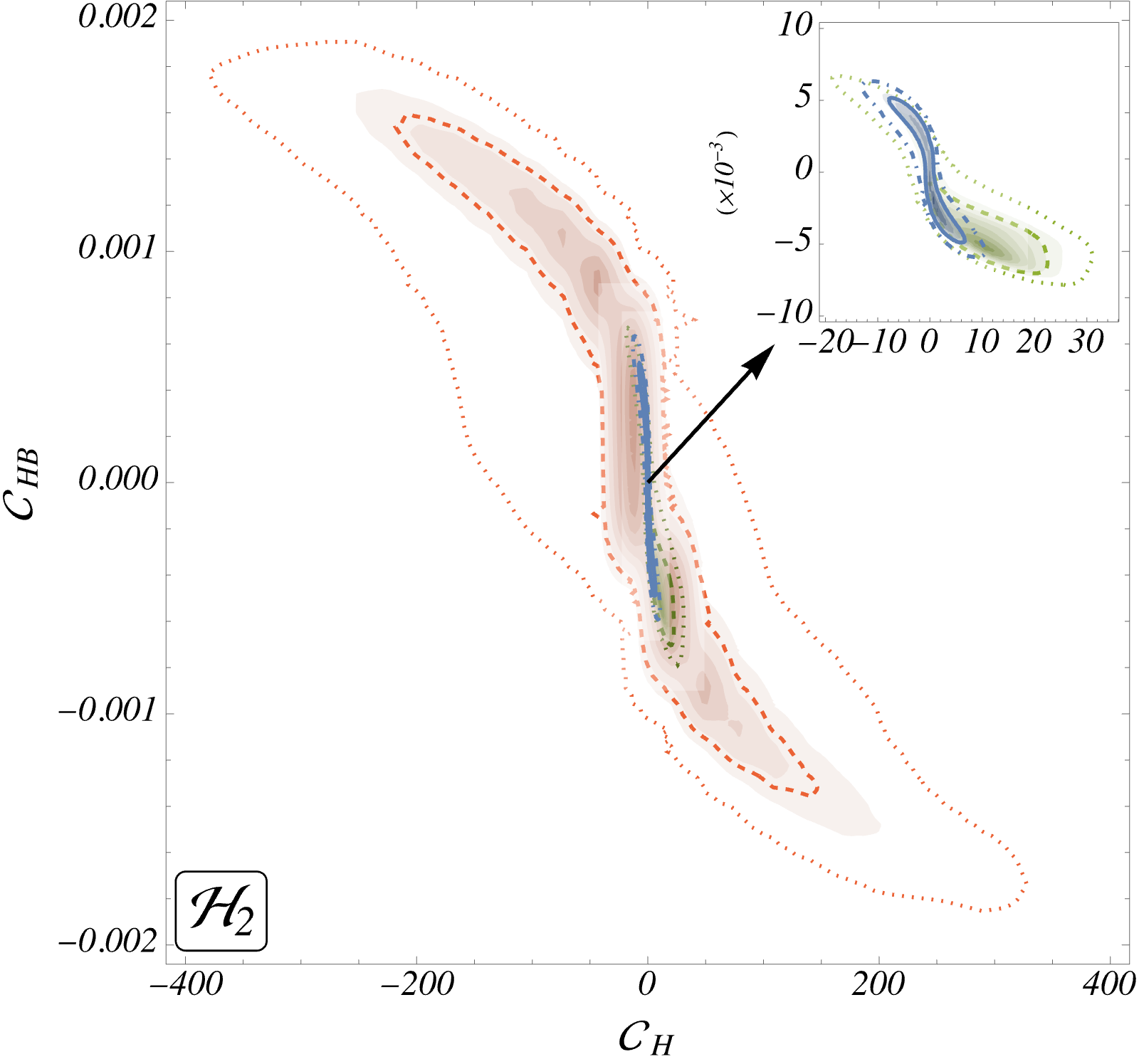}\label{fig:2HDM_CHvsCHB}}~
	\subfloat[$\mathcal{C}_{H\square}$ - $\mathcal{C}_{HD}$]
	{\includegraphics[width=0.325\textwidth, height=5.8cm]{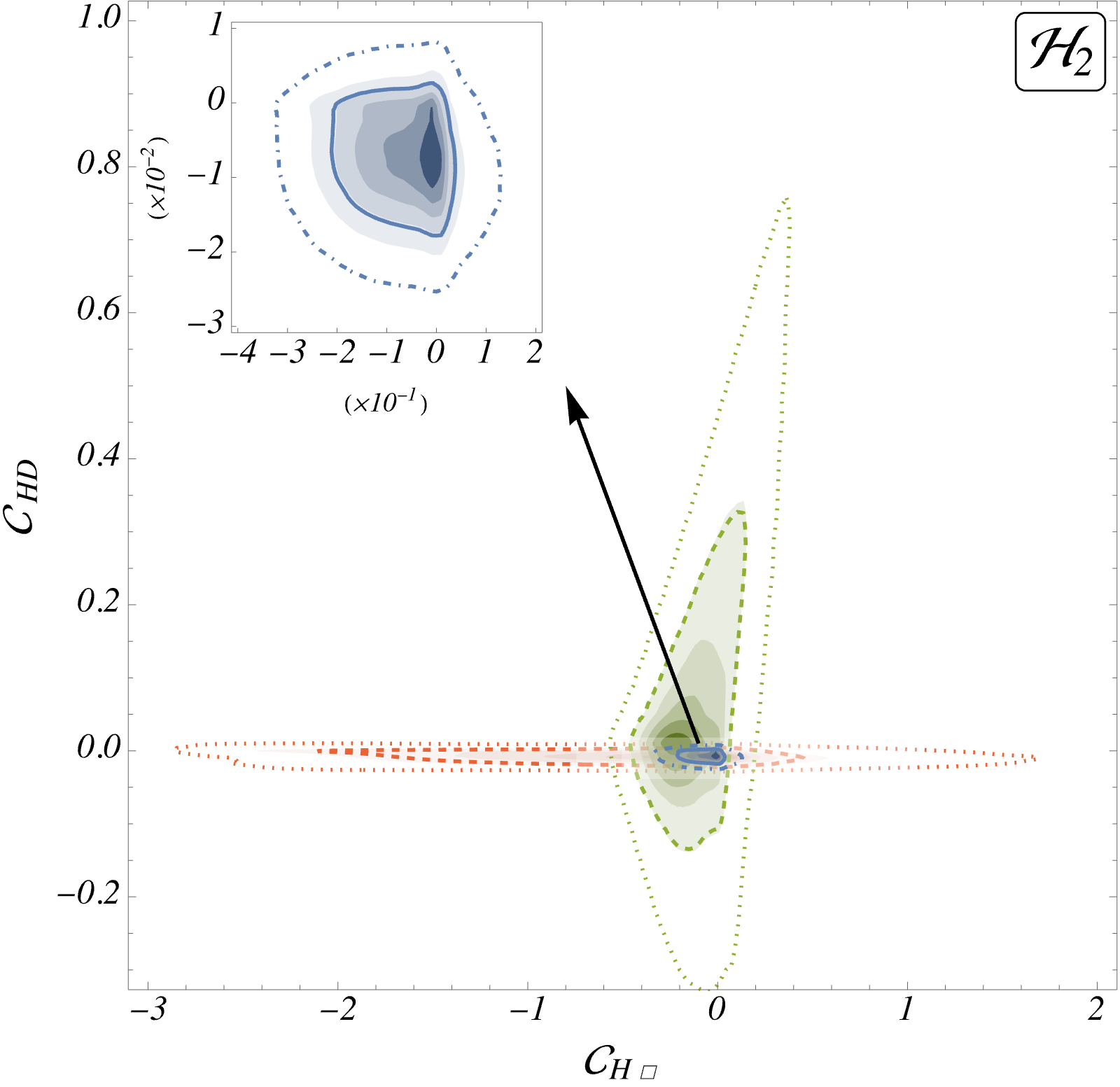}\label{fig:2HDM_CHboxvsCHD}}\\
	\subfloat[$\mathcal{C}_{HW}$ - $\mathcal{C}_{HWB}$]
	{\includegraphics[width=0.325\textwidth, height=5.8cm]{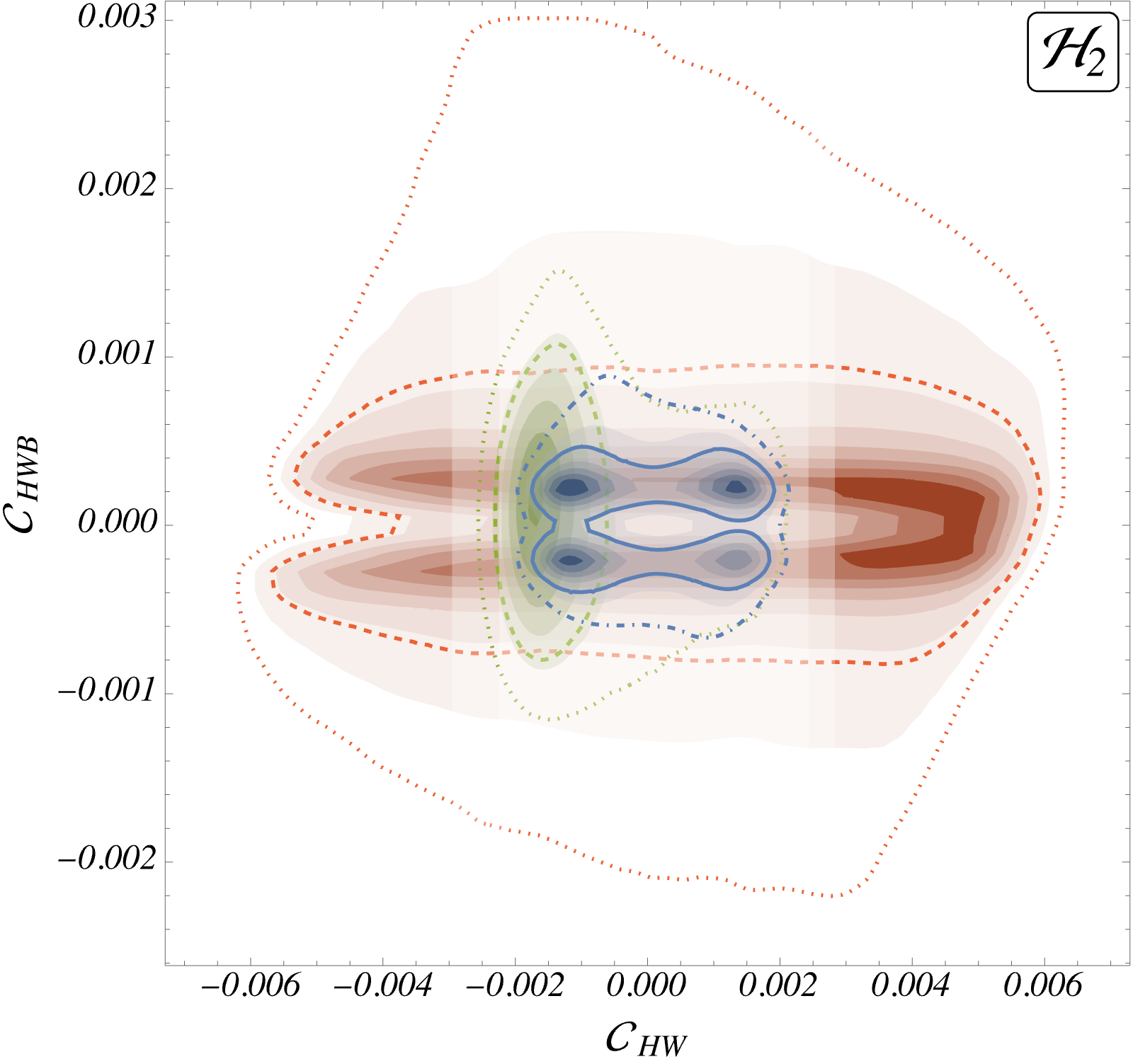}}~
	\subfloat[$\mathcal{C}_{dH}$ - $\mathcal{C}_{uH}$]
	{\includegraphics[width=0.325\textwidth, height=5.8cm]{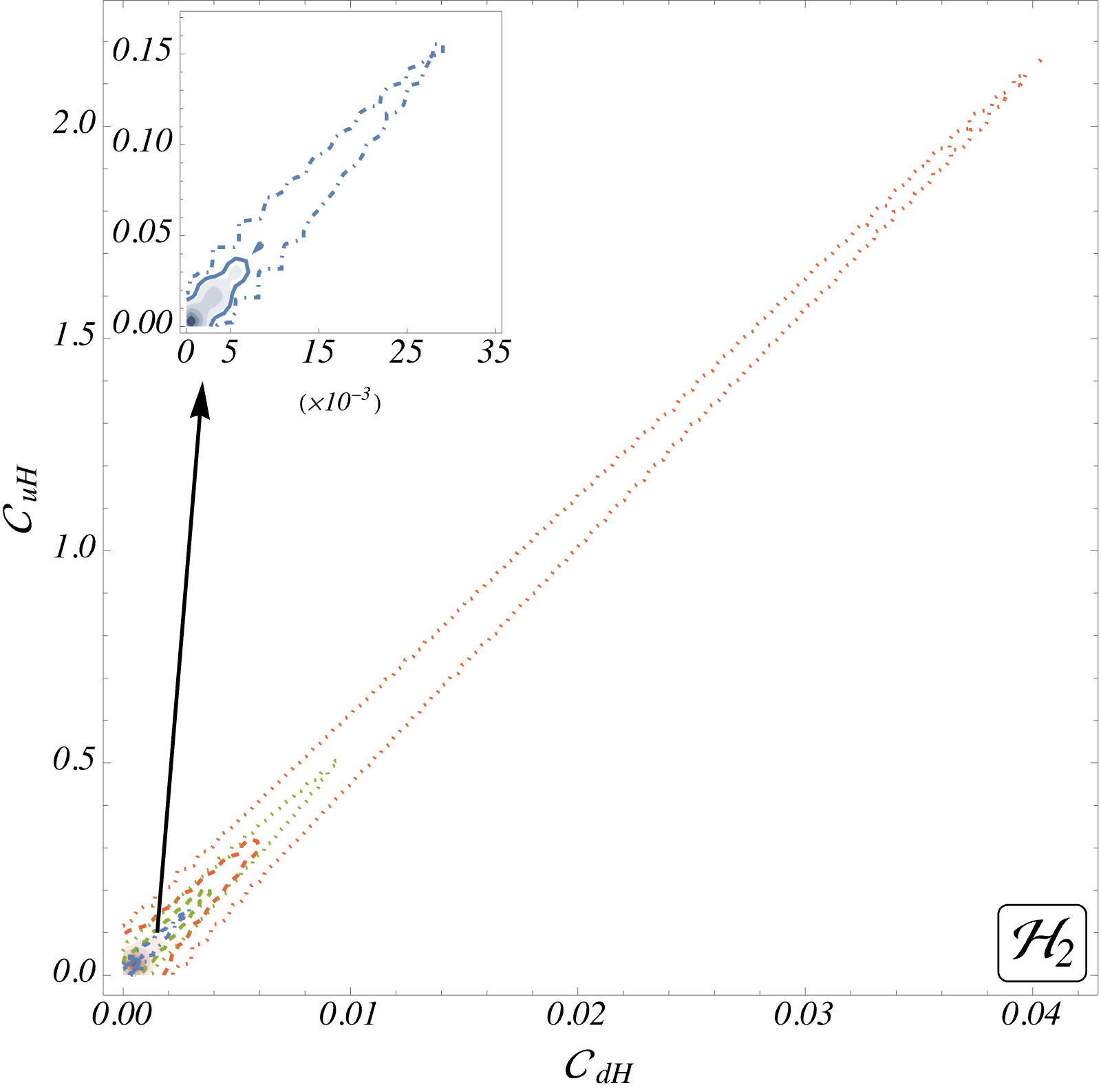}\label{fig:2HDM_CdHvsCuH}}~
	\subfloat[Legend]
	{\includegraphics[width=3.2cm, height=3cm]{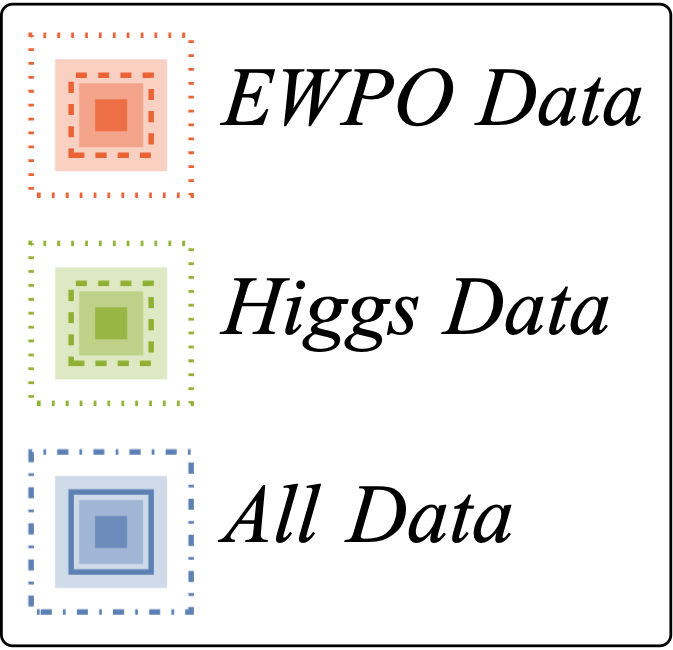}\label{}}
	\caption{\small 
    Two-dimensional posteriors among the relevant WCs induced by $\mathcal{H}_{2}$, listed in black in tab.~\ref{tab:H2}. These regions are obtained from the parameter distributions among $\lambda_{\mathcal{H}_{2},1}$, $\lambda_{\mathcal{H}_{2},2}$ and $\lambda_{\mathcal{H}_{2},3}$ shown in figs.~\ref{fig:2HDM_param12}-\ref{fig:2HDM_param13}. The line contours represent the 68\% and 95\% CIs and the filled contours with changing opacity denote the high-probability regions with decreasing probabilities (darker to lighter). 
	We separate the results from fitting ``EWPO" data only (red), ``Higgs" data only (green) and ``All" data (blue).}
	\label{fig:wc_2HDM}
\end{figure}

We integrate out the heavy BSM fields up to 1-loop and tabulate the complete sets of emerging effective operators and accompanying WCs including the heavy-light mixing contribution from scalars in the loops. 
It is important to note that only those BSMs generate heavy-light mixed WCs in which the heavy field couples to SM fields linearly  \cite{delAguila:2016zcb, Henning:2016lyp, Zhang:2016pja, Ellis:2017jns, DasBakshi:2020pbf}\footnote{A detailed discussion on the heavy-light effective action formulae is given in App.~\ref{subsec:HLaction}.}. This can be visualised by considering one-particle-irreducible 1-loop diagrams where loop propagators are both heavy and light (SM) fields, but external legs are only light (SM) fields.
We further employ the equations of motion, Fierz identities, and integration by parts suitably to ensure the emergence of the exhaustive lists of effective operators up to 1-loop~\footnote{In Ref.~\cite{DasBakshi:2020pbf}, the effective operators are tagged as per their dominant emergence through tree (T), only heavy (HH), and heavy-light (HL) mixing.}. The effective operators are depicted in the Warsaw basis~\cite{Grzadkowski:2010es}. The complete matching results can also be downloaded from the GitHub repository~\cite{Githubresult} where the readers will find the operators and the associated WCs in SILH basis as well for all these BSM scenarios. We require the Wilson coefficients of Warsaw basis effective operators in this analysis, because the observables are parameterised in terms of the SMEFT Warsaw basis operators to perform the global and the model-specific fits. The complete SMEFT dimension-six matching results for some of the SM extensions are already available in literature in Warsaw basis, for example for the real singlet scalar model~\cite{Haisch:2020ahr,Jiang:2018pbd,Cohen:2020fcu}, and scalar leptoquarks~\cite{Gherardi:2020qhc}. However, for the remaining models, to the best of our knowledge, the complete SMEFT dimension-six matching results in the Warsaw basis up to 1-loop including heavy-light mixing are not available yet. A tree-level matching dictionary \cite{deBlas:2017xtg}, and partial results for some models including scalar heavy-light mixing (in SILH basis) \cite{Henning:2014wua,Zhang:2016pja,Ellis:2016enq} are available as well.

\begin{figure}[htb]
	\centering
	\subfloat[$\mathcal{C}_{dH}$ - $\mathcal{C}_{eH}$]
	{\includegraphics[width=0.325\textwidth, height=6.2cm]{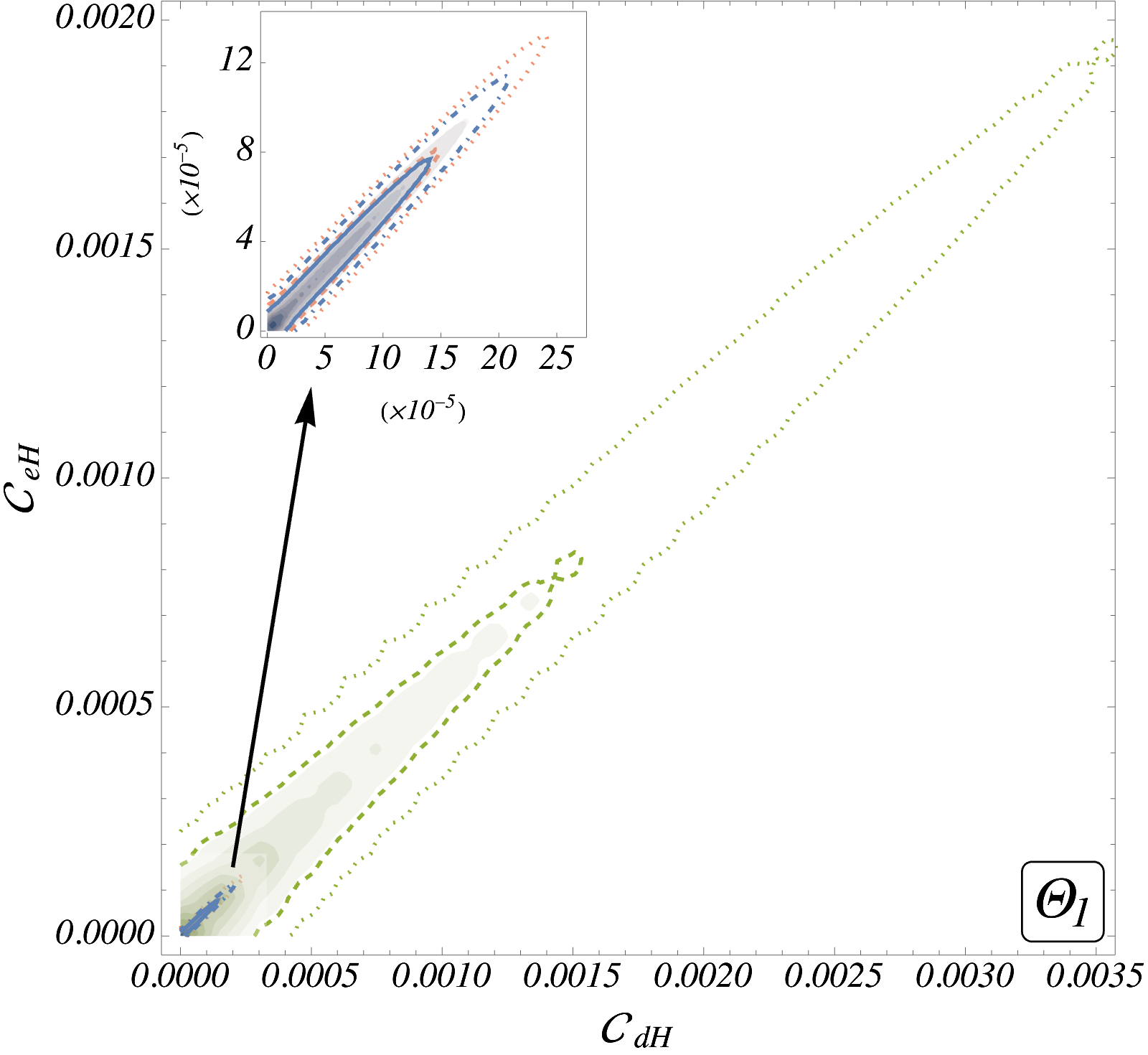}\label{fig:Theta1_CdHvsCeH}}~
	\subfloat[$\mathcal{C}_{H}$ - $\mathcal{C}_{HB}$]
	{\includegraphics[width=0.325\textwidth, height=6.2cm]{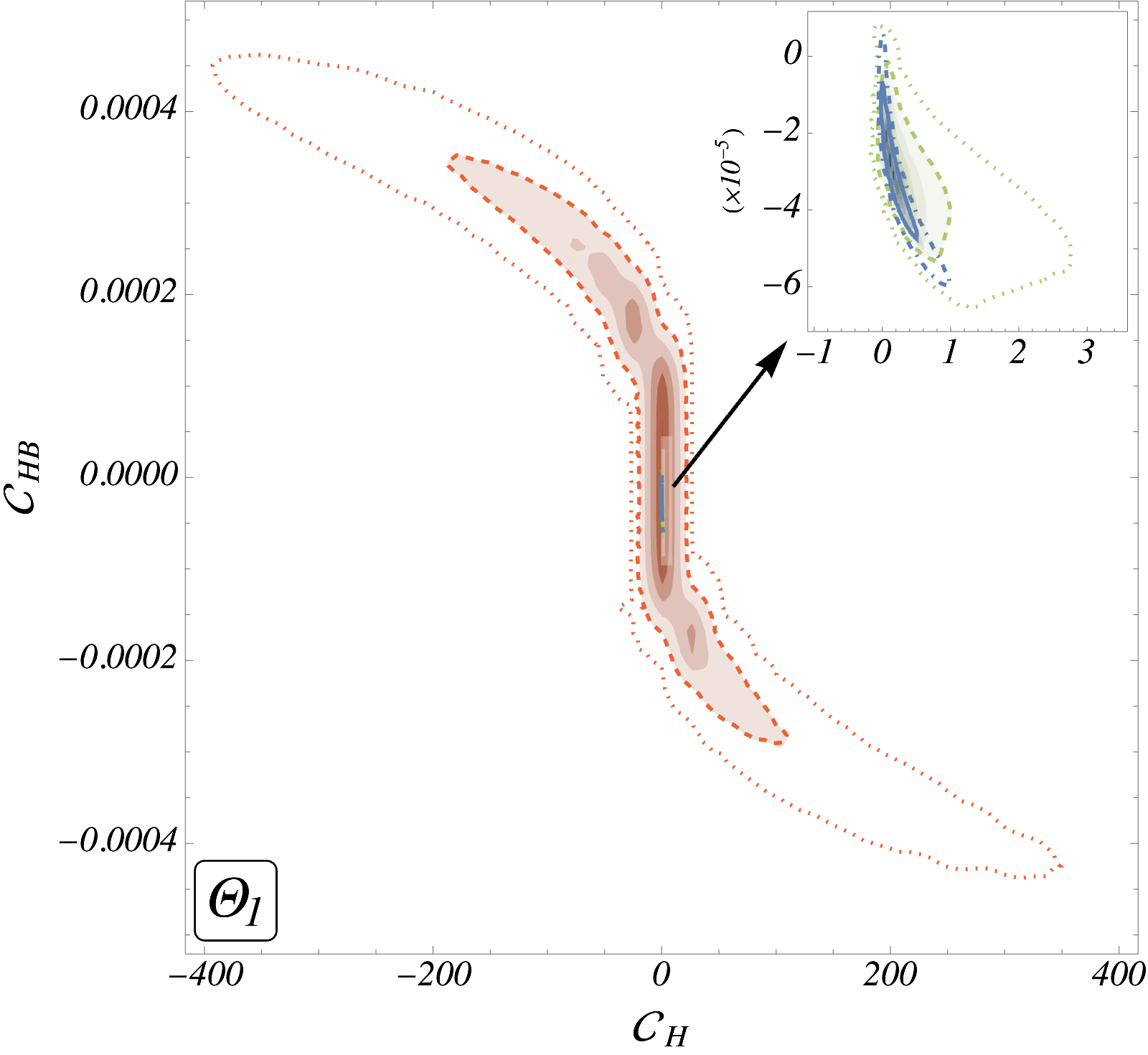}\label{fig:Theta1_CHvsCHB}}~
	\subfloat[$\mathcal{C}_{H\square}$ - $\mathcal{C}_{HD}$]
	{\includegraphics[width=0.325\textwidth, height=6.2cm]{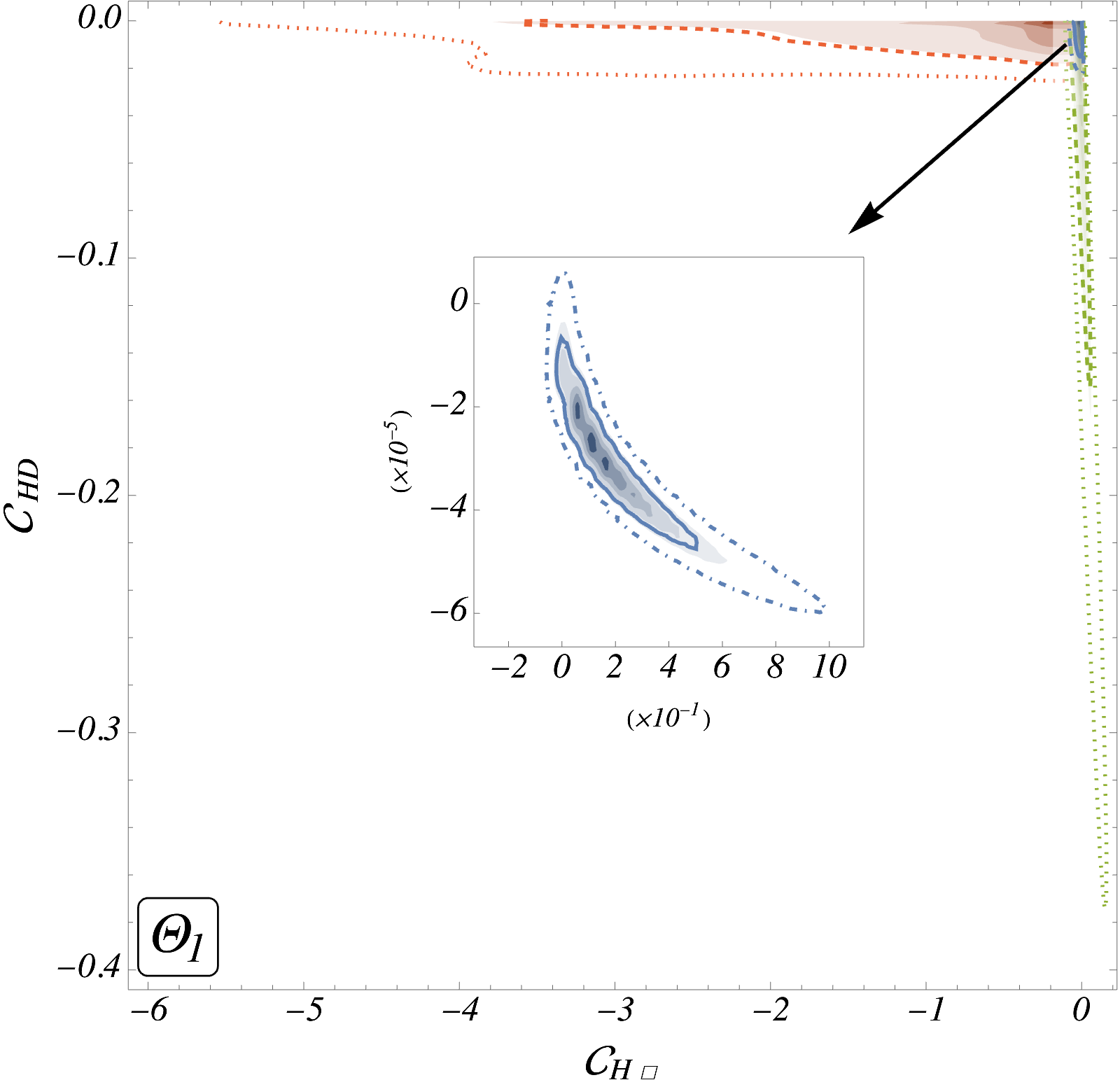}\label{fig:Theta1_CHboxvsCHD}}\\
	\subfloat[$\mathcal{C}_{HW}$ - $\mathcal{C}_{HWB}$]
	{\includegraphics[width=0.325\textwidth, height=6.2cm]{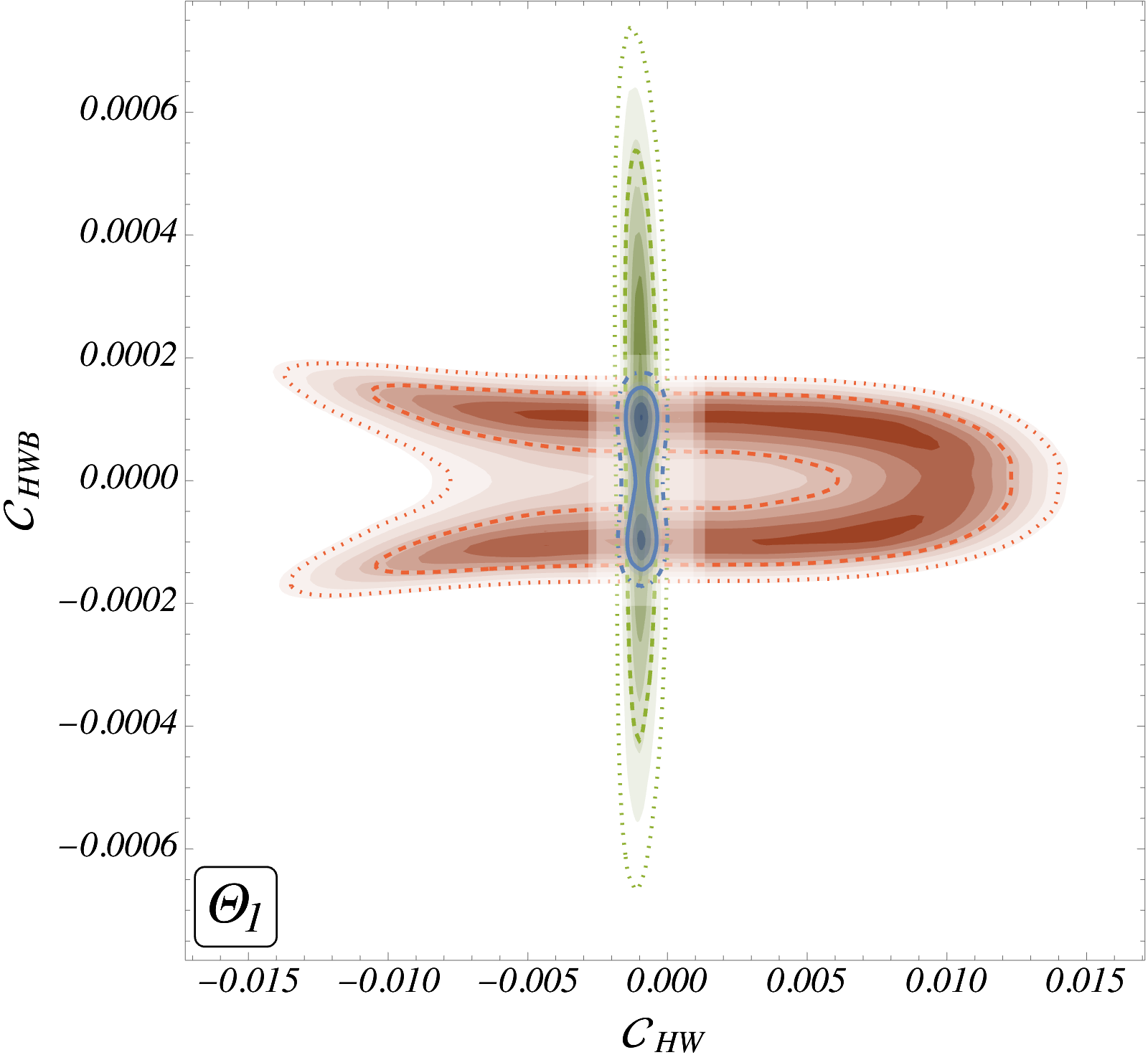}\label{fig:Theta1_CHWvsCHWB}}~
	\subfloat[$\mathcal{C}_{dH}$ - $\mathcal{C}_{uH}$]
	{\includegraphics[width=0.325\textwidth, height=6.2cm]{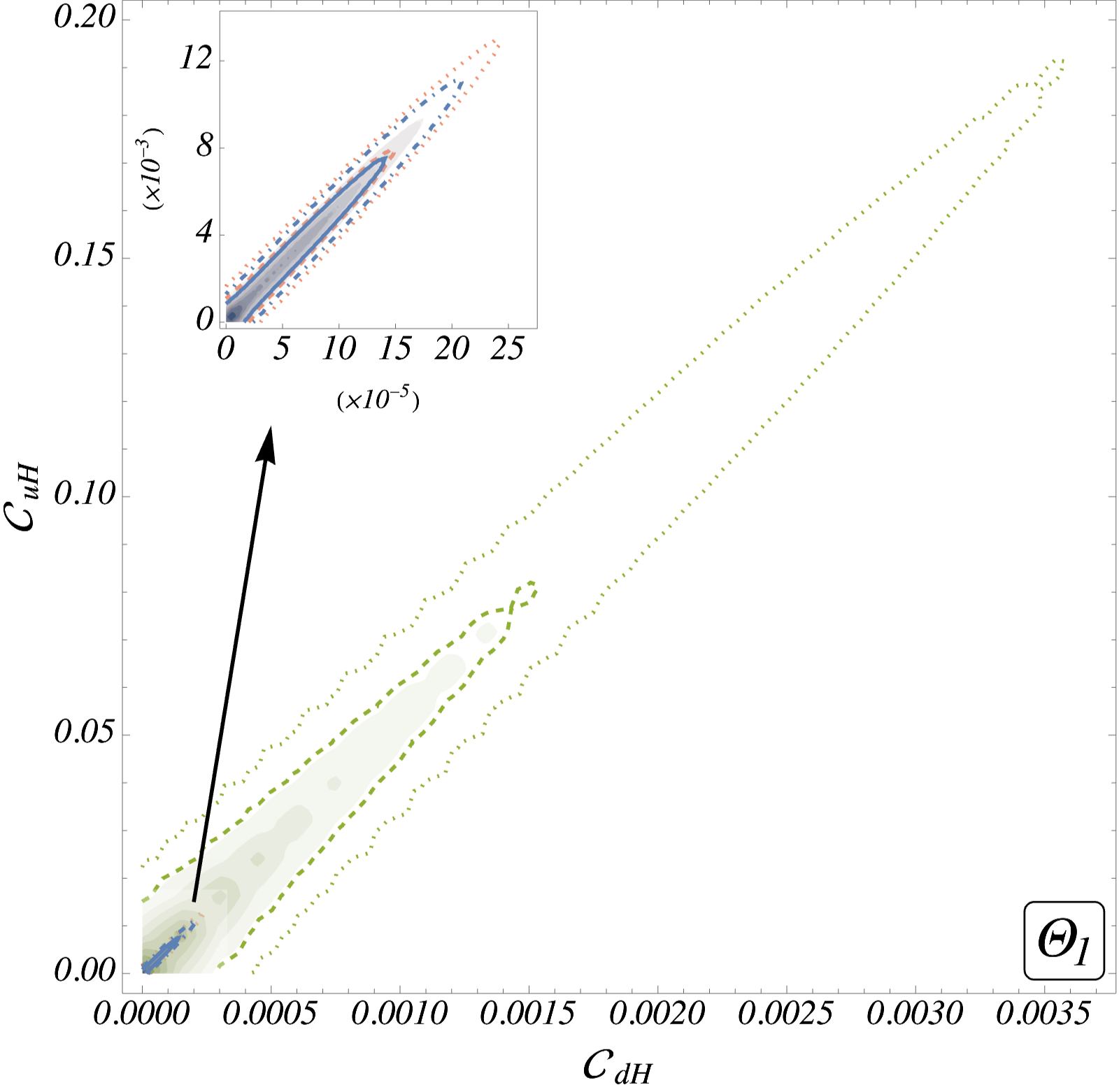}\label{fig:Theta1_CdHvsCuH}}~
	\subfloat[Legend]
	{\includegraphics[width=3.3cm, height=3cm]{figures/wcLegend.png}\label{}}
	\caption{\small Two-dimensional WC posteriors, similar to fig.~\ref{fig:wc_2HDM}, obtained from the parameter distributions of $\eta_{\Theta_1}^{(1)},\; \eta_{\Theta_1}^{(2)}$ for $\Theta_{1}$. }
	\label{fig:wc_Theta1}
\end{figure}

In the following subsections, we tabulate the effective operators generated  and the associated WCs in terms of the model parameters in the Warsaw basis for two representative cases: The two Higgs doublet model, $\mathcal{H}_2$, and a scalar leptoquark, $\Theta_1$. These models encapsulate the individual features of a colour singlet and a non-singlet heavy field, respectively. 
The two models share the exact same set of relevant~\footnote{In our analysis, the operators which are present in our chosen set of observables and are functions of BSM parameters, are relevant. In the result section, we focus on those primarily.} effective operators apart from the additional emergence of $Q_{\text{HG}}$ in the latter model, see tab.~\ref{tab:Theta1}. We tabulate the results for rest of the adopted scenarios in appendix~\ref{sec:remainingmodels}.  

Here, we provide the operators at the scale where they emerge, \textit{i.e.}~the cut-off scale $(\Lambda)$ which we take equal to the mass of heavy BSM fields ($m_{hf}$), in all cases that we discuss. We further assume that the mass of the heavy field $m_{hf} \gg$~EWSB scale. Thus, the heavy fields can be integrated out safely, validating the notion of EFT and all the dimension-6 operators are suppressed by $1/m_{hf}^2$. 
Here, to start with we ignore the running of the effective operators and perform the analysis. Later, in section~\ref{sec:RGE_effects} we for extra Electro-Weak Doublet Scalar(EWDS) scenario, we discuss how the inclusion of running of effective operators leave an impact of parameter space.

In the model-independent analysis in section~\ref{sec:modindep}, we introduce an explicit flavour dependence for the operators of the $\psi^2 \phi^3$ class with the corresponding WCs $Q_{\mu H}$, $Q_{\tau H}$, $Q_{c H}$, $Q_{b H}$, and $Q_{t H}$. As we are using CoDEx results for the model-dependent part of the analysis and CoDEx does not differentiate between flavours (yet), $\mathcal{C}_{\mu H}$ and $\mathcal{C}_{c H}$ are set to be zero for the rest of the analysis and we work with operators consisting of third generation fermions only. For the rest of the paper, the associated WCs will be denoted as $\mathcal{C}_{u H}$, $\mathcal{C}_{d H}$, and $\mathcal{C}_{e H}$.  In the later subsections, we highlight the operators (in red) that do not affect our chosen set of observables and thus are absent from our analysis. The blue coloured operators are functions of SM parameters only and thus, independent of our fit-parameters. Each of the observables can be thought of as a set of effective operators which has been useful to classify different single scalar field extensions of the SM~\cite{DasBakshi:2020pbf}. Relying on that concept, we design the methodology of our analysis to pin down the individual and mutual status of these BSM scenarios in this section.

In section~\ref{sec:modindep}, we obtain the constraints on the set of 23 SMEFT WCs in a model-independent manner, using the observables listed in section~\ref{sec:obs}. In this section, we move a step ahead and connect the relevant observables expressed in terms of the SMEFT dimension-6 operators and their respective WCs to the BSM model parameters. The SMEFT matching results obtained for a particular BSM, mentioned above, allow us to write the WCs in terms of the respective model parameters and $m_{hf}$. Consequently, the bounds on the model parameters of specific BSMs are obtained directly from the relevant experimental data.

The methodology of the statistical analysis is similar to the one discussed in the model-independent part. Fits are performed over the relevant BSM parameters while considering the best-fit values for the SM ones, see eq.~\eqref{eq:SMinputval}. Uniform priors within range $\{-50,50\}$ are chosen for these non-SM parameters and $m_{hf}$ is chosen as to be 1 TeV uniformly in this analysis. The following subsections showcase two example scenarios where the SM is extended by two scalars, $\mathcal{H}_{2}$ and $\Theta_{1}$, based upon the relations among the associated WCs of the emerged effective operators through the model parameters for individual cases.

\subsection{Extra EW Doublet Scalar: $\mathcal{H}_2 \equiv (1_C,2_L,\left.-\frac{1}{2}\right\vert_Y)$}
This model contains an extra isospin-doublet scalar $ (\mathcal{H}_2)$ which is a colour-singlet with hypercharge $ Y=-\frac{1}{2} $. The BSM Lagrangian is given as \cite{Deshpande:1977rw, Henning:2014wua, Nie_1999, Branchina:2018qlf,Pilaftsis:1999qt,Gunion:2002zf},
{\small
	\begin{align}
		\label{2HDMLag}
		\mathcal{L}_{\mathcal{H}_2} & = \mathcal{L}_{_\text{SM}}^{d\leq 4} \; + |\mathcal{D}_{\mu} \mathcal{H}_{2} |^{2} - m_{\mathcal{H}_{2}}^{2} |\mathcal{H}_{2}|^{2} - \ \frac{\lambda_{\mathcal{H}_{2}}}{4} |\mathcal{H}_{2}|^{4} - ( \eta_{H} |\widetilde{H} |^{2} + \eta_{\mathcal{H}_{2}} |\mathcal{H}_{2}|^{2}) (\widetilde{H}^{\dagger} \mathcal{H}_{2} + \mathcal{H}_{2}^{\dagger} \widetilde{H}) \nonumber\\
		& - \lambda_{\mathcal{H}_{2},1} |\widetilde{H}|^{2} |\mathcal{H}_{2}|^{2} - \lambda_{\mathcal{H}_{2},2} |\widetilde{H}^{\dagger} \mathcal{H}_{2} |^{2} - \lambda_{\mathcal{H}_{2},3} \left[ (\widetilde{H}^{\dagger} \mathcal{H}_{2})^{2} + (\mathcal{H}_{2}^{\dagger} \widetilde{H})^{2} \right]\nonumber\\
		& -\left\lbrace Y^{(e)}_{\mathcal{H}_2} \overline{l}_L \, \widetilde{\mathcal{H}}_2 \, e_R + Y^{(u)}_{\mathcal{H}_2} \overline{q}_L \, \mathcal{H}_2 \, u_R + Y^{(d)}_{\mathcal{H}_2} \overline{q}_L\, \widetilde{\mathcal{H}}_2 \, d_R + \text{h.c.} \right\rbrace.
	\end{align}
}	
Here, $m_{\mathcal{H}_2}$ is the mass of the heavy field and serves as the cut-off scale. We assume that the heavy Higgs doublet is decoupled to the SM one, and they do not mix \cite{Gunion:2002zf}. This model contains nine BSM parameters $\eta_{H},\; \eta_{\mathcal{H}_{2}},\;  \lambda_{\mathcal{H}_{2}}$, $\lambda_{\mathcal{H}_{2},1},\; \lambda_{\mathcal{H}_{2},2},\;  \lambda_{\mathcal{H}_{2},3}, \; Y^{(u)}_{\mathcal{H}_2}$,\; $Y^{(d)}_{\mathcal{H}_2}$,\; $Y^{(e)}_{\mathcal{H}_2}$, and the WCs are functions of all these parameters along with the SM ones, see tab.~\ref{tab:H2}. We assume that the BSM Yukawas couple only to the third generation of fermions. In the decoupling limit, these Yukawas do not mix with the SM ones \cite{Gunion:2002zf}. We further note that the WCs of the relevant operators $Q_\text{uH}$, $Q_\text{dH}$, $Q_\text{eH}$, $Q_\text{H}$, $Q_\text{HW}$, $Q_\text{HB}$, $Q_\text{HWB}$,  $Q_{\text{H}\square}$, and $Q_{HD}$ contain all the BSM parameters. In our numerical analysis, we choose to work with a $Z_2$-invariant ($H \rightarrow H$  and $\mathcal{H}_{2} \rightarrow -\mathcal{H}_{2}$) BSM Lagrangian. Thus the quartic couplings $\eta_{H},\; \eta_{\mathcal{H}_{2}}$ and the Yukawa couplings $Y^{(u)}_{\mathcal{H}_2}$,\; $Y^{(d)}_{\mathcal{H}_2}$,\; $Y^{(e)}_{\mathcal{H}_2}$ couplings do not appear in our analysis, where the rest of the three parameters get constrained by our chosen experimental data set. Note that though $\lambda_{\mathcal{H}_{2}}$ corresponds to $Z_2$ invariant term, but it still cannot be constrained as it appears in the WCs with the $Z_2$ violating couplings, always. Thus, to constrain these {\it unconstrained} couplings of the EWDS model, one needs to look for observables beyond the chosen ones in this work that get corrections from the following operators: $Q_{\text{le}}$, $Q_{\text{qd}}{}^{(1)}$, $Q_{\text{qu}}{}^{(1)}$,  $Q_{\text{qu}}{}^{(1)}$,	$Q_{\text{quqd}}{}^{(1)}$, $Q_{\text{lequ}}{}^{(1)}$, $Q_{\text{ledq}}$, see tab.~\ref{tab:H2}. These additional observables will be helpful to constrain $\lambda_{\mathcal{H}_{2}}$ even for the $Z_2$ symmetric Lagrangian. We need to keep in mind that the choice of the new set of observables is guided by the structures of the unconstrained WCs of this particular model only. 

\subsubsection*{Constraints on the model parameters}

Using the relations listed in tab.~\ref{tab:H2}, we obtain the constraints on the BSM parameters directly from the experimental data. The relevant BSM fit parameters for this model are $\lambda_{\mathcal{H}_{2},1},\; \lambda_{\mathcal{H}_{2},2}$ and $\lambda_{\mathcal{H}_{2},3} $ with the mass of the heavy doublet, $m_{hf}$ (cut-off scale), set to 1 TeV. The list of the dimension-6 operators coloured in black and blue in tab.~\ref{tab:H2} are replaced by the corresponding WCs in the SMEFT parameterisation of the observables. The WCs for the operators in blue are functions of only SM parameters (inputs given in eq.~\eqref{eq:SMinputval}). Those for operators in black are functions of the relevant BSM parameters and are thus relevant in constraining them. Uniform distributions within the range $\{-50,50\}$ are chosen as priors for these BSM couplings. 

Using the samples from the un-normalised posteriors, we show the correlations between various BSM parameters in fig.~\ref{fig:model_param_plots} as high-probability contours of two dimensional marginal posteriors. Constant-probability-contours enclose respectively $68\%$ (black solid, red/blue dashed) and $95\%$ (black dot-dashed, red/blue dotted) credible regions. We also show coloured regions with variable-density-contour-shading (black/red/blue) pointing to regions of high-probability. These regions are significant to adjudge the different high-probability regions in the allowed parameter space. This is evident from the posteriors obtained from the ``All" measurements fit, which contain more than one high-probability region. For instance, in fig.~\ref{fig:2HDM_param12}, there are four different high-probability regions within the $68\%$ credible region enclosed by the solid black constant-probability-contour.

In order to demonstrate the constraining power of different datasets, in the top row of the fig.~\ref{fig:model_param_plots}, the posterior distributions of these three parameters obtained are shown from ``All" (black), ``Higgs" (red) and ``EWPO" (blue) sets of experimental measurements. While the constraints from Higgs data are overall a bit stronger, EWPO data add orthogonal information, leading to significantly tightened bounds when combining the data from both sectors. In fig.~\ref{fig:2HDM_param12}, the bound on $\lambda_{\mathcal{H}_{2},1}$ from EWPO only is very relaxed as compared to the others. This happens primarily because in case of EWPO, $\lambda_{\mathcal{H}_{2},1}$ gets constrained through $Q_H$ only whose contribution appears at two-loop in the SMEFT parameterisation of $m_{W}$ and is very small. For other datasets on the other hand, $\lambda_{\mathcal{H}_{2},1}$ gets bounded from other operators $Q_{H\square}$, $Q_{HB}$, $Q_{HW}$ besides $Q_H$. The corresponding WCs receive strong constraints from Higgs signal strengths and in particular from di-Higgs data, as depicted by the corresponding parameter spaces.

\subsubsection*{Model-Dependent constraints on the WCs }

As mentioned in the beginning of this section, we take our model-independent analysis in sec.~\ref{sec:modindep} to the next step, by determining the allowed WC spaces for specific models, using their parameter-posteriors and the WC matching results obtained after integrating out the heavy BSM particle. In this case, the matched WCs are functions of BSM parameters and dependent on one another. The WC-spaces obtained in this way are directed from the constraints of the relevant model parameters and thus are termed as ``model-dependent". Using the large MCMC samples generated from the model-parameter-posterior, we generate the multi-variate distributions of those WCs (now functions of model parameters). We show these distributions in the WC-space with the help of marginal-posteriors of two WCs at a time.

For $\mathcal{H}_{2}$, we generate the distributions of the WCs (black) using relations (obtained after matching) from tab.~\ref{tab:H2} expressed in terms of parameters: $\lambda_{\mathcal{H}_{2},1},\; \lambda_{\mathcal{H}_{2},2}$ and $\lambda_{\mathcal{H}_{2},3}$, and propagating the model-parameter-posteriors. As before, we demonstrate the relative effects of different datasets named ``All", ``EWPO", and ``Higgs" separately. Instead of showing all the possible 2D-marginal contour-plots for the WCs, we choose to show a few sample plots in fig.~\ref{fig:wc_2HDM}.

Solid blue (dashed red/green) constant-probability-contours enclose the $68\%$ and dot-dashed blue (dotted red/green) ones enclose the $95\%$ credible regions, respectively. Coloured regions (blue/red/green) with variable density depict the high probability regions. In some cases, the parameter spaces obtained for the three datasets differ from one another by order(s) of magnitude. For instance, the bluish regions corresponding to ``All" measurements are imperceptibly tiny in comparison to those corresponding to both the ``EWPO" (red) and ``Higgs" (green) datasets. For ease of viewing, we have magnified such regions and shown them as insets.    

As mentioned earlier, these obtained WC-spaces are related by non-linear relations of the model parameters. This is explicitly illustrated from the WC-expressions for $\mathcal{C}_{dH}$, $\mathcal{C}_{eH}$ and $\mathcal{C}_{uH}$ in tab.~\ref{tab:H2} which, after taking the contributions from the relevant BSM parameters, turn out to be positive definite. As a result, the corresponding parameter spaces shown in figs.~\ref{fig:2HDM_CdHvsCeH} and \ref{fig:2HDM_CdHvsCuH} are delimited only in the positive quadrant.

\subsection{Scalar Leptoquark: $\Theta_1\equiv (3_C,2_L,\left.\frac{1}{6}\right\vert_Y)$}
In this model, we extend the SM by a colour-triplet isospin-doublet scalar ($\Theta_1$) with hypercharge $ Y = \frac{1}{6} $. We consider the BSM Lagrangian \cite{Buchmuller:1986zs,Arnold:2013cva},
\begin{align}\label{eq:theta1lag}
	\mathcal{L}_{\Theta_1} & = \mathcal{L}_{_\text{SM}}^{d\leq 4} \;+  \left(D_{\mu} \Theta_1\right)^\dagger \, \left(D^{\mu} \Theta_1\right) - m_{{\Theta_1}}^2 \, \Theta_1^\dagger \Theta_1- \eta_{_{\Theta_1}}^{(1)} H^\dagger H \, \Theta_1^\dagger \Theta_1- \eta_{_{\Theta_1}}^{(2)} \left(H^\dagger \sigma^i H\right) \, \left(\Theta_1^\dagger \sigma^i \Theta_1\right)\nonumber\\
	&-\lambda_{\Theta_1}^{(1)} \left(\Theta_1^\dagger \Theta_1\right)^2-\lambda_{\Theta_1}^{(2)} \left(\Theta_1^\dagger \sigma^i \Theta_1\right)^2+\left\lbrace y_{\Theta_1} \Theta_1^\alpha {\overline{d}_R^\alpha} i \sigma^2 l_L  +\text{h.c.}\right\rbrace.
\end{align}
Here, $m_{\Theta_1}$ is the mass of the heavy field, \textit{i.e.}~the cut-off scale appears in the WCs.  This model contains five BSM parameters, $\eta_{\Theta_1}^{(1)},\; \eta_{\Theta_1}^{(2)},\;  \lambda_{\Theta_1}^{(1)},\; \lambda_{\Theta_1}^{(2)}$, and $y_{\Theta_1}$, and the WCs are functions of these parameters on top of the SM ones, see tab.~\ref{tab:Theta1}. In the BSM Yukawa coupling, we assume that $y_{\Theta_1}$ couples to third generation fermions only. As mentioned earlier, not all the emerged operators affect our selected observables, only $Q_{\text{uH}}$, $Q_{\text{dH}}$, $Q_\text{eH}$, $Q_\text{H}$, $Q_\text{HW}$, $Q_\text{HB}$, $ Q_\text{HWB}$, $Q_\text{HG}$, $Q_{\text{H}\square}$, and $Q_\text{HD}$ are the relevant ones for our analysis. Thus out of these five BSM parameters, only $\eta_{\Theta_1}^{(1)}$, and $\eta_{\Theta_1}^{(2)}$ get constrained by the experimental data while the others play no role. To constrain them, we need to add more observables that get contributions from the $Q_{\text{ld}}$ operators, see tab.~\ref{tab:Theta1}. Note that the unconstrained operator, that are functions of the BSM parameters,  sets belonging to these example models are mutually orthogonal, the requirement of additional observables for these two scenarios also do not overlap. Thus, to constrain all the BSM parameters~\footnote{The corrections to the low-energy observables including the electric dipole moments require an enhancement in the SMEFT operator list to include multiple generations of fermions (see Refs.~\cite{Falkowski:2017pss, Dekens:2018bci}), which is beyond the scope of this paper. In this work, we concern ourselves only with  flavour diagonal operators. We leave the flavour off-diagonal scenario as a future project.}, one may have to look for some of the observables for individual models case by case. 

\subsubsection*{Constraints on the model parameters $\eta_{\Theta_1}^{(1)},\; \eta_{\Theta_1}^{(2)}$}

With uniform priors within the range $\{-50,50\}$ and following the same methodology as mentioned for $\mathcal{H}_{2}$, we showcase the parameter-space of the two BSM couplings ($\eta_{\Theta_1}^{(1)}$ and  $\eta_{\Theta_1}^{(2)}$) for model $\Theta_{1}$ as two-dimensional marginalised posteriors in fig.~\ref{fig:Theta1_param}, for the same datasets as before: ``EWPO" (blue), ``Higgs" (red), and ``All" (black). The figure shows that the ``EWPO" bound on $\eta_{\Theta_1}^{(1)}$ is the most relaxed, owing to the loop-order contribution to $Q_{H}$, whereas those allowed by ``Higgs" and ``All" datasets are of almost same order and have negative limits. This is because apart from $Q_{H}$, there are other operators like $Q_{HB}$, $Q_{H\square}$, $Q_{HG}$ and  $Q_{HW}$ which give relatively stronger contributions to other two datasets. In contrast, the constraints on $\eta_{\Theta_1}^{(2)}$ are the weakest from ``Higgs'' data and are of similar range for ``EWPO'' and ``All''. This is a consequence of the strong contribution from $Q_{HWB}$ and $Q_{HD}$ to ``EWPO'', which is also evident for ``All' measurements.

\subsubsection*{Model-Dependent constraints on the WCs }

In the next part of the analysis, similar to the case of $\mathcal{H}_{2}$, distributions are generated for the 10 WCs of the model $\Theta_1$ (expressions in black in tab.~\ref{tab:Theta1}) for the three different datasets. Similar to fig.~\ref{fig:wc_2HDM}, some chosen two-dimensional marginal distributions are shown in fig.~\ref{fig:wc_Theta1} for $\Theta_{1}$. As already noted, these WC-distributions are generated from the model-parameter-posteriors (i.e.~$\eta_{\Theta_1}^{(1)}$ and  $\eta_{\Theta_1}^{(2)}$), using the expressions listed in tab.~\ref{tab:Theta1}. The WCs $\mathcal{C}_{eH}$, $\mathcal{C}_{dH}$, and $\mathcal{C}_{uH}$, representing Yukawa-type interactions, are functions of the squared power of $\eta_{\Theta_1}^{(2)}$. As a result, these WCs will only take positive values when determined in the $\Theta_1$ model. This is clearly shown in 2D marginalised WC-distributions in figs.~\ref{fig:Theta1_CdHvsCeH} and \ref{fig:Theta1_CdHvsCuH}. The opposite behaviour is visible in the WC-space of $\mathcal{C}_{HD}$ in the fig.~\ref{fig:Theta1_CHboxvsCHD} which yields negative limits. From fig.~\ref{fig:Theta1_param}, it is evident that the constraints for  $\eta_{\Theta_1}^{(1)}$ for ``Higgs" and ``All" datasets have negative limits. This leads to the negative bounds on $Q_{HW}$ which is a linear function of  $\eta_{\Theta_1}^{(1)}$, as depicted in fig.~\ref{fig:Theta1_CHWvsCHWB} for the corresponding datasets.\\

Order-of-magnitude variations in the size of the WC-spaces between models, as shown in figs.~\ref{fig:wc_2HDM} and \ref{fig:wc_Theta1}, point to the significance of the SMEFT matching expressions as well as the BSM parameter-spaces in determining model-dependent WC-constraints.

\begin{table*}[h!]
	\caption{\small Warsaw basis  effective operators and the associated WCs that emerge after integrating-out the heavy field $\Theta_1$: (3,2,$\frac{1}{6}$). See caption of tab.~\ref{tab:H2} for colour coding.}
	\label{tab:Theta1}
	\centering
	\renewcommand{\arraystretch}{1.8}
	\subfloat{
		\begin{tabular}{|*{2}{>{\rowfonttype}c|}}%{|c|c|}
			\hline \hline
			Dim-6 Ops.&Wilson coefficients\\
			\hline \hline
			$Q_{\text{HB}}$  &  $\frac{g_Y^2 \eta _{\Theta _1}^{(1)}}{1152 \pi ^2 m_{\Theta _1}^2}$  \\
			\hline
			$Q_{H\square }$  &  $-\frac{g_W^4}{2560 \pi ^2 m_{\Theta _1}^2}-\frac{\eta _{\Theta _1}^{(1)}{}^2}{32 \pi ^2 m_{\Theta _1}^2}+\frac{\eta _{\Theta _1}^{(2)}{}^2}{512 \pi ^2 m_{\Theta _1}^2}$  \\
			\hline
			$Q_{\text{HD}}$  &  $-\frac{g_Y^4}{5760 \pi ^2 m_{\Theta _1}^2}-\frac{\eta _{\Theta _1}^{(2)}{}^2}{128 \pi ^2 m_{\Theta _1}^2}$  \\
			\hline
			$Q_{\text{HG}}$  &  $\frac{g_S^2 \eta _{\Theta _1}^{(1)}}{192 \pi ^2 m_{\Theta _1}^2}$  \\
			\hline
			$Q_{\text{HW}}$  &  $\frac{g_W^2 \eta _{\Theta _1}^{(1)}}{128 \pi ^2 m_{\Theta _1}^2}$  \\
			\hline
			$Q_{\text{HWB}}$  &  $\frac{g_W g_Y \eta _{\Theta _1}^{(2)}}{768 \pi ^2 m_{\Theta _1}^2}$  \\
			\hline
			$Q_{\text{uH}}$  &  $\frac{\eta _{\Theta _1}^{(2)}{}^2 Y_u^{\text{SM}}}{256 \pi ^2 m_{\Theta _1}^2}$  \\
			\hline
			$Q_{\text{dH}}$  &  $\frac{\eta _{\Theta _1}^{(2)}{}^2 Y_d^{\text{SM}}}{256 \pi ^2 m_{\Theta _1}^2}$  \\
			\hline
			$Q_{\text{eH}}$  &  $\frac{\eta _{\Theta _1}^{(2)}{}^2 Y_e^{\text{SM}}}{256 \pi ^2 m_{\Theta _1}^2}$  \\
			\hline
			$Q_H$  &  $-\frac{\eta _{\Theta _1}^{(1)}{}^3}{16 \pi ^2 m_{\Theta _1}^2}-\frac{3 \eta _{\Theta _1}^{(1)} \eta _{\Theta _1}^{(2)}{}^2}{256 \pi ^2 m_{\Theta _1}^2}+\frac{\eta _{\Theta _1}^{(2)}{}^2 \lambda _H^{\text{SM}}}{128 \pi ^2 m_{\Theta _1}^2}$  \\
			\hline \rowfont{\color{blue}}
			$Q_{\text{ll}}$  &  $-\frac{g_W^4}{2560 \pi ^2 m_{\Theta _1}^2}-\frac{g_Y^4}{23040 \pi ^2 m_{\Theta _1}^2}$  \\
			\hline
			$Q_{\text{Hl}}{}^{(1)}$  &  $\frac{g_Y^4}{11520 \pi ^2 m_{\Theta _1}^2}$  \\
			\hline
			$Q_{\text{Hq}}{}^{(1)}$  &  $-\frac{g_Y^4}{34560 \pi ^2 m_{\Theta _1}^2}$  \\
			\hline
			$Q_{\text{Hl}}{}^{(3)}$  &  $-\frac{g_W^4}{640 \pi ^2 m_{\Theta _1}^2}$  \\
			\hline
			$Q_{\text{Hq}}{}^{(3)}$  &  $-\frac{g_W^4}{640 \pi ^2 m_{\Theta _1}^2}$  \\
			\hline
			$Q_G$  &  $\frac{g_S^3}{2880 \pi ^2 m_{\Theta _1}^2}$  \\
			\hline
			$Q_{\text{Hu}}$  &  $-\frac{g_Y^4}{8640 \pi ^2 m_{\Theta _1}^2}$  \\
			\hline
			$Q_{\text{Hd}}$  &  $\frac{g_Y^4}{17280 \pi ^2 m_{\Theta _1}^2}$  \\
			\hline
			$Q_{\text{He}}$  &  $\frac{g_Y^4}{5760 \pi ^2 m_{\Theta _1}{}^2}$  \\
			\hline
			$Q_W$  &  $\frac{g_W^3}{1920 \pi ^2 m_{\Theta _1}^2}$  \\
			\hline \hline
	\end{tabular}}
	\subfloat{
		\begin{tabular}{|*{2}{>{\rowfonttype}c|}}%{|c|c|}
			\hline \hline
			Dim-6 Ops.&Wilson coefficients\\
			\hline \hline \rowfont{\color{red}}
			$Q_{\text{lq}}{}^{(1)}$  &  $\frac{g_Y^4}{34560 \pi ^2 m_{\Theta _1}^2}$  \\
			\hline
			$Q_{\text{qd}}{}^{(1)}$  &  $\frac{g_Y^4}{51840 \pi ^2 m_{\Theta _1}^2}$  \\
			\hline
			$Q_{\text{qq}}{}^{(1)}$  &  $-\frac{g_Y^4}{207360 \pi ^2 m_{\Theta _1}^2}$  \\
			\hline
			$Q_{\text{qu}}{}^{(1)}$  &  $-\frac{g_Y^4}{25920 \pi ^2 m_{\Theta _1}^2}$  \\
			\hline
			$Q_{\text{ud}}{}^{(1)}$  &  $\frac{g_Y^4}{12960 \pi ^2 m_{\Theta _1}^2}$  \\
			\hline
			$Q_{\text{lq}}{}^{(3)}$  &  $-\frac{g_W^4}{1280 \pi ^2 m_{\Theta _1}^2}$  \\
			\hline
			$Q_{\text{qq}}{}^{(3)}$  &  $-\frac{g_W^4}{2560 \pi ^2 m_{\Theta _1}^2}$  \\
			\hline
			$Q_{\text{dd}}$  &  $-\frac{g_Y^4}{51840 \pi ^2 m_{\Theta _1}^2}$  \\
			\hline
			$Q_{\text{ed}}$  &  $-\frac{g_Y^4}{8640 \pi ^2 m_{\Theta _1}^2}$  \\
			\hline
			$Q_{\text{ee}}$  &  $-\frac{g_Y^4}{5760 \pi ^2 m_{\Theta _1}^2}$  \\
			\hline
			$Q_{\text{ld}}$  &  $-\frac{g_Y^4}{17280 \pi ^2 m_{\Theta _1}^2}-\frac{9 y_{\Theta _1}^2 \left(4\lambda^{(1)}_{\Theta _1}+\lambda^{(2)}_{\Theta _1}\right)}{128 \pi ^2 m_{\Theta _1}^2}-\frac{y_{\Theta _1}^2}{4 m_{\Theta _1}^2}$  \\
			\hline
			$Q_{\text{le}}$  &  $-\frac{g_Y^4}{5760 \pi ^2 m_{\Theta _1}^2}$  \\
			\hline
			$Q_{\text{lu}}$  &  $\frac{g_Y^4}{8640 \pi ^2 m_{\Theta _1}^2}$  \\
			\hline
			$Q_{\text{eu}}$  &  $\frac{g_Y^4}{4320 \pi ^2 m_{\Theta _1}^2}$  \\
			\hline
			$Q_{\text{qu}}^{(8)}$&$-\frac{g_S^4}{480 \pi^2 m_{\Theta_1}^2}$\\
			\hline
			$Q_{\text{qe}}$  &  $\frac{g_Y^4}{17280 \pi ^2 m_{\Theta _1}^2}$  \\
			\hline
			$Q_{\text{uu}}$  &  $-\frac{g_Y^4}{12960 \pi ^2 m_{\Theta _1}^2}$  \\
			\hline
			$Q_{\text{ud}}^{(8)}$&$-\frac{g_S^4}{480 \pi^2 m_{\Theta_1}^2}$\\
			\hline
			$Q_{\text{qd}}^{(8)}$&$-\frac{g_S^4}{480 \pi^2 m_{\Theta_1}^2}$\\
			\hline \hline
	\end{tabular}}
\end{table*}

\begin{figure}[htb]
	\centering
	\subfloat[$\lambda_{\mathcal{H}_{2},1}$ - $\lambda_{\mathcal{H}_{2},2}$]
	{\includegraphics[width=0.325\textwidth, height=6cm]{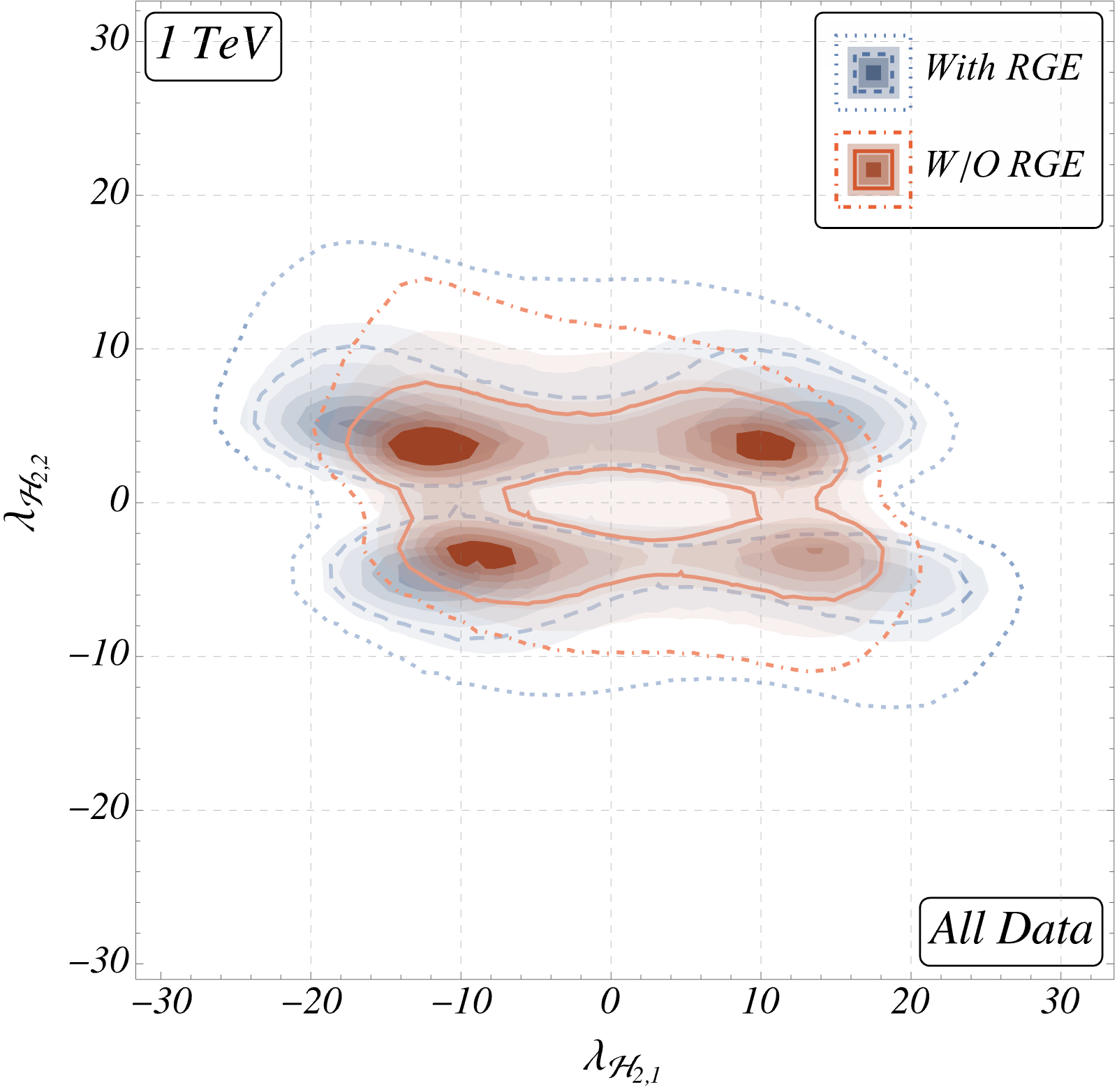}\label{fig:RGE_1TeV_2HDM_param12}}~
	\subfloat[$\lambda_{\mathcal{H}_{2},2}$ - $\lambda_{\mathcal{H}_{2},3}$]
	{\includegraphics[width=0.325\textwidth, height=6cm]{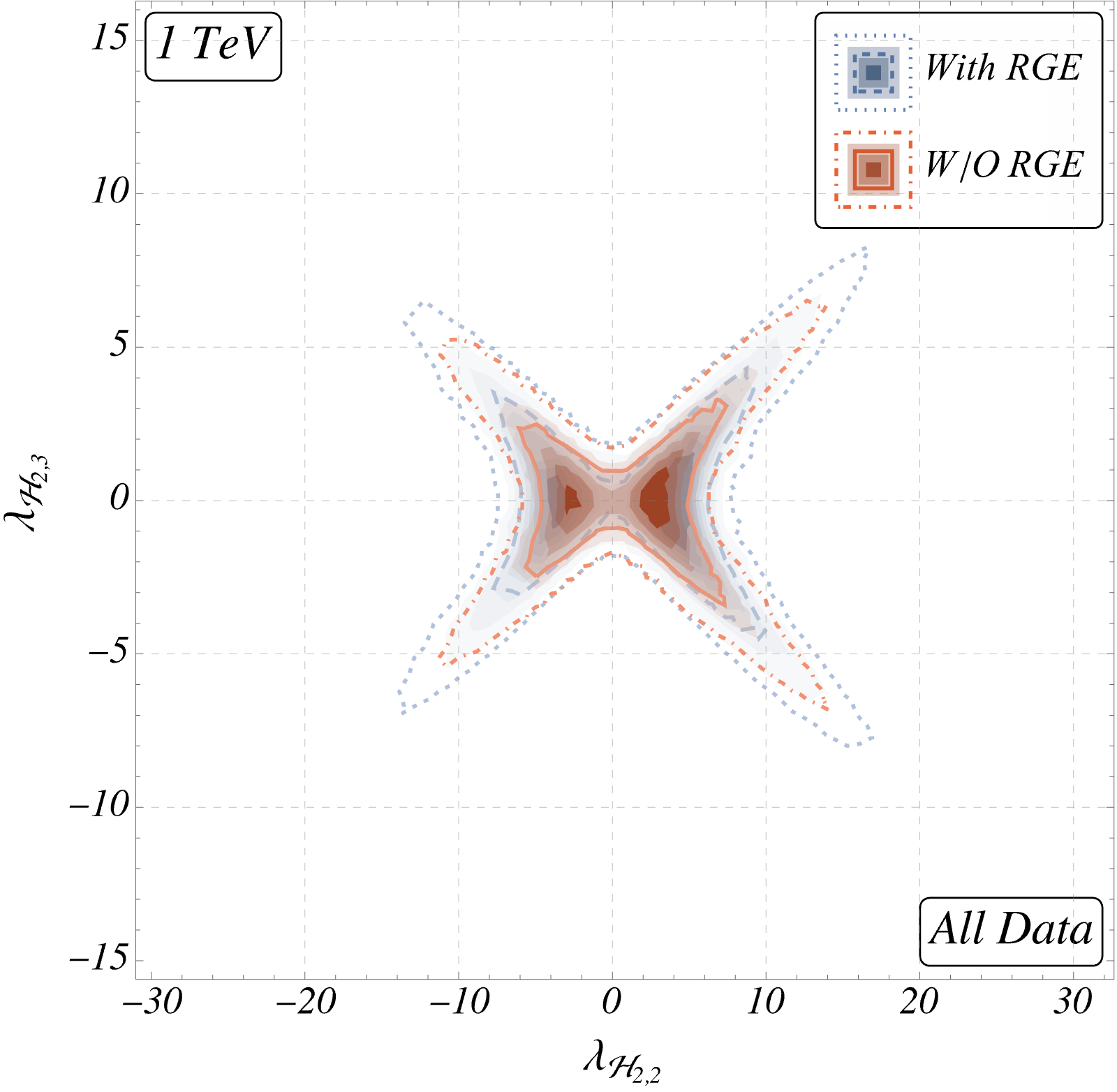}\label{fig:RGE_1TeV_2HDM_param23}}~
	\subfloat[$\lambda_{\mathcal{H}_{2},1}$ - $\lambda_{\mathcal{H}_{2},3}$]
	{\includegraphics[width=0.325\textwidth, height=6cm]{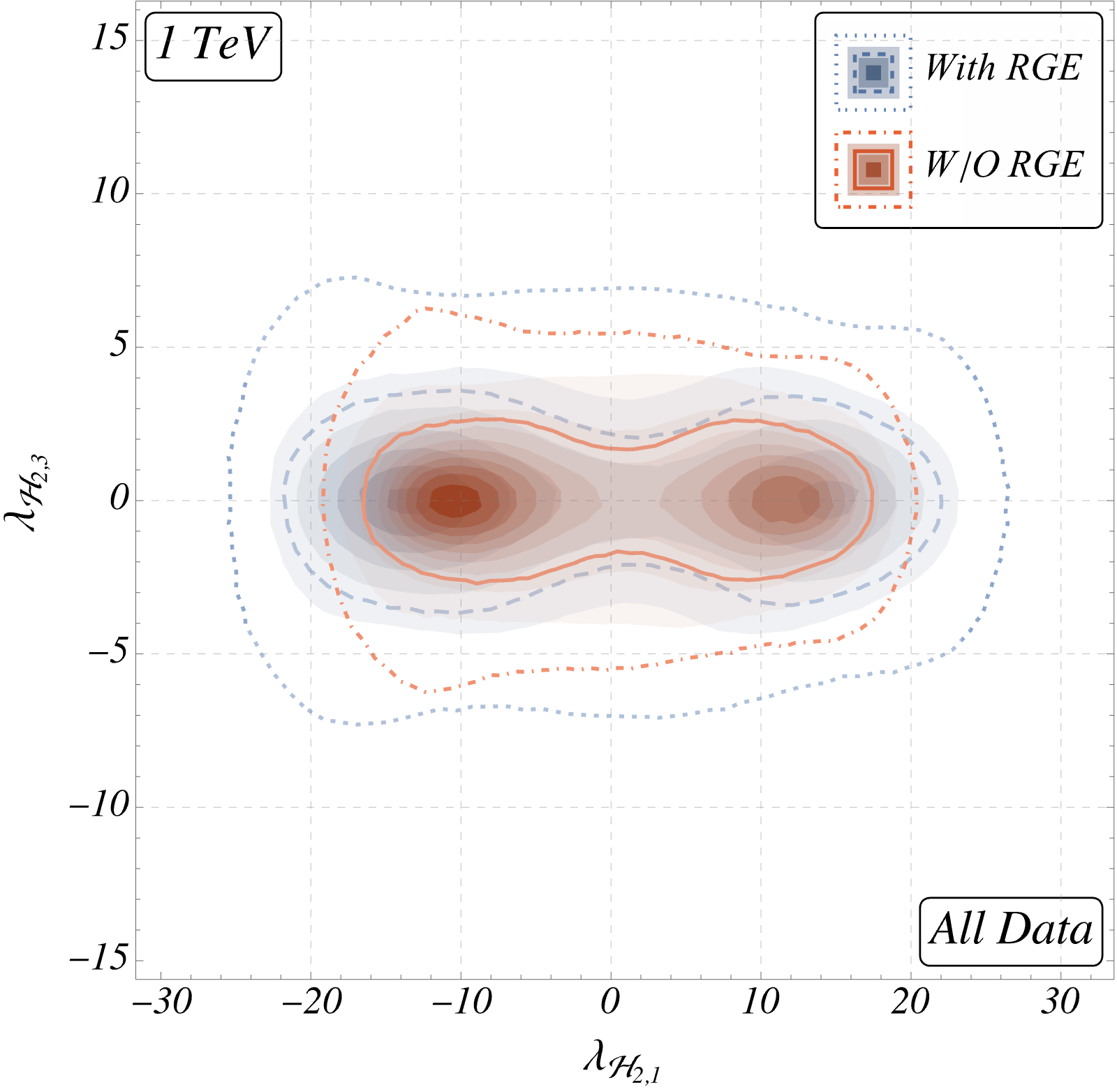}\label{fig:RGE_1TeV_2HDM_param13}}\\
    \subfloat[$\lambda_{\mathcal{H}_{2},1}$ - $\lambda_{\mathcal{H}_{2},2}$]
	{\includegraphics[width=0.325\textwidth, height=6cm]{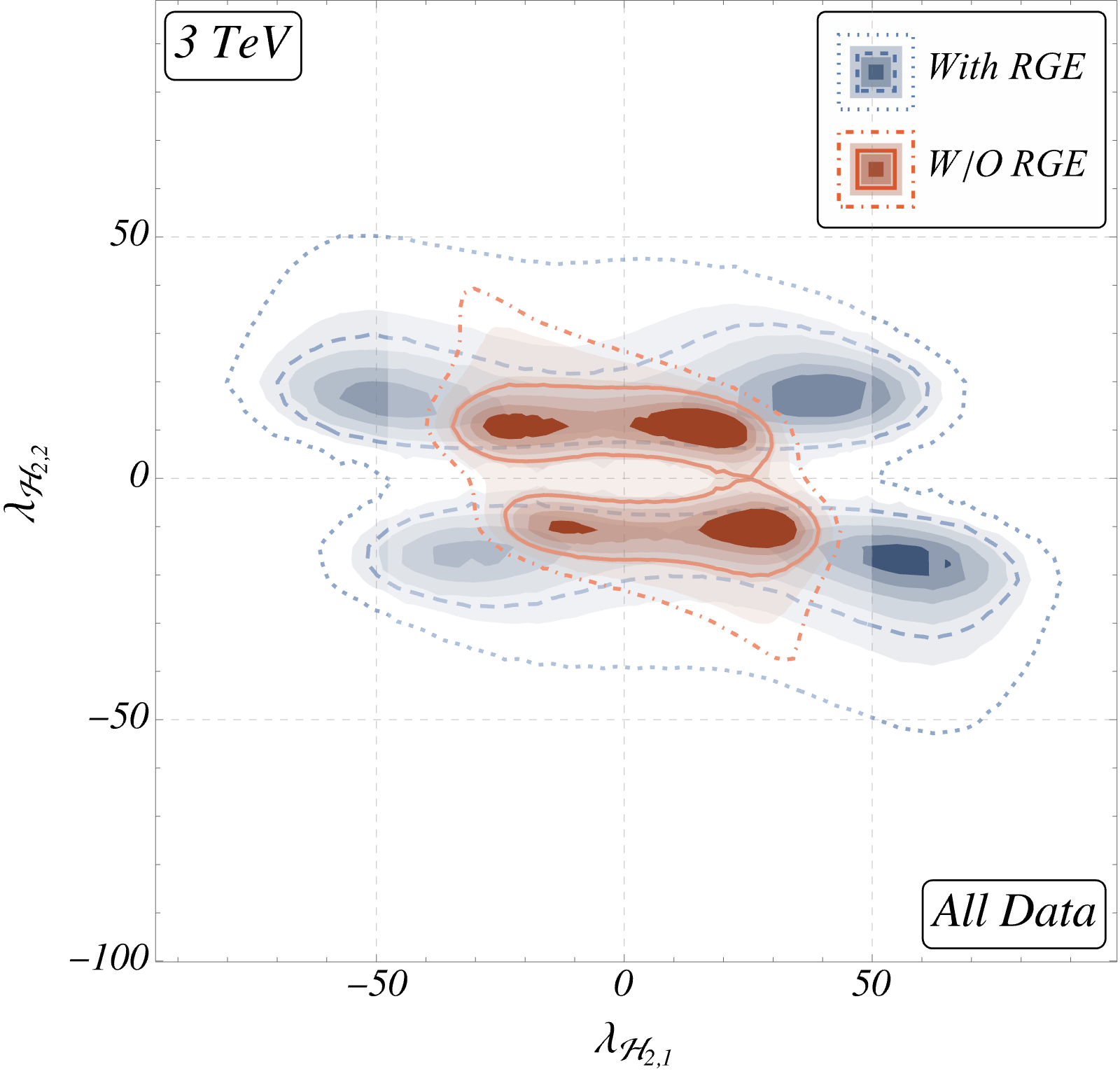}\label{fig:RGE_3TeV_2HDM_param12}}~
	\subfloat[$\lambda_{\mathcal{H}_{2},2}$ - $\lambda_{\mathcal{H}_{2},3}$]
	{\includegraphics[width=0.325\textwidth, height=6cm]{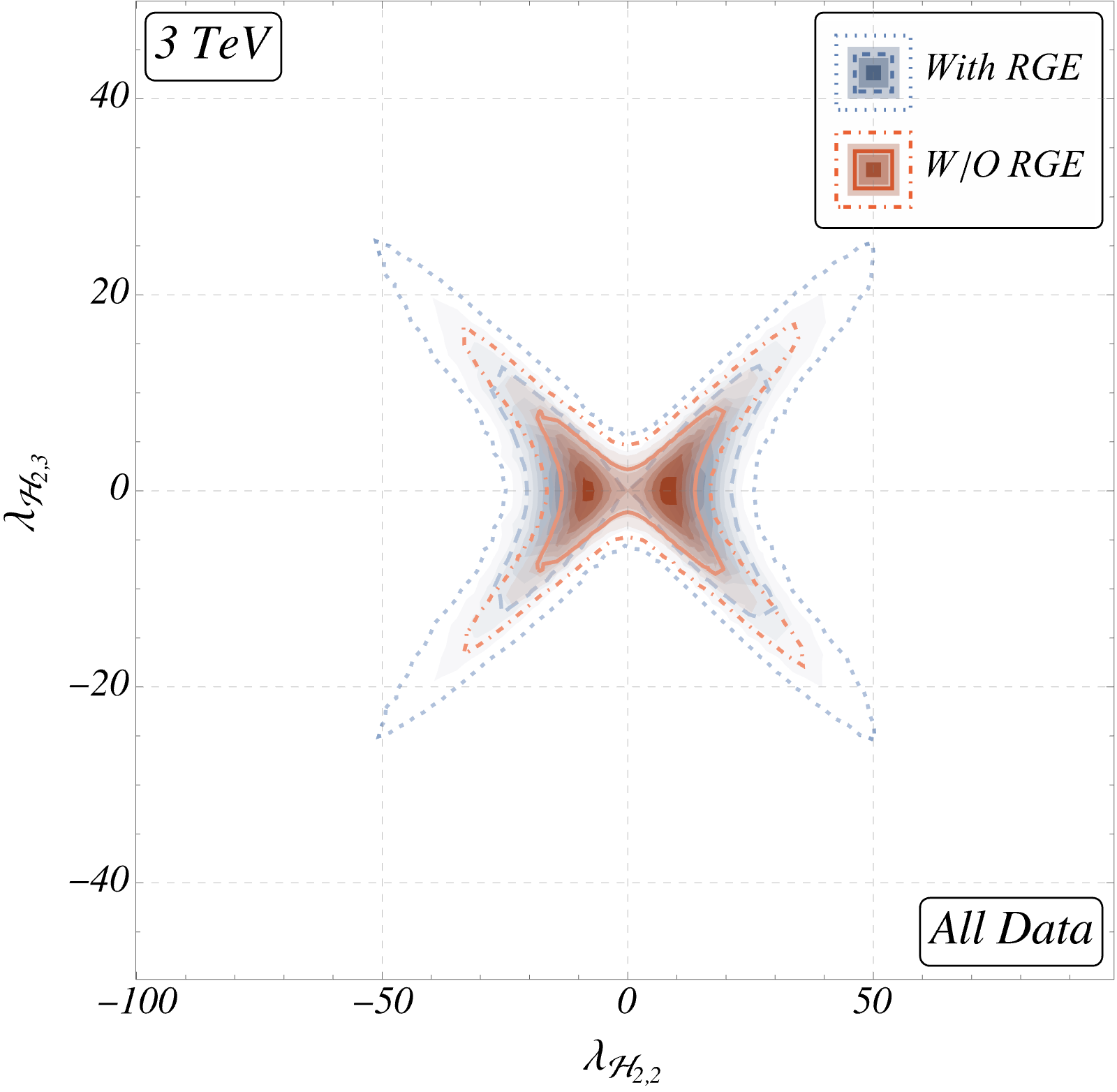}\label{fig:RGE_3TeV_2HDM_param23}}~
	\subfloat[$\lambda_{\mathcal{H}_{2},1}$ - $\lambda_{\mathcal{H}_{2},3}$]
	{\includegraphics[width=0.325\textwidth, height=6cm]{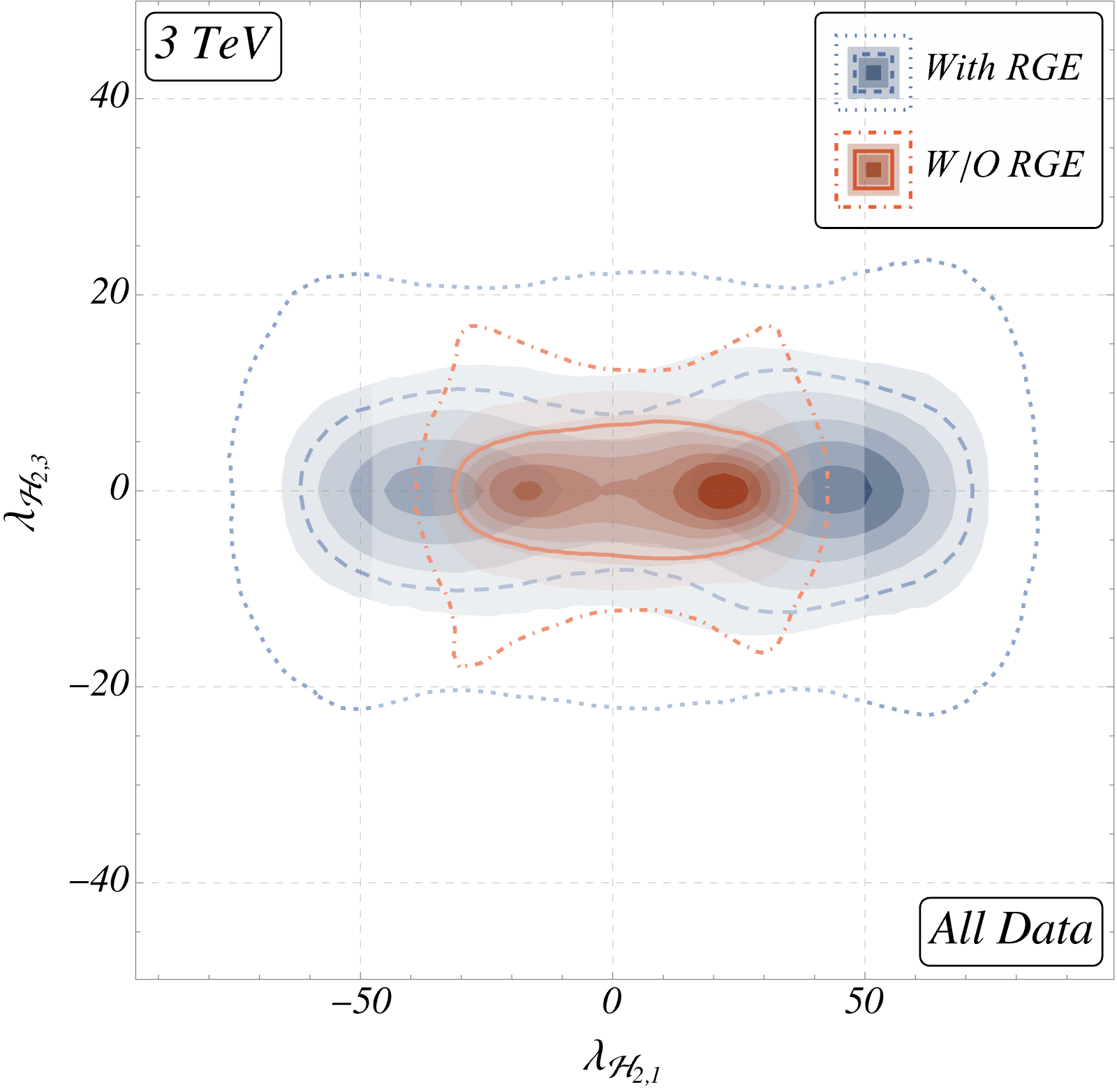}\label{fig:RGE_3TeV_2HDM_param13}}\\
	\caption{\small 
    Two-dimensional marginalised posteriors among the BSM parameters for $\mathcal{H}_{2}$  for $m_{\mathcal{H}_{2}} =$ 1 TeV  (top row) and $m_{\mathcal{H}_{2}} = 3$ TeV  (bottom row). We show the results from a fit with ``All"  data. The credible regions shown in blue correspond to the model parameter spaces obtained after including RG evolved operators and in red ones, these are ignored.  }
	\label{fig:RGE_model_param_plots}
\end{figure}

\section{Effect of Renormalisation Group Equations on the model-dependent analysis}\label{sec:RGE_effects}

In the previous section~\ref{sec:BSMs}, the SMEFT matching of the BSM theory with a heavy scalar field is performed at the high scale $\Lambda$, which is taken as the mass of the heavy scalar $m_{hf}$. There, we have ignored the running of the SMEFT operators, emerged at the high scale $\Lambda$, up to the electroweak scale where they are eventually mapped into the experimental observables. In this section, the operators generated by integrating out the heavy scalar doublet $\mathcal{H}_{2}$ at the scale $\Lambda$ are evolved to the $M_{Z}$ scale using the RGEs encoded in refs.~\cite{Jenkins:2013wua, Jenkins:2013zja, Alonso:2013hga}. The WCs at $M_Z$, $\mathcal{C}_i(M_Z)$, are computed using the matching scale WCs $\mathcal{C}_i(\Lambda)$  and the SMEFT anomalous dimension matrix (ADM)~$\gamma_{ij}$ in the leading-log approximation,

\begin{align}
    \frac{d \, \mathcal{C}_i(\mu)}{d \, \text{log} \mu} &= \sum\limits_j \frac{1}{16 \pi^2} \gamma_{ij} \mathcal{C}_j, \nonumber
\end{align}
and, at leading order,
\begin{align}
    \mathcal{C}_i(M_Z) &= \mathcal{C}_i(\Lambda) + \sum\limits_j \frac{1}{16 \pi^2} \gamma_{ij} \mathcal{C}_j(\Lambda) \text{log}\left[\frac{M_Z}{\Lambda}\right].
\end{align}

For $\mathcal{H}_{2}$, a total of 51 operators are generated, 14 of which are exclusively induced by the RG running, and the RG evolved matching result is available in the Github repository \href{https://github.com/effExTeam/Precision-Observables-and-Higgs-Signals-Effective-passageto-select-BSM}{\faGithub}. Using this RG evolved matching relations of the WCs, the constraints on the BSM parameters $\lambda_{\mathcal{H}_{2},1},\; \lambda_{\mathcal{H}_{2},2}$, and $\lambda_{\mathcal{H}_{2},3} $ are obtained using ``All" experimental measurements listed in tab.~\ref{tab:obset} and the corresponding two-dimensional marginalised posteriors are shown in fig.~\ref{fig:RGE_model_param_plots} in blue colour with $m_{\mathcal{H}_{2}} $ set to 1 TeV. In the same figure, to see the effect of RG running, we show the two-dimensional marginalised model parameter posteriors obtained neglecting RG running in red colour (these are the ones shown in black in the top row of fig.~\ref{fig:model_param_plots}). It is visible that the obtained parameter spaces are relaxed after including the RG effects in the analysis.
Further, we also study the effects on the constraints for different choices of  $m_{\mathcal{H}_{2}}$. We observe that the constraints on the BSM parameters weaken with higher values of $m_{\mathcal{H}_{2}}$. We show the allowed parameter spaces for $m_{\mathcal{H}_{2}}$ set to 3 TeV in the bottom row of fig.~\ref{fig:RGE_model_param_plots}. On comparing the red (W/O RGE) and blue (With RGE) regions in the top row (with $m_{\mathcal{H}_{2}}$ = $1$ TeV) with the corresponding ones in the bottom (with $m_{\mathcal{H}_{2}}$ = $3$ TeV), we note that the model parameter constraints relax with increase in $m_{\mathcal{H}_{2}}$.

\begin{figure}[t]
	\centering
	\subfloat[$\mathcal{C}_{H}$ - $\mathcal{C}_{HB}$]
	{\includegraphics[width=0.325\textwidth, height=5.8cm]{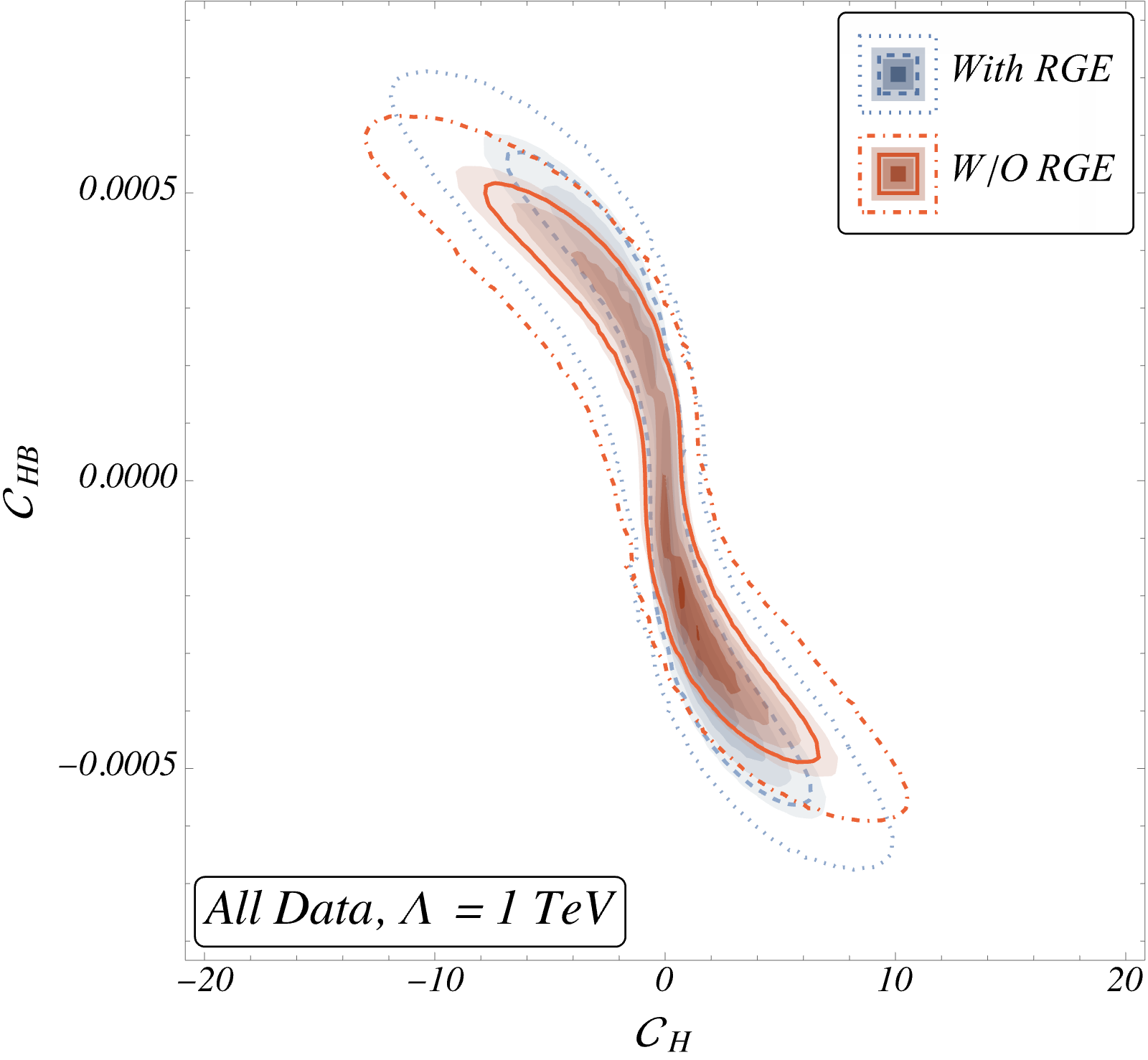}\label{fig:2HDM_CHvsCHB}}~
	\subfloat[$\mathcal{C}_{H\square}$ - $\mathcal{C}_{HD}$]
	{\includegraphics[width=0.325\textwidth, height=5.8cm]{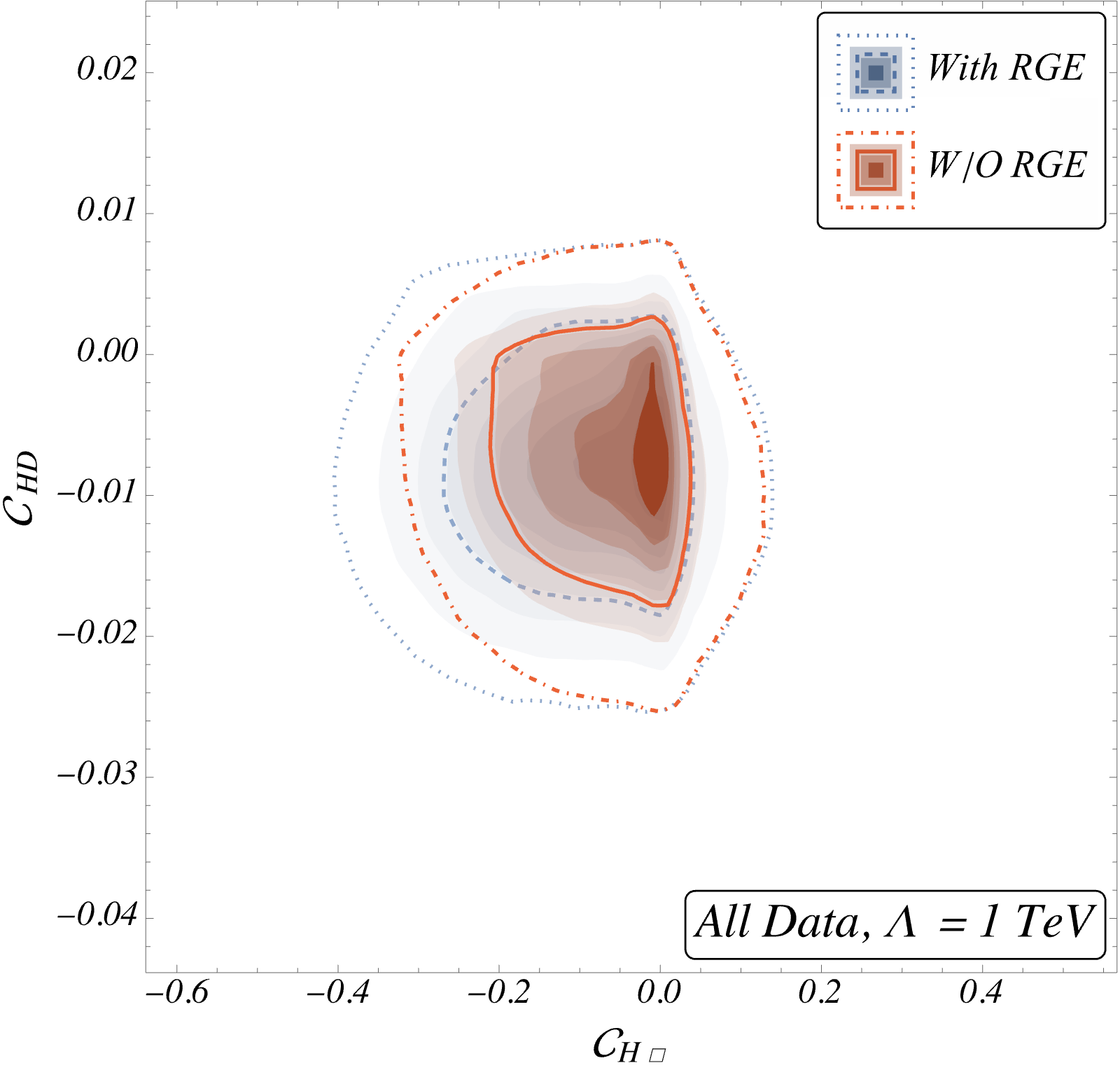}\label{fig:2HDM_CHboxvsCHD}}~
	\subfloat[$\mathcal{C}_{HW}$ - $\mathcal{C}_{HWB}$]
	{\includegraphics[width=0.325\textwidth, height=5.8cm]{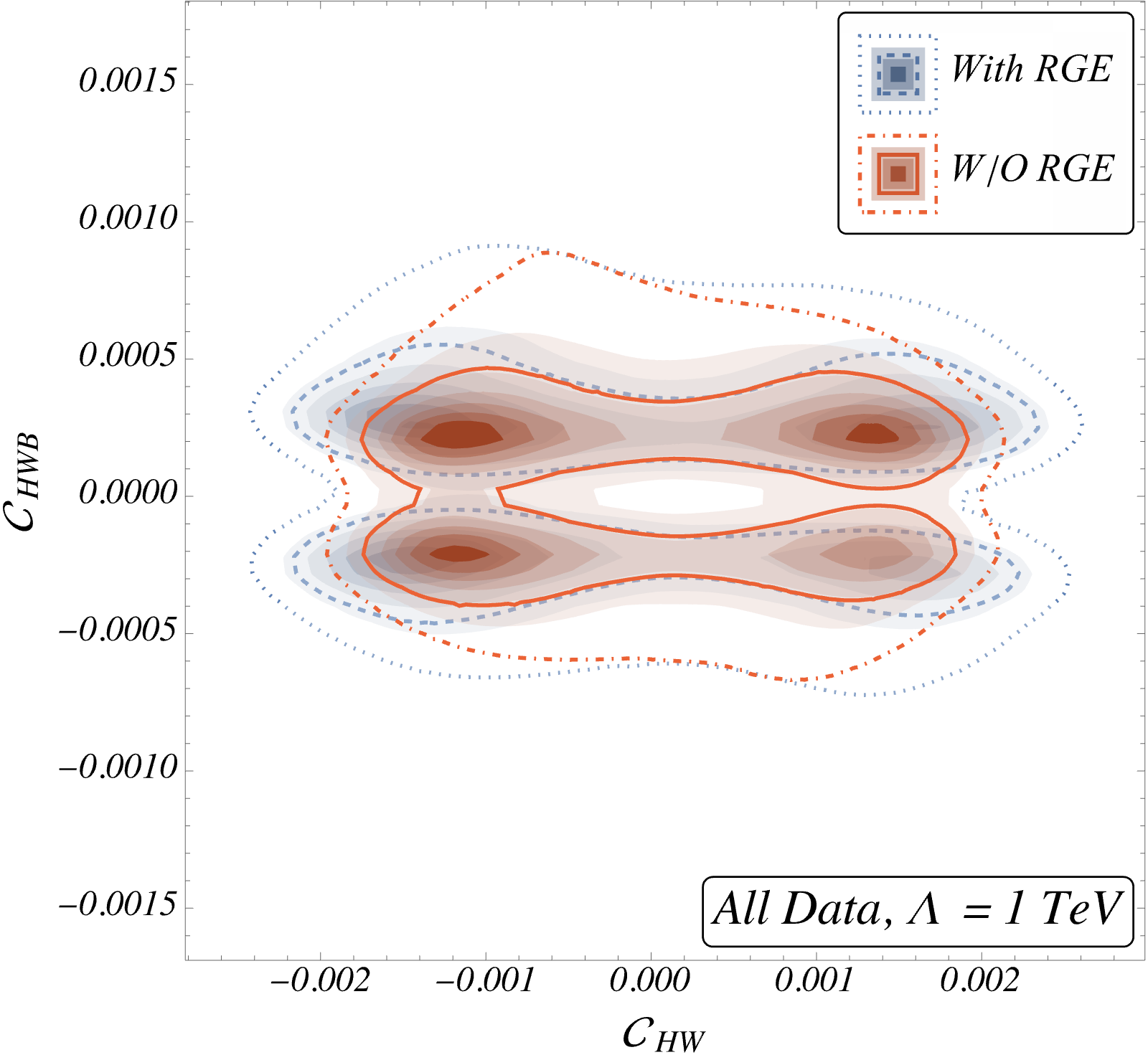}}\\
	\subfloat[$\mathcal{C}_{H}$ - $\mathcal{C}_{HB}$]
	{\includegraphics[width=0.325\textwidth, height=5.8cm]{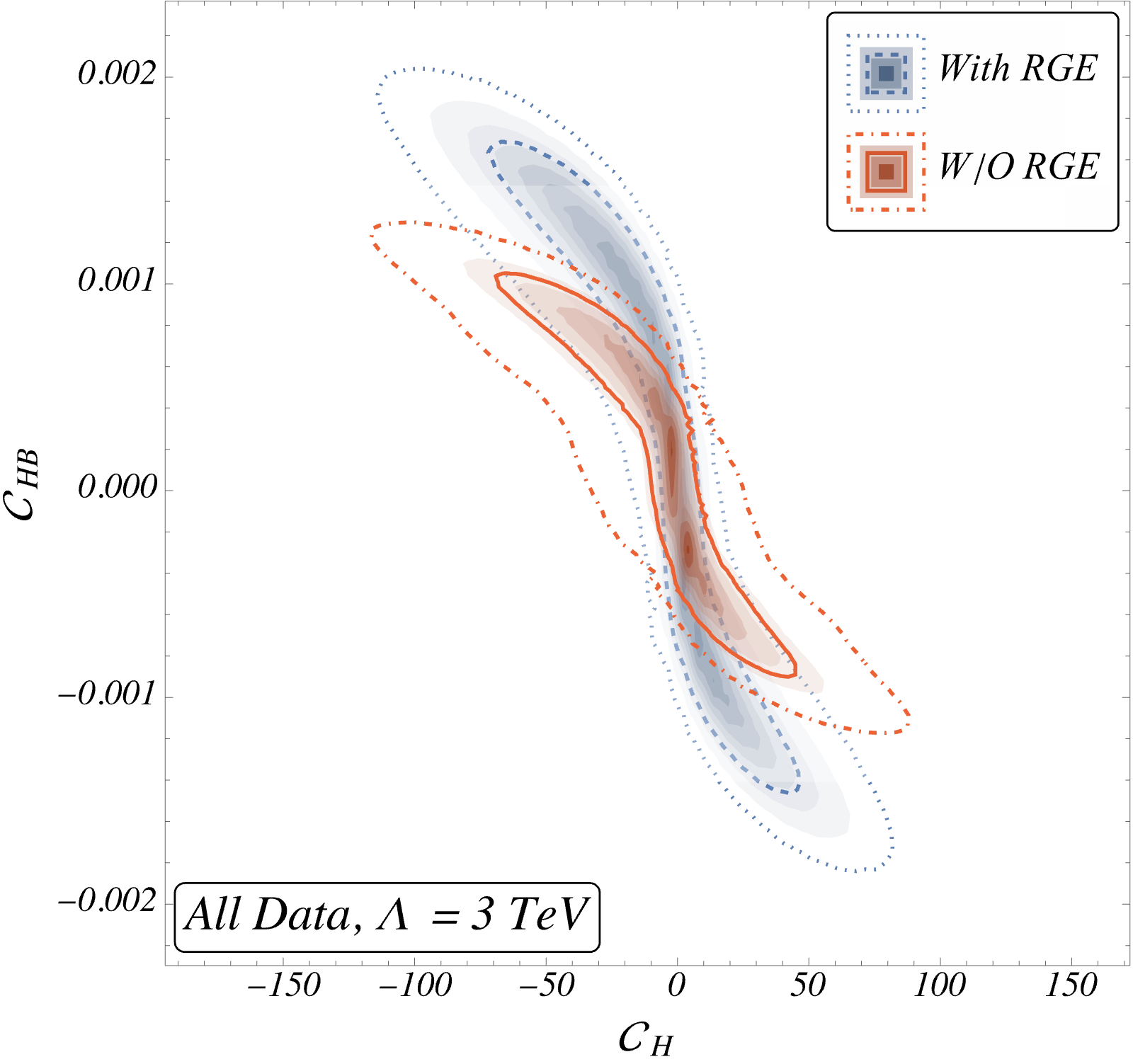}\label{fig:2HDM_CHvsCHB}}~
	\subfloat[$\mathcal{C}_{H\square}$ - $\mathcal{C}_{HD}$]
	{\includegraphics[width=0.325\textwidth, height=5.8cm]{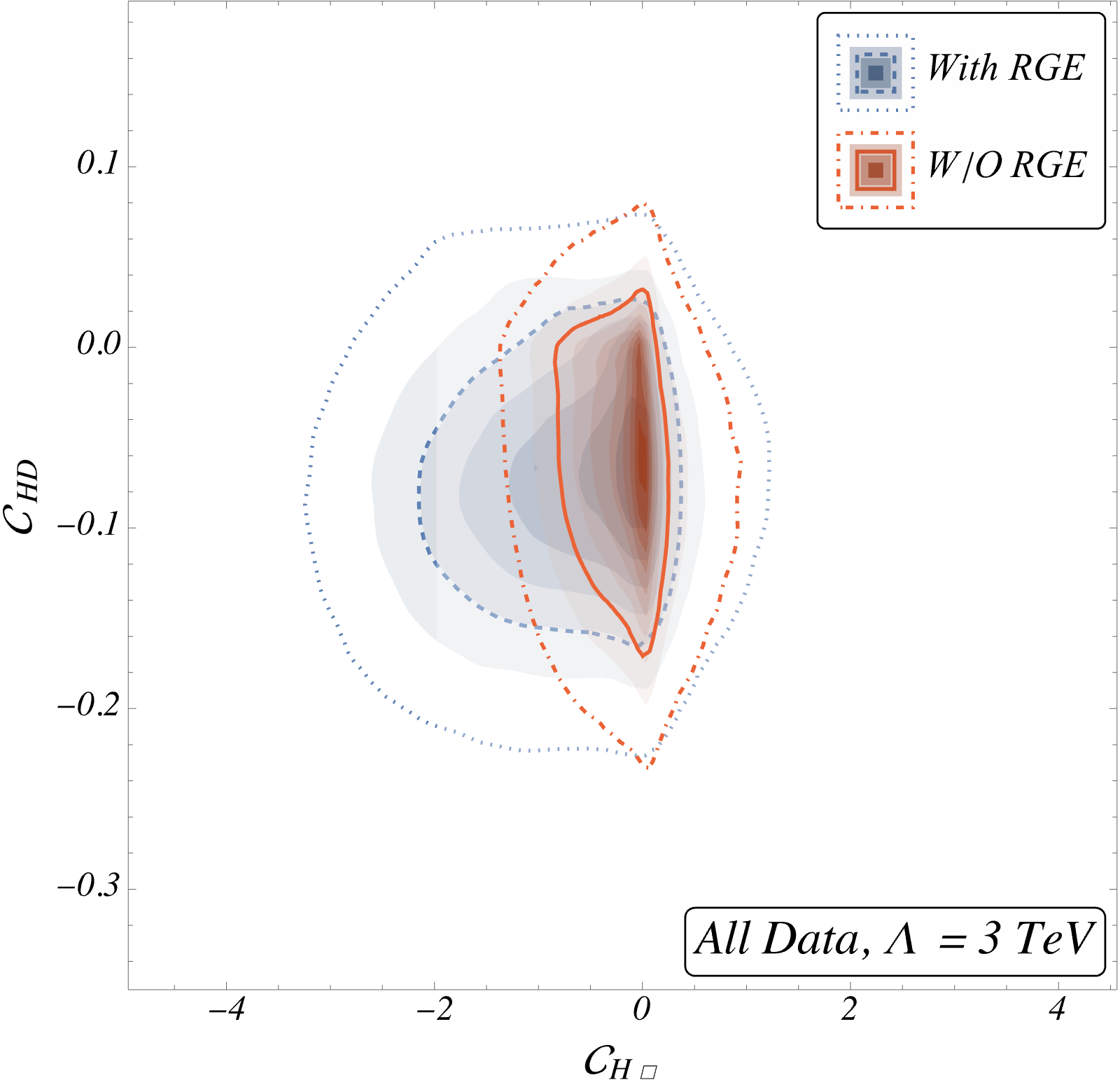}\label{fig:2HDM_CHboxvsCHD}}~
	\subfloat[$\mathcal{C}_{HW}$ - $\mathcal{C}_{HWB}$]
	{\includegraphics[width=0.325\textwidth, height=5.8cm]{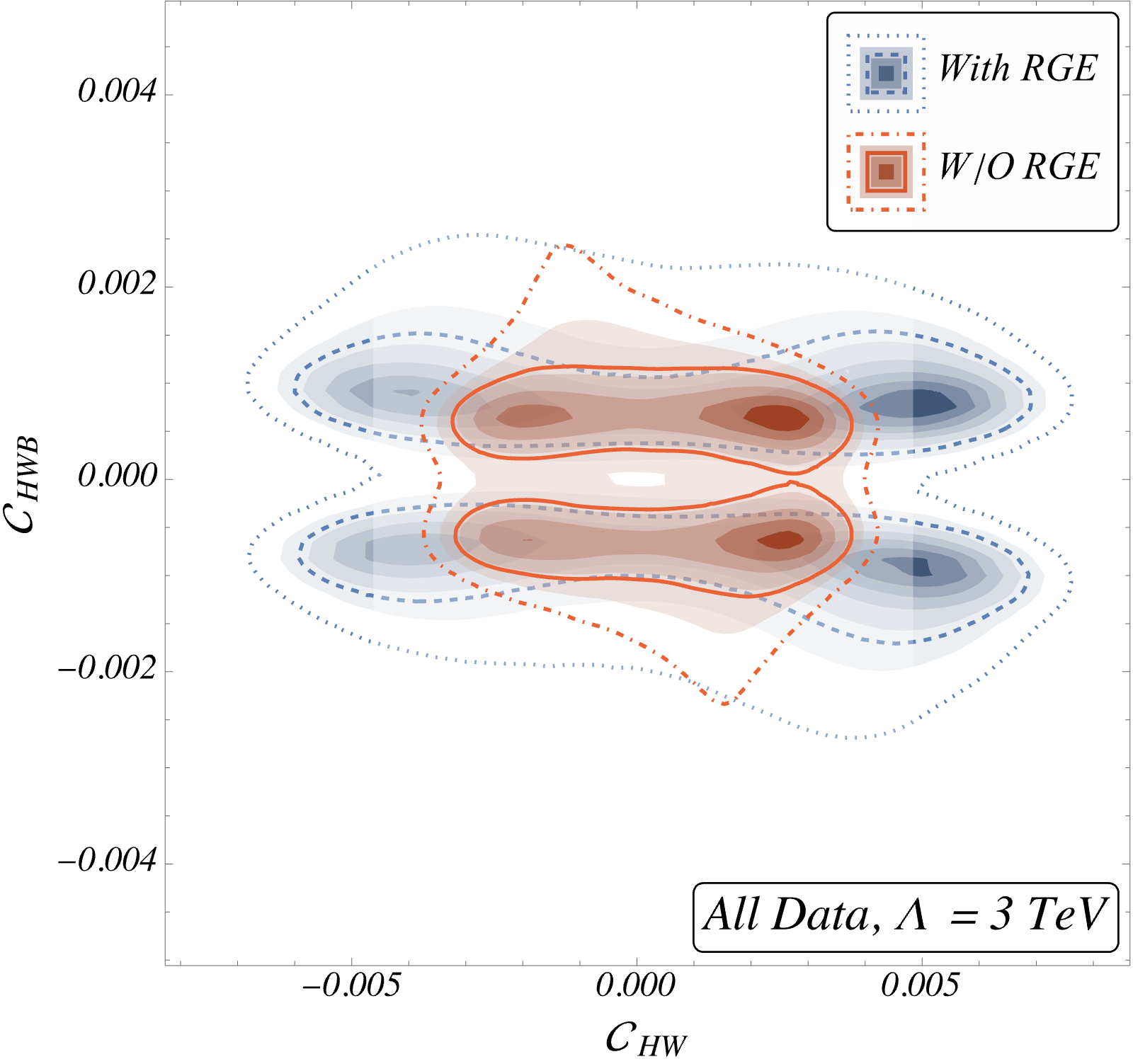}}
	\caption{\small 
     Two-dimensional posteriors among the WCs induced by $\mathcal{H}_{2}$ for $m_{\mathcal{H}_{2}} =$ 1 TeV  (top row) and $m_{\mathcal{H}_{2}} = 3$ TeV  (bottom row). Similar to fig.~\ref{fig:RGE_model_param_plots}, the credible regions shown in red correspond to scenario without RGE and the blue ones with RGE.}
	\label{fig:RGE_wc_2HDM}
\end{figure}

The model-dependent WCs' credible regions obtained from the model parameter posteriors and RG evolved matching results are shown in fig.~\ref{fig:RGE_wc_2HDM} in blue colour. We show two-dimensional marginalised sample plots. Similar to fig.~\ref{fig:RGE_model_param_plots}, we also show the WCs' regions obtained when ignoring the RG running in red. These plots depict similar results that with the inclusion of RGE and with a higher value of $m_{\mathcal{H}_{2}}$, the allowed parameter spaces become comparatively relaxed. 

%%%%%%%%%%%%%%%%%%%%%%%%%%%%%%%%%%
%%%%%%%%%%%%%%%%%%%%%%%%%%%%%%%%%%%%%%%

\begin{figure}[t]
	\centering
	\subfloat[$\mathcal{C}_{uB}$ - $\mathcal{C}_{uW}$]
	{\includegraphics[width=0.325\textwidth, height=5.8cm]{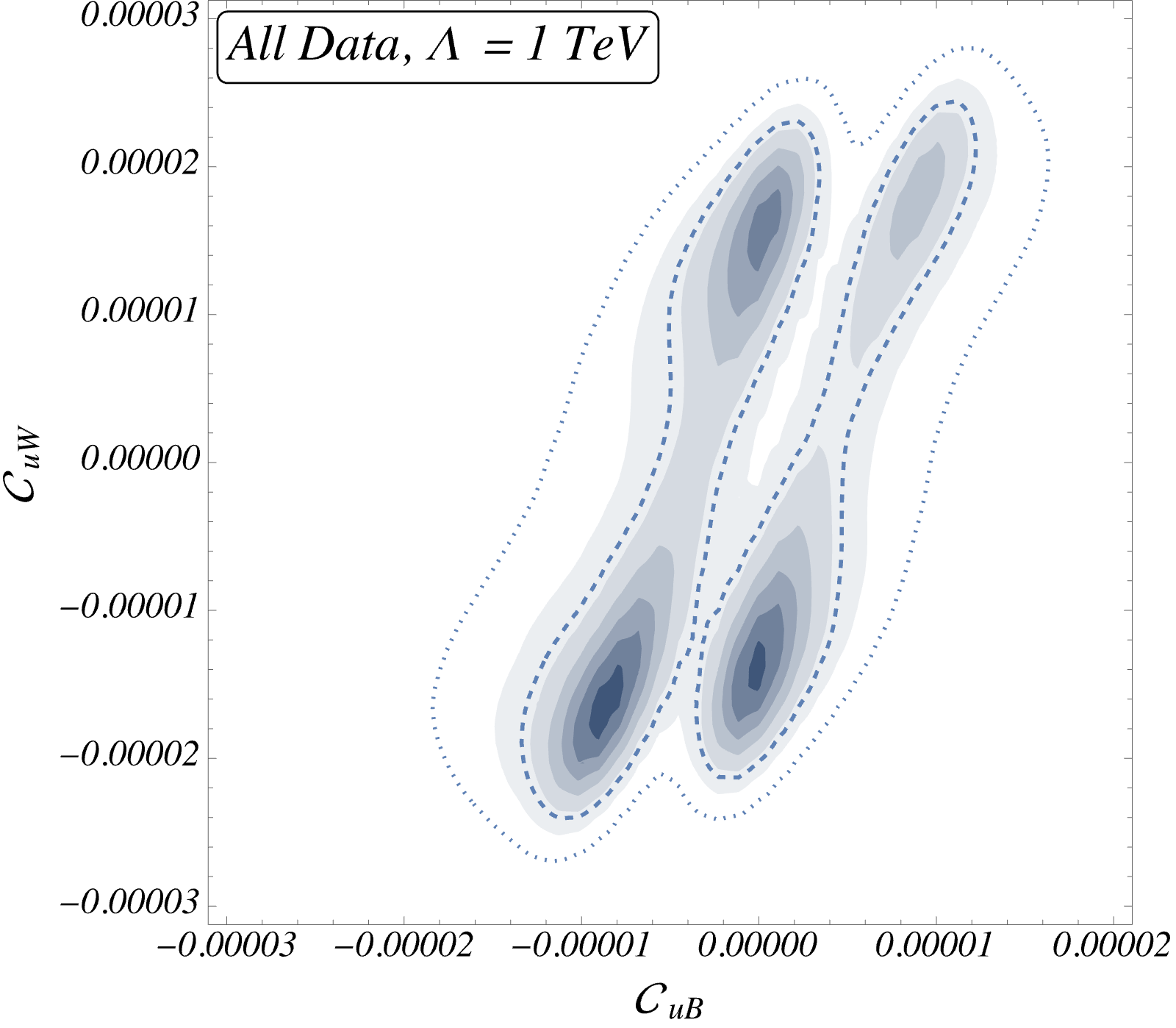}\label{fig:2HDM_CuBvsCuW}}~~
	\subfloat[$\mathcal{C}_{eB}$ - $\mathcal{C}_{Hud}$]
	{\includegraphics[width=0.325\textwidth, height=5.8cm]{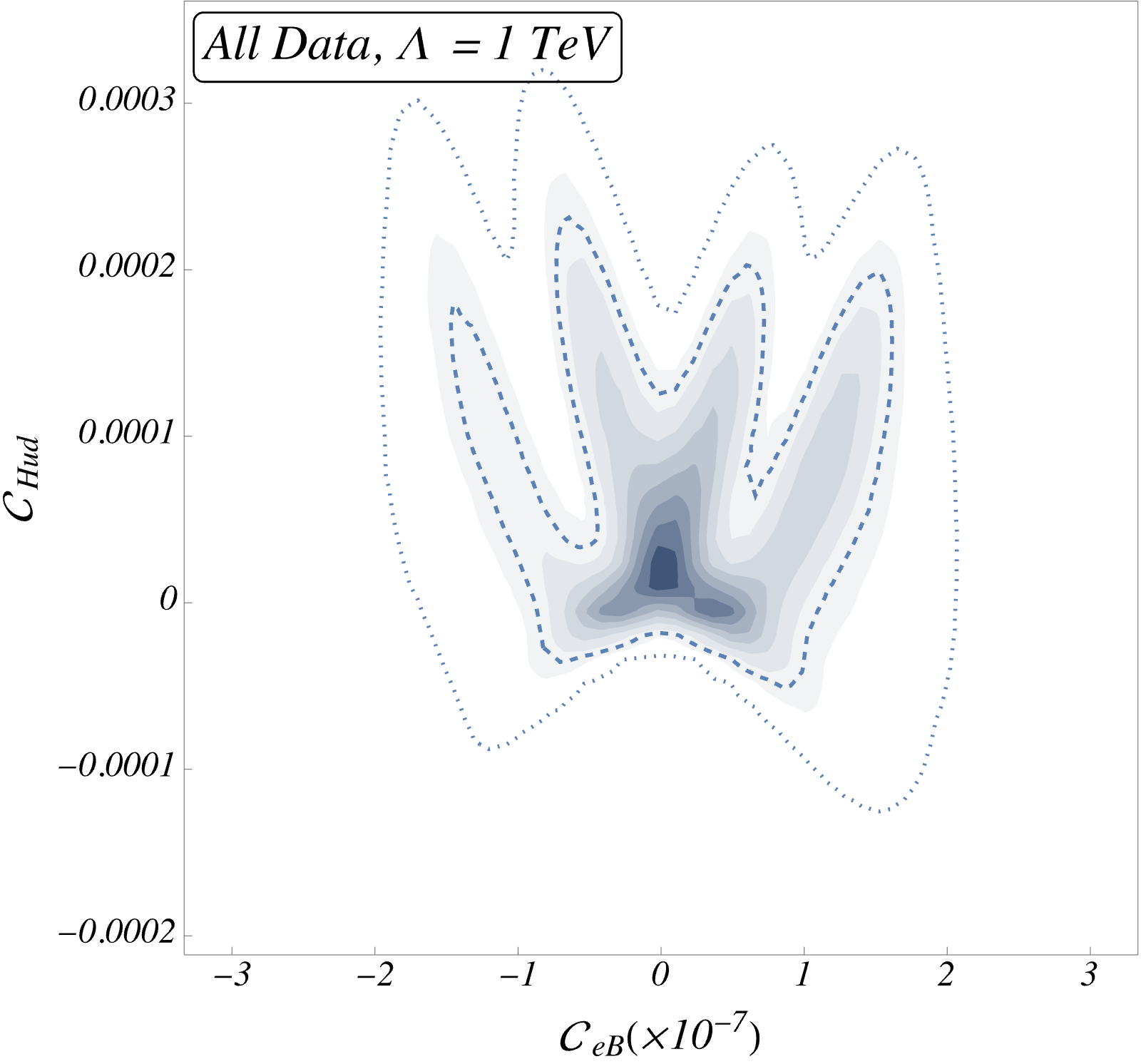}\label{fig:2HDM_CeBvsCHud}}
	\caption{\small 
     Two-dimensional marginalised WCs posterior for $m_{\mathcal{H}_{2}}$ ($\Lambda$) = 1 TeV with RG evolved matching results.}
	\label{fig:RGE_wc2_2HDM}
\end{figure}

After including the RG running of the SMEFT operators, the WC relations of 23 SMEFT operators are given as functions of the model parameters $\lambda_{\mathcal{H}_{2},1},\; \lambda_{\mathcal{H}_{2},2}$, and $\lambda_{\mathcal{H}_{2},3}$. Thus, using the model parameter posterior distributions in the RG evolved matching relations, the multivariate distributions are obtained for these 23 WCs . This list includes some operators (for eg. $Q_{uB}$, $Q_{uW}$, $Q_{dB}$, $Q_{dW}$, $Q_{ueB}$, $Q_{eW}$) which are not constrained in the model-independent analysis by the experimental measurements. We show the example plots of such model-dependent WCs' regions in fig.~\ref{fig:RGE_wc2_2HDM}. The operators $Q_{uB}$, $Q_{uW}$, $Q_{eB}$ and $Q_{Hud}$ are RGE generated, and the corresponding WCs include an extra suppression of $1/16 \pi^2$. The effect of this small factor is visible in the strictly constrained WC space.

%%%%%%%%%%%%%%%%%%%%%%%%%%%%%%%%%%
%%%%%%%%%%%%%%%%%%%%%%%%%%%%%%%%%%%%%%%

\section{Conclusions}\label{sec:conclusion}

In this paper, we have analysed and constrained potential directions of beyond Standard Model physics under two themes: first, in a model-independent way using  dimension-6 SMEFT and second for $16$ single scalar particle extensions of the SM using a similar statistical methodology and datasets. 

We have performed a global analysis of $23$~Wilson coefficients in the Warsaw basis based on an up-to-date set of observables of the Higgs and di-boson sectors plus electroweak precision observables. Our fit includes di-Higgs measurements for the first time which improves the bounds on a modified Higgs potential through $\wc_H$.
The addition of recent datasets, and in particular STXS data up to high energies, strengthens the limits on the Wilson coefficients by up to a factor $9$ compared to previous analyses. The biggest improvements are obtained for operators describing Higgs-top and Higgs-gluon interactions as well as for $\wc_{H \square}$. We have analysed in detail the role of different STXS measurements in constraining the different Wilson coefficients. 

Moreover, we have tabulated the complete sets of  dimension-6 effective operators and the associated WCs in the Warsaw basis after integrating out the single heavy non-SM scalars for $16$ different new physics scenarios up to 1-loop level incorporating the scalar heavy-light mixing contributions from the loops. The WCs are expressed in terms of the model parameters. We have employed a similar analysis strategy for each of the BSM scenarios and constrained the model parameters. We have highlighted the individual impact of different observables (EWPO and Higgs data) through two-dimensional posteriors in model parameter as well as WC planes. We have explicitly discussed the results for the EWDS~$\mathcal{H}_2$ and a scalar leptoquark extension~$\Theta_1$. Plots for the remaining 14 models considered are available in the Supplementary material~\cite{SupMat}\footnote{These plots can also be downloaded from the GitHub repository~\cite{Githubresult}.}.

We have investigated the effects of RG running on the allowed space of the model parameters and the WCs for two choices of the cut-off scale, 1~TeV and 3~TeV. We have performed the RG running (in leading-log approximation) on the matching results for an extra Electroweak scalar doublet model for two different cut-off scales 1~TeV and 3~TeV. We have noted that once the RG effects are taken into consideration, the allowed ranges of BSM parameters are relaxed compared to no RG case. We have also displayed the allowed ranges of WCs associated with the radiatively generated effective operators.

Currently, the considered dataset leaves some of the effective operators that arise from the single scalar extensions studied unconstrained and therefore fails to encapsulate the impact of all new physics interactions. Specifically, this concerns the WCs associated with 4-fermion operators. For the future, we are planning to add observables constraining 4-fermion operators~\cite{Falkowski:2017pss}, \textit{e.g.}~from dilepton production, parity violation and low-energy flavour observables, to constrain so far unbound directions in parameter space. These would be particularly relevant for constraining leptoquark models~\cite{Schmaltz:2018nls}.

\section{Acknowledgement}
We thank Julien Baglio, Silvia Ferrario Ravasio, Stephen Jones, Ken Mimasu, and Tevong You for helpful discussions, and Ennio Salvioni, Gauthier Durieux, and Matthew Mccullough for pointing out an error in matching result of the Quartet model $(\Sigma)$. The work of A., S.D.B.~and J.C.~is supported by the Science and Engineering Research Board, Government of India, under the agreements SERB/PHY/2019501. SDB is also supported in part by SRA (Spain) under Grant No.\ PID2019-106087GB-C21 / 10.13039/501100011033;
by the Junta de Andalucía (Spain) under Grants No.\ FQM- 101, A-FQM-467-UGR18, and P18-FR-4314 (FEDER). M.S.~is supported by the STFC under grant ST/P001246/1. A.B.~gratefully acknowledges support from the Alexander-von-Humboldt foundation as a Feodor Lynen Fellow. We thank the IPPP for access to its computing cluster.

\FloatBarrier
\newpage
\appendix

\section{Functional matching by \texttt{CoDEx}}
\vspace{1em}

\texttt{CoDEx} uses effective action formulae to compute 1-loop level WCs, which involves solving functional ``$Trace-Log$'' of UV action derivatives. We follow the procedures in Ref.~\cite{Henning:2014wua} (for loops with only heavy field propagators), and Refs.~\cite{Zhang:2016pja} and \cite{Ellis:2016enq} for derivation and cross-checks of heavy-light mixed 1-loop effective action formulae, which \texttt{CoDEx} solves for each BSM. Note that there are more than one procedures to calculate the effective action formulae \cite{Henning:2014wua,Drozd:2015rsp,Fuentes-Martin:2016uol,Zhang:2016pja}. However, readers are welcome to study all of them as each method has its own advantages. It is worth mentioning that in this work we calculate matching of BSMs with single heavy scalar extension of the SM particle content, and therefore, the functional traces have a relatively simple form, which is not the case in most generic scenarios assuming arbitrary number of heavy fields with non-degenerate masses and arbitrary spins \cite{Fuentes-Martin:2020udw,Cohen:2020qvb,Kramer:2019fwz}. 

In the following text, we elaborate on the implementation of effective action formulae on BSMs to integrate out the heavy field up to 1-loop. The effective action $ S_{\text{eff}}[\phi]$ is defined

\begin{align}\label{eq:effactiondefn}
	e^{i\, S_{\text{eff}}[\phi](\mu)} = \int \mathcal{D}\Phi\, e^{i\, S_{\text{UV}}[\phi,\Phi](\mu)},
\end{align}

where, $\Phi$ and $\phi$ represent are a heavy real scalar and a light field respectively, and $ S_{\text{UV}}[\phi,\Phi] (\mu) $ represents the UV action defines at scale $\mu$. In the following discussion, we set $ \mu $ equal to be the cut-off scale $ \Lambda $, which is also the matching scale of the UV theory to the EFT. We expand the UV action around the classical configuration of the heavy field $ \Phi_c $ defined by,

\begin{align}\label{eq:classsoldefn}
	\left.\frac{\delta S_{\text{UV}}[\phi,\Phi]}{\delta \Phi}\right\rvert_{\Phi_c} = 0.
\end{align}

Now, $ S_{\text{UV}}[\phi,\Phi] \rightarrow S_{\text{UV}}[\phi,\Phi_c + \eta]$ in Eq.~\ref{eq:effactiondefn}, where $ \eta  $ represents the fluctuation field defined as $ \Phi = \Phi_c + \eta$. Using Eq.~\ref{eq:classsoldefn}, we have

\begin{align}\label{eq:heavyloopeffactiondef}
	S_{\text{eff}}[\phi] = S_{\text{UV}}[\phi,\Phi_c] + \frac{i}{2}\, \text{Tr}\, \text{Log} \left[\left.-\frac{\delta^2 S_{\text{UV}}[\phi,\Phi]}{\delta\Phi^2}\right\rvert_{\Phi_c}\right]+ \mathcal{O}(\eta^2).
\end{align}

In Eq.~\ref{eq:heavyloopeffactiondef}, we recognize the first and second terms on the RHS as the tree-level and the pure heavy-loop processes in the effective action,  respectively. Similarly, we can capture the contribution from the mixed heavy-light loop processes by expanding the UV action around the classical configuration of the light field(s), about which we discuss in detail at a later stage in this section. One can determine the classical configuration of the heavy field by using the Euler-Lagrange equation on the UV Lagrangian $ \mathcal{L}_{\text{UV}} $. Then substituting the heavy field solution back into the UV Lagrangian leads to tree-level matching. Subsequently, by imposing a mass-dimensions cut-off and Standard Model field Equation of Motions, Fierz identities, and Integration-by-Parts relations, one gets the Wilson coefficients of effective operators of a given mass dimension.
\vspace{2em}

\subsection{1-Loop matching: only heavy field as loop propagators}
\vspace{1em}

The functional ``$Trace-Log$'' of action derivatives can be expanded in terms of covariant derivatives, quadratic matrix `$U$' and heavy field mass. The quadratic matrix `$U$' contains the interaction of heavy fields and light fields. We define

\begin{align}
	S_{\text{eff}}^{\text{1loop(H)}}(\phi)&=\frac{i}{2}\, \text{Tr}\, \text{log} \left[\left.-\frac{\delta^2 S_{\text{UV}}[\phi,\Phi]}{\delta\Phi^2}\right\rvert_{\Phi_c}\right] 
	= i c_s \text{Tr } \text{log} \left[ -P^2 +M^2 +U(\phi) \right].
\end{align}

Here, $ M $ is the mass of the heavy field $ \Phi $. $ c_{s} $ takes numerical values $ +\frac{1}{2}$ and $+1$ for real and complex scalars respectively. Now, by expanding the functional trace by inserting a complete set of momentum and spatial states, and applying Baker-Campbell-Hausdorff formula \cite{Henning:2014wua}, we get

\begin{align}
	S_{\text{eff}}^{\text{1loop(H)}}(\phi)&= i c_s \int d^4x \int  \frac{d^4k}{(2 \pi)^4} \tr \, \log \left[ -\left(k_{\mu} + \widetilde{G}_{\nu\mu}\frac{\partial}{\partial k_{\nu}}\right)^2 + M^2 + \widetilde{U}(\phi) \right],
\end{align}

where,

\begin{align}
 	\widetilde{G}_{\nu\mu} = - \sum_{n=0}^{\infty} \frac{n+1}{(n+2)!} \left[P_{\alpha_1},...,\left[P_{\alpha_n},\left[P_{\nu},P_{\mu}\right]\right]\right] \frac{\partial^n}{\partial k_{\alpha_1} ... \partial k_{\alpha_n}} ,
\end{align}

and,

\begin{align}
	\widetilde{U} = \sum_{n=0}^{\infty} \frac{1}{n!} \left[P_{\alpha_1},...,\left[P_{\alpha_n},U\right]\right] \frac{\partial^n}{\partial k_{\alpha_1} ... \partial k_{\alpha_n}}.
\end{align}

This transformation puts all the covariant derivatives in commutators and hence, the resulting effective operators are manifestly gauge invariant. Now, we expand the argument of `$log$' and replace by an integral (on $ M^2 $),

{\footnotesize
\begin{align}
	S_{\text{eff}}^{\text{1loop(H)}}(\phi)
	&= -i c_s \int d^4x \int  \frac{d^4k}{(2 \pi)^4} \int dM^2 \, \tr \left[\Delta_F\left[1+ \Delta_F\left(\left\lbrace k_{\mu}, \widetilde{G}_{\nu\mu} \frac{\partial}{\partial k_{\nu}}\right\rbrace+\widetilde{G}_{\sigma\mu}\widetilde{G}^{\sigma}_{\nu}\frac{\partial}{\partial k_{\mu}}\frac{\partial}{\partial k_{\nu}}-\widetilde{U}\right)\right] \right]^{-1},
\end{align}}

where, $ \Delta_F = \frac{1}{k^2 - M^2} $. The above equation is the master formula for gauge invariant effective operators up to arbitrary mass-dimension. In order to truncate at dimension-six, we expand the argument of `tr' and replace the momentum integrals in terms of heavy field mass. Then, the effective action formula for pure-heavy-loop up to mass-dimension six is given by,

{\footnotesize
	\begin{align}\label{eq:leff1loop}
	\mathcal{L}^{(dim-6)}_{1-loop}[\phi] =&\frac{c_s}{(4\pi)^2} \, \tr \, \Bigg\{ M^2 U +\frac{1}{M^2}\Bigg[ -\frac{1}{60} \, \big(P_{\m}G_{\m\n}'\big)^2 - \frac{1}{90} \, G_{\m\n}'G_{\n\s}'G_{\s\m}' -\frac{1}{12} \, (P_{\m}U)^2
	- \frac{1}{6}\, U^3 - \frac{1}{12}\, U G_{\m\n}'G_{\m\n}' \Bigg] \nonumber \\ 
	&+ \frac{1}{M^4} \Bigg[\frac{1}{24} \, U^4 + \frac{1}{12}\, U \big(P_{\m}U\big)^2 + \frac{1}{120}\, \big(P^2U\big)^2 +\frac{1}{24} \, \Big( U^2 G'_{\m\n}G'_{\m\n} \Big) 
	- \frac{1}{120} \, \big[(P_{\m}U),(P_{\n}U)\big] G'_{\m\n} \nonumber \\ 
	& - \frac{1}{120}\, \big[U[U,G'_{\m\n}]\big] G'_{\m\n} \Bigg] + \frac{1}{M^6} \Bigg[-\frac{1}{60} \, U^5 - \frac{1}{20} \, U^2\big(P_{\m}U\big)^2 - \frac{1}{30} \, \big(UP_{\m}U\big)^2 \Bigg] 
	+ \frac{1}{M^8} \bigg[ \frac{1}{120} \, U^6 \bigg]\Bigg\} \,.
	\end{align}}
	
Here, $P_{\mu} = i \ D_{\mu} \ \text{and} \ G'_{\mu \nu}=[D_{\mu} , D_{\nu}]$. It is important to note down that `$\tr$' in the above equation is the trace performed over the internal symmetry indices. Note that the loop-level WCs are renomalization scheme dependent and are defined in $\overline{\text{MS}}$ scheme.

\texttt{CoDEx} builds the covariant derivative operator for the heavy field from its SM quantum numbers, and constructs `$U$' from interactions present in the UV Lagrangian. Then, the field strength tensors in `$G_{\m\n}'$' are defined from the covariant derivative. For cases where, the heavy field transforms under both color and isospin groups, `$U$' would contain two pairs of indices, one each for $SU(3)_C$ and $SU(2)_L$ space of the field multiplet. Using all these information, the package evaluates the trace in Eq.~\ref{eq:leff1loop}, and gets the effective Lagrangian in an off-shell basis, which is reduced to Warsaw (or other user-defined onshell) basis using onshell relations.
\vspace{2em}

\subsection{1-Loop matching: heavy-light loops}\label{subsec:HLaction}
\vspace{1em}

The mixed heavy-light contribution is derived by expanding the UV action around the classical configuration of light fields, similar to the pure heavy-loop approach. The 1-loop effective action $\Delta S^{\text{1-loop}}_{\text{eff}}$ is defined as,

\begin{align}\label{eq:eadef}
\Delta S^{\text{1-loop}}_{\text{eff}}[\phi,\Phi_c]=\frac{i}{2}\text{Tr} \ \text{log} \left[ \left.  -\frac{\delta^2 S_{\text{UV}}[\phi,\Phi]}{\delta(\phi, \Phi)^2} \right|_{\Phi = \Phi_c(\phi)} \right].
\end{align}

And we define,

\begin{align}\label{eq:qdef}
Q_{\text{UV}}[\phi, \Phi] \equiv    -\frac{\delta^2 S_{\text{UV}}[\phi,\Phi]}{\delta(\phi, \Phi)^2}.
\end{align}

This contains quantum corrections from both heavy and light fields. Firstly, we block-diagonalize the quadratic matrix $Q_{\text{UV}}$, which is a square matrix in the field space. We define $\Delta$'s following Eq.~\ref{eq:qdef},

\begin{align}\label{eq:qmat}
Q_{\text{UV}}[\phi, \Phi] = 
\begin{pmatrix}
\Delta_{\text{H}} & \Delta_{\text{HL}}\\\Delta_{\text{LH}} & \Delta_{\text{L}}
\end{pmatrix}
\equiv{\Large \begin{pmatrix}
	-\frac{\delta^2 S_{\text{UV}}[\phi,\Phi]}{\delta \Phi^2}&-\frac{\delta^2 S_{\text{UV}}[\phi,\Phi]}{\delta\Phi\delta\phi}  \\
	-\frac{\delta^2 S_{\text{UV}}[\phi,\Phi]}{\delta\phi\delta\Phi}&-\frac{\delta^2 S_{\text{UV}}[\phi,\Phi]}{\delta\phi^2}
	\end{pmatrix}}.
\end{align}

Then, we define,

\begin{align}\label{eq:qdiag}
V=\begin{pmatrix}
\mathbb{1}  & \mathbb{0}\\
-\Delta_{\text{L}}^{-1}\Delta_{\text{LH}}&\mathbb{1}
\end{pmatrix} \hspace{3em}  \Rightarrow V^{\dagger} Q_{\text{UV}} V=
\begin{pmatrix}
\Delta_{\text{H}}-\Delta_{\text{HL}}\Delta_{\text{L}}^{-1}\Delta_{\text{LH}} & \mathbb{0}\\ \mathbb{0} & \Delta_{\text{L}}
\end{pmatrix}.
\end{align}

We use the distributive property of determinants,

\begin{align}\label{eq:ea1loop}
\Delta S^{\text{1-loop}}_{\text{eff}}[\phi,\Phi_c]&=\left.\left(\frac{i}{2}\text{Tr} \ \text{log} \left[ \Delta_{\text{H}}-\Delta_{\text{HL}}\Delta_{\text{L}}^{-1}\Delta_{\text{LH}} \right] + \frac{i}{2}\text{Tr} \ \text{log} \left[ \Delta_{\text{L}}\right]\right)\right|_{\Phi = \Phi_c}.
\end{align}

Now, we define $\Delta$'s from the UV Lagrangian $S_{\text{UV}}$ in terms of the kinetic and interaction terms,

\begin{align}
S_{\text{UV}}[\phi, \Phi] &= \int d^4x \ \mathcal{L}_{\text{UV}}[\phi,\Phi]\nonumber\\
\Rightarrow -\Delta_{\text{H}} &= \frac{\delta^2 S_{\text{UV}}[\phi, \Phi]}{\delta \Phi^2} = \int d^4x \ \frac{\delta^2 \mathcal{L}_{\text{UV}}[\phi,\Phi]}{\delta \Phi^2}=\int d^4x \ \left(\mathcal{P}^2 - M^2 + U_{\text{H}}\right),
\end{align}

where, $\mathcal{P}_{\mu} \equiv i \mathcal{D}_{\mu}$, and, $M^2$ is the squared mass matrix of $\Phi$. $U_\text{H}$ accommodates the light-field and self-interaction terms of $\Phi$ in the UV Lagrangian. Similarly,

\begin{align}
-\Delta_{\text{L}} &= \frac{\delta^2 S_{\text{UV}}[\phi, \Phi]}{\delta \phi^2} = \int d^4x \ \frac{\delta^2 \mathcal{L}_{\text{UV}}[\phi,\Phi]}{\delta \phi^2}=\int d^4x \ \left(\mathcal{P}^2 - m^2 + U_{\text{L}}\right).
\end{align}

where, $ m $ is the mass of $\phi$. $U_\text{L}$ carries the heavy-field and self-interaction of the light field. The off-diagonal elements of $Q_{\text{UV}}$ are defined,

\begin{align}
-\Delta_{\text{HL}} &= \frac{\delta^2 S_{\text{UV}}[\phi, \Phi]}{\delta\Phi \delta\phi} = \int d^4x \ \frac{\delta^2 \mathcal{L}_{\text{UV}}[\phi,\Phi]}{\delta\Phi \delta\phi}=\int d^4x \  U_{\text{HL}},
\end{align}

and,

\begin{align}
-\Delta_{\text{LH}} &= \frac{\delta^2 S_{\text{UV}}[\phi, \Phi]}{\delta\phi \delta\Phi} = \int d^4x \ \frac{\delta^2 \mathcal{L}_{\text{UV}}[\phi,\Phi]}{\delta\phi \delta\Phi}=\int d^4x \  U_{\text{LH}}.
\end{align}

$U_\text{HL}$ and $U_\text{LH}$  carry the interactions between heavy $\Phi$ and light $\phi$ fields. Now, we have defined R.H.S. of Eq.~\ref{eq:ea1loop} completely in terms of interactions present in the Lagrangian. Then the next step is to extract the heavy-light mixed loop contribution from the R.H.S. We proceed by implementing the matching condition of a UV theory and its EFT:

\begin{align}\label{eq:matchcond}
\Gamma_{\text{L,UV}}[\phi]&=\Gamma_{\text{EFT}}[\phi],
\end{align}

where, $\Gamma_{\text{L,UV}}[\phi]$ is the 1-light-particle-irreducible(1LPI) effective action calculated in the UV theory, and $\Gamma_{\text{EFT}}[\phi]$ is the 1-particle-irreducible(1PI) effective action calculated in the EFT. This matching condition is imposed order by order in perturbation. At 1-loop order in UV,

\begin{align}\label{eq:GammaUV1loop}
\Gamma_{\text{L,UV}}^{\text{1-loop}}[\phi] = \Delta S^{\text{1-loop}}_{\text{eff}}[\phi,\Phi_c]=\frac{i}{2}\text{Tr}\ \text{log} \left[\left.Q_{\text{UV}}[\phi,\Phi]\right|_{\Phi=\Phi_c}\right].
\end{align}

Whereas at 1-loop order in EFT,

\begin{align}\label{eq:eftea}
\Gamma^{\text{1-loop}}_{\text{EFT}}[\phi]=\int d^4x \ \mathcal{L}^{\text{1-loop}}_{\text{EFT}}[\phi]+\frac{i}{2}\text{Tr}\ \text{log}\left[-\frac{\delta^2 \mathcal{S_{\text{EFT}}}[\phi]}{\delta \phi^2}\right],
\end{align}

where,
\begin{align}
S_{\text{EFT}}[\phi] \equiv S_{\text{UV}}[\phi,\Phi_c]=\left.S_{\text{UV}}[\phi,\Phi]\right|_{\Phi=\Phi_c(\phi)}.
\end{align}

The 1-loop effective action in EFT contains contribution from (a) 1-loop generated operators contributing at tree-level, and (b) tree-level operators inserted at 1-loop. The R.H.S. of Eq.~\ref{eq:eftea} contains these two contributions. Substituting Eqs.~\ref{eq:GammaUV1loop} and \ref{eq:eftea} in Eq.~\ref{eq:matchcond}, we get

\begin{align}\label{eq:Left1loopreduction}
&\frac{i}{2}\text{Tr}\ \text{log} \left[\left.Q_{\text{UV}}[\phi,\Phi]\right|_{\Phi=\Phi_c}\right]=\int d^4x \ \mathcal{L}^{\text{1-loop}}_{\text{EFT}}[\phi]+\frac{i}{2}\text{Tr}\ \text{log}\left[-\frac{\delta^2 \mathcal{S_{\text{EFT}}}[\phi]}{\delta \phi^2}\right]\nonumber\\
\implies&\int d^4x \ \mathcal{L}^{\text{1-loop}}_{\text{EFT}}[\phi]=\frac{i}{2}\text{Tr}\ \text{log} \left[\left.Q_{\text{UV}}[\phi,\Phi]\right|_{\Phi=\Phi_c}\right]-\frac{i}{2}\text{Tr}\ \text{log}\left[-\frac{\delta^2 \mathcal{S_{\text{EFT}}}[\phi]}{\delta \phi^2}\right].
\end{align}

Now, we solve the second term in the R.H.S. of Eq.~\ref{eq:Left1loopreduction},

\begin{align}
-\frac{\delta^2 \mathcal{S_{\text{EFT}}}[\phi]}{\delta \phi^2} &= -\frac{\delta^2}{\delta \phi^2}\left(\left.S_{\text{UV}}[\phi,\Phi]\right|_{\Phi=\Phi_c}\right)=-\frac{\delta}{\delta \phi}\left\lbrace\left.\left(\frac{\delta S_{\text{UV}}[\phi,\Phi]}{\delta \phi} + \frac{\delta \Phi}{\delta \phi} \frac{\delta S_{\text{UV}}[\phi,\Phi]}{\delta \Phi}\right)\right|_{\Phi=\Phi_c}\right\rbrace\nonumber\\
&=-\frac{\delta}{\delta \phi}\left(\left.\frac{\delta S_{\text{UV}}[\phi,\Phi]}{\delta \phi}\right|_{\Phi=\Phi_c}\right)=\left.\left(-\frac{\delta^2 S_{\text{UV}}[\phi,\Phi]}{\delta \phi^2} - \frac{\delta \Phi}{\delta \phi} \frac{\delta^2 S_{\text{UV}}[\phi,\Phi]}{\delta \Phi \delta \phi}\right)\right|_{\Phi=\Phi_c}\nonumber\\
&=\left.\left( \Delta_{\text{L}} - \Delta_{\text{LH}}\hat{\Delta}_{\text{H}}^{-1}\Delta_{\text{HL}} \right)\right|_{\Phi=\Phi_c}.
\end{align}
	
We used the definition of $\Phi_c$ on $1^\text{st}$ line to come to $2^\text{nd}$ line, and on $2^\text{nd}$ line, we used

\begin{align}
&\frac{\delta}{\delta \phi}\left( \left.\frac{\delta S_{\text{UV}}[\phi,\Phi]}{\delta \Phi}\right|_{\Phi=\Phi_c} \right)=0\nonumber\\
\implies&\left.\left(\frac{\delta^2 S_{\text{UV}}[\phi,\Phi]}{\delta\phi \delta\Phi}+\frac{\delta \Phi}{\delta \phi} \frac{\delta^2 S_{\text{UV}}[\phi,\Phi]}{\delta \Phi^2}\right)\right|_{\Phi=\Phi_c}=0\nonumber\\
\implies&\left.\left(\frac{\delta \Phi}{\delta \phi} \Delta_{\text{H}} \right)\right|_{\Phi=\Phi_c}=\left.\left(-\Delta_{\text{LH}}\right)\right|_{\Phi=\Phi_c}\nonumber\\
\implies&\left.\left(\frac{\delta \Phi}{\delta \phi} \right)\right|_{\Phi=\Phi_c}= \left.\left(-\Delta_{\text{LH}}\right)\right|_{\Phi=\Phi_c}\left(\left.\Delta_{\text{H}}\right|_{\Phi=\Phi_c}\right)^{-1}.
\end{align}

to arrive at the $3^\text{rd}$ line of Eq.~\ref{eq:Left1loopreduction}. The hat is put on $\Delta_{\text{H}}$ to indicate that it is a local operator in the EFT. We rewrite Eq.~\ref{eq:Left1loopreduction},

\begin{align}
&\int d^4x \ \mathcal{L}^{\text{1-loop}}_{\text{EFT}}[\phi]\nonumber\\
=&\left.\left(\frac{i}{2}\text{Tr} \ \text{log} \left[ \Delta_{\text{H}}-\Delta_{\text{HL}}\Delta_{\text{L}}^{-1}\Delta_{\text{LH}} \right] + \frac{i}{2}\text{Tr} \ \text{log} \left[ \Delta_{\text{L}}\right]-\frac{i}{2}\text{Tr}\ \text{log}\left[\Delta_{\text{L}} - \Delta_{\text{LH}}\hat{\Delta}_{\text{H}}^{-1}\Delta_{\text{HL}}\right]\right)\right|_{\Phi = \Phi_c}\nonumber\\
=&\left.\left(\frac{i}{2}\text{Tr} \ \text{log} \left[ \Delta_{\text{H}}-\Delta_{\text{HL}}\Delta_{\text{L}}^{-1}\Delta_{\text{LH}} \right] + \frac{i}{2}\text{Tr} \ \text{log} \left[ \Delta_{\text{L}}\right]\right.\right.\nonumber\\
&\left.\left.-\frac{i}{2}\text{Tr}\ \text{log}\left[\Delta_{\text{L}}\right]-\frac{i}{2}\text{Tr}\ \text{log}\left[\mathbb{1} - \Delta_{\text{L}}^{-1}\Delta_{\text{LH}}\hat{\Delta}_{\text{H}}^{-1}\Delta_{\text{HL}}\right]\right)\right|_{\Phi = \Phi_c}\nonumber\\
=&\left.\left(\frac{i}{2}\text{Tr} \ \text{log} \left[ \Delta_{\text{H}}-\Delta_{\text{HL}}\Delta_{\text{L}}^{-1}\Delta_{\text{LH}} \right]-\frac{i}{2}\text{Tr}\ \text{log}\left[\mathbb{1} - \Delta_{\text{L}}^{-1}\Delta_{\text{LH}}\hat{\Delta}_{\text{H}}^{-1}\Delta_{\text{HL}}\right]\right)\right|_{\Phi = \Phi_c}.
\end{align}

The contribution from loops with light propagators only ($\text{Tr}\ \text{log}\ \left[ \Delta_{\text{L}} \right]$) , cancels while matching the UV theory to its EFT, as expected. We further reduce using  Sylvester's determinant identity on the second term,

\begin{align}\label{eq:heavylightea}
&\left.\left(\frac{i}{2}\text{Tr} \ \text{log} \left[ \Delta_{\text{H}}-\Delta_{\text{HL}}\Delta_{\text{L}}^{-1}\Delta_{\text{LH}} \right]-\frac{i}{2}\text{Tr}\ \text{log}\left[\hat{\Delta}_{\text{H}} - \Delta_{\text{HL}}\Delta_{\text{L}}^{-1}\Delta_{\text{LH}}\right]+\frac{i}{2}\text{Tr}\ \text{log}\left[\hat{\Delta}_{\text{H}}\right]\right)\right|_{\Phi = \Phi_c}\nonumber\\
=&\int d^4x \ \mathcal{L}^{\text{1-loop}}_{\text{EFT}}[\phi].
\end{align}

Now, we have the contribution from both the heavy-light mixed loops and loops containing only heavy field propagators. 
One may cross-check the R.H.S. by substituting $\Delta_{\text{L}}=\Delta_{\text{HL}}=\Delta_{\text{LH}}=0$ and check that we get $\left.\left(\frac{i}{2}\text{Tr} \ \text{log} \left[ \Delta_{\text{H}}\right]\right)\right|_{\Phi = \Phi_c}$ , which contains loops with heavy field propagators only. The next step is to expand the R.H.S. of Eq.~\ref{eq:heavylightea} in terms of the covariant derivative operator, mass matrices and the quadratic matrix U's, with a goal to construct an 1-loop effective action formula. We use the key observation made in Ref.~\cite{Fuentes-Martin:2016uol}, that the effective action can be split into ``hard'' and ``soft'' regions, and these regions produce the IR- and UV-divergent integrals, respectively. The hard region is captured when the loop integral is calculated in the limit $ M^2 \gg m^2 $. So we solve,

\begin{align}\label{eq:lefthard}
&\int d^4x \ \mathcal{L}^{\text{1-loop}}_{\text{EFT}}[\phi]=\left.\frac{i}{2}\text{Tr} \ \text{log} \left[ \Delta_{\text{H}}-\Delta_{\text{HL}}\Delta_{\text{L}}^{-1}\Delta_{\text{LH}} \right]\right|_{\text{hard}}\nonumber\\
=&\left.\frac{i}{2} \int \frac{d^dq}{(2\pi)^d} \, \text{tr} \ \text{log} \left[ \Delta_{\text{H}}-\Delta_{\text{HL}}\Delta_{\text{L}}^{-1}\Delta_{\text{LH}} \right]\right|_{\text{hard}}\nonumber\\
=&\left.\frac{i}{2} \int \frac{d^dq}{(2\pi)^d} \, \text{tr} \ \text{log} \left[-(P^2 -2P.q +q^2) +M^2 + U_{\text{H}}-\Delta_{\text{HL}}\Delta_{\text{L}}^{-1}\Delta_{\text{LH}} \right]\right|_{\text{hard}}\nonumber\\
=&-\frac{i}{2} \sum_{n=1}^{\infty}\frac{1}{n} \int \frac{d^dq}{(2\pi)^d} \, \text{tr} \ \text{log} \left[(q^2 - m^2)^{-1}\right.\nonumber\\
&\left.\left. \left\lbrace-P^2 +2P.q+ \left.U_{\text{H}}\right|_{P\rightarrow P-q}-\left.\Delta_{\text{HL}}\Delta_{\text{L}}^{-1}\Delta_{\text{LH}}\right|_{P\rightarrow P-q}\right\rbrace \right]^n\right|_{\text{hard}}.
\end{align}

The last equality is true up to an additive constant because we factored out $ (q^2 -M^2 ) $ and implemented the logarithmic expansion on the last step. Here, we expand the sum and keep terms encompassing all the dimension-6 (and lower) effective operators. To reduce these further, we follow the method of covariant diagrams as in Ref.~\cite{Zhang:2016pja} to arrive at the master formulae. The master trace formulae and their corresponding integration factors are tabulated in tabs.~\ref{tab:allu}-\ref{tab:fourp2}. These formulae are derived for single heavy field or multiple mass-degenerate heavy fields extension to the SM. We have compared these with that presented in Ref.~\cite{Ellis:2017jns} for mass-degenerate cases and we have found complete agreement.

\begin{table}[h]
		\caption{Effective action formulae with only $ U $'s. See tab.~\ref{tab:factor} for integration factor. The formula terms tabulated here, similar to the pure-heavy-loop effective action, are calculated using dimensional regularization and in $\overline{MS}$ scheme. The matching scale is set equal to heavy field mass here.}
		\label{tab:allu}
		\centering
		\renewcommand*{\arraystretch}{1.6}
		\subfloat{
			\begin{tabular}{|c|c|l|}
				\hline \hline
				Factors & Formulae \\ \hline \hline
				$- i c_{s} \mathcal{I}^{11}$  & \text{tr}$\left(U_{HL} U_{LH}\right)$                 \\ \hline
				$- i c_{s} \mathcal{I}^{21} $ & \text{tr}$\left(U_H U_{HL} U_{LH}\right)$             \\ \hline
				$- i c_{s} \mathcal{I}^{31}$  & \text{tr}$\left(U_H U_H U_{HL} U_{LH}\right)$         \\ \hline
				$- i c_{s} \mathcal{I}^{41}$  & \text{tr}$\left(U_H U_H U_H U_{HL} U_{LH}\right)$     \\ \hline
				$- i c_{s} \mathcal{I}^{51}$  & \text{tr}$\left(U_H U_H U_H U_H U_{HL} U_{LH}\right)$ \\ \hline
				$- i c_{s} \mathcal{I}^{12} $ & \text{tr}$\left(U_{HL} U_L U_{LH}\right)$             \\ \hline
				$- i c_{s} \mathcal{I}^{22}$  & \text{tr}$\left(U_H U_{HL} U_L U_{LH}\right)$         \\ \hline
				$- i c_{s} \mathcal{I}^{32}$  & \text{tr}$\left(U_H U_H U_{HL} U_L U_{LH}\right)$     \\ \hline
				$- i c_{s} \mathcal{I}^{42}$  & \text{tr}$\left(U_H U_H U_H U_{HL} U_L U_{LH}\right)$ \\ \hline
				$- i c_{s} \mathcal{I}^{13}$  & \text{tr}$\left(U_{HL} U_L U_L U_{LH}\right)$         \\ \hline
				$- i c_{s} \mathcal{I}^{23}$  & \text{tr}$\left(U_H U_{HL} U_L U_L U_{LH}\right)$     \\ \hline
				$- i c_{s} \mathcal{I}^{33}$  & \text{tr}$\left(U_H U_H U_{HL} U_L U_L U_{LH}\right)$ \\ \hline
				$- i c_{s} \mathcal{I}^{14}$  & \text{tr}$\left(U_{HL} U_L U_L U_L U_{LH}\right)$     \\ \hline \hline
			\end{tabular} 
		}
		\subfloat{
			\begin{tabular}{|c|l|}
				\hline \hline
				Factors & Formulae \\ \hline \hline
				$- i c_{s} \mathcal{I}^{24}$       & \text{tr}$\left(U_H U_{HL} U_L U_L U_L U_{LH}\right)$             \\ \hline
				$- i c_{s} \mathcal{I}^{15}$       & \text{tr}$\left(U_{HL} U_L U_L U_L U_L U_{LH}\right)$             \\ \hline
				$- i c_{s} \frac{1}{2} \mathcal{I}^{22}$ & \text{tr}$\left(U_{HL} U_{LH} U_{HL} U_{LH}\right)$               \\ \hline
				$- i c_{s} \mathcal{I}^{32}$       & \text{tr}$\left(U_H U_{HL} U_{LH} U_{HL} U_{LH}\right)$           \\ \hline
				$- i c_{s} \frac{1}{2} \mathcal{I}^{42}$ & \text{tr}$\left(U_H U_{HL} U_{LH} U_H U_{HL} U_{LH}\right)$       \\ \hline
				$- i c_{s} \mathcal{I}^{42}$       & \text{tr}$\left(U_H U_H U_{HL} U_{LH} U_{HL} U_{LH}\right)$       \\ \hline
				$- i c_{s} \mathcal{I}^{23}$       & \text{tr}$\left(U_{HL} U_{LH} U_{HL} U_L U_{LH}\right)$           \\ \hline
				$- i c_{s} \mathcal{I}^{33}$       & \text{tr}$\left(U_H U_{HL} U_{LH} U_{HL} U_L U_{LH}\right)$       \\ \hline
				$- i c_{s} \mathcal{I}^{33}$       & \text{tr}$\left(U_H U_{HL} U_L U_{LH} U_{HL} U_{LH}\right)$       \\ \hline
				$- i c_{s} \frac{1}{2}\mathcal{I}^{24}$  & \text{tr}$\left(U_{HL} U_L U_{LH} U_{HL} U_L U_{LH}\right)$       \\ \hline
				$- i c_{s} \mathcal{I}^{24}$       & \text{tr}$\left(U_{HL} U_{LH} U_{HL} U_L U_L U_{LH}\right)$       \\ \hline
				$- i c_{s} \frac{1}{3} \mathcal{I}^{33}$ & \text{tr}$\left(U_{HL} U_{LH} U_{HL} U_{LH} U_{HL} U_{LH}\right)$ \\ \hline \hline
			\end{tabular}	
		}
	\end{table}

\begin{table}[h]
	\caption{Effective action formulae with two $ P $'s  (Part-I).}
	\label{tab:twop}
	\centering
	\renewcommand*{\arraystretch}{1.8}
	\begin{tabular}{|l|l|}
		\hline \hline
				Factors & Formulae \\ \hline \hline
		$f^{^{2}}_{_{PU}}=- i c_{s} \ 2 \, \mathcal{I}[q^2]^{22} $                                                                               & $\text{tr}\left([\mathcal{P}_{\mu}, U_{HL}] [\mathcal{P}^{\mu}, U_{LH}] \right)$                                       \\ \hline\hline
		$f^{^{3(H)}}_{_{PU,a}}=- i c_{s} \,   2\left(\mathcal{I}[q^2]^{32} + \mathcal{I}[q^2]^{41}  \right)  $& $\text{tr}\left([\mathcal{P}_{\mu},U_{H}] [\mathcal{P}^{\mu},U_{HL}] U_{LH} \right)$                                   \\ \cline{1-2}
		$f^{^{3(H)}}_{_{PU,b}}=- i c_{s} \,   2\left(\mathcal{I}[q^2]^{32} + \mathcal{I}[q^2]^{41}  \right)  $& $\text{tr}\left([\mathcal{P}_{\mu},U_{H}] U_{HL} [\mathcal{P}^{\mu},U_{LH}]\right)$                                    \\ \cline{1-2}
		$f^{^{3(H)}}_{_{PU,c}}=- i c_{s} \,  4 \, \mathcal{I}[q^2]^{32} $                                                                        & $\text{tr}\left(U_{H}[\mathcal{P}_{\mu},U_{HL}][\mathcal{P}_{\mu},U_{LH}]\right)$                                      \\ \hline\hline
		$f^{^{3(L)}}_{_{PU,a}}=- i c_{s} \,   2\left(\mathcal{I}[q^2]^{14} + \mathcal{I}[q^2]^{23}  \right)   $& $\text{tr}\left([\mathcal{P}_{\mu},U_{HL}] [\mathcal{P}^{\mu},U_{L}]U_{LH}\right)$                                     \\ \cline{1-2}
		$f^{^{3(L)}}_{_{PU,b}}=- i c_{s} \,   4 \, \mathcal{I}[q^2]^{23}  $                                                                      & $\text{tr}\left([\mathcal{P}_{\mu},U_{HL}] U_{L} [\mathcal{P}^{\mu},U_{LH}]\right)$                                    \\ \cline{1-2}
		$f^{^{3(L)}}_{_{PU,c}}=- i c_{s} \,   2\left(\mathcal{I}[q^2]^{14} + \mathcal{I}[q^2]^{23}  \right)  $  & $\text{tr}\left(U_{HL}[\mathcal{P}_{\mu},U_{L}][\mathcal{P}^{\mu},U_{LH}]\right)$                                      \\ \hline\hline
	\end{tabular}
\end{table}

\begin{table}[h]
	\caption{Effective action formulae with four $ P $'s.}
	\label{tab:fourp1}
	\centering
	\renewcommand*{\arraystretch}{2}
	\begin{tabular}{|l|l|}
		\hline \hline
		Factors & Formulae \\ \hline \hline
		$f^{^{2}}_{_{PPU,a}}=- i c_{s} \,   4 \left(\mathcal{I}[q^4]^{33}+2\mathcal{I}[q^4]^{42} + 2 \, \mathcal{I}[q^4]^{51} \right)  $ & $\text{tr}\left(G'_{\mu\nu} G'^{\mu\nu} U_{HL} U_{LH}\right)$   \\ \cline{1-2}
		$f^{^{2}}_{_{PPU,b}}=- i c_{s} \,   4 \left(\mathcal{I}[q^4]^{33}+2\mathcal{I}[q^4]^{24} + 2 \, \mathcal{I}[q^4]^{15} \right)  $ & $\text{tr}\left(G'_{\mu\nu} G'^{\mu\nu} U_{LH} U_{HL}\right)$   \\ \cline{1-2}
		$f^{^{2}}_{_{PPU,c}}=- i c_{s} \,   8\,\mathcal{I}[q^4]^{33}  $            & $\text{tr}\left(G'_{\nu\mu} [\mathcal{P}^{\mu}, U_{HL}] [\mathcal{P}^{\nu},U_{LH}]\right)$ \\ \cline{1-2}
		$f^{^{2}}_{_{PPU,d}}=- i c_{s} \,   8\,\mathcal{I}[q^4]^{33}  $           & $\text{tr}\left(G'_{\nu\mu} [\mathcal{P}^{\mu}, U_{LH}] [\mathcal{P}^{\nu},U_{HL}]\right)$ \\ \cline{1-2}
		$f^{^{2}}_{_{PPU,e}}=- i c_{s} \,  8\,\mathcal{I}[q^4]^{33}   $ & $\text{tr}\left(\left[\mathcal{P}_{\mu}, \left[\mathcal{P}_{\mu} ,U_{HL}\right]\right]  \left[\mathcal{P}^{\nu}, \left[\mathcal{P}^{\nu}, U_{LH}\right]\right]\right)$  \\ \cline{1-2}
		$f^{^{2}}_{_{PPU,f}}=- i c_{s} \, 4 \, \left(\mathcal{I}[q^4]^{33} + \mathcal{I}[q^4]^{42} \right)$ & $\text{tr}\left([\mathcal{P}^{\mu},U_{HL}]U_{LH} [\mathcal{P}^{\nu}, G'_{\mu\nu}] \right)$   \\ \cline{1-2}
		$f^{^{2}}_{_{PPU,g}}=- i c_{s} \, 4 \, \left(\mathcal{I}[q^4]^{33} + \mathcal{I}[q^4]^{42} \right)$ & $\text{tr}\left(U_{HL}\left[\mathcal{P}^{\mu},U_{LH}\right] [\mathcal{P}^{\nu}, G'_{\nu\mu}] \right)$   \\ \cline{1-2}
		$f^{^{2}}_{_{PPU,h}}=- i c_{s} \, 4 \, \left(\mathcal{I}[q^4]^{33} + \mathcal{I}[q^4]^{24} \right)$ & $\text{tr}\left(U_{LH}\left[\mathcal{P}^{\mu},U_{HL}\right] [\mathcal{P}^{\nu}, G'_{\nu\mu}] \right)$   \\ \cline{1-2}
		$f^{^{2}}_{_{PPU,i}}=- i c_{s} \, 4 \, \left(\mathcal{I}[q^4]^{33} + \mathcal{I}[q^4]^{24} \right)$ & $\text{tr}\left(\left[\mathcal{P}^{\mu},U_{LH}\right]U_{HL} [\mathcal{P}^{\nu}, G'_{\mu\nu}] \right)$   \\ \cline{1-2}
		\hline \hline
	\end{tabular}
\end{table}

\begin{table}[h]
	\caption{Effective action formulae with two $ P $'s  (Part-II).}
	\label{tab:fourp2}
	\centering
	\renewcommand*{\arraystretch}{1.6}
	\begin{tabular}{|l|l|}
		\hline \hline
		Factors & Formulae \\ \hline \hline
		$ f^{^{4(HH)}}_{_{PU,a}}=- i c_{s} \,   2 \, \left( \mathcal{I}[q^2]^{42} + 2 \, \mathcal{I}[q^2]^{51}  \right)  $  &  $\text{tr}\left([\mathcal{P}_{\mu},U_{H}][\mathcal{P}^{\mu},U_{H}] U_{HL} U_{LH}\right)$  \\\cline{1-2}
		$ f^{^{4(HH)}}_{_{PU,b}}=- i c_{s} \,   4 \, \left( \mathcal{I}[q^2]^{42} +  \mathcal{I}[q^2]^{51}  \right)  $  &  $\text{tr}\left([\mathcal{P}_{\mu},U_{H}]U_{H} [\mathcal{P}^{\mu},U_{HL}] U_{LH}\right)$  \\\cline{1-2}
		$ f^{^{4(HH)}}_{_{PU,c}}=- i c_{s} \,   2 \, \left( \mathcal{I}[q^2]^{42} + 2 \, \mathcal{I}[q^2]^{51}  \right)  $  &  $\text{tr}\left([\mathcal{P}_{\mu},U_{H}]U_{H} U_{HL}[\mathcal{P}^{\mu},U_{LH}]\right)$  \\\cline{1-2}
		$ f^{^{4(HH)}}_{_{PU,d}}=- i c_{s} \,   4 \, \left( \mathcal{I}[q^2]^{42} + \mathcal{I}[q^2]^{51}  \right)  $  &  $\text{tr}\left(U_{H}[\mathcal{P}_{\mu},U_{H}]U_{HL} [\mathcal{P}^{\mu},U_{LH}]\right)$  \\\cline{1-2}
		$ f^{^{4(HH)}}_{_{PU,e}}=- i c_{s} \,   6 \,  \mathcal{I}[q^2]^{42}  $  &  $\text{tr}\left(U_{H} U_{H} [\mathcal{P}_{\mu},U_{HL}] [\mathcal{P}^{\mu},U_{LH}]\right)$  \\\cline{1-2}
		$ f^{^{4(HH)}}_{_{PU,f}}=- i c_{s} \,  2 \, \left( \mathcal{I}[q^2]^{42} + 2 \, \mathcal{I}[q^2]^{51}  \right)   $	&$\text{tr}\left(U_{H} [\mathcal{P}_{\mu},U_{H}]  [\mathcal{P}^{\mu}, U_{HL}]U_{LH}\right)$\\
		\hline \hline
		$ f^{^{4(HL)}}_{_{PU,a}}=- i c_{s} \,  2 \, \left( \mathcal{I}[q^2]^{33} +  \, \mathcal{I}[q^2]^{42}  \right)   $&$\text{tr}\left([\mathcal{P}_{\mu},U_{H}] [\mathcal{P}^{\mu}, U_{HL}]U_{L} U_{LH}\right)$\\\cline{1-2}
		$ f^{^{4(HL)}}_{_{PU,b}}=- i c_{s} \,  2 \, \left( \mathcal{I}[q^2]^{33} + 2 \, \mathcal{I}[q^2]^{42}  \right)   $&$\text{tr}\left([\mathcal{P}_{\mu},U_{LH}]  [\mathcal{P}^{\mu}, U_{H}]U_{HL}U_{L}\right)$\\\cline{1-2}
		$ f^{^{4(HL)}}_{_{PU,c}}=- i c_{s} \,  2 \, \left( \mathcal{I}[q^2]^{24} + 2 \, \mathcal{I}[q^2]^{33}  \right)   $&$\text{tr}\left(U_{H} U_{HL}[\mathcal{P}_{\mu},U_{L}]  [\mathcal{P}^{\mu}, U_{LH}]\right)$\\\cline{1-2}
		$ f^{^{4(HL)}}_{_{PU,d}}=- i c_{s} \,  2 \, \left( \mathcal{I}[q^2]^{24} + 2 \, \mathcal{I}[q^2]^{33}  \right)   $&$\text{tr}\left(U_{H} [\mathcal{P}_{\mu},U_{HL}]  [\mathcal{P}^{\mu}, U_{L}]U_{LH}\right)$\\\cline{1-2}
		$ f^{^{4(HL)}}_{_{PU,e}}=- i c_{s} \,  2 \, \left( \mathcal{I}[q^2]^{24} + 2 \, \mathcal{I}[q^2]^{33} +\mathcal{I}[q^2]^{42}  \right)   $&$\text{tr}\left( [\mathcal{P}_{\mu},U_{H}]U_{HL}  [\mathcal{P}^{\mu}, U_{L}]U_{LH}\right)$\\\cline{1-2}
		$ f^{^{4(HL)}}_{_{PU,f}}=- i c_{s} \,  8 \, \left( \mathcal{I}[q^2]^{33} \right)   $&$\text{tr}\left(U_{H} [\mathcal{P}_{\mu},U_{HL}] U_{L} [\mathcal{P}^{\mu}, U_{LH}] \right)$\\\cline{1-2}
		\hline \hline
		$ f^{^{4(LL)}}_{_{PU,a}}=- i c_{s} \,   2 \left( 2 \mathcal{I}[q^2]^{15} + \mathcal{I}[q^2]^{24} \right)  $  &  $\text{tr}\left([\mathcal{P}_{\mu},U_{HL}][\mathcal{P}^{\mu},U_{L}]U_{L}U_{LH}\right)$  \\\cline{1-2}
		$ f^{^{4(LL)}}_{_{PU,b}}=- i c_{s} \,  4 \left(\mathcal{I}[q^2]^{15} + \mathcal{I}[q^2]^{24} \right)  $  &  $\text{tr}\left([\mathcal{P}_{\mu},U_{HL}]U_{L}[\mathcal{P}^{\mu},U_{L}]U_{LH}\right)$  \\\cline{1-2}
		$ f^{^{4(LL)}}_{_{PU,c}}=- i c_{s} \,   6 \mathcal{I}[q^2]^{24}  $  &  $\text{tr}\left([\mathcal{P}_{\mu},U_{HL}]U_{L}U_{L}[\mathcal{P}^{\mu},U_{LH}]\right)$  \\\cline{1-2}
		$ f^{^{4(LL)}}_{_{PU,d}}=- i c_{s} \,   2 \, \left( 2 \mathcal{I}[q^2]^{15} + \mathcal{I}[q^2]^{24} \right)  $  &  $\text{tr}\left(U_{HL}[\mathcal{P}_{\mu},U_{L}][\mathcal{P}^{\mu},U_{L}]U_{LH}\right)$  \\\cline{1-2}
		$ f^{^{4(LL)}}_{_{PU,e}}=- i c_{s} \,   4 \left(\mathcal{I}[q^2]^{15} + \mathcal{I}[q^2]^{24} \right)  $  &  $\text{tr}\left(U_{HL}[\mathcal{P}_{\mu},U_{L}]U_{L}[\mathcal{P}^{\mu},U_{LH}]\right)$  \\\cline{1-2}
		$ f^{^{4(LL)}}_{_{PU,f}}=- i c_{s} \,   2 \left( 2 \mathcal{I}[q^2]^{15} + \mathcal{I}[q^2]^{24} \right)  $  &  $\text{tr}\left(U_{HL}U_{L}[\mathcal{P}_{\mu},U_{L}][\mathcal{P}^{\mu},U_{LH}]\right)$  \\
		\hline \hline
		$f^{^{4(00)}}_{_{PU,a}}=- i c_{s} \,   \left(\mathcal{I}[q^2]^{24} + 2 \, \mathcal{I}[q^2]^{33} + \mathcal{I}[q^2]^{42} \right)  $  &  $\text{tr}\left([\mathcal{P}_{\mu},U_{HL}]U_{LH}[\mathcal{P}^{\mu},U_{HL}]U_{LH}\right)$  \\\cline{1-2}
		$f^{^{4(00)}}_{_{PU,b}}=- i c_{s} \,   2\, \left(\mathcal{I}[q^2]^{24} +2 \, \mathcal{I}[q^2]^{33} \right)  $  &  $\text{tr}\left([\mathcal{P}_{\mu},U_{HL}][\mathcal{P}^{\mu},U_{LH}]U_{HL}U_{LH}\right)$  \\\cline{1-2}
		$f^{^{4(00)}}_{_{PU,c}}=- i c_{s} \,   2\, \left(2\,\mathcal{I}[q^2]^{33} +  \mathcal{I}[q^2]^{42}\right)  $  &  $\text{tr}\left([\mathcal{P}_{\mu},U_{HL}]U_{LH}U_{HL}[\mathcal{P}^{\mu},U_{LH}]\right)$  \\\cline{1-2}
		$f^{^{4(00)}}_{_{PU,d}}=- i c_{s} \,   \left(\mathcal{I}[q^2]^{24} + 2 \, \mathcal{I}[q^2]^{33} + \mathcal{I}[q^2]^{42} \right)  $  &  $\text{tr}\left(U_{HL}[\mathcal{P}_{\mu},U_{LH}]U_{HL}[\mathcal{P}^{\mu},U_{LH}]\right)$  \\
		\hline \hline
	\end{tabular}
\end{table}

\begin{table}[h]
	\caption{Table of Integration factor ($\mathcal{I}$). Matching scale is equal to the heavy field mass $ \mu = M $.}
	\label{tab:factor}
	\centering
	\renewcommand*{\arraystretch}{1.6}
	\begin{tabular}{|c|c||c|c|}
		\hline \hline
		$\left.\mathcal{I}\left[q^{2n}\right]^{\alpha\beta}\right|_{n=0}$& Factor $ \times \frac{i}{16 \pi^2} $ &				$\left.\mathcal{I}\left[q^{2n}\right]^{\alpha\beta}\right|_{n=1,2}$& 			Factor     $ \times \frac{i}{16 \pi^2} $             \\ \hline \hline
		$ \mathcal{I}^{11} $&					 $ 1 $ &													$ \mathcal{I}\left[q^{2}\right]^{13} $ &							 $ \frac{3}{8 M^2} $ \\ \hline
		$ \mathcal{I}^{12} $& 					$ \frac{1}{M^2} $ 	&						  $ \mathcal{I}\left[q^{2}\right]^{22} $ & 						$ -\frac{1}{4 M^2} $ \\ \hline		
		$ \mathcal{I}^{21} $& 					$ -\frac{1}{M^2} $ & 						  $ \mathcal{I}\left[q^{2}\right]^{31} $&						 $ -\frac{1}{8 M^2} $ \\ \hline	
		$ \mathcal{I}^{13} $&					 $ \frac{1}{M^4} $ &						  $ \mathcal{I}\left[q^{2}\right]^{14} $ &							$ \frac{3}{8 M^4} $ \\ \hline	
		$ \mathcal{I}^{22} $&					 $ -\frac{2}{M^4} $ &						 $ \mathcal{I}\left[q^{2}\right]^{23} $ &							$ -\frac{5}{8 M^4} $ \\ \hline
		$ \mathcal{I}^{31} $& 					 $ \frac{1}{2 M^4} $ &						$ \mathcal{I}\left[q^{2}\right]^{32} $ & 							$ \frac{1}{8 M^4} $ \\ \hline		
		$ \mathcal{I}^{14} $& 					 $ \frac{1}{M^6} $ &							$ \mathcal{I}\left[q^{2}\right]^{41} $&					 $ \frac{1}{24 M^4} $ \\ \hline	
		$ \mathcal{I}^{23} $&					 $ -\frac{3}{M^6} $ &						 $ \mathcal{I}\left[q^{2}\right]^{15} $ &							 $ \frac{3}{8 M^6} $ \\ \hline	
		$ \mathcal{I}^{32} $&					 $ \frac{5}{2 M^6} $ &						 $ \mathcal{I}\left[q^{2}\right]^{24} $ &							 $ -\frac{1}{M^6} $ \\ \hline
		$ \mathcal{I}^{41} $& 					$ -\frac{1}{3 M^6} $&					 $ \mathcal{I}\left[q^{2}\right]^{33} $ & 							$ \frac{3}{4 M^6} $ \\ \hline		
		$ \mathcal{I}^{15} $& 					$ \frac{1}{M^8} $ & 							$ \mathcal{I}\left[q^{2}\right]^{42} $&					 $ -\frac{1}{12 M^6} $ \\ \hline	
		$ \mathcal{I}^{24} $&					 $ -\frac{4}{M^8} $ &							$ \mathcal{I}\left[q^{2}\right]^{51} $ &						$ -\frac{1}{48 M^6} $ \\ \hline	
		$ \mathcal{I}^{33} $&					 $ \frac{11}{2 M^8} $ &						 $ \mathcal{I}\left[q^{2}\right]^{51} $ &						$ -\frac{1}{48 M^6} $ \\ \hline	
		$ \mathcal{I}^{42} $&					 $ -\frac{17}{6 M^8} $ &					 $ \mathcal{I}\left[q^{4}\right]^{15} $ &						$ \frac{11}{144 M^4} $ \\ \hline	
		$ \mathcal{I}^{51} $&					 $ \frac{1}{4 M^8} $ &						 $ \mathcal{I}\left[q^{4}\right]^{24} $ &						$ -\frac{17}{144 M^4} $ \\ \hline	
		&&																																	   $ \mathcal{I}\left[q^{4}\right]^{33} $ &						$ \frac{1}{48 M^4} $ \\ \cline{3-4}
		&&   																																	$ \mathcal{I}\left[q^{4}\right]^{42} $ &						$ \frac{1}{144 M^4} $ \\ \cline{3-4}
		&&   																																	$ \mathcal{I}\left[q^{4}\right]^{42} $ &						$ \frac{1}{288 M^4} $ \\ \hline	\hline
	\end{tabular}
\end{table}

\clearpage

\vspace{1cm}

\section{Matching results for scalar extended SM scenarios}\label{sec:remainingmodels}

\subsection{Colour Singlet Heavy Scalars}
Here, we have given the exhaustive sets of effective operators and the associated WCs that are emerged after integrating out the colour singlet heavy scalars up to 1-loop including the heavy-light mixing.
\subsubsection{Real Singlet: $\mathcal{S} \equiv (1_C,1_L, \left. 0\right\vert_Y$) }

Here, we have extended the SM by a real gauge singlet scalar ($\mathcal{S} $), and the modified Lagrangian involving this heavy field is written as \cite{Zhang:2016pja,Jiang:2018pbd,Dawson:2017vgm,Haisch:2020ahr}:
{\small
	\begin{align}
	\label{RSSLag}
	\mathcal{L}_{\mathcal{S}} & = \mathcal{L}_{_\text{SM}}^{d\leq 4} \; + \ \frac{1}{2} \ (\mathcal{\partial}_{\mu} \mathcal{S} )^{2} \ - \ \frac{1}{2} \ m_{\mathcal{S}}^{2} \, \mathcal{S}^{2} - c_{\mathcal{S}} |H|^{2} \mathcal{S}  - \frac{1}{2} \kappa_{\mathcal{S}}|H|^{2}  \mathcal{S}^{2} - \frac{1}{3!} \, \mu_{\mathcal{S}} \, \mathcal{S}^{3}  - \frac{1}{4!} \, \lambda_{\mathcal{S}} \, \mathcal{S}^{4}\,.
	\end{align}
}
Here, $m_{\mathcal{S}}$ is  mass of the heavy field ($\mathcal{S} $).  This model contains four BSM parameters $c_{\mathcal{S}}, \kappa_{\mathcal{S}}, \mu_{\mathcal{S}},  \lambda_{\mathcal{S}}$, and the WCs are functions of these parameters along with the SM ones, see tab.~\ref{tab:RSS}.

\begin{table*}[h!]
	\caption{\small Warsaw basis  effective operators and the associated WCs that emerge after integrating-out the heavy field $\mathcal{S} : (1,1,0)$. See caption of tab.~\ref{tab:H2} for colour coding.}
	\label{tab:RSS}
	\scriptsize
	\centering
	\renewcommand{\arraystretch}{1.8}
	\subfloat{
		\begin{tabular}{|c|c|}
			\hline \hline
			Dim-6 Ops.&Wilson coefficients\\
			\hline \hline
			$Q_{\text{dH}}$  &  $-\frac{3 Y_d^{\text{SM}} c_{\mathcal{S}}^2 \kappa _{\mathcal{S}}}{64 \pi ^2 m_{\mathcal{S}}^4}+\frac{Y_d^{\text{SM}} c_{\mathcal{S}}^3 \mu _{\mathcal{S}}}{64 \pi ^2 m_{\mathcal{S}}^6}-\frac{9 Y_d^{\text{SM}} c_{\mathcal{S}}^4}{64 \pi ^2 m_{\mathcal{S}}^6}$  \\
			&$+\frac{29 \lambda _H^{\text{SM}} Y_d^{\text{SM}} c_{\mathcal{S}}^2}{192 \pi ^2 m_{\mathcal{S}}^4}$\\
			\hline
			$Q_{\text{eH}}$  &  $-\frac{3 Y_e^{\text{SM}} c_{\mathcal{S}}^2 \kappa _{\mathcal{S}}}{64 \pi ^2 m_{\mathcal{S}}^4}+\frac{Y_e^{\text{SM}} c_{\mathcal{S}}^3 \mu _{\mathcal{S}}}{64 \pi ^2 m_{\mathcal{S}}^6}-\frac{9 Y_e^{\text{SM}} c_{\mathcal{S}}^4}{64 \pi ^2 m_{\mathcal{S}}^6}$  \\
			&$+\frac{29 \lambda _H^{\text{SM}} Y_e^{\text{SM}} c_{\mathcal{S}}^2}{192 \pi ^2 m_{\mathcal{S}}^4}$\\
			\hline
			$Q_{\text{uH}}$  &  $-\frac{3 Y_u^{\text{SM}} c_{\mathcal{S}}^2 \kappa _{\mathcal{S}}}{64 \pi ^2 m_{\mathcal{S}}^4}+\frac{Y_u^{\text{SM}} c_{\mathcal{S}}^3 \mu _{\mathcal{S}}}{64 \pi ^2 m_{\mathcal{S}}^6}-\frac{9 Y_u^{\text{SM}} c_{\mathcal{S}}^4}{64 \pi ^2 m_{\mathcal{S}}^6}$  \\
			&$+\frac{29 \lambda _H^{\text{SM}} Y_u^{\text{SM}} c_{\mathcal{S}}^2}{192 \pi ^2 m_{\mathcal{S}}^4}$\\
			\hline
			$Q_H$  &  $\frac{43 c_{\mathcal{S}}^6}{48 \pi ^2 m_{\mathcal{S}}^8}-\frac{7 g_W^2 \lambda _H^{\text{SM}} c_{\mathcal{S}}^2}{288 \pi ^2 m_{\mathcal{S}}^4}+\frac{41 \lambda _H^{\text{SM}}{}^2 c_{\mathcal{S}}^2}{48 \pi ^2 m_{\mathcal{S}}^4}$  \\
			&  $+\frac{37 c_{\mathcal{S}}^4 \kappa _{\mathcal{S}}}{32 \pi ^2 m_{\mathcal{S}}^6}+\frac{11 c_{\mathcal{S}}^2 \kappa _{\mathcal{S}}^2}{32 \pi ^2 m_{\mathcal{S}}^4}-\frac{c_{\mathcal{S}}^2 \kappa _{\mathcal{S}}}{2 m_{\mathcal{S}}^4}$  \\
			&  $-\frac{c_{\mathcal{S}}^2 \kappa _{\mathcal{S}} \lambda _{\mathcal{S}}}{32 \pi ^2 m_{\mathcal{S}}^4}-\frac{c_{\mathcal{S}}^4 \lambda _{\mathcal{S}}}{32 \pi ^2 m_{\mathcal{S}}^6}-\frac{15 c_{\mathcal{S}}^5 \mu _{\mathcal{S}}}{32 \pi ^2 m_{\mathcal{S}}^8}$  \\
			&  $-\frac{5 c_{\mathcal{S}}^3 \kappa _{\mathcal{S}} \mu _{\mathcal{S}}}{16 \pi ^2 m_{\mathcal{S}}^6}-\frac{c_{\mathcal{S}} \kappa _{\mathcal{S}}^2 \mu _{\mathcal{S}}}{64 \pi ^2 m_{\mathcal{S}}^4}+\frac{c_{\mathcal{S}}^3 \mu _{\mathcal{S}}}{6 m_{\mathcal{S}}^6}$  \\
			&  $+\frac{c_{\mathcal{S}}^2 \kappa _{\mathcal{S}} \mu _{\mathcal{S}}^2}{32 \pi ^2 m_{\mathcal{S}}^6}+\frac{c_{\mathcal{S}}^3 \lambda _{\mathcal{S}} \mu _{\mathcal{S}}}{48 \pi ^2 m_{\mathcal{S}}^6}+\frac{c_{\mathcal{S}}^4 \mu _{\mathcal{S}}^2}{16 \pi ^2 m_{\mathcal{S}}^8}$  \\
			&$-\frac{57 \lambda _H^{\text{SM}} c_{\mathcal{S}}^4}{32 \pi ^2 m_{\mathcal{S}}^6}+\frac{13 \lambda _H^{\text{SM}} c_{\mathcal{S}}^3 \mu _{\mathcal{S}}}{32 \pi ^2 m_{\mathcal{S}}^6}-\frac{c_{\mathcal{S}}^3 \mu _{\mathcal{S}}^3}{96 \pi ^2 m_{\mathcal{S}}^8}$\\
			&$-\frac{27 \lambda _H^{\text{SM}} c_{\mathcal{S}}^2 \kappa _{\mathcal{S}}}{32 \pi ^2 m_{\mathcal{S}}^4}-\frac{\kappa _{\mathcal{S}}^3}{192 \pi ^2 m_{\mathcal{S}}^2}$\\
			\hline
			$Q_{H\square }$  &  $\frac{13 c_{\mathcal{S}}^4}{192 \pi ^2 m_{\mathcal{S}}^6}-\frac{c_{\mathcal{S}}^2}{2 m_{\mathcal{S}}^4}-\frac{7 g_W^2 c_{\mathcal{S}}^2}{384 \pi ^2 m_{\mathcal{S}}^4}$  \\
			&  $+\frac{25 c_{\mathcal{S}}^2 \kappa _{\mathcal{S}}}{192 \pi ^2 m_{\mathcal{S}}^4}-\frac{c_{\mathcal{S}}^2 \lambda _{\mathcal{S}}}{32 \pi ^2 m_{\mathcal{S}}^4}-\frac{13 c_{\mathcal{S}}^3 \mu _{\mathcal{S}}}{192 \pi ^2 m_{\mathcal{S}}^6}$  \\
			&  $-\frac{\kappa _{\mathcal{S}}^2}{384 \pi ^2 m_{\mathcal{S}}^2}+\frac{11 c_{\mathcal{S}}^2 \mu _{\mathcal{S}}^2}{384 \pi ^2 m_{\mathcal{S}}^6}-\frac{5 c_{\mathcal{S}} \kappa _{\mathcal{S}} \mu _{\mathcal{S}}}{192 \pi ^2 m_{\mathcal{S}}^4}$  \\
			&$-\frac{7 g_Y^2 c_{\mathcal{S}}^2}{1152 \pi ^2 m_{\mathcal{S}}^4}$\\
			\hline \hline
	\end{tabular}}
	\subfloat{
		\begin{tabular}{|*{2}{>{\rowfonttype}c|}}%{|c|c|}
			\hline \hline
			Dim-6 Ops.&Wilson coefficients\\
			\hline \hline
			$Q_{\text{HB}}$  &  $\frac{g_Y^2 c_{\mathcal{S}}^2}{256 \pi ^2 m_{\mathcal{S}}^4}$  \\
			\hline
			$Q_{\text{HD}}$  &  $-\frac{7 g_Y^2 c_{\mathcal{S}}^2}{288 \pi ^2 m_{\mathcal{S}}^4}$  \\
			\hline
			$Q_{\text{Hd}}$  &  $\frac{7 g_Y^2 c_{\mathcal{S}}^2}{1728 \pi ^2 m_{\mathcal{S}}^4}$  \\
			\hline
			$Q_{\text{He}}$  &  $\frac{7 g_Y^2 c_{\mathcal{S}}^2}{576 \pi ^2 m_{\mathcal{S}}^4}$  \\
			\hline
			$Q_{\text{Hu}}$  &  $-\frac{7 g_Y^2 c_{\mathcal{S}}^2}{864 \pi ^2 m_{\mathcal{S}}^4}$  \\
			\hline
			$Q_{\text{HW}}$  &  $\frac{g_W^2 c_{\mathcal{S}}^2}{256 \pi ^2 m_{\mathcal{S}}^4}$  \\
			\hline
			$Q_{\text{HWB}}$  &  $\frac{g_W g_Y c_{\mathcal{S}}^2}{128 \pi ^2 m_{\mathcal{S}}^4}$  \\
			\hline
			$Q_{\text{Hl}}{}^{(1)}$  &  $\frac{7 g_Y^2 c_{\mathcal{S}}^2}{1152 \pi ^2 m_{\mathcal{S}}^4}$  \\
			\hline
			$Q_{\text{Hq}}{}^{(1)}$  &  $-\frac{7 g_Y^2 c_{\mathcal{S}}^2}{3456 \pi ^2 m_{\mathcal{S}}^4}$  \\
			\hline
			$Q_{\text{Hl}}{}^{(3)}$  &  $-\frac{7 g_W^2 c_{\mathcal{S}}^2}{1152 \pi ^2 m_{\mathcal{S}}^4}$  \\
			\hline
			$Q_{\text{Hq}}{}^{(3)}$  &  $-\frac{7 g_W^2 c_{\mathcal{S}}^2}{1152 \pi ^2 m_{\mathcal{S}}^4}$  \\
			\hline \rowfont{\color{red}} 
			$Q_{\text{lequ}}{}^{(1)}$  &  $\frac{Y_e^{\text{SM}} Y_u^{\text{SM}} c_{\mathcal{S}}^2}{96 \pi ^2 m_{\mathcal{S}}^4}$  \\
			\hline 
			$Q_{\text{qd}}{}^{(1)}$  &  $-\frac{Y_d^{\text{SM}} Y_d^{\text{SM$\dagger $}} c_{\mathcal{S}}^2}{192 \pi ^2 m_{\mathcal{S}}^4}$  \\
			\hline 
			$Q_{\text{qu}}{}^{(1)}$  &  $-\frac{Y_u^{\text{SM}} Y_u^{\text{SM$\dagger $}} c_{\mathcal{S}}^2}{192 \pi ^2 m_{\mathcal{S}}^4}$  \\
			\hline 
			$Q_{\text{quqd}}{}^{(1)}$  &  $-\frac{Y_d^{\text{SM}} Y_u^{\text{SM}} c_{\mathcal{S}}^2}{96 \pi ^2 m_{\mathcal{S}}^4}$  \\
			\hline 
			$Q_{\text{le}}$  &  $-\frac{Y_e^{\text{SM}} Y_e^{\text{SM$\dagger $}} c_{\mathcal{S}}^2}{192 \pi ^2 m_{\mathcal{S}}^4}$  \\
			\hline 
			$Q_{\text{ledq}}$  &  $\frac{Y_d^{\text{SM$\dagger $}} Y_e^{\text{SM}} c_{\mathcal{S}}^2}{96 \pi ^2 m_{\mathcal{S}}^4}$  \\
			\hline \hline
	\end{tabular}}
\end{table*}

\subsubsection{Real Triplet: $\Delta\equiv(1_C,3_L, \left. 0\right\vert_Y)$}
In this model, we have extended the SM by a real colour-singlet isospin-triplet scalar ($\Delta$). The Lagrangian involving the heavy field is written as \cite{Henning:2014wua,Ellis:2017jns},
{\small
	\begin{align}\label{RTSLag}
		\mathcal{L}_{\Delta} & = \mathcal{L}_{_\text{SM}}^{d\leq 4} \; + \, \frac{1}{2} \, (\mathcal{D}_{\mu} \Delta )^{2} \, - \, \frac{1}{2} \ m_{\Delta}^{2} \, \Delta^{a} \, \Delta^{a} + 2 \, \kappa_{\Delta} \, H^{\dagger} \tau^{a} H \, \Delta^{a} - \, \eta_{\Delta} \, |H|^{2} \, \Delta^{a} \, \Delta^{a} - \frac{1}{4} \, \lambda_{\Delta} \, (\Delta^{a} \Delta^{a})^{2}.
	\end{align}
}
Here, $m_{\Delta}$ is  mass of the heavy field.  This model contains three BSM parameters $\kappa_{\Delta}, \eta_{\Delta},  \lambda_{\Delta}$, and the WCs are functions of these parameters along with the SM ones, see tab.~\ref{tab:130}. 

\begin{table*}[h!]
	\caption{\small Warsaw basis  effective operators and the associated WCs that emerge after integrating-out the heavy field $\Delta : (1,3,0)$. See caption of tab.~\ref{tab:H2} for colour coding.}
	\scriptsize
	\label{tab:130}
	\centering
	\renewcommand{\arraystretch}{1.8}
	\subfloat{
		\begin{tabular}{|c|c|}
			\hline \hline
			Dim-6 Ops.&Wilson coefficients\\
			\hline \hline
			$Q_{\text{dH}}$  &  $-\frac{21 \eta _{\Delta } \kappa _{\Delta }^2 Y_d^{\text{SM}}}{32 \pi ^2 m_{\Delta }^4}-\frac{21 \kappa _{\Delta }^4 Y_d^{\text{SM}}}{64 \pi ^2 m_{\Delta }^6}+\frac{\kappa _{\Delta }^2 Y_d^{\text{SM}}}{m_{\Delta }^4}$  \\
			&  $+\frac{5 \kappa _{\Delta }^2 \lambda _{\Delta } Y_d^{\text{SM}}}{8 \pi ^2 m_{\Delta }^4}+\frac{27 \kappa _{\Delta }^2 \lambda _H^{\text{SM}} Y_d^{\text{SM}}}{64 \pi ^2 m_{\Delta }^4}$  \\
			\hline
			$Q_{\text{eH}}$  &  $-\frac{21 \eta _{\Delta } \kappa _{\Delta }^2 Y_e^{\text{SM}}}{32 \pi ^2 m_{\Delta }^4}-\frac{21 \kappa _{\Delta }^4 Y_e^{\text{SM}}}{64 \pi ^2 m_{\Delta }^6}+\frac{\kappa _{\Delta }^2 Y_e^{\text{SM}}}{m_{\Delta }^4}$  \\
			&  $+\frac{5 \kappa _{\Delta }^2 \lambda _{\Delta } Y_e^{\text{SM}}}{8 \pi ^2 m_{\Delta }^4}+\frac{27 \kappa _{\Delta }^2 \lambda _H^{\text{SM}} Y_e^{\text{SM}}}{64 \pi ^2 m_{\Delta }^4}$  \\
			\hline
			$Q_{\text{uH}}$  &  $-\frac{21 \eta _{\Delta } \kappa _{\Delta }^2 Y_u^{\text{SM}}}{32 \pi ^2 m_{\Delta }^4}-\frac{21 \kappa _{\Delta }^4 Y_u^{\text{SM}}}{64 \pi ^2 m_{\Delta }^6}+\frac{\kappa _{\Delta }^2 Y_u^{\text{SM}}}{m_{\Delta }^4}$  \\
			&  $+\frac{5 \kappa _{\Delta }^2 \lambda _{\Delta } Y_u^{\text{SM}}}{8 \pi ^2 m_{\Delta }^4}+\frac{27 \kappa _{\Delta }^2 \lambda _H^{\text{SM}} Y_u^{\text{SM}}}{64 \pi ^2 m_{\Delta }^4}$  \\
			\hline
			$Q_H$  &  $-\frac{\eta _{\Delta }^3}{8 \pi ^2 m_{\Delta }^2}-\frac{\eta _{\Delta } \kappa _{\Delta }^2}{m_{\Delta }^4}+\frac{2 \kappa _{\Delta }^2 \lambda _H^{\text{SM}}}{m_{\Delta }^4}$  \\
			&  $+\frac{13 \eta _{\Delta }^2 \kappa _{\Delta }^2}{8 \pi ^2 m_{\Delta }^4}+\frac{47 \eta _{\Delta } \kappa _{\Delta }^4}{16 \pi ^2 m_{\Delta }^6}+\frac{19 \kappa _{\Delta }^6}{16 \pi ^2 m_{\Delta }^8}$  \\
			&  $-\frac{g_W^2 \kappa _{\Delta }^2 \lambda _H^{\text{SM}}}{288 \pi ^2 m_{\Delta }^4}-\frac{5 \eta _{\Delta } \kappa _{\Delta }^2 \lambda _{\Delta }}{8 \pi ^2 m_{\Delta }^4}-\frac{5 \kappa _{\Delta }^4 \lambda _{\Delta }}{16 \pi ^2 m_{\Delta }^6}$  \\
			&  $-\frac{53 \eta _{\Delta } \kappa _{\Delta }^2 \lambda _H^{\text{SM}}}{16 \pi ^2 m_{\Delta }^4}+\frac{5 \kappa _{\Delta }^2 \lambda _{\Delta } \lambda _H^{\text{SM}}}{4 \pi ^2 m_{\Delta }^4}-\frac{85 \kappa _{\Delta }^4 \lambda _H^{\text{SM}}}{32 \pi ^2 m_{\Delta }^6}$  \\
			&$+\frac{3 \kappa _{\Delta }^2 \lambda _H^{\text{SM}}{}^2}{2 \pi ^2 m_{\Delta }^4}$\\
			\hline
			$Q_{H\square }$  &  $-\frac{g_W^4}{3840 \pi ^2 m_{\Delta }^2}-\frac{\eta _{\Delta }^2}{32 \pi ^2 m_{\Delta }^2}+\frac{\kappa _{\Delta }^2}{ 2 m_{\Delta }^4}$  \\
			&  $-\frac{g_W^2 \kappa _{\Delta }^2}{384 \pi ^2 m_{\Delta }^4}-\frac{7 g_Y^2 \kappa _{\Delta }^2}{384 \pi ^2 m_{\Delta }^4}-\frac{7 \eta _{\Delta } \kappa _{\Delta }^2}{32 \pi ^2 m_{\Delta }^4}$  \\
			&  $+\frac{5 \kappa _{\Delta }^2 \lambda _{\Delta }}{16 \pi ^2 m_{\Delta }^4}+\frac{3 \kappa _{\Delta }^2 \lambda _H^{\text{SM}}}{16 \pi ^2 m_{\Delta }^4}-\frac{13 \kappa _{\Delta }^4}{64 \pi ^2 m_{\Delta }^6}$  \\
			\hline
			$Q_{\text{HD}}$  &  $-\frac{7 g_Y^2 \kappa _{\Delta }^2}{96 \pi ^2 m_{\Delta }^4}+\frac{\eta _{\Delta } \kappa _{\Delta }^2}{\pi ^2 m_{\Delta }^4}-\frac{2 \kappa _{\Delta }^2}{m_{\Delta }^4}$  \\
			&  $-\frac{5 \kappa _{\Delta }^2 \lambda _{\Delta }}{4 \pi ^2 m_{\Delta }^4}-\frac{3 \kappa _{\Delta }^2 \lambda _H^{\text{SM}}}{16 \pi ^2 m_{\Delta }^4}-\frac{\kappa _{\Delta }^4}{16 \pi ^2 m_{\Delta }^6}$  \\
			\hline
			$Q_{\text{HW}}$  &  $\frac{\eta _{\Delta } g_W^2}{96 \pi ^2 m_{\Delta }^2}+\frac{25 g_W^2 \kappa _{\Delta }^2}{768 \pi ^2 m_{\Delta }^4}$  \\
			\hline
			$Q_{\text{Hl}}{}^{(3)}$  &  $-\frac{g_W^2 \kappa _{\Delta }^2}{1152 \pi ^2 m_{\Delta }^4}-\frac{g_W^4}{960 \pi ^2 m_{\Delta }^2}$  \\
			\hline
			$Q_{\text{Hq}}{}^{(3)}$  &  $-\frac{g_W^2 \kappa _{\Delta }^2}{1152 \pi ^2 m_{\Delta }^4}-\frac{g_W^4}{960 \pi ^2 m_{\Delta }^2}$  \\
			\hline \hline
	\end{tabular}}
	\subfloat{
		\begin{tabular}{|*{2}{>{\rowfonttype}c|}}%{|c|c|}
			\hline \hline
			Dim-6 Ops.&Wilson coefficients\\
			\hline \hline
			$Q_{\text{Hl}}{}^{(1)}$  &  $\frac{7 g_Y^2 \kappa _{\Delta }^2}{384 \pi ^2 m_{\Delta }^4}$\\
			\hline
			$Q_{\text{Hq}}{}^{(1)}$  &  $-\frac{7 g_Y^2 \kappa _{\Delta }^2}{1152 \pi ^2 m_{\Delta }^4}$  \\
			\hline
			$Q_{\text{Hd}}$  &  $\frac{7 g_Y^2 \kappa _{\Delta }^2}{576 \pi ^2 m_{\Delta }^4}$  \\
			\hline
			$Q_{\text{He}}$  &  $\frac{7 g_Y^2 \kappa _{\Delta }^2}{192 \pi ^2 m_{\Delta }^4}$  \\
			\hline
			$Q_{\text{Hu}}$  &  $-\frac{7 g_Y^2 \kappa _{\Delta }^2}{288 \pi ^2 m_{\Delta }^4}$  \\
			\hline
			$Q_{\text{HB}}$  &  $\frac{3 g_Y^2 \kappa _{\Delta }^2}{256 \pi ^2 m_{\Delta }^4}$  \\
			\hline
			$Q_{\text{HWB}}$  &  $-\frac{g_W g_Y \kappa _{\Delta }^2}{128 \pi ^2 m_{\Delta }^4}$  \\
			\hline \rowfont{\color{blue}}
			$Q_{\text{ll}}$  &  $-\frac{g_W^4}{3840 \pi ^2 m_{\Delta }^2}$  \\
			\hline
			$Q_W$  &  $\frac{g_W^3}{2880 \pi ^2 m_{\Delta }^2}$  \\
			\hline \rowfont{\color{red}}
			$Q_{\text{lequ}}{}^{(1)}$  &  $\frac{\kappa _{\Delta }^2 Y_e^{\text{SM}} Y_u^{\text{SM}}}{32 \pi ^2 m_{\Delta }^4}$  \\
			\hline  
			$Q_{\text{qd}}{}^{(1)}$  &  $-\frac{\kappa _{\Delta }^2 Y_d^{\text{SM}} Y_d^{\text{SM$\dagger $}}}{64 \pi ^2 m_{\Delta }^4}$  \\
			\hline  
			$Q_{\text{qq}}{}^{(1)}$  &  $\frac{g_W^4}{3840 \pi ^2 m_{\Delta }^2}$  \\
			\hline  
			$Q_{\text{qu}}{}^{(1)}$  &  $-\frac{\kappa _{\Delta }^2 Y_u^{\text{SM}} Y_u^{\text{SM$\dagger $}}}{64 \pi ^2 m_{\Delta }^4}$  \\
			\hline  
			$Q_{\text{quqd}}{}^{(1)}$  &  $-\frac{\kappa _{\Delta }^2 Y_d^{\text{SM}} Y_u^{\text{SM}}}{32 \pi ^2 m_{\Delta }^4}$  \\
			\hline  
			$Q_{\text{lq}}{}^{(3)}$  &  $-\frac{g_W^4}{1920 \pi ^2 m_{\Delta }^2}$  \\
			\hline  
			$Q_{\text{qq}}{}^{(3)}$  &  $-\frac{g_W^4}{3840 \pi ^2 m_{\Delta }^2}$  \\
			\hline  
			$Q_{\text{le}}$  &  $-\frac{\kappa _{\Delta }^2 Y_e^{\text{SM}} Y_e^{\text{SM$\dagger $}}}{64 \pi ^2 m_{\Delta }^4}$  \\
			\hline  
			$Q_{\text{ledq}}$  &  $\frac{\kappa _{\Delta }^2 Y_d^{\text{SM$\dagger $}} Y_e^{\text{SM}}}{32 \pi ^2 m_{\Delta }^4}$  \\
			\hline \hline
	\end{tabular}}
\end{table*}

\subsubsection{Complex Singlet: $\mathcal{S}_1\equiv(1_C,1_L, \left. 1\right\vert_Y)$}
Here, we have extended the SM by a colour-singlet isospin-singlet scalar ($\mathcal{S}_1$) with hypercharge $ Y=1 $. The Lagrangian involving the heavy field is written as \cite{Bilenky:1993bt,Bilenky:1994kt},
\begin{align}\label{eq:model-S1}
	\mathcal{L}_{\mathcal{S}_1} & = \mathcal{L}_{_\text{SM}}^{d\leq 4} \; +   \left(D_{\mu}\mathcal{S}_{1}\right)^\dagger  \left(D^{\mu}\mathcal{S}_{1}\right) - m_{\mathcal{S}_{1}}^2 \mathcal{S}_{1}^\dagger \mathcal{S}_{1}  - \eta_{{\mathcal{S}_{1}}} |H|^2  |\mathcal{S}_{1}|^2  - \lambda_{{\mathcal{S}_{1}}} |\mathcal{S}_{1}|^4\nonumber\\
	&-\left\lbrace y_{\mathcal{S}_1} l_{L}^{T} \, C \,i \sigma^2  l_{L} \, \mathcal{S}_1 + \text{h.c.}\right\rbrace.
\end{align}
Here, $m_{\mathcal{S}_1}$ is  mass of the heavy field.  This model contains three BSM parameters $\eta_{\mathcal{S}_1},  \lambda_{\mathcal{S}_1}, y_{\mathcal{S}_1}$, and the WCs are functions of these parameters along with the SM ones, see tab.~\ref{tab:111}. 

\begin{table*}[h!]
	\caption{\small Warsaw basis  effective operators and the associated WCs that emerge after integrating-out the heavy field $\mathcal{S}_1 : (1,1,1)$. See caption of tab.~\ref{tab:H2} for colour coding.}
	\label{tab:111}
	\scriptsize
	\centering
	\renewcommand{\arraystretch}{1.8}
	\subfloat{
		\begin{tabular}{|*{2}{>{\rowfonttype}c|}}%{|c|c|}
			\hline \hline
			Dim-6 Ops.&Wilson coefficients\\
			\hline \hline
			$Q_H$  &  $\frac{\eta _{\mathcal{S}_1}^3}{96 \pi ^2 m_{\mathcal{S}_1}^2}$  \\
			\hline
			$Q_{\text{HB}}$  &  $-\frac{\eta _{\mathcal{S}_1} g_Y^2}{192 \pi ^2 m_{\mathcal{S}_1}^2}$  \\
			\hline
			$Q_{H\square }$  &  $-\frac{\eta _{\mathcal{S}_1}^2}{192 \pi ^2 m_{\mathcal{S}_1}^2}$  \\
			\hline
			$Q_{\text{ll}}$  &  $-\frac{g_Y^4}{3840 \pi ^2 m_{\mathcal{S}_1}^2}+\frac{y_{\mathcal{S}_1}^2}{4 m_{\mathcal{S}_1}^2}-\frac{y_{\mathcal{S}_1}^2 \lambda_{{\mathcal{S}_{1}}}}{16 \pi^2 m_{\mathcal{S}_1}^2 }$  \\
			\hline \rowfont{\color{blue}}
			$Q_{\text{Hl}}{}^{(1)}$  &  $\frac{g_Y^4}{1920 \pi ^2 m_{\mathcal{S}_1}^2}$  \\
			\hline
			$Q_{\text{Hq}}{}^{(1)}$  &  $-\frac{g_Y^4}{5760 \pi ^2 m_{\mathcal{S}_1}^2}$  \\
			\hline
			$Q_{\text{HD}}$  &  $-\frac{g_Y^4}{960 \pi ^2 m_{\mathcal{S}_1}^2}$  \\
			\hline
			$Q_{\text{Hd}}$  &  $\frac{g_Y^4}{2880 \pi ^2 m_{\mathcal{S}_1}^2}$  \\
			\hline
			$Q_{\text{He}}$  &  $\frac{g_Y^4}{960 \pi ^2 m_{\mathcal{S}_1}^2}$  \\
			\hline
			$Q_{\text{Hu}}$  &  $-\frac{g_Y^4}{1440 \pi ^2 m_{\mathcal{S}_1}^2}$  \\
			\hline \rowfont{\color{red}}
			$Q_{\text{lq}}{}^{(1)}$  &  $\frac{g_Y^4}{5760 \pi ^2 m_{\mathcal{S}_1}^2}$  \\
			\hline
			$Q_{\text{qd}}{}^{(1)}$  &  $\frac{g_Y^4}{8640 \pi ^2 m_{\mathcal{S}_1}^2}$  \\
			\hline \hline
	\end{tabular}}
	\subfloat{
		\begin{tabular}{|*{2}{>{\rowfonttype}c|}}%{|c|c|}
			\hline \hline
			Dim-6 Ops.&Wilson coefficients\\
			\hline \hline \rowfont{\color{red}} 
			$Q_{\text{ed}}$  &  $-\frac{g_Y^4}{1440 \pi ^2 m_{\mathcal{S}_1}^2}$  \\
			\hline
			$Q_{\text{ee}}$  &  $-\frac{g_Y^4}{960 \pi ^2 m_{\mathcal{S}_1}^2}$  \\
			\hline
			$Q_{\text{eu}}$  &  $\frac{g_Y^4}{720 \pi ^2 m_{\mathcal{S}_1}^2}$  \\
			\hline
			$Q_{\text{ld}}$  &  $-\frac{g_Y^4}{2880 \pi ^2 m_{\mathcal{S}_1}^2}$  \\
			\hline
			$Q_{\text{le}}$  &  $-\frac{g_Y^4}{960 \pi ^2 m_{\mathcal{S}_1}^2}$  \\
			\hline			
			$Q_{\text{lu}}$  &  $\frac{g_Y^4}{1440 \pi ^2 m_{\mathcal{S}_1}^2}$  \\
			\hline
			$Q_{\text{qe}}$  &  $\frac{g_Y^4}{2880 \pi ^2 m_{\mathcal{S}_1}^2}$  \\
			\hline
			$Q_{\text{uu}}$  &  $-\frac{g_Y^4}{2160 \pi ^2 m_{\mathcal{S}_1}^2}$  \\
			\hline
			$Q_{\text{dd}}$  &  $-\frac{g_Y^4}{8640 \pi ^2 m_{\mathcal{S}_1}^2}$  \\			
			\hline
			$Q_{\text{qu}}{}^{(1)}$  &  $-\frac{g_Y^4}{4320 \pi ^2 m_{\mathcal{S}_1}^2}$  \\
			\hline
			$Q_{\text{qq}}{}^{(1)}$  &  $-\frac{g_Y^4}{34560 \pi ^2 m_{\mathcal{S}_1}^2}$  \\
			\hline
			$Q_{\text{ud}}{}^{(1)}$  &  $\frac{g_Y^4}{2160 \pi ^2 m_{\mathcal{S}_1}^2}$  \\
			\hline \hline
	\end{tabular}}
\end{table*}

\subsubsection{Complex Singlet: $\mathcal{S}_2 \equiv (1_C,1_L, \left. 2\right\vert_Y)$}
In this model, we have extended the SM by a colour-singlet isospin-triplet scalar ($\mathcal{S}_2$) with hypercharge $ Y = 2 $. The Lagrangian involving the heavy field is written as \cite{deBlas:2014mba,deBlas:2017xtg},
{\small
\begin{align}\label{eq:model-S2}
	\mathcal{L}_{\mathcal{S}_2} & = \mathcal{L}_{_\text{SM}}^{d\leq 4} \; +   \left(D_{\mu}\mathcal{S}_{2}\right)^\dagger  \left(D^{\mu}\mathcal{S}_{2}\right) - m_{\mathcal{S}_{2}}^2 \mathcal{S}_{2}^\dagger \mathcal{S}_{2}  - \eta_{{\mathcal{S}_{2}}} |H|^2  |\mathcal{S}_{2}|^2  - \lambda_{{\mathcal{S}_{2}}} |\mathcal{S}_{2}|^4\nonumber\\
	&-\left\lbrace y_{\mathcal{S}_2} e_{R}^{T} \, C \,  e_{R} \, \mathcal{S}_2 + \text{h.c.}\right\rbrace.
\end{align}}
Here, $m_{\mathcal{S}_2}$ is  mass of the heavy field.  This model contains three BSM parameters $\eta_{\mathcal{S}_2},  \lambda_{\mathcal{S}_2}, y_{\mathcal{S}_2}$, and the WCs are functions of these parameters along with the SM ones, see tab.~\ref{tab:112}. 

\begin{table*}[h!]
	\caption{\small Warsaw basis  effective operators and the associated WCs that emerge after integrating-out the heavy field $\mathcal{S}_2 : (1,1,2)$. See caption of tab.~\ref{tab:H2} for colour coding.}
	\label{tab:112}
	\scriptsize
	\centering
	\renewcommand{\arraystretch}{2.0}
	\subfloat{
		\begin{tabular}{|*{2}{>{\rowfonttype}c|}}%{|c|c|}
			\hline \hline
			Dim-6 Ops.&Wilson coefficients\\
			\hline \hline
			$Q_H$  &  $\frac{\eta _{\mathcal{S}_2}^3}{96 \pi ^2 m_{\mathcal{S}_2}^2}$  \\
			\hline
			$Q_{\text{HB}}$  &  $-\frac{\eta _{\mathcal{S}_2} g_Y^2}{48 \pi ^2 m_{\mathcal{S}_2}^2}$  \\
			\hline
			$Q_{H\square }$  &  $-\frac{\eta _{\mathcal{S}_2}^2}{192 \pi ^2 m_{\mathcal{S}_2}^2}$  \\
			\hline \rowfont{\color{blue}}
			$Q_{\text{Hl}}{}^{(1)}$  &  $\frac{g_Y^4}{480 \pi ^2 m_{\mathcal{S}_2}^2}$  \\
			\hline
			$Q_{\text{Hq}}{}^{(1)}$  &  $-\frac{g_Y^4}{1440 \pi ^2 m_{\mathcal{S}_2}^2}$  \\
			\hline
			$Q_{\text{HD}}$  &  $-\frac{g_Y^4}{240 \pi ^2 m_{\mathcal{S}_2}^2}$  \\
			\hline
			$Q_{\text{Hd}}$  &  $\frac{g_Y^4}{720 \pi ^2 m_{\mathcal{S}_2}^2}$  \\
			\hline
			$Q_{\text{He}}$  &  $\frac{g_Y^4}{240 \pi ^2 m_{\mathcal{S}_2}^2}$  \\
			\hline
			$Q_{\text{Hu}}$  &  $-\frac{g_Y^4}{360 \pi ^2 m_{\mathcal{S}_2}^2}$  \\
			\hline  
			$Q_{\text{ll}}$  &  $-\frac{g_Y^4}{960 \pi ^2 m_{\mathcal{S}_2}^2}$  \\
			\hline \rowfont{\color{red}} 
			$Q_{\text{ld}}$  &  $-\frac{g_Y^4}{720 \pi ^2 m_{\mathcal{S}_2}^2}$  \\
			\hline
			$Q_{\text{le}}$  &  $-\frac{g_Y^4}{240 \pi ^2 m_{\mathcal{S}_2}^2}$  \\
			\hline\hline
	\end{tabular}}
	\subfloat{
		\begin{tabular}{|*{2}{>{\rowfonttype}c|}}%{|c|c|}
			\hline \hline 
			Dim-6 Ops.&Wilson coefficients\\
			\hline\hline  \rowfont{\color{red}} 
			$Q_{\text{lu}}$  &  $\frac{g_Y^4}{360 \pi ^2 m_{\mathcal{S}_2}^2}$  \\
			\hline
			$Q_{\text{lq}}{}^{(1)}$  &  $\frac{g_Y^4}{1440 \pi ^2 m_{\mathcal{S}_2}^2}$  \\
			\hline  
			$Q_{\text{qd}}{}^{(1)}$  &  $\frac{g_Y^4}{2160 \pi ^2 m_{\mathcal{S}_2}^2}$  \\
			\hline  
			$Q_{\text{qq}}{}^{(1)}$  &  $-\frac{g_Y^4}{8640 \pi ^2 m_{\mathcal{S}_2}^2}$  \\
			\hline  
			$Q_{\text{qu}}{}^{(1)}$  &  $-\frac{g_Y^4}{1080 \pi ^2 m_{\mathcal{S}_2}^2}$  \\
			\hline  
			$Q_{\text{ud}}{}^{(1)}$  &  $\frac{g_Y^4}{540 \pi ^2 m_{\mathcal{S}_2}^2}$  \\
			\hline  
			$Q_{\text{dd}}$  &  $-\frac{g_Y^4}{2160 \pi ^2 m_{\mathcal{S}_2}^2}$  \\
			\hline  
			$Q_{\text{ed}}$  &  $-\frac{g_Y^4}{360 \pi ^2 m_{\mathcal{S}_2}^2}$  \\
			\hline  
			$Q_{\text{eu}}$  &  $\frac{g_Y^4}{180 \pi ^2 m_{\mathcal{S}_2}^2}$  \\
			\hline  
			$Q_{\text{uu}}$  &  $-\frac{g_Y^4}{540 \pi ^2 m_{\mathcal{S}_2}^2}$  \\
			\hline  
			$Q_{\text{qe}}$  &  $\frac{g_Y^4}{720 \pi ^2 m_{\mathcal{S}_2}^2}$  \\
			\hline  
			$Q_{\text{ee}}$  &  $-\frac{g_Y^4}{240 \pi ^2 m_{\mathcal{S}_2}^2}-\frac{\lambda _{\mathcal{S}_2} y_{\mathcal{S}_2}^2}{16 \pi ^2 m_{\mathcal{S}_2}^2}+\frac{y_{\mathcal{S}_2}^2}{4 m_{\mathcal{S}_2}^2}$\\
			\hline \hline
	\end{tabular}}
\end{table*}

\subsubsection{Complex Triplet: $\Delta_1\equiv(1_C,3_L, \left. 1\right\vert_Y)$}
Here, we have extended the SM by a colour-singlet isospin-triplet scalar ($\Delta_1$) with hypercharge $ Y=1 $. This model is also known as `Type-II seesaw'. The Lagrangian involving the heavy field is written as \cite{Arhrib:2011uy},
{\begin{align}
		\label{lbsmCTS}
		\mathcal{L}_{\Delta_1} & = \mathcal{L}_{_\text{SM}}^{d\leq 4} \; + Tr [ (D_{\mu} \Delta_1)^{\dagger} (D^{\mu} \Delta_1 ) ] - m_{\Delta_1}^{2} Tr [ \Delta_1^{\dagger} \Delta_1 ] - \mathcal{L}_{Y} - V(H,\Delta_1) ,
	\end{align}
}
where,
{\begin{align}
		V(H,\Delta_1) =& \lambda_{\Delta_1,1} (H^{\dagger} H) Tr[ \Delta_1^{\dagger} \Delta_1 ] + \lambda_{\Delta_1,2} (Tr[ \Delta_1^{\dagger} \Delta_1 ])^2 + \lambda_{\Delta_1,3}Tr[ (\Delta_1^{\dagger} \Delta_1)^2 ] \nonumber\\
		& + \lambda_{\Delta_1,4}Tr[ H^\dagger \Delta_1 {\Delta_1}^\dagger H \} ] + [ \mu_{\Delta_1} ( H^{T} i \sigma^{2} \Delta_1^{\dagger} H ) + \text{h.c.}], \\
		\text{and,}~~~~~\mathcal{L}_{Y} =& y_{\Delta_1} l_{L}^{T} C i \sigma^2 \Delta_1 l_{L} + \text{h.c.} 
	\end{align}
}
Here, $m_{\Delta_1}$ is  mass of the heavy field.  This model contains six BSM parameters $\lambda_{\Delta_1,1}, \lambda_{\Delta_1,2}, \lambda_{\Delta_1,3},$  $ \lambda_{\Delta_1,4}, \mu_{\Delta_1},  y_{\Delta_1}$, and the WCs are functions of these parameters along with the SM ones, see tab.~\ref{tab:131}.

\begin{table*}[h!]
	\caption{\small Warsaw basis  effective operators and the associated WCs that emerge after integrating-out the heavy field  $\Delta_1 : (1,3,1)$. See caption of tab.~\ref{tab:H2} for colour coding.}
	\label{tab:131}
	\centering
	\scriptsize
	\renewcommand{\arraystretch}{1.7}
	\subfloat{
		\begin{tabular}{|c|c|}
			\hline \hline
			Dim-6 Ops.&Wilson coefficients\\
			\hline \hline
			$Q_{\text{uH}}$  &  $\frac{3 \mu _{\Delta _1}^2 \lambda _H^{\text{SM}} Y_u^{\text{SM}}}{2 \pi ^2 m_{\Delta _1}^4}-\frac{37 \mu _{\Delta _1}^4 Y_u^{\text{SM}}}{6 \pi ^2 m_{\Delta _1}^6}+\frac{2 \mu _{\Delta _1}^2 Y_u^{\text{SM}}}{m_{\Delta _1}^4}$  \\
			&  $-\frac{5 \mu _{\Delta _1}^2 \lambda _{\Delta _1,1} Y_u^{\text{SM}}}{8 \pi ^2 m_{\Delta _1}^4}-\frac{37 \mu _{\Delta _1}^2 \lambda _{\Delta _1,4} Y_u^{\text{SM}}}{48 \pi ^2 m_{\Delta _1}^4}+\frac{\lambda _{\Delta _1,4}{}^2 Y_u^{\text{SM}}}{48 \pi ^2 m_{\Delta _1}^2}$  \\
			\hline
			$Q_{\text{dH}}$  &  $\frac{3 \mu _{\Delta _1}^2 \lambda _H^{\text{SM}} Y_d^{\text{SM}}}{2 \pi ^2 m_{\Delta _1}^4}-\frac{37 \mu _{\Delta _1}^4 Y_d^{\text{SM}}}{6 \pi ^2 m_{\Delta _1}^6}+\frac{2 \mu _{\Delta _1}^2 Y_d^{\text{SM}}}{m_{\Delta _1}^4}$  \\
			&  $-\frac{5 \mu _{\Delta _1}^2 \lambda _{\Delta _1,1} Y_d^{\text{SM}}}{8 \pi ^2 m_{\Delta _1}^4}-\frac{37 \mu _{\Delta _1}^2 \lambda _{\Delta _1,4} Y_d^{\text{SM}}}{48 \pi ^2 m_{\Delta _1}^4}+\frac{\lambda _{\Delta _1,4}{}^2 Y_d^{\text{SM}}}{48 \pi ^2 m_{\Delta _1}^2}$  \\
			\hline
			$Q_{\text{eH}}$  &  $\frac{3 \mu _{\Delta _1}^2 \lambda _H^{\text{SM}} Y_e^{\text{SM}}}{2 \pi ^2 m_{\Delta _1}^4}-\frac{37 \mu _{\Delta _1}^4 Y_e^{\text{SM}}}{6 \pi ^2 m_{\Delta _1}^6}+\frac{2 \mu _{\Delta _1}^2 Y_e^{\text{SM}}}{m_{\Delta _1}^4}$  \\
			&  $-\frac{5 \mu _{\Delta _1}^2 \lambda _{\Delta _1,1} Y_e^{\text{SM}}}{8 \pi ^2 m_{\Delta _1}^4}-\frac{37 \mu _{\Delta _1}^2 \lambda _{\Delta _1,4} Y_e^{\text{SM}}}{48 \pi ^2 m_{\Delta _1}^4}+\frac{\lambda _{\Delta _1,4}{}^2 Y_e^{\text{SM}}}{48 \pi ^2 m_{\Delta _1}^2}$  \\
			\hline
			$Q_H$  &  $\frac{4 \mu _{\Delta _1}^2 \lambda _H^{\text{SM}}}{m_{\Delta _1}^4}-\frac{4 \mu _{\Delta _1}^2 \lambda _{\Delta _1,1}}{m_{\Delta _1}^4}-\frac{4 \mu _{\Delta _1}^2 \lambda _{\Delta _1,4}}{m_{\Delta _1}{}^4}$  \\
			&  $-\frac{g_W^2 \mu _{\Delta _1}^2 \lambda _H^{\text{SM}}}{72 \pi ^2 m_{\Delta _1}^4}-\frac{130 \mu _{\Delta _1}^4 \lambda _H^{\text{SM}}}{3 \pi ^2 m_{\Delta _1}^6}+\frac{74 \mu _{\Delta _1}^6}{\pi ^2 m_{\Delta _1}^8}$  \\
			&  $+\frac{45 \mu _{\Delta _1}^2 \lambda _H^{\text{SM}}{}^2}{8 \pi ^2 m_{\Delta _1}^4}-\frac{37 \mu _{\Delta _1}^2 \lambda _H^{\text{SM}} \lambda _{\Delta _1,1}}{4 \pi ^2 m_{\Delta _1}^4}+\frac{48 \mu _{\Delta _1}^4 \lambda _{\Delta _1,1}}{\pi ^2 m_{\Delta _1}^6}$  \\
			&  $+\frac{13 \mu _{\Delta _1}^2 \lambda _{\Delta _1,1}{}^2}{2 \pi ^2 m_{\Delta _1}^4}-\frac{20 \mu _{\Delta _1}^4 \lambda _{\Delta _1,2}}{\pi ^2 m_{\Delta _1}^6}-\frac{16 \mu _{\Delta _1}^2 \lambda _{\Delta _1,1} \lambda _{\Delta _1,2}}{\pi ^2 m_{\Delta _1}^4}$  \\
			&  $-\frac{12 \mu _{\Delta _1}^2 \lambda _{\Delta _1,1} \lambda _{\Delta _1,3}}{\pi ^2 m_{\Delta _1}^4}-\frac{20 \mu _{\Delta _1}^4 \lambda _{\Delta _1,3}}{\pi ^2 m_{\Delta _1}^6}$  \\
			&  $-\frac{\lambda _{\Delta _1,1}{}^3}{4 \pi ^2 m_{\Delta _1}^2}+\frac{25 \mu _{\Delta _1}^2 \lambda _{\Delta _1,1} \lambda _{\Delta _1,4}}{2 \pi ^2 m_{\Delta _1}^4}+\frac{97 \mu _{\Delta _1}^4 \lambda _{\Delta _1,4}}{2 \pi ^2 m_{\Delta _1}^6}$  \\
			&  $-\frac{229 \mu _{\Delta _1}^2 \lambda _H^{\text{SM}} \lambda _{\Delta _1,4}}{24 \pi ^2 m_{\Delta _1}^4}+\frac{39 \mu _{\Delta _1}^2 \lambda _{\Delta _1,4}{}^2}{8 \pi ^2 m_{\Delta _1}^4}+\frac{\lambda _H^{\text{SM}} \lambda _{\Delta _1,4}{}^2}{24 \pi ^2 m_{\Delta _1}^2}$  \\
			&  $-\frac{3 \lambda _{\Delta _1,1}{}^2 \lambda _{\Delta _1,4}}{8 \pi ^2 m_{\Delta _1}^2}-\frac{5 \lambda _{\Delta _1,1} \lambda _{\Delta _1,4}{}^2}{16 \pi ^2 m_{\Delta _1}^2}-\frac{3 \lambda _{\Delta _1,4}{}^3}{32 \pi ^2 m_{\Delta _1}^2}$  \\
			\hline
			$Q_{H\square }$  &  $-\frac{g_W^2 \mu _{\Delta _1}^2}{96 \pi ^2 m_{\Delta _1}^4}-\frac{g_W^4}{1920 \pi ^2 m_{\Delta _1}^2}+\frac{2\mu _{\Delta _1}^2}{m_{\Delta _1}^4}$  \\
			&  $+\frac{11 g_Y^2 \mu _{\Delta _1}^2}{96 \pi ^2 m_{\Delta _1}^4}+\frac{21 \mu _{\Delta _1}^2 \lambda _H^{\text{SM}}}{32 \pi ^2 m_{\Delta _1}{}^4}-\frac{41 \mu _{\Delta _1}^4}{12 \pi ^2 m_{\Delta _1}^6}$  \\
			&  $+\frac{\mu _{\Delta _1}^2 \lambda _{\Delta _1,1}}{8 \pi ^2 m_{\Delta _1}^4}-\frac{\mu _{\Delta _1}^2 \lambda _{\Delta _1,4}}{48 \pi ^2 m_{\Delta _1}^4}-\frac{\lambda _{\Delta _1,1}{}^2}{16 \pi ^2 m_{\Delta _1}^2}$  \\
			&  $+\frac{\lambda _{\Delta _1,4}{}^2}{192 \pi ^2 m_{\Delta _1}^2}-\frac{\lambda _{\Delta _1,1} \lambda _{\Delta _1,4}}{16 \pi ^2 m_{\Delta _1}^2}$  \\
			\hline
			$Q_{\text{HD}}$  &  $\frac{11 g_Y^2 \mu _{\Delta _1}^2}{24 \pi ^2 m_{\Delta _1}^4}-\frac{g_Y^4}{320 \pi ^2 m_{\Delta _1}^2}+\frac{4 \mu _{\Delta _1}^2}{m_{\Delta _1}^4}$  \\
			&  $+\frac{3 \mu _{\Delta _1}^2 \lambda _H^{\text{SM}}}{8 \pi ^2 m_{\Delta _1}^4}+\frac{\mu _{\Delta _1}^2 \lambda _{\Delta _1,4}}{6 \pi ^2 m_{\Delta _1}^4}-\frac{8 \mu _{\Delta _1}^4}{3 \pi ^2 m_{\Delta _1}^6}$  \\
			&$-\frac{\lambda _{\Delta _1,4}{}^2}{24 \pi ^2 m_{\Delta _1}^2}$\\
			\hline
			$Q_{\text{HB}}$  &  $\frac{11 g_Y^2 \mu _{\Delta _1}^2}{64 \pi ^2 m_{\Delta _1}^4}+\frac{g_Y^2 \lambda _{\Delta _1,1}}{32 \pi ^2 m_{\Delta _1}^2}+\frac{g_Y^2 \lambda _{\Delta _1,4}}{64 \pi ^2 m_{\Delta _1}^2}$  \\
			\hline
			$Q_{\text{HW}}$  &  $\frac{25 g_W^2 \mu _{\Delta _1}^2}{192 \pi ^2 m_{\Delta _1}^4}+\frac{g_W^2 \lambda _{\Delta _1,1}}{48 \pi ^2 m_{\Delta _1}^2}+\frac{g_W^2 \lambda _{\Delta _1,4}}{96 \pi ^2 m_{\Delta _1}^2}$  \\
			\hline
			$Q_{\text{HWB}}$  &  $-\frac{13 g_W g_Y \mu _{\Delta _1}^2}{96 \pi ^2 m_{\Delta _1}^4}-\frac{g_W g_Y \lambda _{\Delta _1,4}}{48 \pi ^2 m_{\Delta _1}^2}$  \\
			\hline
			$Q_{\text{ll}}$  &  $-\frac{g_W^4}{1920 \pi ^2 m_{\Delta _1}^2}-\frac{g_Y^4}{1280 \pi ^2 m_{\Delta _1}^2}+\frac{y_{\Delta _1}^2}{2 m_{\Delta _1}^2}$  \\
			&  $+\frac{\lambda _{\Delta _1,2} y_{\Delta _1}^2}{\pi ^2 m_{\Delta _1}^2}+\frac{3 \lambda _{\Delta _1,3} y_{\Delta _1}^2}{4 \pi ^2 m_{\Delta _1}^2}$  \\
			\hline \hline
	\end{tabular}}
	\subfloat{
		\begin{tabular}{|*{2}{>{\rowfonttype}c|}}%{|c|c|}{|c|c|}{|c|c|}
			\hline \hline
			Ops.&Wilson coefficients\\
			\hline \hline
			$Q_{\text{Hq}}{}^{(1)}$  &  $\frac{11 g_Y^2 \mu _{\Delta _1}^2}{288 \pi ^2 m_{\Delta _1}^4}-\frac{g_Y^4}{1920 \pi ^2 m_{\Delta _1}^2}$  \\
			\hline
			$Q_{\text{Hd}}$  &  $\frac{g_Y^4}{960 \pi ^2 m_{\Delta _1}^2}-\frac{11 g_Y^2 \mu _{\Delta _1}^2}{144 \pi ^2 m_{\Delta _1}^4}$  \\
			\hline
			$Q_{\text{He}}$  &  $\frac{g_Y^4}{320 \pi ^2 m_{\Delta _1}^2}-\frac{11 g_Y^2 \mu _{\Delta _1}^2}{48 \pi ^2 m_{\Delta _1}^4}$  \\
			\hline
			$Q_{\text{Hu}}$  &  $\frac{11 g_Y^2 \mu _{\Delta _1}^2}{72 \pi ^2 m_{\Delta _1}^4}-\frac{g_Y^4}{480 \pi ^2 m_{\Delta _1}^2}$  \\
			\hline
			$Q_{\text{Hl}}{}^{(3)}$  &  $-\frac{g_W^2 \mu _{\Delta _1}^2}{288 \pi ^2 m_{\Delta _1}^4}-\frac{g_W^4}{480 \pi ^2 m_{\Delta _1}^2}$  \\
			\hline
			$Q_{\text{Hq}}{}^{(3)}$  &  $-\frac{g_W^2 \mu _{\Delta _1}^2}{288 \pi ^2 m_{\Delta _1}^4}-\frac{g_W^4}{480 \pi ^2 m_{\Delta _1}^2}$  \\
			\hline
			$Q_{\text{Hl}}{}^{(1)}$  &  $\frac{g_Y^4}{640 \pi ^2 m_{\Delta _1}^2}-\frac{11 g_Y^2 \mu _{\Delta _1}^2}{96 \pi ^2 m_{\Delta _1}^4}$  \\
			\hline \rowfont{\color{blue}}
			$Q_W$  &  $\frac{g_W^3}{1440 \pi ^2 m_{\Delta _1}^2}$  \\
			\hline \rowfont{\color{red}}
			$Q_{\text{le}}$  &  $-\frac{g_Y^4}{320 \pi ^2 m_{\Delta _1}^2}-\frac{\mu _{\Delta _1}^2 Y_e^{\text{SM}} Y_e^{\text{SM$\dagger $}}}{16 \pi ^2 m_{\Delta _1}^4}$  \\
			\hline
			$Q_{\text{qu}}{}^{(1)}$  &  $-\frac{g_Y^4}{1440 \pi ^2 m_{\Delta _1}^2}-\frac{\mu _{\Delta _1}^2 Y_u^{\text{SM}} Y_u^{\text{SM$\dagger $}}}{16 \pi ^2 m_{\Delta _1}^4}$  \\
			\hline
			$Q_{\text{qd}}{}^{(1)}$  &  $\frac{g_Y^4}{2880 \pi ^2 m_{\Delta _1}^2}-\frac{\mu _{\Delta _1}^2 Y_d^{\text{SM}} Y_d^{\text{SM$\dagger $}}}{16 \pi ^2 m_{\Delta _1}^4}$  \\
			\hline
			$Q_{\text{lq}}{}^{(1)}$  &  $\frac{g_Y^4}{1920 \pi ^2 m_{\Delta _1}^2}$  \\
			\hline
			$Q_{\text{qq}}{}^{(1)}$  &  $-\frac{g_Y^4}{11520 \pi ^2 m_{\Delta _1}^2}$  \\
			\hline
			$Q_{\text{quqd}}{}^{(1)}$  &  $-\frac{\mu _{\Delta _1}^2 Y_d^{\text{SM}} Y_u^{\text{SM}}}{8 \pi ^2 m_{\Delta _1}^4}$  \\
			\hline
			$Q_{\text{ld}}$  &  $-\frac{g_Y^4}{960 \pi ^2 m_{\Delta _1}^2}$  \\
			\hline
			$Q_{\text{ledq}}$  &  $\frac{\mu _{\Delta _1}^2 Y_d^{\text{SM$\dagger $}} Y_e^{\text{SM}}}{8 \pi ^2 m_{\Delta _1}^4}$  \\
			\hline
			$Q_{\text{ed}}$  &  $-\frac{g_Y^4}{480 \pi ^2 m_{\Delta _1}^2}$  \\
			\hline
			$Q_{\text{lu}}$  &  $\frac{g_Y^4}{480 \pi ^2 m_{\Delta _1}^2}$  \\
			\hline
			$Q_{\text{qe}}$  &  $\frac{g_Y^4}{960 \pi ^2 m_{\Delta _1}^2}$  \\
			\hline
			$Q_{\text{uu}}$  &  $-\frac{g_Y^4}{720 \pi ^2 m_{\Delta _1}^2}$  \\
			\hline
			$Q_{\text{ee}}$  &  $-\frac{g_Y^4}{320 \pi ^2 m_{\Delta _1}^2}$  \\
			\hline
			$Q_{\text{eu}}$  &  $\frac{g_Y^4}{240 \pi ^2 m_{\Delta _1}^2}$  \\
			\hline
			$Q_{\text{lequ}}{}^{(1)}$  &  $\frac{\mu _{\Delta _1}^2 Y_e^{\text{SM}} Y_u^{\text{SM}}}{8 \pi ^2 m_{\Delta _1}^4}$  \\
			\hline
			$Q_{\text{ud}}{}^{(1)}$  &  $\frac{g_Y^4}{720 \pi ^2 m_{\Delta _1}^2}$  \\
			\hline
			$Q_{\text{lq}}{}^{(3)}$  &  $-\frac{g_W^4}{960 \pi ^2 m_{\Delta _1}^2}$  \\
			\hline
			$Q_{\text{qq}}{}^{(3)}$  &  $-\frac{g_W^4}{1920 \pi ^2 m_{\Delta _1}^2}$  \\
			\hline
			$Q_{\text{dd}}$  &  $-\frac{g_Y^4}{2880 \pi ^2 m_{\Delta _1}^2}$  \\
			\hline \hline
	\end{tabular}}
\end{table*}

\subsubsection{Complex Quartet: $\Sigma \equiv (1_C,4_L,\left.\frac{1}{2}\right\vert_Y)$}
Here, we have extended the SM by a colour-singlet isospin-quartet scalar ($\Sigma$) with hypercharge $ Y= \frac{1}{2} $. The Lagrangian involving the heavy field is written as \cite{Babu:2009aq,Bambhaniya:2013yca,Dawson:2017vgm},
\begin{align}\label{eq:quartetlag}
	\mathcal{L}_{\Sigma} & = \mathcal{L}_{_\text{SM}}^{d\leq 4} \; +   \left(D_{\mu}\Sigma\right)^\dagger \, \left(D^{\mu}\Sigma\right) - m_{_\Sigma}^2 \, \Sigma^\dagger \Sigma - \mu_{_\Sigma} \, \left[\left(\Sigma^\dagger H\right)^2 + \text{h.c.}\right] - \kappa_{_\Sigma} \left[\Sigma^\dagger B_\Sigma + \text{h.c.}\right] \nonumber\\
	& - \zeta_{\Sigma}^{(1)} \left(H^{\dagger}H\right) \left(\Sigma^{\dagger}\Sigma\right) - \zeta_{\Sigma}^{(2)} \left(H^{\dagger} \tau^I  H\right) \left(\Sigma^{\dagger} \, T_{_4}^I \, \Sigma\right) - \lambda_{_{\Sigma}}^{(1)} \left(\Sigma^\dagger \Sigma\right)^2  - \lambda_{_{\Sigma}}^{(2)} \left(\Sigma^\dagger \, T_{_4}^I \, \Sigma\right)^2.
\end{align}
Here, $m_{\Sigma}$ is  mass of the heavy field. $T_4^I$'s are the $SU(2)$ generators in 4-dimensional representation, and $B_\Sigma = \left(H_1^2 \widetilde{H}_1, \frac{1}{\sqrt{3}}H_1^2 \widetilde{H}_2 + \frac{2}{\sqrt{3}} H_1 H_2 \widetilde{H}_1 , \frac{1}{\sqrt{3}}H_2^2 \widetilde{H}_1 + \frac{2}{\sqrt{3}} H_1 H_2 \widetilde{H}_2, H_2^2 \widetilde{H}_2\right)^T$. This model contains six BSM parameters $\kappa_{_\Sigma}, \mu_{_\Sigma},  \zeta_{\Sigma}^{(1)}, \zeta_{\Sigma}^{(2)}, \lambda_{_{\Sigma}}^{(1)}, \lambda_{_{\Sigma}}^{(2)}$, and the WCs are functions of these parameters along with the SM ones, see tab.~\ref{tab:Sigma}.

\begin{table*}[!hbt]
	\caption{\small Warsaw basis  effective operators and the associated WCs that emerge after integrating-out the heavy field $ \Sigma : (1,4,\frac{1}{2}) $. See caption of tab.~\ref{tab:H2} for colour coding.}
	\label{tab:Sigma}
	\scriptsize
	\centering
	\renewcommand{\arraystretch}{1.8}
	\subfloat{
		\begin{tabular}{|*{2}{>{\rowfonttype}c|}}%{|*{2}{c|}}
			\hline \hline
			Dimension-6 Ops.& Wilson coefficients \\
			\hline \hline
			$Q_{\text{HW}}$  &  $-\frac{5 g_W^2 \zeta _{\Sigma}^{(1)}}{192 \pi ^2 m_{\Sigma}^2}$  \\
			\hline
			$Q_{\text{HB}}$  &  $-\frac{g_Y^2 \zeta _{\Sigma}^{(1)}}{192 \pi ^2 m_{\Sigma}^2}$  \\
			\hline
			$Q_{\text{HWB}}$  &  $-\frac{5 g_W g_Y \zeta _{\Sigma}^{(2)}}{384 \pi ^2 m_{\Sigma}^2}$  \\
			\hline
			$Q_H$  &  $\frac{\zeta _{\Sigma}^{(1)}{}^3}{24 \pi ^2 m_{\Sigma}^2}+\frac{5 \zeta _{\Sigma}^{(1)} \zeta _{\Sigma}^{(2)}{}^2}{128 \pi ^2 m_{\Sigma}^2}+\frac{5 \zeta _{\Sigma}^{(2)}{}^2 \lambda _H^{\text{SM}}}{192 \pi ^2 m_{\Sigma}^2}$  \\
			&  $+\frac{5 \mu _{\Sigma}^2 \zeta _{\Sigma}^{(1)}}{36 \pi ^2 m_{\Sigma}^2}+\frac{5 \mu _{\Sigma}^2 \zeta _{\Sigma}^{(2)}}{144 \pi ^2 m_{\Sigma}^2}+\frac{5 \mu _{\Sigma}^2 \lambda _H^{\text{SM}}}{108 \pi ^2 m_{\Sigma}^2}$  \\
			&$+\frac{\kappa^2_\Sigma}{3 m_\Sigma^2}-\frac{5\kappa^2_\Sigma \lambda_{\Sigma_1}}{24 \pi^2 m_\Sigma^2}-\frac{5\kappa^2_\Sigma \lambda_{\Sigma_2}}{32 \pi^2 m_\Sigma^2}$\\
			\hline
			$Q_{\text{HD}}$  &  $-\frac{g_Y^4}{960 \pi ^2 m_{\Sigma}^2}+\frac{5 \mu _{\Sigma}^2}{108 \pi ^2 m_{\Sigma}^2}-\frac{5 \zeta _{\Sigma}^{(2)}{}^2}{192 \pi ^2 m_{\Sigma}^2}$  \\
			\hline
			$Q_{\text{uH}}$  &  $\frac{5 \mu _{\Sigma}^2 Y_u^{\text{SM}}}{216 \pi ^2 m_{\Sigma}^2}+\frac{5 \zeta _{\Sigma}^{(2)}{}^2 Y_u^{\text{SM}}}{384 \pi ^2 m_{\Sigma}^2}$  \\
			\hline
			$Q_{\text{dH}}$  &  $\frac{5 \mu _{\Sigma}^2 Y_d^{\text{SM}}}{216 \pi ^2 m_{\Sigma}^2}+\frac{5 \zeta _{\Sigma}^{(2)}{}^2 Y_d^{\text{SM}}}{384 \pi ^2 m_{\Sigma}^2}$  \\
			\hline
			$Q_{\text{eH}}$  &  $\frac{5 \mu _{\Sigma}^2 Y_e^{\text{SM}}}{216 \pi ^2 m_{\Sigma}^2}+\frac{5 \zeta _{\Sigma}^{(2)}{}^2 Y_e^{\text{SM}}}{384 \pi ^2 m_{\Sigma}^2}$  \\
			\hline
			$Q_{H\square }$  &  $-\frac{g_W^4}{768 \pi ^2 m_{\Sigma}^2}-\frac{\zeta _{\Sigma}^{(1)}{}^2}{48 \pi ^2 m_{\Sigma}{}^2}+\frac{5 \zeta _{\Sigma}^{(2)}{}^2}{768 \pi ^2 m_{\Sigma}{}^2}$  \\
			&  $+\frac{5 \mu _{\Sigma}^2}{216 \pi ^2 m_{\Sigma}^2} + \frac{\kappa_{\Sigma}^2}{8 \pi^2 m_{\Sigma}^2} $  \\
			\hline \rowfont{\color{blue}}
			$Q_{\text{Hl}}{}^{(1)}$  &  $\frac{g_Y^4}{1920 \pi ^2 m_{\Sigma}^2}$  \\
			\hline
			$Q_{\text{Hq}}{}^{(1)}$  &  $-\frac{g_Y^4}{5760 \pi ^2 m_{\Sigma}^2}$  \\
			\hline
			$Q_{\text{Hl}}{}^{(3)}$  &  $-\frac{g_W^4}{192 \pi ^2 m_{\Sigma}^2}$  \\
			\hline
			$Q_{\text{Hq}}{}^{(3)}$  &  $-\frac{g_W^4}{192 \pi ^2 m_{\Sigma}^2}$  \\
			\hline
			$Q_{\text{Hd}}$  &  $\frac{g_Y^4}{2880 \pi ^2 m_{\Sigma}^2}$  \\
			\hline
			$Q_{\text{He}}$  &  $\frac{g_Y^4}{960 \pi ^2 m_{\Sigma}^2}$  \\
			\hline
			$Q_{\text{Hu}}$  &  $-\frac{g_Y^4}{1440 \pi ^2 m_{\Sigma}^2}$  \\
			\hline \hline
	\end{tabular}}
	\subfloat{
		\begin{tabular}{|*{2}{>{\rowfonttype}c|}}%
			\hline \hline
			Dimension-6 Ops.& Wilson coefficients \\
			\hline \hline \rowfont{\color{blue}}
			$Q_W$  &  $\frac{g_W^3}{576 \pi ^2 m_{\Sigma}^2}$  \\
			\hline
			$Q_{\text{ll}}$  &  $-\frac{g_W^4}{768 \pi ^2 m_{\Sigma}^2}-\frac{g_Y^4}{3840 \pi ^2 m_{\Sigma}^2}$  \\
			\hline \rowfont{\color{red}} 
			$Q_{\text{eu}}$  &  $\frac{g_Y^4}{720 \pi ^2 m_{\Sigma}^2}$  \\
			\hline
			$Q_{\text{lq}}{}^{(1)}$  &  $\frac{g_Y^4}{5760 \pi ^2 m_{\Sigma}^2}$  \\
			\hline
			$Q_{\text{qq}}{}^{(1)}$  &  $-\frac{g_Y^4}{34560 \pi ^2 m_{\Sigma}^2}$  \\
			\hline
			$Q_{\text{le}}$  &  $-\frac{g_Y^4}{960 \pi ^2 m_{\Sigma}^2}$  \\
			\hline
			$Q_{\text{lu}}$  &  $\frac{g_Y^4}{1440 \pi ^2 m_{\Sigma}^2}$  \\
			\hline
			$Q_{\text{qe}}$  &  $\frac{g_Y^4}{2880 \pi ^2 m_{\Sigma}^2}$  \\
			\hline
			$Q_{\text{uu}}$  &  $-\frac{g_Y^4}{2160 \pi ^2 m_{\Sigma}^2}$  \\
			\hline
			$Q_{\text{ld}}$  &  $-\frac{g_Y^4}{2880 \pi ^2 m_{\Sigma}^2}$  \\
			\hline
			$Q_{\text{qd}}{}^{(1)}$  &  $\frac{g_Y^4}{8640 \pi ^2 m_{\Sigma}^2}$  \\
			\hline
			$Q_{\text{qu}}{}^{(1)}$  &  $-\frac{g_Y^4}{4320 \pi ^2 m_{\Sigma}^2}$  \\
			\hline
			$Q_{\text{ud}}{}^{(1)}$  &  $\frac{g_Y^4}{2160 \pi ^2 m_{\Sigma}^2}$  \\
			\hline
			$Q_{\text{lq}}{}^{(3)}$  &  $-\frac{g_W^4}{384 \pi ^2 m_{\Sigma}^2}$  \\
			\hline
			$Q_{\text{qq}}{}^{(3)}$  &  $-\frac{g_W^4}{768 \pi ^2 m_{\Sigma}^2}$  \\
			\hline
			$Q_{\text{dd}}$  &  $-\frac{g_Y^4}{8640 \pi ^2 m_{\Sigma}^2}$  \\
			\hline
			$Q_{\text{ed}}$  &  $-\frac{g_Y^4}{1440 \pi ^2 m_{\Sigma}^2}$  \\
			\hline
			$Q_{\text{ee}}$  &  $-\frac{g_Y^4}{960 \pi ^2 m_{\Sigma}^2}$  \\
			\hline\hline
	\end{tabular}}
\end{table*}

\subsection{Colour non-Singlet Heavy Scalar Leptoquarks}
Here, we have given the exhaustive sets of effective operators and the associated WCs that are emerged after integrating out the heavy colour non-singlet heavy scalars up to 1-loop.

\subsubsection{Complex colour triplet, isospin singlet: $\varphi_1 \equiv (3_C,1_L,\left.-\frac{1}{3}\right\vert_Y) $}
In this model, we have extended the SM by a colour-triplet isospin-singlet scalar ($\varphi_1$) with hypercharge $ Y = -\frac{1}{3} $. The Lagrangian involving the heavy field is written as \cite{Bauer:2015knc,Bandyopadhyay:2016oif},
\begin{align}\label{eq:varphi1lag}
	\mathcal{L}_{\varphi_1} & = \mathcal{L}_{_\text{SM}}^{d\leq 4} \;+  \left(D_{\mu} \varphi_1\right)^\dagger \, \left(D^{\mu} \varphi_1\right) - m_{{\varphi_1}}^2 \, \varphi_1^\dagger \varphi_1- \eta_{\varphi_{_1}} H^\dagger H \, \varphi_{_1}^\dagger \varphi_{_1}  -\lambda_{\varphi_{_1}} \left(\varphi_{_1}^\dagger \varphi_{_1}\right)^2\nonumber\\
	&+\left\lbrace y_{\varphi_1}^{(i)} {\varphi_1^\alpha}^\dagger {q_L^\alpha}^T C i \sigma^2 l_L + y_{\varphi_1}^{(ii)} {\varphi_1^\alpha}^\dagger {u_R^\alpha}^T C  e_R  +\text{h.c.}\right\rbrace.
\end{align}
Here, $m_{\varphi_1}$ is  mass of the heavy field.  This model contains four BSM parameters $\eta_{\varphi_1},  \lambda_{\varphi_1}, y_{\varphi_1}^{(i)}, y_{\varphi_1}^{(ii)}$, and the WCs are functions of these parameters along with the SM ones, see tab.~\ref{tab:Phi1}. 

\begin{table*}[h!]
	\caption{\small Warsaw basis  effective operators and the associated WCs that emerge after integrating-out the heavy field $\varphi_1 : (3,1,-\frac{1}{3})$. See caption of tab.~\ref{tab:H2} for colour coding.}
	\label{tab:Phi1}
	\centering
	\scriptsize
	\renewcommand{\arraystretch}{1.8}
	\subfloat{
		\begin{tabular}{|*{2}{>{\rowfonttype}c|}}
			\hline \hline
			Dim-6 Ops.&Wilson coefficients\\
			\hline \hline
			$Q_H$  &  $-\frac{\eta _{\varphi _1}^3}{32 \pi ^2 m_{\varphi _1}^2}$  \\
			\hline
			$Q_{\text{HB}}$  &  $\frac{\eta _{\varphi _1} g_Y^2}{576 \pi ^2 m_{\varphi _1}^2}$  \\
			\hline
			$Q_{\text{HG}}$  &  $\frac{g_S^2 \eta _{\varphi _1}}{384 \pi ^2 m_{\varphi _1}^2}$  \\
			\hline
			$Q_{H\square }$  &  $-\frac{\eta _{\varphi _1}^2}{64 \pi ^2 m_{\varphi _1}^2}$  \\
			\hline \rowfont{\color{blue}}
			$Q_{\text{Hl}}{}^{(1)}$  &  $\frac{g_Y^4}{5760 \pi ^2 m_{\varphi _1}^2}$  \\
			\hline
			$Q_{\text{Hq}}{}^{(1)}$  &  $-\frac{g_Y^4}{17280 \pi ^2 m_{\varphi _1}^2}$  \\
			\hline
			$Q_G$  &  $\frac{g_S^3}{5760 \pi ^2 m_{\varphi _1}^2}$  \\
			\hline
			$Q_{\text{HD}}$  &  $-\frac{g_Y^4}{2880 \pi ^2 m_{\varphi _1}^2}$  \\
			\hline
			$Q_{\text{Hu}}$  &  $-\frac{g_Y^4}{4320 \pi ^2 m_{\varphi _1}^2}$  \\
			\hline
			$Q_{\text{Hd}}$  &  $\frac{g_Y^4}{8640 \pi ^2 m_{\varphi _1}^2}$  \\
			\hline
			$Q_{\text{He}}$  &  $\frac{g_Y^4}{2880 \pi ^2 m_{\varphi _1}^2}$  \\
			\hline
			$Q_{\text{ll}}$  &  $-\frac{g_Y^4}{11520 \pi ^2 m_{\varphi _1}^2}$  \\
			\hline \rowfont{\color{red}}
			$Q_{\text{ld}}$  &  $-\frac{g_Y^4}{8640 \pi ^2 m_{\varphi _1}^2}$  \\
			\hline
			$Q_{\text{le}}$  &  $-\frac{g_Y^4}{2880 \pi ^2 m_{\varphi _1}^2}$  \\
			\hline
			$Q_{\text{lu}}$  &  $\frac{g_Y^4}{4320 \pi ^2 m_{\varphi _1}^2}$  \\
			\hline \hline
	\end{tabular}}
	\subfloat{
		\begin{tabular}{|*{2}{>{\rowfonttype}c|}}
			\hline \hline
			Dim-6 Ops.&Wilson coefficients\\
			\hline \hline \rowfont{\color{red}} 
			$Q_{\text{qe}}$  &  $\frac{g_Y^4}{8640 \pi ^2 m_{\varphi _1}^2}$  \\
			\hline
			$Q_{\text{ed}}$  &  $-\frac{g_Y^4}{4320 \pi ^2 m_{\varphi _1}^2}$  \\
			\hline
			$Q_{\text{ee}}$  &  $-\frac{g_Y^4}{2880 \pi ^2 m_{\varphi _1}^2}$  \\
			\hline
			$Q_{\text{lq}}{}^{(1)}$  &  $\frac{g_Y^4}{17280 \pi ^2 m_{\varphi _1}^2}$  \\
			\hline
			$Q_{\text{qq}}{}^{(1)}$  &  $-\frac{g_Y^4}{103680 \pi ^2 m_{\varphi _1}^2}$  \\
			\hline
			$Q_{\text{ud}}{}^{(1)}$  &  $\frac{g_Y^4}{6480 \pi ^2 m_{\varphi _1}^2}$  \\
			\hline
			$Q_{\text{ud}}^{(8)}$&$-\frac{g_S^4}{960 \pi^2 m_{\varphi_1}^2}$\\
			\hline
			$Q_{\text{qu}}^{(8)}$&$-\frac{g_S^4}{960 \pi^2 m_{\varphi_1}^2}$\\
			\hline
			$Q_{\text{qd}}^{(8)}$&$-\frac{g_S^4}{960 \pi^2 m_{\varphi_1}^2}$\\
			\hline
			$Q_{\text{qu}}{}^{(1)}$  &  $-\frac{g_Y^4}{12960 \pi ^2 m_{\varphi _1}^2}$  \\
			\hline
			$Q_{\text{qd}}{}^{(1)}$  &  $\frac{g_Y^4}{25920 \pi ^2 m_{\varphi _1}^2}$  \\
			\hline
			$Q_{\text{lq}}{}^{(3)}$  &  $-\frac{3 \lambda _{\varphi _1} y_{\varphi _1}^{(i)}{}^2}{16 \pi ^2 m_{\varphi _1}^2}-\frac{3 y_{\varphi _1}^{(i)}{}^2}{8 m_{\varphi _1}^2}$  \\
			\hline
			$Q_{\text{dd}}$  &  $-\frac{g_Y^4}{25920 \pi ^2 m_{\varphi _1}^2}$  \\
			\hline
			$Q_{\text{eu}}$  &  $\frac{g_Y^4}{2160 \pi ^2 m_{\varphi _1}^2}+\frac{3 \lambda _{\varphi _1} y_{\varphi _1}^{(\text{ii})}{}^2}{8 \pi ^2 m_{\varphi _1}^2}+\frac{3 y_{\varphi _1}^{(\text{ii})}{}^2}{4 m_{\varphi _1}^2}$  \\
			\hline
			$Q_{\text{uu}}$  &  $-\frac{g_Y^4}{6480 \pi ^2 m_{\varphi _1}^2}$  \\
			\hline \hline
	\end{tabular}}
\end{table*}

\subsubsection{Complex colour triplet, isospin singlet: $\varphi_2 \equiv (3_C,1_L,\left.-\frac{4}{3}\right\vert_Y)$}
The heavy field $ (\varphi_2) $ in this model is a scalar leptoquark, similar to the model discussed above, but with a different hypercharge $ Y= -\frac{4}{3} $. Consider the BSM Lagrangian \cite{Davidson_2010,Arnold:2013cva},
\begin{align}\label{eq:varphi2lag}
	\mathcal{L}_{\varphi_2} & = \mathcal{L}_{_\text{SM}}^{d\leq 4} \;+  \left(D_{\mu} \varphi_2\right)^\dagger \, \left(D^{\mu} \varphi_2\right) - m_{{\varphi_2}}^2 \, \varphi_2^\dagger \varphi_2- \eta_{\varphi_{_2}} H^\dagger H \, \varphi_{_2}^\dagger \varphi_{_2}  -\lambda_{\varphi_{_2}} \left(\varphi_{_2}^\dagger \varphi_{_2}\right)^2\nonumber\\
	&+\left\lbrace y_{\varphi_2} {\varphi_2^\alpha}^\dagger {d_R^\alpha}^T C  e_R  +\text{h.c.}\right\rbrace.
\end{align}
Here, $m_{\varphi_2}$ is  mass of the heavy field.  This model contains three BSM parameters $\eta_{\varphi_2},  \lambda_{\varphi_2}, y_{\varphi_2}$, and the WCs are functions of these parameters along with the SM ones, see tab.~\ref{tab:Phi2}.

\begin{table*}[h!]
	\caption{\small Warsaw basis  effective operators and the associated WCs that emerge after integrating-out the heavy field $\varphi_2 : (3,1,-\frac{4}{3})$. See caption of tab.~\ref{tab:H2} for colour coding.}
	\label{tab:Phi2}
	\centering
	\scriptsize
	\renewcommand{\arraystretch}{1.8}
	\subfloat{
		\begin{tabular}{|*{2}{>{\rowfonttype}c|}}
			\hline \hline
			Dim-6 Ops.&Wilson coefficients\\
			\hline \hline
			$Q_H$  &  $-\frac{\eta _{\varphi _2}^3}{32 \pi ^2 m_{\varphi _2}^2}$  \\
			\hline
			$Q_{\text{HB}}$  &  $\frac{\eta _{\varphi _2} g_Y^2}{36 \pi ^2 m_{\varphi _2}^2}$  \\
			\hline
			$Q_{H\square }$  &  $-\frac{\eta _{\varphi _2}^2}{64 \pi ^2 m_{\varphi _2}^2}$  \\
			\hline
			$Q_{\text{HG}}$  &  $\frac{g_S^2 \eta _{\varphi _2}}{384 \pi ^2 m_{\varphi _2}^2}$  \\
			\hline \rowfont{\color{blue}}
			$Q_{\text{Hq}}{}^{(1)}$  &  $-\frac{g_Y^4}{1080 \pi ^2 m_{\varphi _2}^2}$  \\
			\hline
			$Q_{\text{HD}}$  &  $-\frac{g_Y^4}{180 \pi ^2 m_{\varphi _2}^2}$  \\
			\hline
			$Q_{\text{Hd}}$  &  $\frac{g_Y^4}{540 \pi ^2 m_{\varphi _2}^2}$  \\
			\hline
			$Q_{\text{He}}$  &  $\frac{g_Y^4}{180 \pi ^2 m_{\varphi _2}^2}$  \\
			\hline
			$Q_{\text{Hl}}{}^{(1)}$  &  $\frac{g_Y^4}{360 \pi ^2 m_{\varphi _2}^2}$  \\
			\hline
			$Q_G$  &  $\frac{g_S^3}{5760 \pi ^2 m_{\varphi _2}^2}$  \\
			\hline
			$Q_{\text{Hu}}$  &  $-\frac{g_Y^4}{270 \pi ^2 m_{\varphi _2}^2}$  \\
			\hline
			$Q_{\text{ll}}$  &  $-\frac{g_Y^4}{720 \pi ^2 m_{\varphi _2}^2}$  \\
			\hline \rowfont{\color{red}}
			$Q_{\text{ee}}$  &  $-\frac{g_Y^4}{180 \pi ^2 m_{\varphi _2}^2}$  \\
			\hline
			$Q_{\text{eu}}$  &  $\frac{g_Y^4}{135 \pi ^2 m_{\varphi _2}^2}$  \\
			\hline
			$Q_{\text{lq}}{}^{(1)}$  &  $\frac{g_Y^4}{1080 \pi ^2 m_{\varphi _2}^2}$  \\
			\hline \hline
	\end{tabular}}
	\subfloat{
		\begin{tabular}{|*{2}{>{\rowfonttype}c|}}
			\hline \hline
			Dim-6 Ops.&Wilson coefficients\\
			\hline \hline \rowfont{\color{red}} 
			$Q_{\text{qq}}{}^{(1)}$  &  $-\frac{g_Y^4}{6480 \pi ^2 m_{\varphi _2}^2}$  \\
			\hline
			$Q_{\text{qd}}{}^{(1)}$  &  $\frac{g_Y^4}{1620 \pi ^2 m_{\varphi _2}^2}$  \\
			\hline
			$Q_{\text{qu}}{}^{(1)}$  &  $-\frac{g_Y^4}{810 \pi ^2 m_{\varphi _2}^2}$  \\
			\hline
			$Q_{\text{dd}}$  &  $-\frac{g_Y^4}{1620 \pi ^2 m_{\varphi _2}^2}$  \\
			\hline
			$Q_{\text{ed}}$  &  $-\frac{g_Y^4}{270 \pi ^2 m_{\varphi _2}^2}+\frac{3 \lambda _{\varphi _2} y_{\varphi _2}^2}{8 \pi ^2 m_{\varphi _2}^2}+\frac{3 y_{\varphi _2}^2}{4 m_{\varphi _2}^2}$  \\
			\hline
			$Q_{\text{uu}}$  &  $-\frac{g_Y^4}{405 \pi ^2 m_{\varphi _2}^2}$  \\
			\hline
			$Q_{\text{qd}}^{(8)}$&$-\frac{g_S^4}{960 \pi^2 m_{\varphi_2}^2}$\\
			\hline
			$Q_{\text{qu}}^{(8)}$&$-\frac{g_S^4}{960 \pi^2 m_{\varphi_2}^2}$\\
			\hline
			$Q_{\text{ld}}$  &  $-\frac{g_Y^4}{540 \pi ^2 m_{\varphi _2}^2}$  \\
			\hline
			$Q_{\text{le}}$  &  $-\frac{g_Y^4}{180 \pi ^2 m_{\varphi _2}^2}$  \\
			\hline
			$Q_{\text{lu}}$  &  $\frac{g_Y^4}{270 \pi ^2 m_{\varphi _2}^2}$  \\
			\hline
			$Q_{\text{qe}}$  &  $\frac{g_Y^4}{540 \pi ^2 m_{\varphi _2}^2}$  \\
			\hline
			$Q_{\text{ud}}{}^{(1)}$  &  $\frac{g_Y^4}{405 \pi ^2 m_{\varphi _2}^2}$  \\
			\hline
			$Q_{\text{ud}}^{(8)}$&$-\frac{g_S^4}{960 \pi^2 m_{\varphi_2}^2}$\\
			\hline \hline
	\end{tabular}}
\end{table*}

\subsubsection{Complex colour triplet, isospin doublet: $\Theta_2 \equiv (3_C,2_L,\left.\frac{7}{6}\right\vert_Y)$}
The heavy field $ (\Theta_2) $ in this model is a scalar leptoquark, similar to the model discussed above, but with a different hypercharge $ Y= \frac{7}{6} $. The Lagrangian involving the heavy field is written as \cite{Buchmuller:1986zs,Arnold:2013cva,Davidson_2010},
\begin{align}\label{eq:theta2lag}
	\mathcal{L}_{\Theta_2} & = \mathcal{L}_{_\text{SM}}^{d\leq 4} \;+  \left(D_{\mu} \Theta_2\right)^\dagger \, \left(D^{\mu} \Theta_2\right) - m_{{\Theta_2}}^2 \, \Theta_2^\dagger \Theta_2- \eta_{_{\Theta_2}}^{(1)} H^\dagger H \, \Theta_2^\dagger \Theta_2- \eta_{_{\Theta_2}}^{(2)} \left(H^\dagger \sigma^i H\right) \, \left(\Theta_2^\dagger \sigma^i \Theta_2\right)\nonumber\\
	&-\lambda_{\Theta_2}^{(1)} \left(\Theta_2^\dagger \Theta_2\right)^2-\lambda_{\Theta_2}^{(2)} \left(\Theta_2^\dagger \sigma^i \Theta_2\right)^2+\left\lbrace y_{\Theta_2}^{(1)} \Theta_2^\alpha {\overline{q}_L^\alpha} e_R +y_{\Theta_2}^{(2)} \Theta_2^\alpha {\overline{u}_R^\alpha} l_L +\text{h.c.}\right\rbrace.
\end{align}
Here, $m_{\Theta_2}$ is  mass of the heavy field.  This model contains six BSM parameters $\eta_{\Theta_2}^{(1)}, \eta_{\Theta_2}^{(2)},  \lambda_{\Theta_2}^{(1)}, \lambda_{\Theta_2}^{(2)},$  $ y_{\Theta_2}^{(1)}, y_{\Theta_2}^{(2)}$, and the WCs are functions of these parameters along with the SM ones, see tab.~\ref{tab:Theta2}. 

\begin{table*}[h!]
	\caption{\small Warsaw basis  effective operators and the associated WCs that emerge after integrating-out the heavy field $\Theta_2$: $(3,2,\frac{7}{6})$. See caption of tab.~\ref{tab:H2} for colour coding.}
	\label{tab:Theta2}
	\centering
	\scriptsize
	\renewcommand{\arraystretch}{1.8}
	\subfloat{
		\begin{tabular}{|*{2}{>{\rowfonttype}c|}}
			\hline \hline
			Dim-6 Ops.&Wilson coefficients\\
			\hline \hline
			$Q_{\text{dH}}$  &  $\frac{\eta _{\Theta _2}^{(2)}{}^2 Y_d^{\text{SM}}}{256 \pi ^2 m_{\Theta _2}^2}$  \\
			\hline
			$Q_{\text{eH}}$  &  $\frac{\eta _{\Theta _2}^{(2)}{}^2 Y_e^{\text{SM}}}{256 \pi ^2 m_{\Theta _2}^2}$  \\
			\hline
			$Q_H$  &  $-\frac{\eta _{\Theta _2}^{(1)}{}^3}{16 \pi ^2 m_{\Theta _2}^2}-\frac{3 \eta _{\Theta _2}^{(1)} \eta _{\Theta _2}^{(2)}{}^2}{256 \pi ^2 m_{\Theta _2}^2}+\frac{\eta _{\Theta _2}^{(2)}{}^2 \lambda _H^{\text{SM}}}{128 \pi ^2 m_{\Theta _2}^2}$  \\
			\hline
			$Q_{\text{HB}}$  &  $\frac{49 g_Y^2 \eta _{\Theta _2}^{(1)}}{1152 \pi ^2 m_{\Theta _2}^2}$  \\
			\hline
			$Q_{H\square }$  &  $-\frac{g_W^4}{2560 \pi ^2 m_{\Theta _2}^2}-\frac{\eta _{\Theta _2}^{(1)}{}^2}{32 \pi ^2 m_{\Theta _2}^2}+\frac{\eta _{\Theta _2}^{(2)}{}^2}{512 \pi ^2 m_{\Theta _2}^2}$  \\
			\hline
			$Q_{\text{HD}}$  &  $-\frac{49 g_Y^4}{5760 \pi ^2 m_{\Theta _2}^2}-\frac{\eta _{\Theta _2}^{(2)}{}^2}{128 \pi ^2 m_{\Theta _2}^2}$  \\
			\hline
			$Q_{\text{HG}}$  &  $\frac{g_S^2 \eta _{\Theta _2}^{(1)}}{192 \pi ^2 m_{\Theta _2}^2}$  \\
			\hline
			$Q_{\text{HW}}$  &  $\frac{g_W^2 \eta _{\Theta _2}^{(1)}}{128 \pi ^2 m_{\Theta _2}^2}$  \\
			\hline
			$Q_{\text{HWB}}$  &  $\frac{7 g_W g_Y \eta _{\Theta _2}^{(2)}}{768 \pi ^2 m_{\Theta _2}^2}$  \\
			\hline \rowfont{\color{blue}}
			$Q_G$  &  $\frac{g_S^3}{2880 \pi ^2 m_{\Theta _2}^2}$  \\
			\hline
			$Q_{\text{Hu}}$  &  $-\frac{49 g_Y^4}{8640 \pi ^2 m_{\Theta _2}^2}$  \\
			\hline
			$Q_{\text{Hd}}$  &  $\frac{49 g_Y^4}{17280 \pi ^2 m_{\Theta _2}^2}$  \\
			\hline
			$Q_{\text{He}}$  &  $\frac{49 g_Y^4}{5760 \pi ^2 m_{\Theta _2}^2}$  \\
			\hline
			$Q_{\text{Hl}}{}^{(1)}$  &  $\frac{49 g_Y^4}{11520 \pi ^2 m_{\Theta _2}^2}$  \\
			\hline
			$Q_{\text{Hq}}{}^{(1)}$  &  $-\frac{49 g_Y^4}{34560 \pi ^2 m_{\Theta _2}^2}$  \\
			\hline
			$Q_{\text{Hl}}{}^{(3)}$  &  $-\frac{g_W^4}{640 \pi ^2 m_{\Theta _2}^2}$  \\
			\hline
			$Q_{\text{Hq}}{}^{(3)}$  &  $-\frac{g_W^4}{640 \pi ^2 m_{\Theta _2}^2}$  \\
			\hline
			$Q_{\text{ll}}$  &  $-\frac{g_W^4}{2560 \pi ^2 m_{\Theta _2}^2}-\frac{49 g_Y^4}{23040 \pi ^2 m_{\Theta _2}^2}$  \\
			\hline \rowfont{\color{red}}
			$Q_{\text{ld}}$  &  $-\frac{49 g_Y^4}{17280 \pi ^2 m_{\Theta _2}^2}$  \\
			\hline\hline
	\end{tabular}}
	\subfloat{
		\begin{tabular}{|*{2}{>{\rowfonttype}c|}}
			\hline \hline
			Dim-6 Ops.&Wilson coefficients\\
			\hline \hline \rowfont{\color{red}}
			$Q_{\text{qe}}$  &  $\frac{49 g_Y^4}{17280 \pi ^2 m_{\Theta _2}^2}-\frac{9 \left(4\lambda _{\Theta _2}^{(1)} + \lambda _{\Theta _2}^{(2)} \right) y_{\Theta _2}^{(1)}{}^2}{128 \pi ^2 m_{\Theta _2}^2}-\frac{y_{\Theta _2}^{(1)}{}^2}{4 m_{\Theta _2}^2}$  \\
			\hline
			$Q_{\text{le}}$  &  $-\frac{49 g_Y^4}{5760 \pi ^2 m_{\Theta _2}^2}$  \\
			\hline
			$Q_{\text{lu}}$  &  $\frac{49 g_Y^4}{8640 \pi ^2 m_{\Theta _2}^2}-\frac{9 \left(4\lambda_{\Theta _2}^{(1)}+\lambda_{\Theta _2}^{(2)} \right) y_{\Theta _2}^{(\text{2})}{}^2}{128 \pi ^2 m_{\Theta _2}^2}-\frac{y_{\Theta _2}^{(\text{2})}{}^2}{4 m_{\Theta _2}^2}$  \\
			\hline
			$Q_{\text{uH}}$  &  $\frac{\eta _{\Theta _2}^{(2)}{}^2 Y_u^{\text{SM}}}{256 \pi ^2 m_{\Theta _2}^2}$  \\
			\hline
			$Q_{\text{uu}}$  &  $-\frac{49 g_Y^4}{12960 \pi ^2 m_{\Theta _2}^2}$  \\
			\hline
			$Q_W$  &  $\frac{g_W^3}{1920 \pi ^2 m_{\Theta _2}^2}$\\
			\hline
			$Q_{\text{qd}}^{(8)}$&$-\frac{g_S^4}{480 \pi^2 m_{\Theta_2}^2}$\\
			\hline
			$Q_{\text{qu}}^{(8)}$&$-\frac{g_S^4}{480 \pi^2 m_{\Theta_2}^2}$\\
			\hline
			$Q_{\text{lq}}{}^{(1)}$  &  $\frac{49 g_Y^4}{34560 \pi ^2 m_{\Theta _2}^2}$  \\
			\hline
			$Q_{\text{qd}}{}^{(1)}$  &  $\frac{49 g_Y^4}{51840 \pi ^2 m_{\Theta _2}^2}$  \\
			\hline
			$Q_{\text{qq}}{}^{(1)}$  &  $-\frac{49 g_Y^4}{207360 \pi ^2 m_{\Theta _2}^2}$  \\
			\hline
			$Q_{\text{qu}}{}^{(1)}$  &  $-\frac{49 g_Y^4}{25920 \pi ^2 m_{\Theta _2}^2}$  \\
			\hline
			$Q_{\text{ud}}{}^{(1)}$  &  $\frac{49 g_Y^4}{12960 \pi ^2 m_{\Theta _2}^2}$  \\
			\hline
			$Q_{\text{lq}}{}^{(3)}$  &  $-\frac{g_W^4}{1280 \pi ^2 m_{\Theta _2}^2}$  \\
			\hline
			$Q_{\text{qq}}{}^{(3)}$  &  $-\frac{g_W^4}{2560 \pi ^2 m_{\Theta _2}^2}$  \\
			\hline
			$Q_{\text{dd}}$  &  $-\frac{49 g_Y^4}{51840 \pi ^2 m_{\Theta _2}^2}$  \\
			\hline
			$Q_{\text{ed}}$  &  $-\frac{49 g_Y^4}{8640 \pi ^2 m_{\Theta _2}^2}$  \\
			\hline
			$Q_{\text{ee}}$  &  $-\frac{49 g_Y^4}{5760 \pi ^2 m_{\Theta _2}^2}$  \\
			\hline
			$Q_{\text{eu}}$  &  $\frac{49 g_Y^4}{4320 \pi ^2 m_{\Theta _2}^2}$  \\
			\hline
			$Q_{\text{ud}}^{(8)}$&$-\frac{g_S^4}{480 \pi^2 m_{\Theta_2}^2}$\\
			\hline \hline
	\end{tabular}}
\end{table*}

\subsubsection{Complex colour triplet, isospin triplet: $\Omega \equiv (3_C,3_L,\left.-\frac{1}{3}\right\vert_Y)$}
In this model, we have extended the SM by a colour-triplet isospin-triplet scalar ($\Omega$) with hypercharge $ Y = -\frac{1}{3} $. The Lagrangian involving the heavy field is written as \cite{Buchmuller:1986zs,Arnold:2013cva},
\begin{align}\label{eq:omegalag}
	\mathcal{L}_{\Omega} & = \mathcal{L}_{_\text{SM}}^{d\leq 4} \; +  \left(D_{\mu} \Omega\right)^\dagger \, \left(D^{\mu} \Omega\right) - m_{{\Omega}}^2 \, \Omega^\dagger \Omega-- \eta_{_{\Omega}}^{(1)} H^\dagger H \, \Omega^\dagger \Omega- \eta_{_{\Omega}}^{(2)} \left(H^\dagger \sigma^i H\right) \, \left(\Omega^\dagger T_{adj}^i \Omega\right)\nonumber\\
	&-\lambda_{\Omega}^{(1)} \left(\Omega^\dagger \Omega\right)^2-\lambda_{\Omega}^{(2)} \left(\Omega^\dagger T_{adj}^i \Omega\right)^2 -\lambda_{\Omega}^{(3)} \left(\Omega^\dagger \lambda^A \Omega\right)^2 +\left\lbrace y_{\Omega} \Omega^\alpha {q_L^\alpha}^T C l_L  +\text{h.c.}\right\rbrace.
\end{align}
Here, $\lambda^A$'s are the $SU(3)_C$ generators, and $m_{\Omega}$ is  mass of the heavy field.  This model contains six BSM parameters $\eta_{\Omega}^{(1)}, \eta_{\Omega}^{(2)}, \lambda_{\Omega}^{(1)}, \lambda_{\Omega}^{(2)}, \lambda_{\Omega}^{(3)},  y_{\Omega}$, and the WCs are functions of these parameters along with the SM ones, see tab.~\ref{tab:Omega}.

\begin{table*}[h!]
	\caption{\small Warsaw basis  effective operators and the associated WCs that emerge after integrating-out the heavy field $\Omega: (3,3,-\frac{1}{3})$. See caption of tab.~\ref{tab:H2} for colour coding.}
	\label{tab:Omega}
	\centering
	\scriptsize
	\renewcommand{\arraystretch}{1.8}
	\subfloat{
		\begin{tabular}{|*{2}{>{\rowfonttype}c|}}
			\hline \hline
			Dim-6 Ops.&Wilson coefficients\\
			\hline \hline
			$Q_{\text{dH}}$  &  $\frac{\eta _{\Omega }^{(2)}{}^2 Y_d^{\text{SM}}}{64 \pi ^2 m_{\Omega }^2}$  \\
			\hline
			$Q_{\text{eH}}$  &  $\frac{\eta _{\Omega }^{(2)}{}^2 Y_e^{\text{SM}}}{64 \pi ^2 m_{\Omega }^2}$  \\
			\hline
			$Q_H$  &  $-\frac{3 \eta _{\Omega }^{(1)}{}^3}{32 \pi ^2 m_{\Omega }^2}-\frac{3 \eta _{\Omega }^{(1)} \eta _{\Omega }^{(2)}{}^2}{64 \pi ^2 m_{\Omega }^2}+\frac{\eta _{\Omega }^{(2)}{}^2 \lambda _H^{\text{SM}}}{32 \pi ^2 m_{\Omega }^2}$  \\
			\hline
			$Q_{\text{HB}}$  &  $\frac{g_Y^2 \eta _{\Omega }^{(1)}}{192 \pi ^2 m_{\Omega }^2}$  \\
			\hline
			$Q_{H\square }$  &  $-\frac{g_W^4}{640 \pi ^2 m_{\Omega }^2}-\frac{3 \eta _{\Omega }^{(1)}{}^2}{64 \pi ^2 m_{\Omega }^2}+\frac{\eta _{\Omega }^{(2)}{}^2}{128 \pi ^2 m_{\Omega }^2}$  \\
			\hline
			$Q_{\text{HD}}$  &  $-\frac{g_Y^4}{960 \pi ^2 m_{\Omega }^2}-\frac{\eta _{\Omega }^{(2)}{}^2}{32 \pi ^2 m_{\Omega }^2}$  \\
			\hline
			$Q_{\text{HG}}$  &  $\frac{g_S^2 \eta _{\Omega }^{(1)}}{128 \pi ^2 m_{\Omega }^2}$  \\
			\hline
			$Q_{\text{HW}}$  &  $\frac{g_W^2 \eta _{\Omega }^{(1)}}{32 \pi ^2 m_{\Omega }^2}$  \\
			\hline
			$Q_{\text{HWB}}$  &  $\frac{g_W g_Y \eta _{\Omega }^{(2)}}{96 \pi ^2 m_{\Omega }^2}$  \\
			\hline
			$Q_{\text{uH}}$  &  $\frac{\eta _{\Omega }^{(2)}{}^2 Y_u^{\text{SM}}}{64 \pi ^2 m_{\Omega }^2}$  \\
			\hline \rowfont{\color{blue}}
			$Q_{\text{Hl}}{}^{(1)}$  &  $\frac{g_Y^4}{1920 \pi ^2 m_{\Omega }^2}$  \\
			\hline
			$Q_{\text{Hq}}{}^{(1)}$  &  $-\frac{g_Y^4}{5760 \pi ^2 m_{\Omega }^2}$  \\
			\hline
			$Q_{\text{Hl}}{}^{(3)}$  &  $-\frac{g_W^4}{160 \pi ^2 m_{\Omega }^2}$  \\
			\hline
			$Q_{\text{Hq}}{}^{(3)}$  &  $-\frac{g_W^4}{160 \pi ^2 m_{\Omega }^2}$  \\
			\hline
			$Q_{\text{ll}}$  &  $-\frac{g_W^4}{640 \pi ^2 m_{\Omega }^2}-\frac{g_Y^4}{3840 \pi ^2 m_{\Omega }^2}$  \\
			\hline
			$Q_{\text{Hu}}$  &  $-\frac{g_Y^4}{1440 \pi ^2 m_{\Omega }^2}$  \\
			\hline
			$Q_{\text{Hd}}$  &  $\frac{g_Y^4}{2880 \pi ^2 m_{\Omega }^2}$  \\
			\hline
			$Q_{\text{He}}$  &  $\frac{g_Y^4}{960 \pi ^2 m_{\Omega }^2}$  \\
			\hline
			$Q_W$  &  $\frac{g_W^3}{480 \pi ^2 m_{\Omega }^2}$  \\
			\hline
			$Q_G$  &  $\frac{g_S^3}{1920 \pi ^2 m_{\Omega }^2}$  \\
			\hline\hline
	\end{tabular}}
	\subfloat{
		\begin{tabular}{|*{2}{>{\rowfonttype}c|}}
			\hline \hline
			Dim-6 Ops.&Wilson coefficients\\
			\hline \hline \rowfont{\color{red}}
			$Q_{\text{lq}}{}^{(3)}$  &  $-\frac{g_W^4}{320 \pi ^2 m_{\Omega }^2}+\frac{\left(15 \lambda^{(1)}_{\Omega }+3\lambda^{(2)}_{\Omega }+2\lambda^{(3)}_{\Omega }\right) y_{\Omega }^2}{32 \pi ^2 m_{\Omega }^2}+\frac{3 y_{\Omega }^2}{8 m_{\Omega }^2}$  \\
			\hline
			$Q_{\text{qq}}{}^{(3)}$  &  $-\frac{g_W^4}{640 \pi ^2 m_{\Omega }^2}$  \\
			\hline
			$Q_{\text{dd}}$  &  $-\frac{g_Y^4}{8640 \pi ^2 m_{\Omega }^2}$  \\
			\hline
			$Q_{\text{ed}}$  &  $-\frac{g_Y^4}{1440 \pi ^2 m_{\Omega }^2}$  \\
			\hline
			$Q_{\text{ee}}$  &  $-\frac{g_Y^4}{960 \pi ^2 m_{\Omega }^2}$  \\
			\hline
			$Q_{\text{eu}}$  &  $\frac{g_Y^4}{720 \pi ^2 m_{\Omega }^2}$  \\
			\hline
			$Q_{\text{ud}}^{(8)}$&$-\frac{g_S^4}{320 \pi^2 m_{\Omega}^2}$\\
			\hline
			$Q_{\text{lq}}{}^{(1)}$  &  $\frac{\left(15 \lambda^{(1)}_{\Omega } +3 \lambda^{(2)}_{\Omega } + 2 \lambda^{(3)}_{\Omega }\right) y_{\Omega }^2}{8 \pi ^2 m_{\Omega }^2}+\frac{3 y_{\Omega }^2}{2 m_{\Omega }^2}$  \\
			&$+\frac{g_Y^4}{5760 \pi ^2 m_{\Omega }^2}$\\
			\hline
			$Q_{\text{qd}}{}^{(1)}$  &  $\frac{g_Y^4}{8640 \pi ^2 m_{\Omega }^2}$  \\
			\hline
			$Q_{\text{qq}}{}^{(1)}$  &  $-\frac{g_Y^4}{34560 \pi ^2 m_{\Omega }^2}$  \\
			\hline
			$Q_{\text{qu}}{}^{(1)}$  &  $-\frac{g_Y^4}{4320 \pi ^2 m_{\Omega }^2}$  \\
			\hline
			$Q_{\text{ud}}{}^{(1)}$  &  $\frac{g_Y^4}{2160 \pi ^2 m_{\Omega }^2}$  \\
			\hline
			$Q_{\text{ld}}$  &  $-\frac{g_Y^4}{2880 \pi ^2 m_{\Omega }^2}$  \\
			\hline
			$Q_{\text{le}}$  &  $-\frac{g_Y^4}{960 \pi ^2 m_{\Omega }^2}$  \\
			\hline
			$Q_{\text{lu}}$  &  $\frac{g_Y^4}{1440 \pi ^2 m_{\Omega }^2}$  \\
			\hline
			$Q_{\text{qe}}$  &  $\frac{g_Y^4}{2880 \pi ^2 m_{\Omega }^2}$  \\
			\hline
			$Q_{\text{uu}}$  &  $-\frac{g_Y^4}{2160 \pi ^2 m_{\Omega }^2}$  \\
			\hline
			$Q_{\text{qu}}^{(8)}$&$-\frac{g_S^4}{320 \pi^2 m_{\Omega}^2}$\\
			\hline
			$Q_{\text{qd}}^{(8)}$&$-\frac{g_S^4}{320 \pi^2 m_{\Omega}^2}$\\
			\hline \hline
	\end{tabular}}
\end{table*}

\subsubsection{Complex colour sextet, isospin triplet: $\chi_{_1}\equiv(6_C,3_L,\left.\frac{1}{3}\right\vert_Y)$}
In this model, we have extended the SM by a colour-sextet isospin-triplet scalar ($\chi_{_1}$) with hypercharge $ Y = \frac{1}{3} $. The Lagrangian involving the heavy field is written as \cite{Chen:2008hh,deBlas:2017xtg},
{\footnotesize
	\begin{align}\label{eq:chi1lag}
	\mathcal{L}_{\chi_1} & = \mathcal{L}_{_\text{SM}}^{d\leq 4} \; +  Tr\left[\left(D_{\mu} \chi_1\right)^\dagger \, \left(D^{\mu} \chi_1\right)\right] - m_{{\chi_1}}^2 \, Tr \left[\chi_1^\dagger \chi_1\right]- \eta_{_{\chi_1}}^{(1)} H^\dagger H \, Tr\left[\chi_1^\dagger \chi_1\right] \nonumber\\ 
	&- \eta_{_{\chi_1}}^{(2)} \left(H^\dagger \sigma^i H\right) \, Tr\left[\left(\chi_1^\dagger \sigma^i \chi_1 \right)\right] -\lambda^{(1)}_{\chi_1} \left(Tr\left[\chi_1^\dagger \chi_1\right] \right)^2-\lambda^{(2)}_{\chi_1} Tr \left[\left(\chi_1^\dagger \chi_1 \right)^2\right]\nonumber\\
	&-\left\lbrace y_{\chi_{_1}} \left(q_L^{\lbrace A|}\right)^T C \left(\chi_{_1}^{AB,I}\right)^\dagger i \sigma^2  q_L^{|B\rbrace}  + \text{h.c.}\right\rbrace.
	\end{align}
}
Here, $m_{\chi_{_1}}$ is  mass of the heavy field.  This model contains five BSM parameters $\eta_{\chi_{_1}}^{(1)}, \eta_{\chi_{_1}}^{(2)},  \lambda_{\chi_{_1}}^{(1)}, \lambda_{\chi_{_1}}^{(2)}, y_{\chi_{_1}}$, and the WCs are functions of these parameters along with the SM ones, see tab.~\ref{tab:Chi1}.

\begin{table*}[h!]
	\caption{\small Warsaw basis  effective operators and the associated WCs that emerge after integrating-out the heavy field $ \chi_1 : (6,3,\frac{1}{3})$. See caption of tab.~\ref{tab:H2} for colour coding.}
	\label{tab:Chi1}
	\centering
	\scriptsize
	\renewcommand{\arraystretch}{1.6}
	\subfloat{
	\begin{tabular}{|*{2}{>{\rowfonttype}c|}}
		\hline \hline
		Dim-6 Ops.&Wilson coefficients\\
		\hline \hline
		$Q_H$  &  $-\frac{{\eta^{(1)}_{\chi _1}}^3}{16 \pi ^2 m_{\chi _1}^2}-\frac{3 \eta _{\chi _1}^{(1)} {\eta _{\chi _1}^{(2)}}^2}{64 \pi ^2 m_{\chi _1}^2}+\frac{{\eta _{\chi _1}^{(2)}}^2 \lambda _H^{\text{SM}}}{16 \pi ^2 m_{\chi _1}^2}$  \\
		\hline
		$Q_{H\square }$  &  $-\frac{g_W^4}{320 \pi ^2 m_{\chi _1}^2}-\frac{{\eta _{\chi _1}^{(1)}}^2}{32 \pi ^2 m_{\chi _1}^2}+\frac{{\eta _{\chi _1}^{(2)}}^2}{64 \pi ^2 m_{\chi _1}^2}$  \\
		\hline
		$Q_{\text{HD}}$  &  $-\frac{g_Y^4}{480 \pi ^2 m_{\chi _1}^2}-\frac{{\eta _{\chi _1}^{(2)}}^2}{16 \pi ^2 m_{\chi _1}^2}$  \\
		\hline
		$Q_{\text{eH}}$  &  $\frac{{\eta _{\chi _1}^{(2)}}^2 Y_e^{\text{SM}}}{32 \pi ^2 m_{\chi _1}^2}$  \\
		\hline
		$Q_{\text{uH}}$  &  $\frac{{\eta _{\chi _1}^{(2)}}^2 Y_u^{\text{SM}}}{32 \pi ^2 m_{\chi _1}^2}$  \\
		\hline
		$Q_{\text{dH}}$  &  $\frac{{\eta _{\chi _1}^{(2)}}^2 Y_d^{\text{SM}}}{32 \pi ^2 m_{\chi _1}^2}$  \\
		\hline 
		$Q_{\text{HG}}$  &  $\frac{g_S^2 {\eta _{\chi _1}^{(1)}}}{192 \pi ^2 m_{\chi _1}^2}$  \\
		\hline
		$Q_{\text{HW}}$  &  $\frac{{\eta _{\chi _1}^{(1)}} g_W^2}{32 \pi ^2 m_{\chi _1}^2}$  \\
		\hline
		$Q_{\text{HWB}}$  &  $\frac{{\eta _{\chi _1}^{(2)}} g_W g_Y}{48 \pi ^2 m_{\chi _1}^2}$  \\
		\hline
		$Q_{\text{HB}}$  &  $\frac{{\eta _{\chi _1}^{(1)}} g_Y^2}{288 \pi ^2 m_{\chi _1}^2}$  \\
		\hline \rowfont{\color{blue}}
		$Q_{\text{Hd}}$  &  $\frac{g_Y^4}{1440 \pi ^2 m_{\chi _1}^2}$  \\
		\hline
		$Q_{\text{He}}$  &  $\frac{g_Y^4}{480 \pi ^2 m_{\chi _1}^2}$  \\
		\hline
		$Q_{\text{Hl}}{}^{(1)}$  &  $\frac{g_Y^4}{960 \pi ^2 m_{\chi _1}^2}$  \\
		\hline
		$Q_{\text{Hq}}{}^{(1)}$  &  $-\frac{g_Y^4}{2880 \pi ^2 m_{\chi _1}^2}$  \\
		\hline
		$Q_{\text{Hl}}{}^{(3)}$  &  $-\frac{g_W^4}{80 \pi ^2 m_{\chi _1}^2}$  \\
		\hline
		$Q_{\text{Hq}}{}^{(3)}$  &  $-\frac{g_W^4}{80 \pi ^2 m_{\chi _1}^2}$  \\
		\hline
		$Q_{\text{ll}}$  &  $-\frac{g_W^4}{320 \pi ^2 m_{\chi _1}^2}-\frac{g_Y^4}{1920 \pi ^2 m_{\chi _1}^2}$  \\
		\hline
		$Q_{\text{Hu}}$  &  $-\frac{g_Y^4}{720 \pi ^2 m_{\chi _1}^2}$  \\
		\hline
		$Q_G$  &  $\frac{g_S^3}{960 \pi ^2 m_{\chi _1}^2}$  \\
		\hline
		$Q_W$  &  $\frac{g_W^3}{240 \pi ^2 m_{\chi _1}^2}$  \\
		\hline\hline
	\end{tabular}}
	\subfloat{
	\begin{tabular}{|*{2}{>{\rowfonttype}c|}}
		\hline \hline
		Dim-6 Ops.&Wilson coefficients\\
		\hline \hline \rowfont{\color{red}}
		$Q_{\text{lq}}{}^{(1)}$  &  $\frac{g_Y^4}{2880 \pi ^2 m_{\chi _1}^2}$  \\
		\hline
		$Q_{\text{qd}}{}^{(1)}$  &  $\frac{g_Y^4}{4320 \pi ^2 m_{\chi _1}^2}$  \\
		\hline
		$Q_{\text{qq}}{}^{(1)}$  &  $-\frac{g_Y^4}{17280 \pi ^2 m_{\chi _1}^2}+ \frac{3 y_{\chi_{_1}}^2}{m_{\chi_{_1}}^2}+\frac{\left(12 \lambda_{\chi_{_1}}^{(1)} + 3 \lambda_{\chi_{_1}}^{(2)}\right) y_{\chi_{_1}}^2}{2 \pi^2  m_{\chi_{_1}}^2}$  \\
		\hline
		$Q_{\text{qu}}{}^{(1)}$  &  $-\frac{g_Y^4}{2160 \pi ^2 m_{\chi _1}^2}$  \\
		\hline
		$Q_{\text{dd}}$  &  $-\frac{g_Y^4}{4320 \pi ^2 m_{\chi _1}^2}$  \\
		\hline
		$Q_{\text{uu}}$  &  $-\frac{g_Y^4}{1080 \pi ^2 m_{\chi _1}^2}$  \\
		\hline
		$Q_{\text{ud}}{}^{(1)}$  &  $\frac{g_Y^4}{1080 \pi ^2 m_{\chi _1}^2}$  \\
		\hline
		$Q_{\text{lq}}{}^{(3)}$  &  $-\frac{g_W^4}{160 \pi ^2 m_{\chi _1}^2}$  \\
		\hline
		$Q_{\text{qq}}{}^{(3)}$  &  $-\frac{g_W^4}{320 \pi ^2 m_{\chi _1}^2}+ \frac{3 y_{\chi_{_1}}^2}{4 m_{\chi_{_1}}^2}+\frac{\left(12 \lambda_{\chi_{_1}}^{(1)} + 3 \lambda_{\chi_{_1}}^{(2)}\right) y_{\chi_{_1}}^2}{8 \pi^2  m_{\chi_{_1}}^2}$  \\
		\hline
		$Q_{\text{qe}}$  &  $\frac{g_Y^4}{1440 \pi ^2 m_{\chi _1}^2}$  \\
		\hline
		$Q_{\text{ed}}$  &  $-\frac{g_Y^4}{720 \pi ^2 m_{\chi _1}^2}$  \\
		\hline
		$Q_{\text{ee}}$  &  $-\frac{g_Y^4}{480 \pi ^2 m_{\chi _1}^2}$  \\
		\hline
		$Q_{\text{eu}}$  &  $\frac{g_Y^4}{360 \pi ^2 m_{\chi _1}^2}$  \\
		\hline
		$Q_{\text{ld}}$  &  $-\frac{g_Y^4}{1440 \pi ^2 m_{\chi _1}^2}$  \\
		\hline
		$Q_{\text{le}}$  &  $-\frac{g_Y^4}{480 \pi ^2 m_{\chi _1}^2}$  \\
		\hline
		$Q_{\text{lu}}$  &  $\frac{g_Y^4}{720 \pi ^2 m_{\chi _1}^2}$  \\
		\hline
		$Q_{qu}^{(8)}$&$-\frac{g_S^4}{160 \pi ^2 m_{\chi _1}^2}$\\
		\hline
		$Q_{qd}^{(8)}$&$-\frac{g_S^4}{160 \pi ^2 m_{\chi _1}^2}$\\
		\hline
		$Q_{ud}^{(8)}$&$-\frac{g_S^4}{160 \pi ^2 m_{\chi _1}^2}$\\
		\hline \hline
	\end{tabular}}
\end{table*}
%%%%%%%%%%%%%%%%%%%%%%%%%%%%%%%%%%%%%%

\subsubsection{Complex colour sextet, isospin singlet: $\chi_{_2}\equiv(6_C,1_L,\left.\frac{4}{3}\right\vert_Y)$}
Here, we have extended the SM by a colour-sextet isospin-singlet scalar ($\chi_{_2}$) with hypercharge $ Y = \frac{4}{3} $. The Lagrangian involving the heavy field is written as \cite{Chen:2008hh,deBlas:2017xtg},
\begin{align}\label{eq:chi2lag}
\mathcal{L}_{\chi_2} & = \mathcal{L}_{_\text{SM}}^{d\leq 4} \; +  \left(D_{\mu} \chi_2\right)^\dagger \, \left(D^{\mu} \chi_2\right) - m_{{\chi_2}}^2 \, \chi_2^\dagger \chi_2 - \eta_{_{\chi_2}} H^\dagger H \, \chi_2^\dagger \chi_2 -\lambda_{\chi_2} \left(\chi_2^\dagger \chi_2 \right)^2\nonumber\\
&-\left\lbrace y_{\chi_{_2}} \left(u_R^{\lbrace A|}\right)^T C \left(\chi_{_2}^{AB}\right)^\dagger u_R^{|B\rbrace}  + \text{h.c.}\right\rbrace.
\end{align}
Here, $m_{\chi_{_2}}$ is  mass of the heavy field.  This model contains three BSM parameters $\eta_{\chi_{_2}},  \lambda_{\chi_{_2}}, y_{\chi_{_2}}$, and the WCs are functions of these parameters along with the SM ones, see tab.~\ref{tab:Chi2}.

\begin{table*}[h!]
	\caption{\small Warsaw basis  effective operators and the associated WCs that emerge after integrating-out the heavy field $ \chi_2 : (6,1,\frac{4}{3})$. See caption of tab.~\ref{tab:H2} for colour coding.}
	\label{tab:Chi2}
	\centering
	\scriptsize
	\renewcommand{\arraystretch}{1.8}
	\begin{tabular}{|*{2}{>{\rowfonttype}c|}}
		\hline \hline
		Dim-6 Ops.&Wilson coefficients\\
		\hline \hline
		$Q_H$  &  $-\frac{\eta _{\chi _2}^3}{16 \pi ^2 m_{\chi _2}^2}$  \\
		\hline
		$Q_{\text{HB}}$  &  $\frac{\eta _{\chi _2} g_Y^2}{18 \pi ^2 m_{\chi _2}^2}$  \\
		\hline
		$Q_{H\square }$  &  $-\frac{\eta _{\chi _2}^2}{32 \pi ^2 m_{\chi _2}^2}$  \\
		\hline
		$Q_{\text{HG}}$  &  $\frac{g_S^2 \eta _{\chi _2}}{192 \pi ^2 m_{\chi _2}^2}$  \\
		\hline \rowfont{\color{blue}}
		$Q_{\text{HD}}$  &  $-\frac{g_Y^4}{90 \pi ^2 m_{\chi _2}^2}$  \\
		\hline
		$Q_{\text{Hd}}$  &  $\frac{g_Y^4}{270 \pi ^2 m_{\chi _2}^2}$  \\
		\hline
		$Q_{\text{He}}$  &  $\frac{g_Y^4}{90 \pi ^2 m_{\chi _2}^2}$  \\
		\hline
		$Q_{\text{Hu}}$  &  $-\frac{g_Y^4}{135 \pi ^2 m_{\chi _2}^2}$  \\
		\hline
		$Q_{\text{ll}}$  &  $-\frac{g_Y^4}{360 \pi ^2 m_{\chi _2}^2}$  \\
		\hline
		$Q_{\text{Hl}}{}^{(1)}$  &  $\frac{g_Y^4}{180 \pi ^2 m_{\chi _2}^2}$  \\
		\hline
		$Q_{\text{Hq}}{}^{(1)}$  &  $-\frac{g_Y^4}{540 \pi ^2 m_{\chi _2}^2}$  \\
		\hline
		$Q_G$  &  $\frac{g_S^3}{2880 \pi ^2 m_{\chi _2}^2}$  \\
		\hline \rowfont{\color{red}}
		$Q_{ud}^{(8)}$ & $-\frac{g_S^4}{480 \pi ^2 m_{\chi _2}^2}$\\
		\hline
		$Q_{qu}^{(8)}$&$-\frac{g_S^4}{480 \pi ^2 m_{\chi _2}^2}$\\
		\hline
		$Q_{qd}^{(8)}$&$-\frac{g_S^4}{480 \pi ^2 m_{\chi _2}^2}$\\
		\hline\hline
	\end{tabular}
	\begin{tabular}{|*{2}{>{\rowfonttype}c|}}
		\hline \hline
		Dim-6 Ops.&Wilson coefficients\\
		\hline \hline \rowfont{\color{red}}
		$Q_{\text{lq}}{}^{(1)}$  &  $\frac{g_Y^4}{540 \pi ^2 m_{\chi _2}^2}$  \\
		\hline
		$Q_{\text{qd}}{}^{(1)}$  &  $\frac{g_Y^4}{810 \pi ^2 m_{\chi _2}^2}$  \\
		\hline
		$Q_{\text{qq}}{}^{(1)}$  &  $-\frac{g_Y^4}{3240 \pi ^2 m_{\chi _2}^2}$  \\
		\hline
		$Q_{\text{qu}}{}^{(1)}$  &  $-\frac{g_Y^4}{405 \pi ^2 m_{\chi _2}^2}$  \\
		\hline
		$Q_{\text{ud}}{}^{(1)}$  &  $\frac{2 g_Y^4}{405 \pi ^2 m_{\chi _2}^2}$  \\
		\hline
		$Q_{\text{dd}}$  &  $-\frac{g_Y^4}{810 \pi ^2 m_{\chi _2}^2}$  \\
		\hline
		$Q_{\text{ed}}$  &  $-\frac{g_Y^4}{135 \pi ^2 m_{\chi _2}^2}$  \\
		\hline
		$Q_{\text{ee}}$  &  $-\frac{g_Y^4}{90 \pi ^2 m_{\chi _2}^2}$  \\
		\hline
		$Q_{\text{eu}}$  &  $\frac{2 g_Y^4}{135 \pi ^2 m_{\chi _2}^2}$  \\
		\hline
		$Q_{\text{ld}}$  &  $-\frac{g_Y^4}{270 \pi ^2 m_{\chi _2}^2}$  \\
		\hline
		$Q_{\text{le}}$  &  $-\frac{g_Y^4}{90 \pi ^2 m_{\chi _2}^2}$  \\
		\hline
		$Q_{\text{lu}}$  &  $\frac{g_Y^4}{135 \pi ^2 m_{\chi _2}^2}$  \\
		\hline
		$Q_{\text{qe}}$  &  $\frac{g_Y^4}{270 \pi ^2 m_{\chi _2}^2}$  \\
		\hline
		$Q_{\text{uu}}$  &  $-\frac{2 g_Y^4}{405 \pi ^2 m_{\chi _2}^2}$  \\
		& $ + \frac{3 y_{\chi_{_2}}^2}{2 m_{\chi_{_2}}^2}+\frac{3 \lambda_{\chi_{_2}}  y_{\chi_{_2}}^2}{8 \pi^2  m_{\chi_{_2}}^2} $ \\
		\hline \hline
	\end{tabular}
\end{table*}
%%%%%%%%%%%%%%%%%%%%%%%%%%%%%%%%%%%%%%
%%%%%%%%%%%%%%%%%%%%%%%%%%%%%%%%%%%%%%

\subsubsection{Complex colour sextet, isospin singlet: $\chi_{_3}\equiv(6_C,1_L,\left.-\frac{2}{3}\right\vert_Y)$}
Here, we have extended the SM by a colour-sextet isospin-singlet scalar ($\chi_{_3}$) with hypercharge $ Y = -\frac{2}{3} $. The Lagrangian involving the heavy field is written as \cite{Chen:2008hh,deBlas:2017xtg},
\begin{align}\label{eq:chi3lag}
\mathcal{L}_{\chi_3} & = \mathcal{L}_{_\text{SM}}^{d\leq 4} \; +  \left(D_{\mu} \chi_3\right)^\dagger \, \left(D^{\mu} \chi_3\right) - m_{{\chi_3}}^2 \, \chi_3^\dagger \chi_3 - \eta_{_{\chi_3}} H^\dagger H \, \chi_3^\dagger \chi_3 -\lambda_{\chi_3} \left(\chi_3^\dagger \chi_3 \right)^2\nonumber\\
&-\left\lbrace y_{\chi_{_3}} \left(d_R^{\lbrace A|}\right)^T C \left(\chi_{_3}^{AB}\right)^\dagger d_R^{|B\rbrace}  + \text{h.c.}\right\rbrace.
\end{align}
Here, $m_{\chi_{_3}}$ is  mass of the heavy field.  This model contains three BSM parameters $\eta_{\chi_{_3}}, \lambda_{\chi_{_3}}, y_{\chi_{_3}}$, and the WCs are functions of these parameters along with the SM ones, see tab.~\ref{tab:Chi3}.

\begin{table*}[h!]
	\caption{\small Warsaw basis  effective operators and the associated WCs that emerge after integrating-out the heavy field $ \chi_3 : (6,1,-\frac{2}{3})$. See caption of tab.~\ref{tab:H2} for colour coding.}
	\label{tab:Chi3}
	\centering
	\scriptsize
	\renewcommand{\arraystretch}{1.8}
	\begin{tabular}{|*{2}{>{\rowfonttype}c|}}
		\hline \hline
		Dim-6 Ops.&Wilson coefficients\\
		\hline \hline
		$Q_H$  &  $-\frac{\eta _{\chi _3}^3}{16 \pi ^2 m_{\chi _3}^2}$  \\
		\hline
		$Q_{\text{HB}}$  &  $\frac{\eta _{\chi _3} g_Y^2}{72 \pi ^2 m_{\chi _3}^2}$  \\
		\hline
		$Q_{\text{HG}}$  &  $\frac{g_S^2 \eta _{\chi _3}}{192 \pi ^2 m_{\chi _3}^2}$  \\
		\hline
		$Q_{H\square }$  &  $-\frac{\eta _{\chi _3}^2}{32 \pi ^2 m_{\chi _3}^2}$  \\
		\hline \rowfont{\color{blue}}
		$Q_{\text{HD}}$  &  $-\frac{g_Y^4}{360 \pi ^2 m_{\chi _3}^2}$  \\
		\hline
		$Q_{\text{Hl}}{}^{(1)}$  &  $\frac{g_Y^4}{720 \pi ^2 m_{\chi _3}^2}$  \\
		\hline
		$Q_{\text{Hq}}{}^{(1)}$  &  $-\frac{g_Y^4}{2160 \pi ^2 m_{\chi _3}^2}$  \\
		\hline
		$Q_G$  &  $\frac{g_S{}^3}{2880 \pi ^2 m_{\chi _3}^2}$  \\
		\hline
		$Q_{\text{Hd}}$  &  $\frac{g_Y^4}{1080 \pi ^2 m_{\chi _3}^2}$  \\
		\hline
		$Q_{\text{He}}$  &  $\frac{g_Y^4}{360 \pi ^2 m_{\chi _3}^2}$  \\
		\hline
		$Q_{\text{Hu}}$  &  $-\frac{g_Y^4}{540 \pi ^2 m_{\chi _3}^2}$  \\
		\hline
		$Q_{\text{ll}}$  &  $-\frac{g_Y^4}{1440 \pi ^2 m_{\chi _3}^2}$  \\
		\hline \rowfont{\color{red}}
		$Q_{qu}^{(8)}$&$-\frac{g_S^4}{480 \pi ^2 m_{\chi _3}^2}$\\
		\hline
		$Q_{qd}^{(8)}$&$-\frac{g_S^4}{480 \pi ^2 m_{\chi _3}^2}$\\
		\hline
		$Q_{ud}^{(8)}$&$-\frac{g_S^4}{480 \pi ^2 m_{\chi _3}^2}$\\
		\hline\hline
	\end{tabular}
	\begin{tabular}{|*{2}{>{\rowfonttype}c|}}
		\hline \hline
		Dim-6 Ops.&Wilson coefficients\\
		\hline \hline \rowfont{\color{red}}
		$Q_{\text{lq}}{}^{(1)}$  &  $\frac{g_Y^4}{2160 \pi ^2 m_{\chi _3}^2}$  \\
		\hline
		$Q_{\text{qd}}{}^{(1)}$  &  $\frac{g_Y^4}{3240 \pi ^2 m_{\chi _3}^2}$  \\
		\hline
		$Q_{\text{qq}}{}^{(1)}$  &  $-\frac{g_Y^4}{12960 \pi ^2 m_{\chi _3}^2}$  \\
		\hline
		$Q_{\text{qu}}{}^{(1)}$  &  $-\frac{g_Y^4}{1620 \pi ^2 m_{\chi _3}^2}$  \\
		\hline
		$Q_{\text{ud}}{}^{(1)}$  &  $\frac{g_Y^4}{810 \pi ^2 m_{\chi _3}^2}$  \\
		\hline
		$Q_{\text{dd}}$  &  $-\frac{g_Y^4}{3240 \pi ^2 m_{\chi _3}^2}$  \\
		&$ + \frac{3 y_{\chi_{_3}}^2}{2 m_{\chi_{_3}}^2}+\frac{3 \lambda_{\chi_{_3}}  y_{\chi_{_3}}^2}{8 \pi^2  m_{\chi_{_3}}^2} $\\
		\hline
		$Q_{\text{ed}}$  &  $-\frac{g_Y^4}{540 \pi ^2 m_{\chi _3}^2}$  \\
		\hline
		$Q_{\text{ee}}$  &  $-\frac{g_Y^4}{360 \pi ^2 m_{\chi _3}^2}$  \\
		\hline
		$Q_{\text{eu}}$  &  $\frac{g_Y^4}{270 \pi ^2 m_{\chi _3}^2}$  \\
		\hline
		$Q_{\text{ld}}$  &  $-\frac{g_Y^4}{1080 \pi ^2 m_{\chi _3}^2}$  \\
		\hline
		$Q_{\text{le}}$  &  $-\frac{g_Y^4}{360 \pi ^2 m_{\chi _3}^2}$  \\
		\hline
		$Q_{\text{lu}}$  &  $\frac{g_Y^4}{540 \pi ^2 m_{\chi _3}^2}$  \\
		\hline
		$Q_{\text{qe}}$  &  $\frac{g_Y^4}{1080 \pi ^2 m_{\chi _3}^2}$  \\
		\hline
		$Q_{\text{uu}}$  & $-\frac{g_Y^4}{810 \pi ^2 m_{\chi _3}^2}$  \\
		\hline \hline
	\end{tabular}
\end{table*}

\subsubsection{Complex colour sextet, isospin singlet: $\chi_{_4}\equiv(6_C,1_L,\left.\frac{1}{3}\right\vert_Y)$}
Here, we have extended the SM by a colour-sextet isospin-singlet scalar ($\chi_{_4}$) with hypercharge $ Y = \frac{1}{3} $. The Lagrangian involving the heavy field is written as \cite{Chen:2008hh,deBlas:2017xtg},
\begin{align}\label{eq:chi4lag}
\mathcal{L}_{\chi_4} & = \mathcal{L}_{_\text{SM}}^{d\leq 4} \; +  \left(D_{\mu} \chi_4\right)^\dagger \, \left(D^{\mu} \chi_4\right) - m_{{\chi_4}}^2 \, \chi_4^\dagger \chi_4- \eta_{_{\chi_4}} H^\dagger H \, \chi_4^\dagger \chi_4 -\lambda_{\chi_4} \left(\chi_4^\dagger \chi_4 \right)^2\nonumber\\
&-\left\lbrace y_{\chi_{_4}}^{(i)} \left(u_R^{\lbrace A|}\right)^T \left(\chi_{_4}^{AB}\right)^\dagger d_R^{|B\rbrace} +y_{\chi_{_4}}^{(ii)} \left(q_L^{\lbrace A|}\right)^T C \left(\chi_{_4}^{AB}\right)^\dagger i \sigma^2 q_L^{|B\rbrace}   + \text{h.c.}\right\rbrace.
\end{align}
Here, $m_{\chi_{_4}}$ is  mass of the heavy field.  This model contains four BSM parameters $\eta_{\chi_{_4}}, \lambda_{\chi_{_4}}, y_{\chi_{_4}}^{(i)}, y_{\chi_{_4}}^{(ii)}$, and the WCs are functions of these parameters along with the SM ones, see tab.~\ref{tab:Chi4}.
\begin{table*}[h!]
	\caption{\small Warsaw basis  effective operators and the associated WCs that emerge after integrating-out the heavy field $ \chi_4 : (6,1,\frac{1}{3})$. See caption of tab.~\ref{tab:H2} for colour coding.}
	\label{tab:Chi4}
	\centering
	\scriptsize
	\renewcommand{\arraystretch}{1.6}
	\begin{tabular}{|*{2}{>{\rowfonttype}c|}}
		\hline \hline
		Dim-6 Ops.&Wilson coefficients\\
		\hline \hline
		$Q_H$  &  $-\frac{\eta _{\chi _4}^3}{16 \pi ^2 m_{\chi _4}^2}$  \\
		\hline
		$Q_{\text{HB}}$  &  $\frac{\eta _{\chi _4} g_Y^2}{288 \pi ^2 m_{\chi _4}^2}$  \\
		\hline
		$Q_{H\square }$  &  $-\frac{\eta _{\chi _4}^2}{32 \pi ^2 m_{\chi _4}^2}$  \\
		\hline
		$Q_{\text{HG}}$  &  $\frac{g_S^2 \eta _{\chi _4}}{192 \pi ^2 m_{\chi _4}^2}$  \\
		\hline \rowfont{\color{blue}}
		$Q_{\text{HD}}$  &  $-\frac{g_Y^4}{1440 \pi ^2 m_{\chi _4}^2}$  \\
		\hline
		$Q_G$  &  $\frac{g_S^3}{2880 \pi ^2 m_{\chi _4}^2}$  \\
		\hline
		$Q_{\text{Hd}}$  &  $\frac{g_Y^4}{4320 \pi ^2 m_{\chi _4}^2}$  \\
		\hline
		$Q_{\text{He}}$  &  $\frac{g_Y^4}{1440 \pi ^2 m_{\chi _4}^2}$  \\
		\hline
		$Q_{\text{Hu}}$  &  $-\frac{g_Y^4}{2160 \pi ^2 m_{\chi _4}^2}$  \\
		\hline
		$Q_{\text{Hl}}{}^{(1)}$  &  $\frac{g_Y^4}{2880 \pi ^2 m_{\chi _4}^2}$  \\
		\hline
		$Q_{\text{Hq}}{}^{(1)}$  &  $-\frac{g_Y^4}{8640 \pi ^2 m_{\chi _4}^2}$  \\
		\hline
		$Q_{\text{ll}}$  &  $-\frac{g_Y^4}{5760 \pi ^2 m_{\chi _4}^2}$  \\
		\hline \rowfont{\color{red}}
		$Q_{\text{ld}}$  &  $-\frac{g_Y^4}{4320 \pi ^2 m_{\chi _4}^2}$  \\
		\hline
		$Q_{\text{le}}$  &  $-\frac{g_Y^4}{1440 \pi ^2 m_{\chi _4}^2}$  \\
		\hline
		$Q_{\text{lu}}$  &  $\frac{g_Y^4}{2160 \pi ^2 m_{\chi _4}^2}$  \\
		\hline
		$Q_{\text{qe}}$  &  $\frac{g_Y^4}{4320 \pi ^2 m_{\chi _4}^2}$  \\
		\hline\hline
	\end{tabular}
	\begin{tabular}{|*{2}{>{\rowfonttype}c|}}
		\hline \hline
		Dim-6 Ops.&Wilson coefficients\\
		\hline \hline \rowfont{\color{red}}
		$Q_{quqd}^{(1)}$ & $\frac{3 {y_{\chi_{_4}}^{(i)}} {y_{\chi_{_4}}^{(ii)}}}{2 m_{\chi_{_4}}^2}+\frac{3 \lambda_{\chi_{_4}}  {y_{\chi_{_4}}^{(i)}}{y_{\chi_{_4}}^{(ii)}}}{8 \pi^2  m_{\chi_{_4}}^2}$\\
		\hline
		$Q_{\text{uu}}$  &  $-\frac{g_Y^4}{3240 \pi ^2 m_{\chi _4}^2}$  \\
		\hline
		$Q_{\text{ee}}$  &  $-\frac{g_Y^4}{1440 \pi ^2 m_{\chi _4}^2}$  \\
		\hline
		$Q_{\text{lq}}{}^{(1)}$  &  $\frac{g_Y^4}{8640 \pi ^2 m_{\chi _4}^2}$  \\
		\hline
		$Q_{\text{qd}}{}^{(1)}$  &  $\frac{g_Y^4}{12960 \pi ^2 m_{\chi _4}^2}$  \\
		\hline
		$Q_{\text{qq}}{}^{(1)}$  &  $-\frac{g_Y^4}{51840 \pi ^2 m_{\chi _4}^2}+ \frac{3 {y_{\chi_{_4}}^{(ii)}}^2}{2 m_{\chi_{_4}}^2}+\frac{3 \lambda_{\chi_{_4}}  {y_{\chi_{_4}}^{(ii)}}^2}{8 \pi^2  m_{\chi_{_4}}^2}$  \\
		\hline
		$Q_{\text{qu}}{}^{(1)}$  &  $-\frac{g_Y^4}{6480 \pi ^2 m_{\chi _4}^2}$  \\
		\hline
		$Q_{\text{ud}}{}^{(1)}$  &  $\frac{g_Y^4}{3240 \pi ^2 m_{\chi _4}^2}  + \frac{3 {y_{\chi_{_4}}^{(i)}}^2}{2 m_{\chi_{_4}}^2}+\frac{3 \lambda_{\chi_{_4}}  {y_{\chi_{_4}}^{(i)}}^2}{8 \pi^2  m_{\chi_{_4}}^2}$  \\
		\hline
		$Q_{\text{dd}}$  &  $-\frac{g_Y^4}{12960 \pi ^2 m_{\chi _4}^2}$  \\
		\hline
		$Q_{\text{ed}}$  &  $-\frac{g_Y^4}{2160 \pi ^2 m_{\chi _4}^2}$  \\
		\hline
		$Q_{\text{eu}}$  &  $\frac{g_Y^4}{1080 \pi ^2 m_{\chi _4}^2}$  \\
		\hline
		$Q_{ud}^{(8)}$&$-\frac{g_S^4}{480 \pi ^2 m_{\chi _4}^2}+ \frac{ {y_{\chi_{_4}}^{(i)}}^2}{2 m_{\chi_{_4}}^2}+\frac{ \lambda_{\chi_{_4}}  {y_{\chi_{_4}}^{(i)}}^2}{8 \pi^2  m_{\chi_{_4}}^2}$\\
		\hline
		$Q_{qu}^{(8)}$&$-\frac{g_S^4}{480 \pi ^2 m_{\chi _4}^2}$\\
		\hline
		$Q_{qd}^{(8)}$ & $-\frac{g_S^4}{480 \pi ^2 m_{\chi _4}^2}$\\
		\hline
		$Q_{quqd}^{(8)}$ & $\frac{{y_{\chi_{_4}}^{(i)}} {y_{\chi_{_4}}^{(ii)}}}{ 2 m_{\chi_{_4}}^2}+\frac{\lambda_{\chi_{_4}}  {y_{\chi_{_4}}^{(i)}}{y_{\chi_{_4}}^{(ii)}}}{8 \pi^2  m_{\chi_{_4}}^2}$\\
		\hline \hline
	\end{tabular}
\end{table*}
\clearpage

\section{More information on Model Independent Bayesian analysis} \label{sec:moreinfo}

\subsection{Priors used} \label{subsec:priors}
In this work, for most of the WCs, the uniform priors have ranges: $\{-10,10\}$. We list only those WCs here which have different ranges:
\begin{enumerate}
    \item For single WC fits 
    \begin{itemize}
        \item For the fit titled ``This Analysis", the prior-range for $\wc_{H}$ is $\{-40,40\}$. 
        \item With ``2020 Data", the WCs $\wc_{H}$, $\wc_{H\square}$, $\wc_{tH}$ and $\wc_{G}$ have prior-ranges $\{-100,100\}$, $\{-70,70\}$, $\{-50,50\}$ and $\{-50,50\}$ respectively.
    \end{itemize}
    \item For global fits
    \begin{itemize}
        \item ``This Analysis" fit intakes priors for $\wc_{H}$, $\wc_{H\square}$, $\wc_{tH}$ and $\wc_{G}$ each with range $\{-40,40\}$.
        
        \item In ``2020 Data" fit, the WCs with distinct priors are  $\wc_{H}$, $\wc_{H\square}$, $\wc_{tH}$ and $\wc_{G}$ with ranges $\{-100,100\}$, $\{-70,70\}$, $\{-50,50\}$ and  $\{-50,50\}$ respectively.
        
        \item In case of ``W/O $ggF$ STXS" fit, priors of range $\{-40,40\}$ are taken for $\wc_{H}$, $\wc_{H\square}$, $\wc_{tH}$ and $\wc_{G}$ each.
        
        \item For ``W/O $Vh$ STXS" fit, priors of range $\{-40,40\}$ are taken for $\wc_{H}$, $\wc_{H\square}$, $\wc_{tH}$ and $\wc_{G}$ each.
        
        \item For ``W/O $WBF$ STXS" measurements, $\wc_{H}$, $\wc_{cH}$, $\wc_{H\square}$, $\wc_{tH}$ and $\wc_{G}$ are with ranges $\{-100,100\}$, $\{-50,50\}$, $\{-200,200\}$, $\{-40,40\}$ and  $\{-40,40\}$ respectively.
        
        \item For ``W/O $ggF$, $t\bar{t}h$ \& $th$ STXS" measurements, the accepted priors of $\wc_{H}$, $\wc_{H\square}$, $\wc_{tH}$, $\wc_{G}$, $\wc_{HG}$ and $\wc_{tG}$ are with ranges $\{-200,200\}$, $\{-40,40\}$, $\{-1500,1500\}$, $\{-2000, 2000\}$, $\{-100,100\}$ and  $\{-150,150\}$ respectively.
    \end{itemize}
     
\end{enumerate}

\newpage
\FloatBarrier
\subsection{Fit Results and  Correlation matrix}

\begin{table}[htb!]
	\centering
	\renewcommand{\arraystretch}{1.7}
	\caption{\small {95\% credible intervals (CI) for one-parameters WC fits and for a global analysis of 23 WCs.  The cut-off scale $\Lambda$ is set to 1~TeV.}}
	\begin{adjustbox}{width=0.6\textwidth}
		\label{tab:fit_results}
		\begin{tabular}{|c|c|c|}
			\hline
			WCs & $95\%$ CI Individual limits  & $95\%$ CI  Global limits \\
			\hline
			$\mathcal{C}^{}_{HWB}$ & $ [-0.0035, 0.0028]  $  & $ [-0.19, 0.15] $ \\
			\hline
			$\mathcal{C}^{}_{HD}$ &  $ [-0.022, 0.0042] $  & $ [-0.40, 0.39] $  \\
			\hline
			$\mathcal{C}^{}_{ll}$ & $ [-0.006, 0.016] $  & $ [-0.10, 0.00] $  \\
			\hline
            $\mathcal{C}^{(1)}_{Hl}$ & $ [-0.005, 0.012] $  & $[-0.08,0.12]$  \\
			\hline
			$\mathcal{C}^{(3)}_{Hl}$ & $ [-0.010, 0.003] $  & $ [-0.054, 0.063] $   \\
			\hline
			$\mathcal{C}^{}_{He}$ & $ [-0.013, 0.008] $  & $ [-0.20, 0.19] $  \\
			\hline
			$\mathcal{C}^{(1)}_{Hq}$ & $ [-0.023, 0.047] $  &  $[-0.057, 0.096]$ \\
			\hline
			$\mathcal{C}^{(3)}_{Hq}$ & $[-0.008, 0.016] $  & $ [-0.033, 0.063] $  \\
			\hline
			$\mathcal{C}^{}_{Hd}$ & $ [-0.15, 0.04] $ & $ [-0.29, 0.11] $ \\
			\hline
			$\mathcal{C}^{}_{Hu}$ & $ [-0.056, 0.081] $  & $ [-0.13, 0.25] $  \\
			\hline
			$\mathcal{C}^{}_{H}$ & $ [-9.6, 6.9] $ & $ [-11., 7.0] $ \\ 
			\hline
			$\mathcal{C}^{}_{H\square}$ & $ [-0.96, -0.13] $  & $ [-1.6, 5.6] $ \\
			\hline            
			$\mathcal{C}^{}_{HG}$ & $ [-0.0038, -0.0002] $  & $ [-0.013, 0.010] $  \\
			\hline
			$\mathcal{C}^{}_{HW}$ & $ [-0.010, 0.005] $  & $ [-0.28, 0.12] $  \\
			\hline
			$\mathcal{C}^{}_{HB}$ & $ [-0.0031, 0.0016] $  &  $ [-0.050, 0.061] $  \\
			\hline
			$\mathcal{C}_{W}$ &  $ [-0.17,0.34]  $  &  $ [-0.18, 0.33]  $  \\
			\hline
			$\mathcal{C}^{}_{G}$ & 	$ [-0.8, 1.2] $ & $ [-1.1, 1.3] $ \\
			\hline
			$\mathcal{C}^{}_{\mu H}$ & $ [-0.0042, 0.0027] $  & $ [-0.0045, 0.0025] $   \\ 
			\hline
			$\mathcal{C}^{}_{\tau H}$ & $ [-0.0040, 0.028] $  & $ [-0.009, 0.029] $  \\				
			\hline
			$\mathcal{C}^{}_{bH}$ & $ [-0.036, 0.004] $  & $ [-0.029, 0.069] $ \\
			\hline
			$\mathcal{C}^{}_{cH}$ & $ [-0.15, -0.01] $ & $ [-1.1, 0.20] $  \\ 
			\hline
			$\mathcal{C}^{}_{tH}$ & $ [0.02, 1.2] $ & $ [-2.6, 2.6] $  \\
			\hline
			$\mathcal{C}^{}_{tG}$ & $ [-0.11, -0.01] $ & $ [-0.28, 0.21] $ \\
			\hline
			\end{tabular}
	\end{adjustbox}
\end{table}

\begin{table}[htb!]
	\centering
	\renewcommand{\arraystretch}{2.4}
	\caption{\small Correlations among the 23 WCs with the fit results shown in column III of tab.~\ref{tab:fit_results}.}
	\Huge{
		\begin{adjustbox}{width=\textwidth}
			\begin{tabular}{|c|*{23}{c}|}
				  \hline 
				  \multirow{2}{*}{{\bf WCs}} & \multicolumn{23}{c|}{{\bf Correlations}} \\ 
				  \cline{2-24}
				  &$\mathcal{C}^{}_{HWB}$ &$\mathcal{C}^{}_{HD}$ & $\mathcal{C}^{}_{ll}$ &$\mathcal{C}^{(1)}_{Hl}$ & $\mathcal{C}^{(3)}_{Hl}$ & $\mathcal{C}^{}_{He}$ & $\mathcal{C}^{(1)}_{Hq}$ &  $\mathcal{C}^{(3)}_{Hq}$ & $\mathcal{C}^{}_{Hd}$ & $\mathcal{C}^{}_{Hu}$ & $\mathcal{C}_{H}$ & $\mathcal{C}^{}_{H\square}$ &   $\mathcal{C}^{}_{HG}$ & 
				  $\mathcal{C}^{}_{HW}$ & $\mathcal{C}^{}_{HB}$ &  $\mathcal{C}_{W}$ &   $\mathcal{C}^{}_{G}$ &  $\mathcal{C}^{}_{\mu H}$ &	 $\mathcal{C}^{}_{\tau H}$ & $\mathcal{C}^{}_{bH}$ &
				  $\mathcal{C}^{}_{cH}$ & $\mathcal{C}^{}_{tH}$ & $\mathcal{C}^{}_{tG}$   \\ 
				  \hline
				  $\mathcal{C}^{}_{HWB}$ &	1&& && && && && && && && && && && \\
				  $\mathcal{C}^{}_{HD}$ &	-0.98 & 1&& && && && && && && && && && &\\
				  	$\mathcal{C}^{}_{ll}$ & -0.03 & 0.06 & 1&& && && && && && && && && && \\
				  	$\mathcal{C}^{(1)}_{Hl}$ &0.96 & -0.98 & -0.22 & 1&& && && && && && && && && &\\
				  	$\mathcal{C}^{(3)}_{Hl}$ &0.09 & -0.24 & 0.31 & 0.17 & 1&& && && && && && && && && \\
				  	$\mathcal{C}^{}_{He}$ &0.98 & -1.00 & -0.07 & 0.98 & 0.24 & 1&& && && && && && && && &\\
				   $\mathcal{C}^{(1)}_{Hq}$ & 	-0.41 & 0.34 & -0.13 & -0.31 & 0.20 & -0.35 & 1&& && && && && && && && \\
				  	 $\mathcal{C}^{(3)}_{Hq}$ & -0.24 & 0.13 & 0.02 & -0.13 & 0.54 & -0.13 & -0.06 & 1&& && && && && && && &\\
				  	$\mathcal{C}^{}_{Hd}$ &-0.01 & 0.02 & -0.05 & -0.02 & -0.08 & -0.02 & 0.37 & 0.09 & 1&& && && && && && && \\
				  $\mathcal{C}^{}_{Hu}$ &	-0.31 & 0.25 & -0.15 & -0.22 & 0.16 & -0.25 & 0.59 & -0.29 & 0.26 & 1&& && && && && && &\\
				  $\mathcal{C}_{H}$ &	-0.10 & 0.09 & -0.02 & -0.09 & 0.01 & -0.10 & 0.08 & -0.01 & 0.03 & 0.12 & 1&& && && && && && \\
				  $\mathcal{C}^{}_{H\square}$ &	-0.60 & 0.58 & -0.03 & -0.56 & 0 & -0.58 & 0.43 & -0.02 & 0.12 & 0.55 & 0.23 & 1&& && && && && &\\
				  $\mathcal{C}^{}_{HG}$ &	0.07 & -0.05 & 0.02 & 0.04 & -0.13 & 0.05 & -0.06 & -0.13 & -0.03 & -0.10 & -0.28 & -0.12 & 1&& && && && && \\
				  $\mathcal{C}^{}_{HW}$ &	0.88 & -0.85 & -0.02 & 0.83 & 0.02 & 0.85 & -0.38 & -0.24 & -0.03 & -0.33 & -0.11 & -0.62 & 0.07 & 1&& && && && &\\
				  $\mathcal{C}^{}_{HB}$ &	0.87 & -0.86 & -0.03 & 0.85 & 0.14 & 0.86 & -0.35 & -0.13 & 0 & -0.26 & -0.09 & -0.54 & 0.07 & 0.53 & 1&& && && && \\
				  $\mathcal{C}_{W}$ &  	0.15 & -0.15 & 0.02 & 0.14 & 0.07 & 0.15 & -0.02 & -0.03 & 0.01 & 0 & -0.01 & -0.07 & 0 & 0.12 & 0.13 & 1&& && && &\\
				  	$\mathcal{C}_{G}$ &  -0.05 & 0.06 & 0 & -0.06 & -0.04 & -0.06 & 0.03 & -0.03 & 0 & 0.03 & 0.01 & 0.02 & -0.11 & -0.03 & -0.07 & -0.01 & 1&& && && \\
				  $\mathcal{C}^{}_{\mu H}$ &	0 & 0 & -0.01 & 0 & 0 & 0 & 0.01 & -0.02 & 0.01 & 0.02 & -0.01 & 0.02 & 0& 0 & 0& 0& 0.04 & 1&& && &\\
				  	$\mathcal{C}^{}_{\tau H}$ &0 & 0 & -0.01 & 0 & -0.01 & 0 & 0.03 & -0.05 & 0.01 & 0.05 & -0.04 & 0.01 & -0.16 & 0.01 & 0.01 & 0 & 0.05 & 0.07 & 1&& && \\
				  $\mathcal{C}^{}_{bH}$ &	0.04 & -0.11 & -0.05 & 0.11 & 0.37 & 0.11 & 0.01 & 0.35 & 0.03 & 0.09 & 0.01 & 0.05 & -0.40 & 0.07 & 0 & 0.02 & -0.01 & 0.05 & 0.28 & 1&& &\\
				  $\mathcal{C}^{}_{cH}$ &	0.51 & -0.48 & 0.04 & 0.45 & -0.08 & 0.48 & -0.37 & -0.06 & -0.12 & -0.51 & -0.22 & -0.95 & 0.15 & 0.52 & 0.48 & 0.06 & 0 & 0 & 0.08 & -0.15 & 1&& \\
				  $\mathcal{C}^{}_{tH}$ &	-0.21 & 0.22 & 0 & -0.21 & -0.07 & -0.22 & 0.15 & -0.08 & 0.03 & 0.15 & -0.19 & 0.21 & 0.37 & -0.24 & -0.14 & -0.03 & -0.39 & -0.02 & 0.09 & -0.01 & -0.08 & 1&\\
				  $\mathcal{C}^{}_{tG}$ &	-0.04 & 0.02 & -0.01 & -0.02 & 0.11 & -0.02 & 0.04 & 0.10 & 0.02 & 0.08 & 0.16 & 0.09 & -0.78 & -0.06 & -0.03 & 0 & -0.17 & -0.05 & 0.05 & 0.27 & -0.12 & 0.14 & 1 \\
				  	\hline
\end{tabular}
\end{adjustbox}}
\label{tab:correlations_23wcs}
\end{table}

\FloatBarrier
\section{Di-Higgs data}
\label{app:diHiggsData}
\begin{table}[ht]
    \centering
    \renewcommand{\arraystretch}{1.7}
    \caption{Considered signal strength measurements for di-Higgs production~\cite{ATLAS:2018dpp,ATLAS:2018rnh,ATLAS:2018uni,CMS:2020tkr, CMS:2021ssj,CMS:2017hea}.}
    \begin{tabular}{|l|cc|}
    \hline
    channel & ATLAS    & CMS \\
    \hline
    $b \bar{b} b \bar{b}$     &  $-12.7 \pm 12.8$ & $-3.9 \pm 3.8$\\
    $b \bar{b} \gamma\gamma$     & $- 6.3^{+9.9}_{-7.5}$ & $2.5 \pm 2.6$ \\
    $b \bar{b} \tau \tau$     &  $-4.1 \pm 8.4 $ & $-5 \pm 15$\\
    \hline
    \end{tabular}
    \label{tab:diHiggs_data}
\end{table}

\section{Relevant dimension-6 SMEFT operators}
\label{app:operators}
\begin{table}[ht]
	\caption{These dimension-6 effective operators (Warsaw basis) contribute to the observables. Here, $\sigma$'s are Pauli matrices, and $\lambda$'s are the Gell-Mann matrices.}
	\centering
	\small
	\renewcommand{\arraystretch}{2.0}
	\begin{tabular}{|c|c|c|c|c|c|}
			\hline \hline
			$Q_H$  &  $\left(H^{\dagger }H )^3\right.$ & $Q_{HG}$  &  $\left(H^{\dagger }H \right)G_{\mu \nu }{}^aG^{a,\mu \nu }$ & $Q_{He}$  &  $\left(H^{\dagger }\it{ i }\overleftrightarrow{\mathcal{D}   }_{\mu }\it{ H }\right) \text{(}\bar{e}_R \gamma ^{\mu }\it{e}_R \text{)}$ \\
			\hline
			$Q_{H\square }$  &  $\left(H^{\dagger }H \text{)$\square $(}H^{\dagger }H \right)$ & $Q_{Hl}^{(1)}$  &  $\left(H^{\dagger }\it{ i }\overleftrightarrow{\mathcal{D}   }_{\mu }\it{ H }\right) \text{(}\bar{l}_L \gamma ^{\mu }\it{l}_L \text{)}$ & $Q_{Hu}$  &  $\left(H^{\dagger }\it{ i }\overleftrightarrow{\mathcal{D}   }_{\mu }\it{ H }\right) \text{(}\bar{u}_R \gamma ^{\mu }\it{u}_R \text{)}$  \\ 
			\hline
			$Q_{HD}$  &  $\left(H^{\dagger }\mathcal{D}_{\mu }H )^*\right(H^{\dagger }\mathcal{D}^{\mu }H )$ & $Q_{Hl}^{(3)}$  &  $\left(H^{\dagger }\it{ i }\sigma ^I \overleftrightarrow{\mathcal{D}   }_{\mu }\it{ H }\right) \text{(}\bar{l}_L \sigma ^I \gamma ^{\mu }\it{l}_L\text{)}$ & $Q_{Hd}$  &  $\left(H^{\dagger }\it{ i }\overleftrightarrow{\mathcal{D}   }_{\mu }\it{ H }\right) \text{(}\bar{d}_R \gamma ^{\mu }\it{d}_R\text{)}$  \\
			\hline
			$Q_{HB}$  &  $\left(H^{\dagger }H \right)B_{\mu \nu }B^{\mu \nu }$ & $Q_{Hq}^{(1)}$  &  $\left(H^{\dagger }\it{ i }\overleftrightarrow{\mathcal{D}   }_{\mu }\it{ H }\right) \text{(}\bar{q}_{_L} \gamma ^{\mu }\it{q}_{_L}\text{)}$ & $Q_{\tau H}$  & $\left(H^{\dagger }H \right) \text{(}\bar{l}_L \ \it{ \tau_R } \ H\text{)+h.c.}$  \\
			\hline
			$Q_{HW}$  &  $\left(H^{\dagger }H \right)W_{\mu \nu }{}^I W^{I,\mu \nu }$ & $Q_{Hq}^{(3)}$  &  $\left(H^{\dagger }\it{ i }\sigma^I\overleftrightarrow{\mathcal{D}   }_{\mu }\it{ H }\right) \text{(}\bar{q}_{_L} \sigma^I \gamma ^{\mu }\it{q}_{_L}\text{)}$ & $Q_{tH}$  & $\left(H^{\dagger }H \right) \text{(}\bar{q}_{_L} \ \it{t}_R \ \widetilde{H}\text{)+h.c.}$  \\
			\hline
			$Q_{HWB}$  &  $\left(H^{\dagger }\sigma^I H \right)W_{\mu \nu }{}^I B^{\mu \nu }$ & $Q_{ll}$  &  $\left(\bar{l}_L \gamma _{\mu }\it{l}_L  \right) \left(\bar{l}_L \gamma ^{\mu }\it{l}_L\text{)}\right.$ & $Q_{bH}$ & $\left(H^{\dagger }H \text{)(}\bar{q}_{_L} \ \it{b}_R \ H \text{)+h.c.}\right.$     \\
			\hline
			$Q_{W}$  & $\epsilon ^{IJK}W_{\rho }{}^{I,\mu }W_{\mu }{}^{J,\nu }W_{\nu }{}^{K,\rho }$  & $Q_{tG}$  & $\left(\bar{q}_{_L} \sigma ^{\mu \nu } \frac{\lambda^a}{2} \it{t}_R \right) \widetilde{H} G_{\mu \nu }{}^a\text{ }$ & $Q_{\mu H}$  & $\left(H^{\dagger }H \right) \text{(}\bar{l}_L \ \it{ \mu_R } \ H\text{)+h.c.}$      \\
			\hline
			 $Q_{G}$  & $f^{abc}G_{\rho }{}^{a,\mu }G_{\mu }{}^{b,\nu }G_{\nu }{}^{c,\rho }$  & &  &   $Q_{cH}$  & $\left(H^{\dagger }H \right) \text{(}\bar{q}_{_L} \ \it{c}_R \ \widetilde{H}\text{)+h.c.}$       \\
			\hline \hline
	\end{tabular}
	\label{table:22operators}
\end{table}

%%%%%%%%%%%%%%%%%%%%%%%%%%%%%%%%%%%%%%%%%%%%%%%%%%%%%%%%%%%%%%%%%
	\FloatBarrier
	\newpage
	\providecommand{\href}[2]{#2}
	\addcontentsline*{toc}{section}{}
	\bibliographystyle{JHEP}
	\bibliography{refs-EFT}

	%%%%%%%%%%%%%%%%%%%%%%%%%%%%%%%%%%%%%%%%%%%%%%%%%%%%%%%%%%%%%%%%%	

\end{document}